\documentclass[12pt,openany]{book}
\usepackage[T1]{fontenc}
\usepackage[utf8]{inputenc}
\usepackage[natbib,style=numeric-comp,refsection=part, sorting=none]{biblatex}
\usepackage[english]{babel}
\usepackage{csquotes}
\usepackage[a4paper,left=2cm,right=2cm,top=3cm,bottom=3cm, twoside]{geometry}
\usepackage{dsfont}
\usepackage{xcolor}
\usepackage[fit, breakall]{truncate}
\usepackage{stmaryrd}
\usepackage{amssymb}
\usepackage{amsmath}
\usepackage{libertine}
\usepackage[pdftex]{graphicx}
\usepackage{caption}
\usepackage{subcaption}
\usepackage{float}
\usepackage{multirow}
\usepackage{apalike}
\usepackage{minitoc}
\usepackage[toc,page]{appendix}
\usepackage{gensymb}
\usepackage[normalem]{ulem}
\usepackage{enumitem}
\usepackage{epigraph}
\usepackage{lscape}
\usepackage{titletoc}
\usepackage{tabularray}
\usepackage{newpxtext,newpxmath}

\usepackage{booktabs}
\usepackage{slashed}
\usepackage[nonumberlist]{glossaries}
\usepackage{calc}
\usepackage{titlesec} 
\usepackage[colorlinks=true,citecolor=green,linkcolor=black,urlcolor=blue]{hyperref}
\usepackage{lipsum}
\usepackage{subcaption}
\usepackage{tabularray}
\newcommand{\jg}[1]{{\color{blue} JG: #1}}

\newcommand{\Tobs}{T_\mathrm{obs}}
\newcommand{\RIN}{\mathrm{RIN}}

\usepackage{fancyhdr}
\renewcommand{\sectionmark}[1]{}
\usepackage{psl-cover}

\title{Exploration of new experimental strategies \\ for the detection of ultralight dark matter : \\ laboratory searches on ground and in space}
\author{Jordan Gué}
\date{01 Octobre 2024}

\institute{SYRTE, Observatoire de Paris}
\doctoralschool{Ecole Doctorale \\Astronomie et Astrophysique\\ d'Ile de France (AAIF)}{127}
\specialty{Astronomie et Astrophysique}

\jurymember{1}{\textbf{Franck Pereira Dos Santos}}{CNRS Research director\\ SYRTE, Observatoire de Paris}{Pr\'esident}
\jurymember{2}{\textbf{Diego Blas}}{Professor\\ IFAE, Universitat Aut\`onoma de Barcelona}{Rapporteur}
\jurymember{3}{\textbf{Hartmut Grote}}{Professor\\
Cardiff University}{Rapporteur}
\jurymember{4}{\textbf{Clare Burrage}}{Professor\\
University of Nottingham}{Examinatrice}
\jurymember{5}{\textbf{Antoine Petiteau}}{CEA Researcher\\
IRFU, CEA Saclay}{Examinateur}
\jurymember{6}{\textbf{Marianna Safronova}}{Professor\\
University of Delaware}{Examinatrice}
\jurymember{7}{\textbf{Peter Wolf}}{CNRS Research director\\SYRTE, Observatoire de Paris}{Directeur de th\`ese}
\jurymember{8}{\textbf{Aurélien Hees}}{CNRS Researcher \\ SYRTE, Observatoire de Paris}{Co-encadrant de th\`ese}
\frabstract{La matière noire ultralégère (avec pour acronyme anglais ULDM), classe de candidats de matière noire de faible masse (< 1 eV), constitue une alternative pertinente aux modèles historiquement dominants, tels que les WIMPs, et a récemment fortement attiré l'attention de la communauté scientifique. Dans cette thèse, nous étudions de nombreuses stratégies expérimentales pour la détection directe d'ULDM, au sol et dans l'espace. Plus précisément, nous proposons une modélisation théorique d'expériences actuelles et futures, et nous dérivons une estimation de leurs sensibilités respectives. Nous nous concentrons principalement sur trois types de phénoménologies. La première consiste en une interaction entre un candidat de matière noire vectoriel, plus communément appelé photon noir (DP), et l'électromagnétisme, qui induit un faible champ électrique oscillant à la fréquence de Compton du DP. Nous proposons d'abord une méthode innovante pour détecter ce petit champ électrique, qui consiste à mesurer l'effet Stark quadratique sur des atomes de Rydberg, piégés au centre d'une cavité micro-onde. Nous montrons qu'une telle expérience aurait une sensibilité lui permettant d'être compétitive avec les expériences de laboratoire existantes. Une autre façon de détecter ce champ électrique consiste en l'utilisation d'une antenne parabolique qui le réfléchit et concentre ainsi la puissance électromagnétique en son centre de courbure, où l'on place une antenne cornet (par exemple, l'expérience \textit{SHUKET} au CEA Saclay). Nous investiguons analytiquement les effets de diffraction et de couplage de mode dans ce type de configuration, et nous montrons que l'intensité du signal attendu peut être significativement réduite comparativement aux estimations habituelles. Dans cette étude, nous proposons aussi une optimisation des paramètres expérimentaux pour augmenter le signal. La seconde phénoménologie d'intérêt pour cette thèse est l'oscillation de la masse et de la fréquence de transition d'atomes ou de masses tests. Ces oscillations pourraient être produites par le couplage non universel entre les particules du Modèle Standard et un candidat scalaire de matière noire, comme le dilaton ou l'axion. Nous étudions en détail l'impact de ces oscillations sur plusieurs variations d'interféromètres atomiques et sur des expériences de tests classiques de l'universalité de la chute libre, et nous démontrons comment ces différentes expériences pourraient sonder des régions inexplorées de l'espace des paramètres. Ces oscillations de masse au repos pourraient aussi être observées à l'aide de détecteurs spatiaux d'ondes gravitationnelles, comme \textit{LISA}, et nous investiguons la possibilité d'une telle détection en utilisant des orbites réalistes pour le détecteur. En particulier, en utilisant des méthodes bayésiennes, nous montrons que \textit{LISA} pourra en effet séparer un signal de matière noire d'un signal d'ondes gravitationnelles. Nous montrons aussi que la faible vitesse de l'onde de matière noire n'est pas résoluble sur la majorité de la bande de fréquences de \textit{LISA}, ce qui induit une diminution de la sensibilité de l'expérience aux constantes de couplage de la matière noire scalaire par rapport aux estimations précédentes.Enfin, nous étudions la biréfringence du vide induite par l'interaction entre les photons et les axions, et nous nous intéressons à sa possible détection à l'aide de cavités optiques, de fibres optiques et de \textit{LISA}. En particulier, nous montrons qu'une légère modification des bancs optiques de \textit{LISA} lui permettrait de devenir l'expérience la plus sensible à ce couplage à faibles masses.}

\enabstract{Ultralight dark matter (ULDM), as a class of low mass (< 1 eV) dark matter (DM) candidates, is a compelling alternative to historically dominant models such as WIMPs and has recently gained significant attention in the scientific community. In this thesis, we study various experimental schemes for the direct detection of ULDM, both on ground and in space. More precisely, we propose a theoretical modeling of current and futuristic experiments, and we derive an estimation of their respective sensitivity. We mainly concentrate on three distinct phenomenologies. The first one is the coupling between a DM $U(1)$ field, known as the dark photon (DP), and electromagnetism, which induces a small electric field oscillating at the DP Compton frequency. We first propose an innovative way of detecting this small electric field by measuring the quadratic Stark shift of Rydberg atoms inside a microwave cavity, and we show that such an experiment could reach competitive constraints compared to existing laboratory experiments. Another possibility of detecting this electric field is to use a spherical mirror, which reflects it and focuses the electromagnetic power at its center of curvature, where a horn antenna is located (e.g. \textit{SHUKET} experiment at CEA Saclay). We analytically investigate the effects of diffraction and mode matching in this type of experiment, and we show that the expected signal intensity can be significantly reduced compared to usual estimates. In this study, we also propose an optimization of the experimental parameters in order to increase the signal. The second main phenomenology considered in this thesis is the oscillation of rest mass and transition frequencies of atoms and test masses. These oscillations could be produced by the non-universal coupling of standard matter with a scalar ULDM candidate (dilaton or axion-like particle). We extensively study the impact of such oscillations on various atom interferometer schemes and classical tests of the universality of free fall, and we demonstrate how these different experiments could probe unconstrained regions of the parameter space. The oscillation of rest mass could also be observed in space-based gravitational wave (GW) detectors, such as \textit{LISA}, and we investigate the possibility of such detection using more realistic orbits of spacecraft compared to previous studies. In particular, using Bayesian methods, we show that \textit{LISA} could disentangle scalar ULDM signals from monochromatic GWs. We also show that the small velocity of the DM wave is not resolvable for most frequencies in the \textit{LISA} band, which induces a decrease in sensitivity to scalar ULDM couplings, with respect to previous studies. Finally, we study the effect of vacuum birefringence and dichroism induced by the coupling between axions and photons, and how it could be detected with optical cavities, fibers, and \textit{LISA}. In particular, we show that a slight modification of \textit{LISA}'s optical benches would make \textit{LISA} the most sensitive experiment to the axion-photon coupling at low axion masses.}

\frkeywords{Matière noire ultralégère, axion, dilaton, photon noir, capteur quantiques, horloge atomique, interferométrie (atomique, optique), LISA, principe d'équivalence, physique fondamentale}
\enkeywords{Ultralight dark matter, axion, dilaton, dark photon, quantum sensors, atomic clock, interferometry (atomic, optical), LISA, equivalence principle, fundamental physics}

\begin{document}

\maketitle

\pagestyle{empty}
\renewcommand{\baselinestretch}{0.95}\normalsize
\tableofcontents
\renewcommand{\baselinestretch}{1.0}\normalsize
\newpage
\pagestyle{fancy}

\part{General context and framework}
\thispagestyle{plain}

At the end of the nineteenth century, classical physics, which includes among others, electromagnetism, statistical mechanics, thermodynamics and classical mechanics, was able to describe accurately many different physical systems, through, for example, Maxwell's equations and Newton's classical laws of motion.
\newline
However, classical physics was lacking explanations for various phenomena, such as Mercury's perihelion, black-body radiation, internal structure of atoms and people were having trouble finding the so-called aether, a postulated medium for the propagation of electromagnetic waves.
\newline
In the twentieth century, a complete change of paradigm appeared, as new fundamental theories emerged, which are the basis of what is known today as modern physics : relativity and quantum mechanics. These theories were able to address many of the unanswered questions from classical physics.
\newline
Based on that, two major fundamental physics theories arose, which respectively describe the infinitely big and the infinitely small : general relativity (GR) and the Standard Model of particle physics (SM). They are considered as the most rigorously and extensively confirmed theories of all time through experiments. 
As imperfect physics theories, it still remains open questions, one of which is the inability to reconcile the two theories into a single \textit{theory of everything}. Another major issue arising from these theories, and relevant for this thesis, is the nature of dark matter, a hypothetical form of matter, abundantly present in the universe, which we can only notice through its gravitational effect.
In this opening part, we will first introduce the theoretical framework, necessary for the understanding of the dark matter puzzle. After reviewing the key concepts related to dark matter relevant for this thesis, we will focus on one specific class of solutions as for the microscopic nature of dark matter, ultralight dark matter. Afterwards, we will discuss the main aspects that characterize ultralight dark matter, which will work as a basis of the next chapters of this thesis. 

\pagebreak

\chapter{Special relativity}

Special relativity is a pillar of modern physics as it is the first theory of space and time, as two parameters of a deeper mathematical object : a four-dimensional spacetime.  
The theory, developed by Einstein in 1905 \cite{Einstein1905}, addressed the inability to reconcile Maxwell's electromagnetism with classical mechanics. It is based on two postulates : the principle of relativity and the invariance of the speed of light. The former states that the laws of physics are the same in every inertial reference frame, i.e with zero net acceleration. The latter states that the speed of light in vacuum is invariant and always equals $c$ in all global inertial reference frames. From these two postulates, one can show that the speed of light $c$ is the maximum velocity allowed by special relativity, from the principle of causality.

The main idea behind this theory is that time is no longer an absolute parameter but depends on the velocity of the reference frame. 
Before the development of special relativity, if one wanted to transform time and space coordinate from one reference frame $R$ to another frame $R'$ where the latter travels with constant velocity $v$ in the $x$ direction, compared to the former, one used the non-relativistic Galilean transformations, i.e
    \begin{align}
        ct' &= ct \: \: , \: \: x' = x-vt \: \: , \: \: y'=y \: \: , \: \: z'=z \, .
    \end{align}
Instead, Lorentz transformations transform the time coordinate as a linear combination of initial time and space coordinates, i.e (for the same situation as before)
\begin{align}\label{Lorentz_transfo}
    ct' &= \gamma\left(ct-\frac{vx}{c}\right)\: \: , \: \: x' = \gamma(x-vt) \, ,
\end{align}
where the $y$ and $z$ coordinates are unchanged, and where $\gamma = 1/\sqrt{1-(v/c)^2}$ is the Lorentz factor. The first equation transforms the quantity $ct$, instead of time $t$ alone, to make sure that all components have the same unit of length. One can see that Lorentz transformations deviate notably from Galilean ones when $v$ approaches the speed of light $c$, in which case effects of time dilation (moving clock ticks slower than resting one) and length contraction (moving observer measures contracted length of objects in the direction of movement) appear. The symmetry group of Lorentz transformations is the Lorentz group which includes space rotations and boosts.

The comprehension of space and time as two sides of the same coin is the basis of the existence of four-vectors $X^\mu = (X^0,\vec X)$, which is a four-dimensional generalization of usual vectors in 3-space $\vec X$. The components of four-vectors transform in a specific way under Lorentz transformations. In a situation similar than previously, the component of a four-vector $X^\nu$ in a given inertial frame transforms in $X'^\mu$ in another inertial frame, which moves in the x-direction with velocity $v$ with respect to the first one, as
\begin{subequations}
\begin{align}
    X'^\mu &= \Lambda^\mu_{\ \nu} X^\nu 
\end{align}
where we introduce the $\Lambda$ matrix of Lorentz group (the group of all Lorentz transformations)
\begin{align}
  \Lambda^\mu_{\ \nu} &= \begin{pmatrix}
        \gamma & -\frac{\gamma v}{c} & 0 & 0 \\
        -\frac{\gamma v}{c} & \gamma & 0 & 0 \\
        0 & 0 & 1 & 0 \\
        0 & 0 & 0 & 1
    \end{pmatrix} \, ,
\end{align}
\end{subequations}
as the transformation matrix of Eq.~\eqref{Lorentz_transfo}. There exist four-vectors describe many properties of objects, in particular the four-position $x^\mu = (ct, \vec x)$, the four-momentum $p^\mu = (E/c,\vec p)$, where $E$ is the energy, or the electromagnetic four-potential $A^\mu=(\phi/c, \vec A)$, where $\phi, \vec A$ are respectively the scalar and vector potentials.

In a (flat and cartesian) four dimensional spacetime (i.e with no energy or gravity involved), distances are computed using the line element 
\begin{align}
    ds^2 &= -c^2 dt^2 + dx^2 + dy^2 + dz^2 \equiv \eta_{\mu\nu} dx^\mu dx^\nu \, ,
\end{align} 
where $\eta_{\mu\nu}$ is the Minkowski metric. This rank-2 tensor is central in special relativity and general relativity (where it will be generalized to $g_{\mu\nu}$ which in general differs from $\eta_{\mu\nu}$ to account for energy content of spacetime) as it is used to describe e.g. time, distances and curvature. 

The theory of special relativity has been confirmed experimentally many times in various systems. In particular, it allowed to understand why one were able to detect cosmic muons on Earth ground. Muons are unstable particles which can be produced in cosmic showers when highly energetic particles enter the Earth atmosphere and recoil with atoms. They are relativistic particles, i.e they travel at a speed close to $c$, but they are unstable and have a lifetime of around $2.2$ $\mu$s, after which they decay to lighter particles. Considering their lifetime and altitude at which they are produced, it is only through time dilation that their detection on ground can be explained.

Special relativity is very important in various fields of physics, and in particular it is a pillar of the two cornerstones of fundamental physics that will be discussed in the next chapters : general relativity and the Standard Model of particle physics.

\chapter{\label{chap:Quantum_mechanics}Quantum mechanics}

Quantum mechanics is the theory that describes phenomena at very small scales (typically at the level of atoms or small molecules, i.e $10^{-10}$ m, and below). 

As its name suggests, it is based on the quantization of physical quantities. This has mainly two consequences. The first one is the discovery of particles, as quanta of energies, such as photons. This allows to explain many experimental results, like the photoelectric effect or the Compton effect \cite{Greiner01}. This is at the origin of the wave-particle duality, i.e the fact that some systems, such as light or atoms, can be described by a wave or by a particle, depending on the specific experiment considered. The second consequence is that the energy, the momentum, the angular momentum or spin of quantum systems can only take discrete values, while classical mechanics allow a continuum spectrum of numerical values for such quantities. 

When shining light through a material, some electrons of the material are ejected, because the energy of light is larger than the binding energy of the electron in the atom. This is the photoelectric effect. If one measures the energy of the electrons $E$ as function of the frequency of the incident light $\omega$, one finds a linear dependence, whose slope is the so-called Planck constant $\hbar$, i.e $E= \hbar \omega$. The Planck constant, also known as the quanta of action, is a fundamental quantity of quantum mechanics, as it is at the basis of quantization. Similarly to the speed of light $c$ in special relativity, quantum mechanical effects start to be relevant when the action of the system is of order $\hbar$. This is the reason why, macroscopically, when the system's action is much greater than $\hbar$, quantum mechanics is negligible and classical mechanics works well.

As a probabilistic theory, it is fundamentally different from classical mechanics which is deterministic. Indeed, in the latter, if the initial position and velocity of an object is known, then it is possible to compute the motion of such object at any point in time. Quantum mechanics, however, describe the state of a particle with a wavefunction, usually denoted $\psi$, a complex function of the state. By the Born rule, the square modulus of the wavefunction $|\psi(s)|^2$ gives the probability of finding the particle in a given eigenstate $s$. 

Quantities that can be measured experimentally are known as observables, and are described by hermitian operators, i.e matrices with real eigenvalues. 
As the eigenvectors of a given observable $A$ form an orthonormal basis, a physical state $\psi$ can be expressed as the superposition of eigenstates of $A$. This is what is known as the superposition principle \cite{Greiner01}.

Assuming we want to measure two quantities of a quantum state $\psi$ that are represented by two observables that are canonically conjugate variables, such as the position $X$ and the momentum $P$, the dispersion of both operators around their expectation values obey $\Delta P \Delta X \geq \hbar/2$ \cite{Greiner01}, which is known as the Heisenberg uncertainty principle. This means that there exists a fundamental limit on the knowledge of both conjugate variables of a system when measuring them simultaneously. This uncertainty exists between any pairs of conjugate variables, for example time and energy, which are canonically related through Schr\"odinger equation (see below). 

The fundamental equation of quantum mechanics is the Schr\"odinger equation \cite{Greiner01}
\begin{align}
    i\hbar \frac{d\psi}{dt} &= H\psi \, ,
\end{align}
which describes the evolution of a quantum state $\psi$ using the Hamiltonian operator $H$, which describes the total energy of the system (kinetic and potential). In a sense, it is the quantum generalization of Newton's second law of motion.

\chapter{\label{GR}General relativity}

10 years after its famous annus mirabilis, Einstein published in 1915 its first paper \cite{Einstein1915} on general relativity (GR) which extends the laws of special relativity to gravitational fields. GR is a theory of gravitation which revolutionized our view of gravity. At that time, Newton's theory of gravitation was the leading framework, and despite explaining various phenomena, lacked explanation of several effects, such as Mercury perihelion precession. In Newton's theory, gravity is a force that acts instantaneously from the emitter to the receiver, which breaks causality and therefore special relativity. Instead, Einstein introduces the same four-dimensional spacetime as special relativity, but which curves under the energy-matter distribution of the manifold. In this sense, gravity is simply the curvature of spacetime, affecting the motion of freely falling massive objects through the curvature of the geodesics they follow. The effects of GR are visible when velocities are close to the speed of light and/or when gravitational field becomes strong (when deviations from flat spacetime are non negligible). When these conditions are not fulfilled, GR is well approximated by Newtonian mechanics.

This theory is one of the most successful theory ever created, as it was tested experimentally in various situations for more than $100$ years now, and has never been disproved.

\section{\label{EP_general}Equivalence principle}

In addition to special relativity, the second pillar of GR is known as the equivalence principle (EP) between gravitation and acceleration. EP includes three different forms : the weak EP (WEP), the Einstein EP (EEP) and the strong EP. Here, we will be interested in the EEP which includes three different facets \cite{Uzan11}. 
The first one is the WEP also known as the universality of free fall (UFF) which states that all bodies fall with the same acceleration in the same gravitational potential, no matter their composition. In other words, gravity is universal. It is satisfied if the gravitational theory is metric (which is the case for GR), which means that all matter fields are universally coupled to gravity. UFF implies that the gravitational mass (which appears in Newton's gravitational law) is equivalent to the inertial mass (which appears in Newton's second law).
The second hypothesis is the Local Lorentz Invariance (LLI) whose principle is that the results of any local experiment, where gravity effects are excluded, do not depend on the position and velocity of the reference frame. 
The third principle is the Local Position Invariance (LPI), stating that results of any experiment do not depend on the spacetime position of the laboratory. This principle is closely related to the steadiness of fundamental constants of nature \cite{Uzan11}. 

\section{Einstein field equations}

The central GR equations are 6 independent metric $g_{\mu\nu}$ field equations, which are known as the Einstein field equations (EFE) \cite{Einstein1915}
\begin{align}\label{EFE}
    G_{\mu\nu} = \frac{8\pi G}{c^4}T_{\mu\nu} - \Lambda g_{\mu\nu}
\end{align}
The left-hand side contains only the Einstein tensor $G_{\mu\nu}$ which contains second order derivatives of the metric. It involves in particular the Ricci scalar, which describes the curvature of spacetime.
The right-hand side represents the matter-energy content. The tensor $T_{\mu\nu}$ is the stress-energy tensor of all matter and energy of spacetime. The constant $8\pi G/c^4$, which in the following will be denoted as $\kappa$ is important on dimensional level (to relate $T_{\mu\nu}$ components in $J/m^3$ units with $G_{\mu\nu}$ components in units of $m^{-2}$), but it also indicates the entity of the phenomena involves, i.e in that case gravitational physics (through $G$) with relativistic effects (through $c$). 
In addition to these terms whose links were theoretically demonstrated by Einstein in his original papers from GR's first principles, $\Lambda g_{\mu\nu}$, where $\Lambda$ stands for the cosmological constant, was included to take into account the acceleration of the expansion of the Universe. This constant can be interpreted as a perfect fluid with negative pressure whose nature is not understood, and which is usually denoted as \textit{dark energy} (see below).

John Archibald Wheeler said "Space tells matter how to move, matter tells space how to curve" \cite{Wheeler98}, which is a very simple statement to describe EFE and to show their non linearity.
The non linearity of Einstein field equations make them very complicated to solve, and only partial solutions or solutions in very easy systems (such as the Schwartzschild solution describing the vacuum solution of EFE outside a spherically symmetric body with no charge or angular momentum) have been found. A whole branch of GR studies aims at solving EFE in peculiar systems using numerical methods, known as numerical relativity. 

Despite being non linear, EFE are very similar in their form to Maxwell's equations of electromagnetism (EM) where distribution of matter (charges and currents in EM) sources curvature of spacetime (electromagnetic fields in EM).

\section{Experimental successes}

Overall, GR is a very successful theory as it predicted a large number of experimental results or explained various misunderstood phenomena.

Most solar system gravitational physics can be accurately described by Newtonian physics because the gravitational field is weak and the velocities are small (compared to speed of light $c$). However, the precession of the orbit of Mercury does not exactly follow Newton's predictions \cite{Weinberg72}. Thanks to GR corrections, which are necessary since the Sun's gravitational potential on Mercury starts to be large, this discrepancy was solved. 

Other effects from GR at Solar system scale were experimentally verified. Some of them are the gravitational redshift \cite{Holberg10} and Shapiro time delay \cite{Kramer06}, where both effects describe how light frequency and geodesics, despite being massless particles, are also affected by massive objects. The former states that two clocks located at different gravitational potential will not tick at the same rate, the one located in a weaker gravitational potential ticking faster. Even though the effect is small, this correction is necessary to ensure the Global Navigation Satellite System (GNSS) to work as expected. The latter, predicted by Irwin Shapiro \cite{Shapiro64}, describes how the propagation time for a light ray depends on the gravitational potential encountered by photons, more precisely how a time delay appears in signal when light travels close to a massive object. It was experimentally confirmed by the emission towards bodies close to the Sun (Mercury or Venus) and measuring the light round trip time, affected by the Sun's gravitational potential \cite{Shapiro68}. 

At galactic scale, GR predicts gravitational lensing effects, which describe how light geodesics are curved around galactic objects, such that the galaxy acts as a lens. As a consequence, several effects can be observed such as the Einstein ring, where the same astrophysical object is observed at all locations around the lens galaxy if source, lens and observer are perfectly aligned, making a light ring around the galaxy.   

A very important field of research resulting from GR is the standard model of cosmology. This model aims at describing the evolution of the universe. In its simplest form, the universe is assumed homogeneous and isotropic on large scales, such that the spacetime metric is the Friedmann-Lemaitre-Robertson-Walker (FLRW) metric $g_{\mu\nu}=\mathrm{det}(-1,a^2(t),a^2(t),a^2(t))$ in our sign convention, where $a(t)$ is the dimensionless scale factor accounting for the expansion of the Universe. This metric essentially describes a flat expanding universe. The standard model of cosmology \cite{Bergstrom99} provides good theoretical account for the Cosmic Microwave Background (CMB), the large scale structures and the accelerating expansion of the Universe. Based on measurements from e.g. \cite{Ade13}, the energy content of the Universe is 5\% of ordinary baryonic matter, around 25\% of non-baryonic matter, more commonly known as cold dark matter (CDM) (see Section \ref{DM_proofs}) and about 70\% of an unknown dark energy ($\Lambda$), responsible for the expansion of the Universe. Following these observations, the standard model of cosmology is referred to as $\Lambda$CDM-model.

Finally, the recent measurement of gravitational waves (GW) by the LIGO/VIRGO collaboration \cite{Abbott16} is an important discovery, as it validates even more GR but it also provides a new way of observing the Universe. GW are currently visible as emitted by extremely massive objects such as black holes or neutron stars, therefore their detection allows to test GR in strong gravitational field regimes. Moreover, the Universe being transparent to GW (there exists no structure that can absorb them or reflect them) which is not the case for EM waves, GW can be used to discover new regions of the Universe.

\section{\label{GR_limits}Limits}

As described earlier, EFE Eq.~\eqref{EFE} contains an heuristical term, related to the acceleration of the expansion of the Universe, by some unknown energy. The simplest form of energy is the cosmological constant, with constant energy density over whole spacetime, but other models assumed e.g scalar fields as the cause of the acceleration of expansion. Some efforts were made to explain dark energy by the vacuum energy of the Universe, but it leads to a discrepancy between theory and observations of more than $120$ orders of magnitude \cite{Weinberg89}. This leads to what is known as the cosmological constant problem.

As shown in EFE through the constant $\kappa = 8\pi G/c^4$, GR is a theory combining gravitation and special relativity. However, it does not include quantum mechanical effects (which are relevant when the action is close to $\hbar$). While GR describes accurately physics of the macroscopic scale, another theory is needed to predict physics of the microscopic scale, namely quantum field theory, whose principles will be detailed in the next section. Running backward in time in the cosmological evolution of the Universe, one should arrive at the conclusion that there was a time in the history of the Universe,  where its full energy content was compressed into a very small spacetime region such that temperature, energy density, etc.. were extremely high. In such a case, our current theories (GR and the Standard Model of particle physics (SM), see next section) are not valid anymore because both gravitational and quantum mechanical effects must be taken into account at the same time, which GR and SM are individually not able to do. A new high energy theoretical framework, sometimes referred to as quantum gravity, including both theories is therefore needed in order to understand the physics of the early universe. In this sense, even though the current standard model of cosmology assumes an initial Big Bang singularity as "birth" of the Universe and then an inflationary epoch that stretched spacetime at an exponential rate, one does not have any complete proof of their existence.

As stated in Section ~\ref{EP_general}, the equivalence principle is very important in GR and is the basis of metric theories, i.e an universal coupling between gravity and all types of matter. However, this principle is only heuristical, i.e it is only supported by observations of test masses which fall at the same rate. In particular, it is not based on an underlying symmetry of the universe \cite{Damour12}, in contrast of e.g gauge principle, as we shall see in the next section. It is also very intriguing that gravitation does not rely on any internal charge, compared to other well known fundamental interactions, such as electromagnetism.  Therefore, in some theoretical scenarios, EP is expected to be broken at some scale, see e.g. \cite{Damour94, damour:2010zr, Fayet19}, which would naturally invalidate GR, as a metric theory.

Last but not least, the problem of dark matter, which is the most relevant limit for this thesis. Together with the Standard Model of particle physics (see Section ~\ref{sec:SM_general}), GR does not predict the existence of dark matter, a new form of matter, different from baryonic matter that we are able to describe microscopically (see Section ~\ref{sec:SM_general}), and which, in our current understanding, only interact gravitationnally with visible matter. As we shall see in Section ~\ref{sec:dark_matter_general}, its introduction allows one to explain various cosmological phenomena. 

\chapter{\label{sec:SM_general}The Standard Model of particle physics}

The second cornerstone of fundamental physics, describing the interactions between elementary particles at the microscopic scale is the Standard Model of particle physics (SM). As its name suggests, it is based on quantum mechanics. Many scientists consider SM as the most successful theory ever created. 

In the fourth century before J.-C., Democritus argued that atoms were the fundamental bricks of matter, and therefore could not be cut. It is only in the nineteenth and twentieth centuries that scientists such as Joseph Thomson and Ernest Rutherford discovered the electron and the proton.

Later in the twentieth century and after the work of Werner Heisenberg, Max born and others on the creation of quantum mechanics on one side and Albert Einstein with its relativistic theory on the other side, the first attempts of quantizing the electromagnetic field were made by Paul Dirac. This is the birth of quantum field theory (QFT) which states that particles are not the most fundamental bricks of the universe, but are simply quantum excitation of something more fundamental : quantum fields. This is the theoretical framework at the basis of the SM.

\section{Symmetries}

As a relativistic quantum theory, QFT enjoys the symmetries of quantum mechanics, i.e space rotations and translations, and time translations, and of special relativity, i.e Lorentz symmetry. Space translations extend Lorentz symmetry to Poincaré symmetry. Another important symmetry of particle interactions is what is known as gauge symmetry. The various quantum fields introduced to describe the interactions between particles are not observables, i.e they cannot be measured. Therefore, the observables do not depend on the way we define those fields. The degree of freedom to shift or rotate the quantum fields is gauge symmetry.

\section{Gauge groups}

The current version of SM is based on three different gauge groups, which form the three fundamental interactions between particles that SM is able to describe accurately : electromagnetism described by quantum electrodynamics (QED), the strong interaction, described by the quantum chromodynamics (QCD) and the weak interaction. 

\subsection{QED}

Quantum electrodynamics \cite{Feynman85, Feynman49,Feynman50} is a quantum generalization of the classical Maxwell's equations for electromagnetism. It is based on a $U(1)$ gauge symmetry. The gauge field associated with this symmetry is the electromagnetic four-potential $A^\mu$ and the gauge particle (force carrier) is the photon.
The gauge symmetry implies that the group transformation is $e^{i q}$, with $q$ the electric charge. This means that QED is an abelian theory in the sense that the group operations are commutative. The consequence of that is the gauge boson of the symmetry, the photon, must be neutral under the symmetry transformation. Therefore, photons cannot self interact. These are the reason why the photon is electrically neutral and that Maxwell's equations (or their relativistic version) are linear.
In addition, as gauge bosons, photons must be massless. 

\subsection{Weak interaction}

Weak interaction \cite{Griffiths87} is based on a $SU(2)$ symmetry, which stands for special unitary matrices of dimension 2. There are three gauge fields associated with this symmetry : they are known as the $W^{\pm}$ bosons, which are electrically charged and the $Z^0$ boson, which is neutral.
All fermions are affected by the weak interaction whose particular effect is to change the flavor of such fermions, i.e their fundamental nature.  
Weak interaction is the only charge ($C$) and charge-parity ($CP$) symmetry breaking interactions (for example, a broken $CP$ symmetry means that a left handed fermion and right handed antifermion do not interact in the same way under the weak interaction). In addition, while they are gauge bosons, $W^{\pm},Z^0$, are not massless (their mass is $\sim 100$ GeV/$c^2$ \cite{Workman22}), and this implies that weak interaction is a very short-range interaction. The generation of mass for $W^{\pm}$ and $Z^0$ bosons is known as the Higgs mechanism and will be discussed in Section ~\ref{sec:Higgs_mech}.

\subsection{QCD}

The strong force (also known as the quantum chromodynamics (QCD))  \cite{Griffiths87} describes the interactions between gluons and quarks, which are the particles components of baryons, such as protons or neutrons. The associated conserved charge of the symmetry is an additional quantum number that only quarks and gluons possess, called the color charge, hence the name of the theory. There exists six different color charges : red, blue, green, anti-red, anti-blue and anti-green. The idea behind this formulation of color is that some combinations of colorful states can be colorless, e.g a mixture of red - green - blue state or red - anti-red does not carry color. Each quark carries one color, while gluons, the QCD force carriers carry two different colors. There exists eight independent gluon color states, which represent the generators of the $SU(3)$ gauge symmetry. 

These bring two important consequences for the theory. The first one is that, contrary to QED, gluons carry the conserved charge of the symmetry, therefore they can interact with themselves, i.e QCD is a non-abelian gauge theory. The second one is the color confinement. There exists no bound states which is colorful, i.e quarks cannot be isolated, and therefore cannot be observed alone. Only colorless bound states composed of two (mesons, like pions) or three (hadrons, like protons and neutrons) quarks can be observed.

\section{\label{sec:Higgs_mech}Higgs boson}

At energies larger than what is known as the weak scale ($E=246$ GeV), weak interaction and electromagnetism are unified into a single interaction, known as the electroweak interaction \cite{Griffiths87}. 
Gauge symmetry is required for any quantum theory to be valid, since it is related to the fact that the definition of the field (up to shift and rotation) should not impact the physical observables. Therefore, at this energy scale, in order for the electroweak theory to be gauged, fermions and gauge bosons, including $W^\pm$, must be massless.

However, we know from various experimental results that fermions are not massless, which means that one needs a mechanism to arise between this highly energetic state and the lower energy state of the current experiments and of the world we are living in. 

In order to address this issue, one can introduce a massive scalar field $\phi$ with a new $U(1)$ gauge symmetry and a quartic potential. When the energy of the system gets below the electroweak scale, the potential of the field changes and gets the form of a Mexican hat, where the vacuum state is not aligned with the zero-particle state $\langle \phi \rangle = 0$. The vacuum states are therefore degenerate, and in order to minimize the energy, the system must choose one of its vacua with a given state $\langle \phi_\mathrm{min} \rangle \neq 0$, which does not correspond to the zero-particle state. One says that the $U(1)$ symmetry has been spontaneously broken. The difference between vacuum state and zero-particle state is known as the vacuum expectation value (vev) of the field, therefore, by spontaneously breaking the symmetry, the field acquires a non-zero vev. 
Since the vacuum state $\phi_\mathrm{min}$ does not correspond to the zero-particle state, it contains a non-zero number of particles. If we now assume that the SM fermionic and bosonic fields, initially massless, interact with the field $\phi$ in its vacuum state $\phi_\mathrm{min}$, they effectively get "weighed down" by those interactions, i.e they behave as massive particles.
In the SM, $\phi$ breaks three degrees of freedom (out of four) of the original electroweak symmetry group $SU(2) \times U(1)$, and therefore only three gauge bosons (out of four), $W^\pm, Z_0$, gain mass after symmetry breaking. This is the reason why the photon remains massless.

This mechanism, i.e fields acquiring mass by interacting with another field that spontaneously break symmetries, is known as the Higgs mechanism. We call $\phi$ the Higgs field, and its associated quantum is the Higgs boson. This idea was developed by three independent groups \cite{Englert64,Higgs64,Guralnik64}, and Peter Higgs and François Englert received the Nobel Prize in 2013 for this discovery.

\section{Matter content}

In addition to the Higgs boson and the interaction carriers described in the previous sections, i.e the photon $\gamma$, the two $W^{\pm}$ bosons, the $Z^0$ boson and the eight gluons $g$, one needs to describe the matter content of the universe, i.e the fermions \cite{Griffiths87}.

There exist three generations of fermions, : the first, second and third generations. From one generation to another, fermions are quite similar, the only relevant parameter to distinguish them is the rest mass increase (or equivalently the lifetime decrease because decay to lighter particles is energetically favored) from generation $N$ to generation $N+1$. 
All fermions are also distinguished by the interactions they are sensitive to, i.e if there are charged over the gauge group associated to that interaction. First, we have the leptons, which carry electric charge and weak charges (weak hypercharge, weak isospin), therefore they can be involved in electromagnetic and weak interactions. Among the leptons, we find the electron $e^-$ and the electron neutrino $\nu_e$ (first generation), the muon $\mu^-$ and the muon neutrino $\nu_\mu$ (second generation), and the tau $\tau^-$ and the tau neutrino $\nu_\tau$ (third generation). 
Second, we have the quarks which, in addition to electric and weak charges, carry color charge, which allows them to be sensitive to the third fundamental interaction, the strong force. There exist in total six different quarks. The first generation contains the up $u$ and down $d$ quarks, the second generation contains the strange $s$ and charm $c$ quarks, and the third generation contain the top $t$ and the bottom $b$ quarks.

Note that for each fermion, there exists an antifermion which has the same mass, but carries opposite charges, implying that a given antifermion is sensitive to the same interactions as its associated fermion.

\begin{figure}
    \centering
    \includegraphics[width=0.5\textwidth]{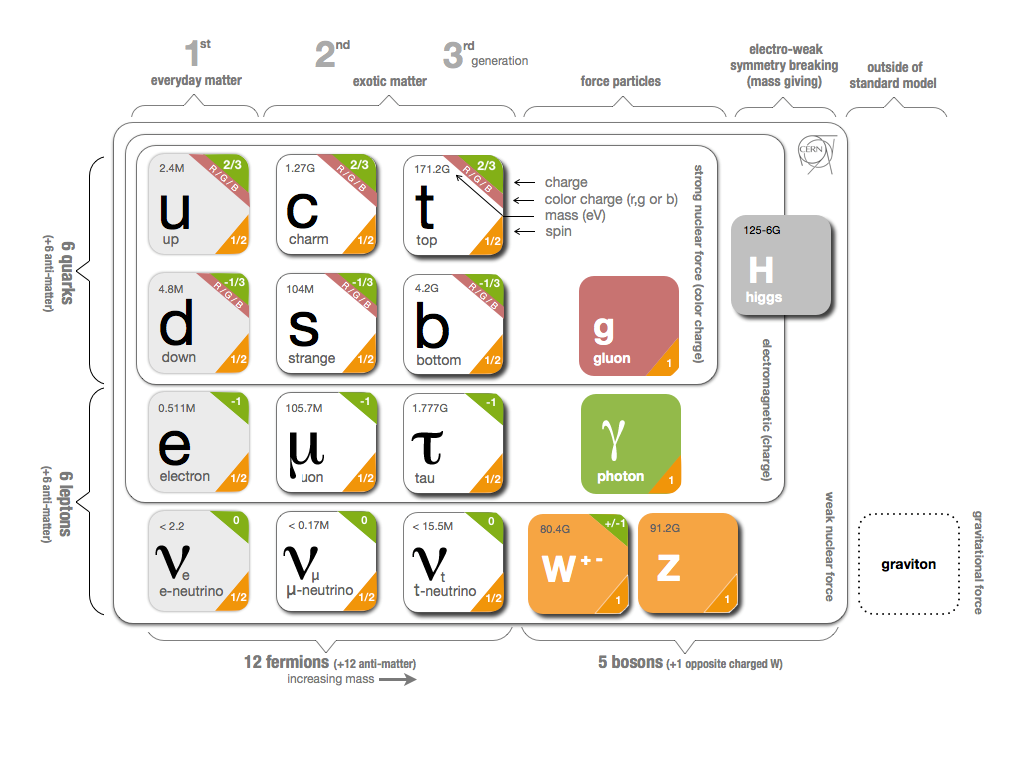}
    \caption{All fundamental particles of the Standard Model (with the hypothetical graviton), with their electric charge, color charge, mass and spin (Credit : CERN)}
    \label{fig:SM_particles}
\end{figure}

The whole zoo of fundamental particles is summarized in Fig.~\ref{fig:SM_particles}, where electric charge, color charge, spin and mass of each particle are provided.

\section{Lagrangian}

The action of the theory is a number and can be calculated by integration of the Lagrangian $\mathcal{L}$ over the four dimensional spacetime measure $d^4x$ on a given manifold. This implies that the Lagrangian must be a scalar under Lorentz transformation, i.e a Lorentz scalar.

The full Lagrangian of SM being extremely lengthy, a reduced version can be used for our practical purposes \cite{Woithe17}
\begin{align}
    \mathcal{L} &= -\frac{1}{4}F^{\mu\nu}F_{\mu\nu} + i\bar \Psi \slashed D \Psi + \Psi_i y_{ij} \Psi_j \phi +\mathrm{h.c.} + |D_\mu \phi|^2 - V(\phi)
\end{align}
The first term, represented by the contraction of the interaction carriers fields strength tensors $F^{\mu\nu}$ with themselves (to construct a Lorentz scalar), gathers the kinetic energies of all interaction carriers. Depending on the particle, this term allows self interaction (e.g for gluons) or simply kinetic energy (e.g for photons).

The second term includes the interaction between force carriers and the fermionic matter fields (leptons and quarks), represented by the spinor $\Psi$, which is a mathematical object used to describe half-integer spin particles, i.e fermions. The $\slashed D$ is the mathematical operator of gauged covariant derivative, which is a generalization of the partial derivative $\partial$ to which we incorporate the gauge fields to account for fermion-boson interactions.

The third (respectively fourth $\mathrm{h.c.}$ for hermitian conjugate) term account for the interaction between fermionic (respectively antifermionic) matter fields and the Higgs field $\phi$, which gives rise to the mass of the fermions. $y_{ij}$ represents the various components of the Yukawa matrix and describes the coupling constant of a given fermion to the Higgs field. 

The fifth term describes the interaction of the weak force bosons with the Higgs field, such that they acquire mass.

Finally, the sixth term represents the potential energy of the Higgs field, which has the form of the so-called "Mexican-hat". As mentioned before, this implies an infinite number of different potential minima, leading to spontaneous symmetry breaking when the field chooses a particular one. This idea is at the basis of the Higgs mechanism for the generation of mass for many of the gauge and fermions fields.

\section{Experimental successes and limits}

After the theoretical completion of the SM, several particle accelerators were built, in particular at CERN, in Geneva, Switzerland, to test experimentally this theory. It consists of beams of particles with large kinetic energy which collide in order to produce large mass particles, to observe them through decay products. More than 40 years ago, the first particle accelerator built at CERN was the Super-Proton-Antiproton-Synchrotron ($Sp\overline{p}S$), then came the Large Electron-Positron Collider (LEP) and finally the Large Hadron Collider (LHC); and the main difference between these accelerators (except the nature of particles colliding) is the kinetic energies reached by the colliding particles, and therefore one is able to produce particles with higher and higher masses.

Since their launch, they allowed scientists to observe the $W$ and $Z$ bosons, gluons, the $\tau$ lepton and the top and bottom quarks. More recently, in 2012, the Higgs boson was found \cite{Aad2012} at LHC, which implied that all particles theoretically predicted by the SM have been experimentally detected.

SM has still issues explaining various physical phenomena. First, as explained in the previous chapter about GR, there is still no unification between QFT and GR, i.e no one has shown how to quantize gravity, i.e explain gravitation interaction completely from quantum fields and particles. 

Additionally, some fine-tuned problems arise in SM, such as the strong CP problem. In short, QCD is theoretically allowed to break the Charge-Parity symmetry, which would be visible experimentally by measuring a non-zero electric dipole moment of the neutron. However, this measurement is consistent with zero with high accuracy \cite{Abel20}, leading us to think that QCD preserve CP symmetry. This problem will be more deeply studied in Chapter ~\ref{axion_pheno} on axions. Finally, one of the major issues in fundamental physics is the microscopic nature of dark matter, which is still unknown, and which is the main topic of this thesis.

\chapter{\label{sec:dark_matter_general}Dark matter}

As quickly stated in Section ~\ref{GR}, a non baryonic matter, more commonly known as dark matter (DM) was first introduced to explain astrophysical observations at galactic scale, as we shall see in the following. Then, it was added into the cosmological equations in order to match various other observations, at the cosmological scale. In the first part of this section, we review the various observations that led to the conjecture of DM, based on the gravitational interaction between DM and the rest of the content of the Universe, and which essentially implies that DM, as its name suggests, is massive. Then, in the second part, we review the various possibilities as for the microscopic nature of DM, and in particular we introduce ultralight dark matter (ULDM) candidates, which will as serve as a foundation for the following chapters.

\section{\label{DM_proofs}Some smoking guns of existence of dark matter}

In this section, we will discuss some astrophysical and cosmological hints for the existence of DM.

\subsection{Galaxy rotation curves}

The first historical hint of DM was made in the $1930$s by Fritz Zwicky. He measured the velocity $v$ of stars of the Coma galaxy clusters as function of their distance to the galactic center $R$ and by using the virial theorem, he deduced the total mass $M$ of the galaxy cluster \cite{Zwicky1937}
\begin{align}\label{virial_theorem}
    Mv^2 &\propto \frac{GM^2}{R} \Rightarrow M \propto \frac{v^2R}{G} \, .
\end{align}
However, the total mass ($\sim 3 \times 10^{14}$ solar masses), deduced by the dynamics of the cluster is much larger than the total luminous mass ($\sim 8 \times 10^{11}$ solar masses), deduced by the number of galaxies inside the cluster. DM was first introduced to account for this missing, non luminous, mass.

\begin{figure}
    \centering
    \includegraphics[width=0.4\textwidth]{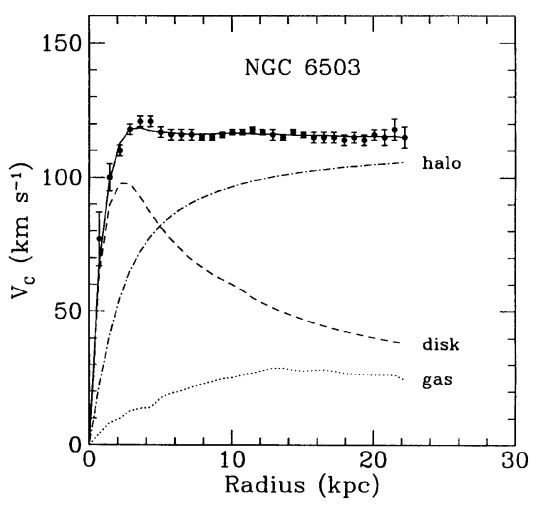}
    \caption{Velocity profile of the NGC 6503 galaxy (from \cite{Freese}). The measured velocity are shown with dots and corresponding error bars. The velocity curve that fits the amount of luminous matter and gas do not fit the observations, and one needs to add a "halo" component, which corresponds to dark matter.}
    \label{fig:rotation_curve}
\end{figure}

Afterwards, several observations of different galaxy rotation curves, i.e the measurement of the velocity of stars as function of their distance to the galactic center, indicate that non luminous mass has to be added. As it can be seen from Fig.~\ref{fig:rotation_curve}, the discrepancy between the measurement of the mass from luminous objects and the measurement of mass through stellar dynamics becomes significant at large distance to the galactic center, which implies that DM would be spherically distributed, encompassing the whole galaxy.

\subsection{Gravitational lensing}

\begin{figure}
    \centering
    \includegraphics[width=0.4\textwidth]{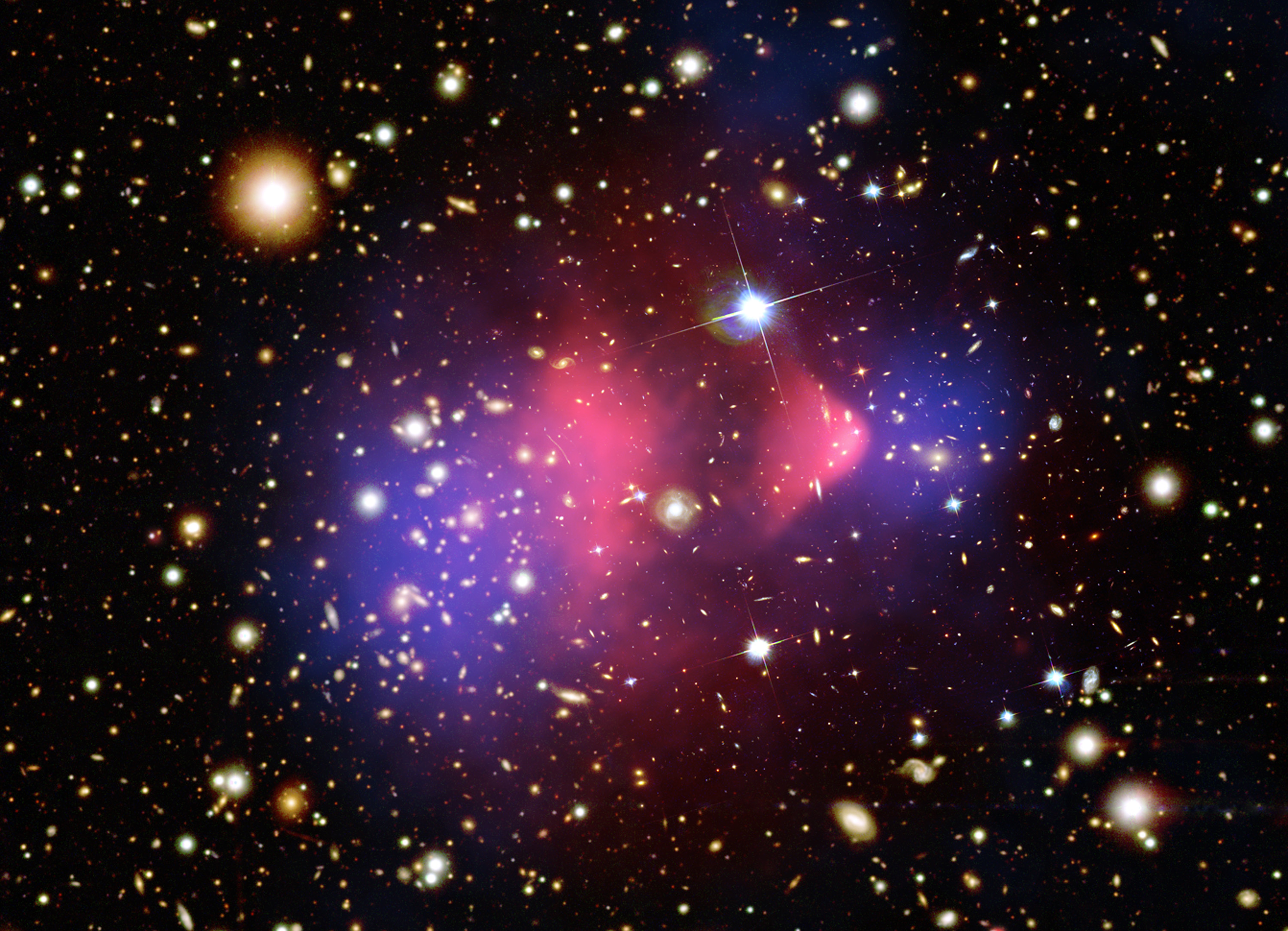}
    \caption{Composite image of the bullet cluster, where the blue area shows the gravitational lensing measurement of the mass and the pink area  is the X-ray measurement of the visible mass in the cluster. X-ray: NASA/CXC/CfA/ M.Markevitch et al.; Lensing Map: NASA/STScI; ESO WFI; Magellan/U.Arizona/ D.Clowe et al. Optical image: NASA/STScI; Magellan/U.Arizona/D.Clowe et al.}
    \label{fig:bullet_cluster}
\end{figure}

Gravitational lensing, another observation at galactic scale, reveals the presence of an invisible mass. As detailed in Section ~\ref{GR}, GR allows massive objects to curve the geodesics of light such that they act as a lens. By detecting the deviated photons on Earth, one is able to reconstruct the mass distribution of the massive object that bent their trajectories. Such measurement was done using the bullet cluster as lens \cite{Clowe06} in addition to the measurement of the visible mass distribution inside the cluster. As it is shown in Fig.~\ref{fig:bullet_cluster}, the measurement of the cluster mass reconstructed from gravitational lensing is shown in blue, while the X-ray measurement, reveals in pink the location of the "visible" mass, which is mostly made of gas. One can clearly see that both distributions do not coincide, which leads us to the conclusion that most of the mass in the cluster is invisible, i.e is DM.

\subsection{Cosmic microwave background}

At cosmic scale, the measurement of the Cosmic Microwave Background (CMB) suggests the existence of DM. In the Standard Model of Cosmology, about $300 \: 000$ years after the presupposed Big-Bang, while the temperature of the universe fell down to $\sim 3000$K, electrons and baryons were able to recombine and form atoms. This epoch, known as recombination, also marks the time of decoupling between such neutral atoms and photons, such that the latter were able to freely propagate through space, thus making the universe transparent. At that moment, the whole universe was immersed in a $3000$ K photon bath. With the expansion of the universe, the wavelength of those photons were stretched such that their temperature is roughly $2.7$ K today, which corresponds to a microwave wavelength. This phenomena is known as the CMB, and was first measured by Penzias and Wilson in 1964, which valued them the Nobel prize in 1978. 
The Planck satellite measured with great accuracy the CMB temperature map \cite{Aghanim20}, and the collaboration was able to measure its power spectrum which is with exceptional agreement with the presence of non baryonic matter, i.e DM, which accounts for $\sim 26.5\%$ of the total energy density of the universe \cite{Aghanim20}. In short, temperature anisotropies of the CMB are linked to the density anisotropies at the time of recombination, which depends on the densities of baryons, DM and photons and their interactions. As DM mainly interact gravitationally with the rest, it collapsed and formed dense regions before decoupling. Through gravitational redshift, photons imprinted those different gravitational potentials and this is why we see temperature anisotropies of those photons in the CMB. 

\section{Dark matter candidates}

Despite the current gravitational evidence of the presence of DM in the universe, through the various observations, described in the previous section, we still have no clue on the microscopic nature of DM, i.e the fundamental particle behind it, its intrinsic properties and its interactions with SM fields. Our current understanding of the microscopic world, in particular from QFT, requires the introduction of an underlying field, which will be denoted in the following by the DM field.

In the Standard Model of Cosmology, the DM field decouples quickly from the rest of the energy, i.e baryons and photons, during the early times of the Universe. Before recombination, from the observation of the CMB, one can conclude that the energy density is not exactly the same everywhere in space, i.e there are some underdense and overdense regions. Overdense regions, containing more DM, baryonic matter and photons, attract more mass, therefore the interaction between baryonic matter and photons increase locally, creating pressure force. This creates sound waves propagating outwards from the overdensities. At the time of recombination, baryons and photons decoupled, which led the photons to propagate freely and which relieved the pressure away. At this point, the left behind DM and baryons stopped propagating but they still formed an overdensed regions of matter, which many believe is the seed of galaxies that we see today. The distance travelled by the sound wave is known as the sound horizon and by measuring the two-point correlation function of the distance between galaxies, we expect to see a bump in the distribution, corresponding to the sound horizon (magnified by the expansion of the universe), what is known as the Baryonic Acoustic Oscillations (BAO) peak.

In this process, DM and baryonic matter attract each other gravitationally to form the galaxies we see today. We define the de Broglie wavelength of the field, as the typical wavelength at which one can describe DM as a matter wave (see Section ~\ref{sec:DM_classical_field})
\begin{align}\label{de_Broglie_wavelength_DM}
    \lambda^\mathrm{dB}_\mathrm{DM} = \frac{2\pi \hbar}{m_\mathrm{DM} v_\mathrm{DM}} \sim 10^{-3}   \left(\frac{\mathrm{eV}}{m_\mathrm{DM}c^2}\right) \: \mathrm{m}\, .
\end{align}
This wavelength is associated to the momentum of the DM wave $m_\mathrm{DM} v_\mathrm{DM} = \hbar k_\mathrm{DM}$, where $k_\mathrm{DM}$ is its wavenumber.
Since DM allowed the formation of large scale structures, in particular galaxies, we require the DM de Broglie wavelength to be at most of the typical size of dwarf galaxies \cite{Battaglieri17}, corresponding to the primordial galaxies to form and which are roughly 1000 light years diameter long. This makes a lower bound on the DM mass to be $m_\mathrm{DM} c^2 \gtrapprox 10^{-22}$ eV \cite{Battaglieri17}. 

There is no observational evidence for an higher bound on the DM mass candidate. However, out of all the possible candidates, the most massive objects which could explain DM would be primordial black holes (PBH) with maximum mass of about $10^2$ solar masses \cite{Carr16} (which is equivalent to $\sim 10^{35}$ g). This corresponds roughly to a mass of $10^{68}$ eV/c$^2$, such that, overall, the bounds on the mass of the DM candidates are 
\begin{align}\label{DM_mass_bounds}
    10^{-22} \: \mathrm{ eV} \leq m_\mathrm{DM}c^2 < 10^{68} \:  \mathrm{ eV} \, .
\end{align}
This means that the range of possible mass for the DM candidate covers 90 orders of magnitude, implying that its search constitutes an experimental challenge.

\begin{figure}
    \centering
    \includegraphics{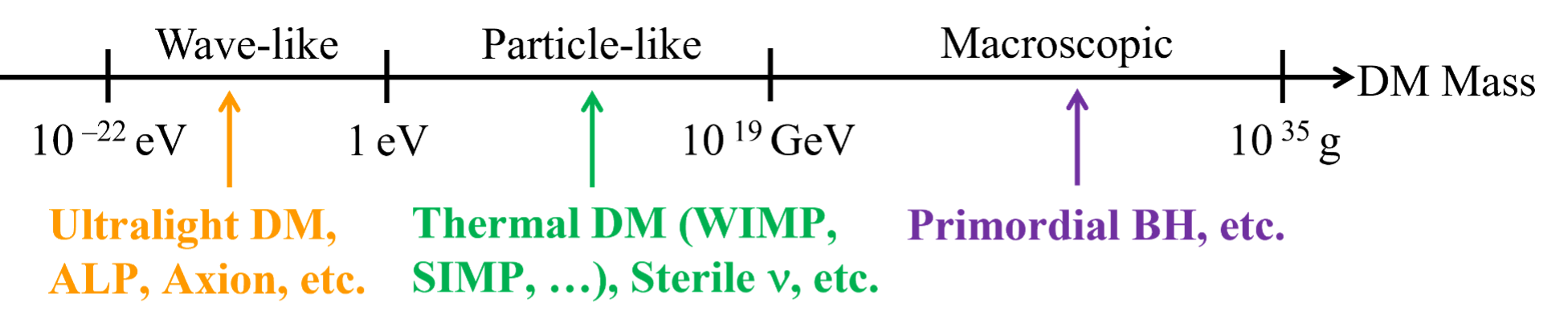}
    \caption{Mass scale of DM candidates, from massive primordial black holes to the light fields, from \cite{matsumoto}.}
    \label{fig:DM_mass_scale}
\end{figure}

Multiple DM candidates exist, depending on the mass, and are summarized in Fig.~\ref{fig:DM_mass_scale}.

\subsection{Macroscopic mass scale}

As stated previously, DM with very large mass, typically larger than  $10^{16}$ g \cite{Carr16} (which corresponds to $\sim 10^{50}$ eV/c$^2$) would be constituted of PBH. This kind of black holes would theoretically be generated by inflation \cite{Carr16}. 

\subsection{Particles mass scale}

At lower mass, typically from $1$ eV/$c^2$ to $10^{19}$ GeV/$c^2$ (which corresponds to the Planck mass), DM would be made of fermionic particles. As shown in Fig.~\ref{fig:DM_mass_scale}, several theoretical models exist, such as sterile neutrinos, but the most studied ones are weakly interacting massive particles (WIMPs), which arise in beyond the Standard Model scenarios. As their name suggests, they are weakly interacting with SM particles and their mass is large, in the GeV-TeV scale, therefore they can be detected in particle accelerators, such as at LHC. Historically, WIMPs were thought to be the most promising DM candidates but lacking a detection at LHC, other models arose, in particular with particles with lower mass.

\subsection{Classical fields mass scale\label{sec:DM_classical_field}}

In the $10^{-22}$ eV/c$^2$ to $1$ eV/$c^2$ mass range, we find DM candidates denoted as ultralight dark matter (ULDM) candidates.
At such low masses, by computing the number density, i.e the average number of particles occupying a box with phase space volume $(\lambda^\mathrm{dB}_\mathrm{DM})^3$ as $\langle N\rangle=(\lambda_\mathrm{dB}/\ell)^3$ where $\ell$ is the distance between particles, we find that 
\begin{align}\label{N_DM_phase_space}
    \langle N\rangle = \frac{\rho_\mathrm{DM}(2\pi \hbar c^2)^3}{(m_\mathrm{DM}c^2)^4 v^3_\mathrm{DM}} \simeq 10^6 \left(\frac{1 \mathrm{eV}}{m_\mathrm{DM}c^2}\right)^4 \, ,
\end{align}
where we used the de Broglie wavelength of the field Eq.~\eqref{de_Broglie_wavelength_DM} and 
inserting DM parameters, especially DM energy density in the Milky Way $\rho_\mathrm{DM} = 0.4$ GeV/cm$^3$ \cite{McMillan11} and galactic velocity $v_\mathrm{DM} = 10^{-3}$ c.
Eq.~\eqref{N_DM_phase_space} means that for DM particles with masses lower than roughly 10 eV, the average occupation number is much larger than 1, implying that the underlying particle is necessarily a boson, due to the Pauli exclusion principle. For sub-eV masses, this occupation number is so large, there is no need to study particles individually, and a standard classical field theory is enough to describe the field. 

An important requirement for the field to be a suitable DM candidate is that it behaves as cold dark matter (CDM) at cosmological scales \cite{Arias}, i.e as pressureless matter ($P \sim 0$) and that its energy density redshiftes as $a^{-3}$, with $a$ the cosmological scale factor. As we shall see in Chapter ~\ref{Cosmo_evolution}, all ULDM models (i.e with mass between $10^{-22}$ to $1$ eV.) fulfill this requirement since the angular frequency of the field exceeds the Hubble constant today.

ULDM includes a large number of models, which depend on the nature of the field, scalar, pseudo-scalar, vector and tensor, and each model has different couplings to SM fields, leading to different observable phenomenology. As we shall see in Chapter ~\ref{DM_pheno}, a various number of ULDM models are studied in this thesis, namely scalar, pseudo-scalar and vector fields, and their respective phenomenology is derived.

\section{Ultralight dark matter intrinsic characteristics}

In this section, we discuss some of the ULDM characteristics that are central for its detection. In Table ~\ref{Table_DM_parameters}, we summarize the experimental values for those ULDM parameters.

\begin{table}[h]
\caption{ULDM intrinsic parameters}
\centering
\begin{tabular}{c c c c}
\hline \hline
Parameters & Symbol & Numerical value & Unit \\
\hline \hline
Local energy density & $\rho_\mathrm{DM}$ & 0.4  \cite{McMillan11} & GeV/cm$^3$ \\
\hline
Rest mass & $m_\mathrm{DM}$ & $[10^{-22}\:;\:1]$  & eV/c$^2$ \\
\hline
Coherence time & $\tau(f_\mathrm{DM})$ & $10^{6}f^{-1}_\mathrm{DM}$ & s \\
\hline
DM mean velocity in heliocentric frame & $v_\mathrm{DM}$ & $3 \times 10^{5}$ \cite{Evans19,Cirelli24} & m/s \\
\hline
DM velocity dispersion & $\sigma_{v}$ & $1.5\times 10^{5}$ \cite{Evans19} & m/s \\
\hline
\end{tabular}
\label{Table_DM_parameters}
\end{table}
\subsection{Rest mass}

As it was mentioned in the last section, the mass of ULDM candidates is contained in 
\begin{align}
    10^{-22} \: \mathrm{eV} \leq m_\mathrm{DM}c^2 \lessapprox 1  \: \mathrm{eV} \, .
\end{align}

\subsection{Local energy density}

Several methods exist to estimate the dark matter energy density, and this, at different scales (e.g. galactic or local) \cite{Cirelli24}. Here, we are only interested in the DM energy density in the vicinity of the Sun, as it is what experiments on Earth will be sensitive too. Such density can be estimated by a parametric fit of the entire rotation curves of the galaxy (global method) $\rho_\mathrm{DM}(r)$ as function of the distance $r$ to the galactic center. Recent determinations of $\rho_\mathrm{DM}(r=r_\odot) \equiv \rho_\mathrm{DM}$ points towards \cite{Cirelli24}
\begin{align}\label{eq:DM_energy_density}
    \rho_\mathrm{DM} &= 0.4 \: \mathrm{GeV/cm}^3 \, .
\end{align}
Eq.~\eqref{eq:DM_energy_density} means that the total DM energy contained in a sphere of radius $\sim 30$ astronomical units (AU) (which corresponds approximately to the Sun-Neptune distance) is $\sim 10^{-13} M_\odot$. This shows how DM is locally subdominant, and therefore, how its gravitational impact on the orbit of the surrounding bodies is completely negligible \cite{Cirelli24}.

\subsection{\label{sec:coherence_DM}Velocity distribution and coherence time}

Galactic DM models assume that DM follows a spherical distribution around galaxies, making the so-called DM halo, as explained in Section ~\ref{DM_proofs}. During its formation, the DM halo virialized, and therefore, acquired a non-zero velocity dispersion $\sigma_{v}$. More precisely, in the heliocentric reference frame, we assume the galactic DM follows a Maxwellian velocity distribution $\mathfrak{F}(\vec v)$ \cite{Derevianko18}
\begin{subequations}
\begin{align}\label{DM_vel_distrib}
    \mathfrak{F}(\vec v) &= \frac{1}{\left(2\pi \sigma^2_{v}\right)^{3/2}} e^{-\frac{\left(\vec v - \vec v_\odot \right)^2}{2\sigma^2_{v}}}
\end{align}
where $\vec v_\odot \equiv \vec v_\mathrm{DM}$ is the mean velocity of the Solar system in the galactic frame and $\sigma_v$ is the dispersion (virial) velocity.
Eq.~\eqref{DM_vel_distrib} can be integrated over a full sphere to become a distribution over the magnitude of the velocity, i.e \cite{savallePhD}
\begin{align}
    \mathfrak{F}(v) &= \sqrt{\frac{2}{\pi}}\frac{v}{\sigma_{v}v_\mathrm{DM}} e^{-\frac{v^2 + v^2_\mathrm{DM}}{2\sigma^2_{v}}}\sinh\left(\frac{vv_\mathrm{DM}}{\sigma^2_{v}}\right) \label{DM_vel_distrib_scalar}
\end{align}
\end{subequations} 
Since the DM frame moves compared to any laboratory frame on Earth with a velocity following the distribution Eq.~\eqref{DM_vel_distrib_scalar}, the DM field itself acquires a kinetic energy $E_k = m_\mathrm{DM}v^2/2$ in addition to its rest energy $E=m_\mathrm{DM}c^2$ such that the measured DM frequency $f^\mathrm{meas.}_\mathrm{DM}$ is in reality
\begin{subequations}
\begin{align}\label{DM_freq_vel}
    f^\mathrm{meas.}_\mathrm{DM}(v) &= f_\mathrm{DM}\left(1+\frac{v^2}{2c^2}\right) \, ,
\end{align}
where 
\begin{align}
    f_\mathrm{DM} &= \frac{m_\mathrm{DM}c^2}{2\pi \hbar} \, ,
\end{align}
\end{subequations}
is the intrinsic DM Compton frequency. In the following of this manuscript, when deriving sensitivities of experiments to ULDM fields at a given frequency $f_\mathrm{DM}$, the kinetic energy correction in Eq.~\eqref{DM_freq_vel} will be neglected at leading order, such that we will consider the observed DM frequency $f^\mathrm{meas}_\mathrm{DM}$ to be equal to the intrinsic DM frequency $f_\mathrm{DM}$. 

Nonetheless, using Eqs.~\eqref{DM_vel_distrib_scalar} and \eqref{DM_freq_vel}, we can now construct a frequency distribution which reads \cite{Savalle21}
\begin{align}\label{DM_freq_distrib}
    \mathfrak{F}(f^\mathrm{meas.}_\mathrm{DM}) &= \sqrt{\frac{2}{\pi}}\frac{c^2}{f_\mathrm{DM} v_\mathrm{DM} \sigma_v}e^{\frac{-v^2_\mathrm{DM}+2c^2\left(1-\frac{f^\mathrm{meas.}_\mathrm{DM}}{f_\mathrm{DM}}\right)}{2\sigma^2_v}}\sinh\left(\frac{v_\mathrm{DM} c}{\sigma^2_v}\sqrt{2\left(\frac{f^\mathrm{meas.}_\mathrm{DM}}{f_\mathrm{DM}}-1\right)}\right) \, .
\end{align}

This equation essentially means that, a more rigorous modeling of galactic DM implies that we cannot model the field as monochromatic : it is a stochastic sum of $N$ different fields oscillating at different frequencies, which all follow the distribution Eq.~\eqref{DM_freq_distrib} \cite{foster:2018aa}.

The frequency broadening Eq.~\eqref{DM_freq_distrib} induces a characteristic coherence time of the field $\tau(\omega_\mathrm{DM})$. For periods shorter than the coherence time, the field behaves as monochromatic (i.e with only one amplitude and phase, but which are still stochastic), while for times longer than the coherence time, the field is a superposition of plane waves with different frequencies. One can compute the first (mean $\mu_{f^\mathrm{meas.}_\mathrm{DM}}$) and second (standard deviation $\delta f^\mathrm{meas.}_\mathrm{DM}$) moments of the frequency distribution by integration over its domain of definition ($\left[f_\mathrm{DM},+\infty\right[$). They are given by
\begin{subequations}
    \begin{align}
    \mu_{f^\mathrm{meas.}_\mathrm{DM}} &= f_\mathrm{DM}\left(1+ \frac{v^2_\mathrm{DM}+3\sigma^2_v}{2c^2}\right) \approx f_\mathrm{DM} \,\\
    \delta f^\mathrm{meas.}_\mathrm{DM} &= \frac{f_\mathrm{DM}}{2c^2}\sqrt{v^4_\mathrm{DM} + 10 v^2_\mathrm{DM} \sigma^2_v + 15 \sigma^4_v} \label{eq:freq_DM_std}
    \end{align}
\end{subequations}
such that we can define the coherence time of the field as 
\begin{align}
    \tau(f_\mathrm{DM}) &= \frac{1}{\delta f^\mathrm{meas.}_\mathrm{DM}} = \frac{2 c^2}{f_\mathrm{DM} \sqrt{v^4_\mathrm{DM} + 10 v^2_\mathrm{DM} \sigma^2_v + 15 \sigma^4_v}} \approx \frac{10^6}{f_\mathrm{DM}}\, ,
    \label{coherence_DM}
\end{align}
where we used $v_\mathrm{DM} \sim 2.33 \times 10^5 \: \mathrm{m/s} \approx 10^{-3}\: c$ \cite{Evans19} and $\sigma_v \sim v_\mathrm{DM}/\sqrt{2} \approx 5\times 10^{-4}\: c$ \cite{Evans19}. 
Note that for these calculations, we neglected the truncation of the velocity distribution Eq.~\eqref{DM_vel_distrib} at the galactic escape velocity $v_\mathrm{esc} \sim 5.5 \times 10^5$ m/s as it was done in \cite{Evans19}. However, it can be shown analytically that it would lead to a slight change in mean and standard deviation of the frequency distribution (of less than 10\%), i.e it will not alter significantly our estimates.

While, in the following of this manuscript, we will model DM as a single monochromatic field, and not as a superposition of fields oscillating at different frequencies following Eq.~\eqref{DM_freq_distrib}, we will still take into account the coherence time of the field when deriving the various sensitivities of experiments to DM, and we show here why.
Let us assume a given (linear) coupling $\zeta$ between DM and any SM sector (electromagnetic, fermionic, etc..) which induces a given signal $s(t)$ in our apparatus. Very generically, the signal searched for $s(t)$ is 
\begin{align}\label{eq:s(t)}
    s(t)= \left[X_s\right] \zeta \cos \left(2\pi f_\mathrm{DM} t+ \Phi \right) \, ,
\end{align}
where we factorize the coupling $\zeta$ from the rest of the amplitude of the signal $[X_s]$.
If the total time of integration of the experiment $T_\mathrm{obs}$ is much shorter than the coherence time, i.e $T_\mathrm{obs} \ll \tau(f_\mathrm{DM})$, Eq.~\eqref{eq:s(t)} is fully valid; i.e the signal is monochromatic, and the experiment sensitivity on the coupling at frequency $f_\mathrm{DM}$ is simply
\begin{subequations}\label{coupling_constraint_general_DM}
\begin{align}
    \zeta(f_\mathrm{DM}) &= \frac{1}{[X_s]}\sqrt{\frac{S_s(f_\mathrm{DM})}{T_\mathrm{obs}}} \equiv \frac{\sqrt{\mathrm{SNR}}}{[X_s]}\sqrt{\frac{S_{n}(f_\mathrm{DM})}{T_{\mathrm{obs}}}} \, ,
\end{align}
where $S_s,S_n$ are respectively the signal and noise power spectral densities (PSD) of the experiment and the SNR (signal-to-noise ratio) is defined as the ratio of signal to noise PSD, i.e $S_s/S_n$.
The SNR is used an an estimator of the signal strength, and it is useful in data analysis because it fixes a detection threshold in frequency domain, for possible discovery of new physics, compared to statistical anomaly. 

Note that in this regime, where $T_\mathrm{obs} \ll \tau(f_\mathrm{DM})$, a correction factor to the sensitivity arises due to the stochastic nature of the amplitude of the field \cite{Centers21}. In our case where we will always consider a 68\% detection threshold (i.e. SNR = 1), this correction factor induces a loss in signal of $\sim 1.5$\footnote{As it is pointed out in \cite{Centers21}, this correction factor depends on the nature of the signal, i.e if the apparatus couples to the field itself or e.g. to its gradient. However, we will neglect this subtlety, because for gradient coupling, the exact value of this correction factor highly depends on how the sensitive axis of the experiment evolves with time, and therefore it must be calculated case by case \cite{Centers21}.}.

On the other hand, when the integration time is much longer than the coherence time of the field, i.e $T_\mathrm{obs} \gg \tau(f_\mathrm{DM})$, this means that the signal searched for and parameterized by Eq.~(\ref{eq:s(t)}) is no longer coherent, i.e. it should be modeled as a sum of several stochastic harmonics, as we discussed it before. Another method to analyse the data is to cut the dataset in fragments with duration smaller than $\tau(f_\mathrm{DM})$ and search for a coherent signal in each of these blocks of data. In such a case, the experimental sensitivity to the coupling is reduced and becomes \cite{Budker14}
\begin{align}
    \zeta(f_\mathrm{DM}) &= \frac{\sqrt{\mathrm{SNR}}}{[X_s]}\sqrt{\frac{S_n(f_\mathrm{DM})}{\sqrt{T_{\mathrm{obs}}\tau(f_\mathrm{DM})}}} \, .
\end{align}
\end{subequations}

\chapter{Alternatives to dark matter}

As we saw in Section ~\ref{DM_proofs}, the various indicators of DM arise by using GR as our theory of gravitation and Newtonian mechanics as its small velocity and weak field limit. Another possibility to overcome the DM problem would be somehow to modify those theories of gravitation, to match again theory with observations. The most famous alternative theory to Newton's second law is MOND (MOdified Newtonian Dynamics) introduced by Milgrom in 1983 \cite{Milgrom83}. The main idea of MOND is to explain the galaxy rotation curves by introducing an universal acceleration $a_0 \sim 10^{-10}$ m/$s^2$ \cite{Milgrom83} in order to avoid smaller acceleration for stars very far away from center of galaxies. For a body of mass m, Newton's second law then becomes
\begin{subequations}
\begin{align}\label{MOND_acceleration}
    F = m \mu\left(\frac{g}{a_0}\right)g \, ,
\end{align}
where $g$ is the gravitational field and $\mu(x)$ is an asymptotic function that tends to 1 for $x \gg 1$ and to $x$ when $x \ll 1$. Therefore, in extreme weak acceleration environment such that $g \ll a_0$, Eq.~\eqref{MOND_acceleration} becomes
\begin{align}
    F &= m \frac{g^2}{a_0} \, .
\end{align}
\end{subequations}
A relativistic version of the MOND paradigm called TeVeS for Tensor-Vector-Scalar gravity was developed in 2004 by Bekenstein \cite{Bekenstein04}. 

While the MOND theory and its extensions suffered for a long time to explain various phenomena, in particular in cluster of galaxies \cite{McGaugh15}, and therefore would still require DM, recent studies (see e.g. \cite{Clifton12, Skordis20, Blanchet24}) were able to solve these problems.
Nonetheless, in this thesis, we will be interested in solutions for dark matter involving new particles, and not through modified gravity.

\clearpage
\pagestyle{plain}
\printbibliography[heading=none]
\clearpage
\pagestyle{fancy}

\part{Phenomenology of some ultralight dark matter candidates}
\chapter{\label{Cosmo_evolution}Cosmological evolution}

\section{Action}

In this section, we will show how a light bosonic field can act as CDM at cosmological scales. In the following of this manuscript, we will be interested in several types of light fields : a scalar field $\phi$, a pseudo-scalar field $a$ and a vector field $\phi^\mu$. 

The dynamics of such fields are encoded in their action which we define as
\begin{subequations}\label{action_fields}
\begin{align}
    S_\phi &= \frac{1}{c}\int d^4x \sqrt{-g}\left[\frac{R}{2\kappa} - \frac{1}{2\kappa}g^{\mu\nu}\partial_\mu \phi \partial_\nu \phi -\frac{V(\phi)}{2\kappa}+ \mathcal{L}_\mathrm{SM}[\Psi_i] + \mathcal{L}_\mathrm{int}[\Psi_i,\phi]\right] \label{scalar_field_action} \,\\
    S_a &= \frac{1}{c}\int d^4x \sqrt{-g}\left[\frac{R}{2\kappa} - \frac{1}{2\kappa}g^{\mu\nu}\partial_\mu a \partial_\nu a -\frac{V(a)}{2\kappa}+ \mathcal{L}_\mathrm{SM}[\Psi_i] + \mathcal{L}_\mathrm{int}[\Psi_i,a]\right] \label{axion_field_action}\,\\
    S_{\phi^\mu} &= \frac{1}{c}\int d^4x \sqrt{-g}\left[\frac{R}{2\kappa} - \frac{1}{4\mu_0}g^{\mu\nu}g^{\alpha\beta}\phi_{\mu\alpha} \phi_{\nu\beta} -\frac{V(\phi^\mu)}{2\mu_0}+ \mathcal{L}_\mathrm{SM}[\Psi_i] + \mathcal{L}_\mathrm{int}[\Psi_i,\phi^\mu]\right] \label{vector_field_action}
\end{align}
\end{subequations}
for the scalar, pseudo-scalar and vector fields, where $\kappa=8\pi G/c^4$ is the Einstein gravitational constant, $\mu_0$ is the vacuum permeability, $g=\mathrm{det}(g_{\mu\nu})$ and $\phi_{\mu\nu}=\partial_\mu\phi_\nu - \partial_\nu\phi_\mu$ is the $\phi^\mu$ field strength tensor. We choose a convention where the scalar $\phi$ and pseudo scalar $a$ fields are dimensionless, while the vector field $\phi^\mu$ has units of $\mathrm{V.s/m}$, as the usual EM potential. In each expression, the first three terms represent the gravitational action which contains GR (through the Ricci scalar $R$) in addition to the kinetic and potential energy of the new field, while the last term represents the coupling of the new field with existing SM fields. In this section, we will focus on the former while the latter will be considered further away as it is responsible for the possible observed phenomenology for direct DM detection.

In Eq.~\eqref{action_fields}, we choose a convention on the scalar field normalization factor in front of the kinetic and potential terms, such that it matches the $1/4$ normalization of the vector kinetic term\footnote{For those respective normalization factors, one can show that integrating by parts the vector field Lagrangian leads to a term $\partial_\mu \phi^\nu \partial^\mu \phi_\nu/2$, which is similar to the scalar field kinetic term.}. Other conventions exist with a different normalization factor, e.g in \cite{hees18}, which can be recovered by a simple redefinition of the field $\phi \rightarrow \sqrt{2} \phi$. 

In general, the potential for the various fields is a polynomial expansion of the field. Without loss of generality, considering a scalar field $\phi$, it has the form 
\begin{align}
    V(\phi) = \frac{m^2_\phi c^2}{\hbar^2}\phi^2 + 
    \mu \phi^3 + \lambda\phi^4 + ...
\end{align}
where the first term is the mass term with $m_\phi$ the mass of the field. The second and third terms represent respectively cubic and quartic self interaction of the field with strength $\mu, \lambda$, in units of $[L^{-2}]$. Higher order self interactions are not represented.

In all models of interest in this manuscript, self interactions will not be considered, such that the potential of the field reduces to the mass term, with mass $m_\phi,m_a,m_{U}$ respectively for the scalar, pseudo-scalar and vector fields. Therefore, the GR action is invariant under $\{\phi,a,\phi^\mu\} \rightarrow \{-\phi,-a,-\phi^\mu\}$ parity transformation, such that both scalar and pseudo-scalar fields have the exact same dynamics at the GR level (i.e neglecting their respective interactions with SM fields). In addition, each vector field spatial component $\phi^i$ (which are the only relevant degrees of freedom\footnote{Solving the equations of motion for the vector field, one can show that there is no $\ddot \phi^0$ term, therefore the temporal component of the field is non dynamical. This naturally yields that its three degrees of freedom (the mass and the two polarization states) can be described by its spatial components only.}) can be described by a scalar field $\phi_U$, which obeys the following action
\begin{align}\label{vector_scalar_equiv_action}
    S^\mathrm{Grav.}_{\phi_U} &= \frac{1}{c}\int d^4x \sqrt{-g}\left[- \frac{1}{2\mu_0}g^{\mu\nu}\partial_\mu \phi_U \partial_\nu \phi_U - \frac{m^2_{U}c^2}{2\mu_0\hbar^2}\phi^2_U\right]
\end{align}
in the Coulomb gauge, up to a total derivative, and neglecting GR for practical purposes. Eq.~\eqref{vector_scalar_equiv_action} has a very similar form of the action of a pure scalar field $\phi$, described in Eq.~\eqref{scalar_field_action} (up to a constant due to the difference in units between scalar and vector fields). Therefore, we will only derive the cosmological evolution of a pure scalar field, whose result will be easily extended to pseudo-scalar and vector fields.

\section{Klein Gordon equation in an expanding Universe}

Let us first discuss scalar fields. As noted in the end of the last section, neglecting their respective interactions with SM fields, both scalar and pseudo-scalar fields have the same dynamics, such that we will only make the calculations in the case of a pure scalar field $\phi$.
From Eq.~\eqref{action_fields}, the Lagrangian describing the kinetic and potential energies of the new scalar field is
    \begin{align}
        \mathcal{L}_\phi &= -\frac{\sqrt{-g}}{2\kappa}\left(g^{\mu\nu}\partial_\mu \phi \partial_\nu \phi + \frac{m^2_\phi c^2}{\hbar^2}\phi^2\right) \label{scalar_lagrangian}\,
    \end{align}
We represent the universe as flat, homogeneous and isotropic and described using the Friedmann-Lemaitre-Robertson-Walker (FLRW) metric defined as 
\begin{align}
    g_{\mu\nu} &= \mathrm{diag}\left(-1,a^2(t),a^2(t),a^2(t)\right)
\end{align}
with $a$ the scale factor, such that $\sqrt{-g}=a^3(t)$. Using the Euler-Lagrange equation, we obtain the equation of motion of the field $\phi$
\begin{subequations}
\begin{align}
    \partial_\mu\left(-a^3(t)\partial^\mu \phi\right) + \frac{m^2_\phi c^2}{\hbar^2}a^3(t)\phi &= 0 \,\\
    \Rightarrow \square \phi - \frac{3H(t)}{c^2}\dot \phi - \frac{m^2_\phi c^2}{\hbar^2}\phi &= 0 \label{KG_eq_exp_Univ}
\end{align}
\end{subequations}
where $\square = \eta^{\mu\nu}\partial_\mu \partial_\nu$ is the d'Alembertian operator, with $\eta_{\mu\nu}=\mathrm{diag(-1,1,1,1)}$, the Minkowski metric, the dot represents derivative with respect to time and where we defined $H(t) = \dot a(t)/a(t)$, the Hubble constant. As it can be seen from Eq.~\eqref{KG_eq_exp_Univ}, the unit of the field does not impact the equation of motion, since it is a linear differential equation. Therefore, the vector field will follow the exact same Klein-Gordon equation, despite its different unit.

Eq.~\eqref{KG_eq_exp_Univ} describes a damped harmonic oscillator. At early times of the universe, when $H \gg m_\phi c^2/\hbar$, the solution of this equation is a constant $\phi \rightarrow \phi_0$, i.e the field is frozen at $\phi_0$ and behaves effectively as massless with a flat potential. As soon as $H \sim m_\phi c^2/\hbar$, the potential of the field gets curved by the mass, and as long as $\phi_0$ does not exactly correspond to the local minimum of the potential, the field rolls down the potential and gets massive. When $H \ll m_\phi c^2/\hbar$, such that we can neglect the field friction, the solution of the Klein-Gordon equation is an oscillating solution of the form 
\begin{subequations}
\begin{align}\label{scalar_sol_KG}
    \phi(t,\vec x) &= \phi_0\cos(\omega_\phi t - \vec k_\phi \cdot \vec x +\Phi)
\end{align}
where $\phi_0, \Phi$ are respectively the amplitude of oscillation and a phase and where 
\begin{align}\label{dispersion_relation_mass_scalar}
    |\vec k_\phi| = \sqrt{\omega^2_\phi/c^2-m^2_\phi c^2/\hbar^2}
\end{align}
\end{subequations}
is the wavevector of the field. A vector field will oscillate in the exact same manner, but with a vectorial amplitude, i.e 
\begin{align}\label{vector_sol_KG}
    \phi^\mu(t,\vec x) &= Y^\mu \cos(\omega_U t - \vec k_U \cdot \vec x +\Phi)
\end{align}
This generation of field oscillation, with non-zero amplitude, which as we shall see in the next section leads to a non-zero energy density for the field, is more commonly known as the misalignment mechanism. Indeed, if by any chance, $\phi_0$ corresponded to the exact potential minimum, the field would not oscillate, and as a consequence would still be effectively massless.

In the following of this manuscript, we require the field to oscillate, such that the necessary condition $H \ll m_\phi c^2/\hbar$ equates to 
\begin{align}
    m_\phi \gg 10^{-33} \: \mathrm{eV/c}^2
\end{align}
with the current value of the Hubble constant $H_0 \sim 10^{-18}$ s$^{-1}$. Note that this condition is looser than the bound of DM mass from observations Eq.~\eqref{DM_mass_bounds}.

\section{\label{Osc_field_DM}Oscillating field as cold dark matter}

We now focus on the energy density and pressure of such fields, and we show that they behave at cosmological scale as cold dark matter, i.e pressureless matter with equation of state $\omega = P/\rho \sim 0$. 

\subsection{Scalar field}

Still neglecting interaction, the stress-energy tensor of the scalar field $T^{\mu\nu}$ can be defined from the field Lagrangian Eq.~\eqref{scalar_lagrangian}, as 
\begin{subequations}
\begin{align}
    T^{\mu\nu} &=\frac{2}{\sqrt{-g}}\frac{\delta (\sqrt{-g}\mathcal{L}_\phi)}{\delta g_{\mu\nu}} \,\\
    &= \frac{1}{\sqrt{-g}\kappa \delta g_{\mu\nu}}\left(\delta(\sqrt{-g})\left(-g^{\mu\nu}\partial_\mu \phi \partial_\nu \phi - \frac{m^2_\phi c^2}{\hbar^2}\phi^2\right)+\sqrt{-g}\left(-\delta g^{\mu\nu}\partial_\mu \phi \partial_\nu\phi\right)\right)
\end{align}
\end{subequations}
where we only vary the metric parameters. Using Jacobi's formula, we have 
\begin{subequations}
    \begin{align}
        \delta \sqrt{-g} &= \frac{1}{2}\sqrt{-g}g^{\mu\nu}\delta g_{\mu\nu} \,\\
        \delta g^{\mu\nu} &= -g^{\mu\alpha}g^{\nu \beta}\delta g_{\alpha\beta}
    \end{align}
\end{subequations}
which simplifies the stress-energy tensor expression to
\begin{align}
    T^{\mu\nu} &= \frac{-g^{\mu\nu}\left(\frac{1}{2}\partial_\alpha \phi \partial^\alpha \phi + \frac{m^2_\phi c^2}{2\hbar^2}\phi^2\right)+\partial^\mu \phi \partial^\nu \phi}{\kappa} \label{stress_tensor_scalar}
\end{align}
For simplicity, we assume the field behaves as a perfect fluid at cosmological scales, such that its stress-energy tensor has the form
\begin{align}\label{Stress_energy_perfect_fluid}
    T^{\mu\nu} &= \left(\rho+\frac{P}{c^2}\right)u^\mu u^\nu + Pg^{\mu\nu}
\end{align}
where $P, \rho$ are respectively the pressure and energy density of the fluid and $u^\mu$ its 4-velocity. By identification with Eq.~\eqref{stress_tensor_scalar}, we find the average pressure over several field oscillations to be 
\begin{subequations}
\begin{align}
   \langle P \rangle &= -\frac{1}{\kappa}\left\langle\frac{1}{2}\partial_\alpha \phi \partial^\alpha \phi + \frac{m^2_\phi c^2}{2\hbar^2}\phi^2\right\rangle \equiv 0\,
\end{align}
while the average energy density is defined as the $00$ component of the stress-energy tensor, i.e 
\begin{align}
\langle\rho \rangle &= \langle T_{00}\rangle = \frac{1}{\kappa}\left\langle\frac{1}{2}\partial_\alpha \phi \partial^\alpha \phi + \frac{m^2_\phi c^2}{2\hbar^2}\phi^2+\left(\frac{1}{c}\frac{\partial \phi}{\partial t}\right)^2\right\rangle \equiv \frac{\omega^2_\phi \phi^2_0}{2 \kappa c^2}
\end{align}
\end{subequations}
where we used Eqs.\eqref{scalar_sol_KG},  \eqref{dispersion_relation_mass_scalar}, and $\langle A+B\cos(2\omega t+\Phi)\rangle = A$.
The observed equation of state of the field is defined as 
\begin{align}
    \frac{\langle P \rangle}{\langle \rho \rangle} = 0\, ,
\end{align}
Therefore, the scalar field behaves cosmologically as a pressureless fluid, i.e cold dark matter with energy density
\begin{align}
    \langle \rho \rangle = \frac{\omega^2_\phi \phi^2_0}{2\kappa c^2} \equiv \rho_\mathrm{DM} 
\end{align}
where $\rho_\mathrm{DM}$ is the local DM energy density, such that we define the amplitude of oscillation of all scalar ULDM candidates to be
\begin{align}\label{energy_density_scalar_DM}
    \phi_0 &= \frac{\sqrt{16 \pi G \rho_\mathrm{DM}}}{\omega_\phi c}
\end{align}
In conventions where the scalar field Lagrangian Eq.~\eqref{scalar_lagrangian} is multiplied by 2 compared to ours (e.g in \cite{hees18}, with the redefinition of the field $\phi\rightarrow \sqrt{2}\phi$), the associated amplitude of the scalar field is $\phi_0 = \sqrt{8 \pi G \rho_\mathrm{DM}}/\omega_\phi c$. We shall see in Chapter ~\ref{DM_pheno} that the observables of interest remain unchanged, as these fields redefinitions are unphysical, as expected. 

\subsection{\label{vector_field_DM}Vector field}

From Eq.~\eqref{vector_field_action}, we can define the vector field Lagrangian as
\begin{align}\label{vector_lagrangian}
  \mathcal L_{\phi^\mu} =& -\sqrt{-g}\left(\frac{1}{4\mu_0}g^{\mu\nu}g^{\alpha\beta}\phi_{\mu\alpha}\phi_{\nu\beta} - \frac{m^2_Uc^2}{2\mu_0\hbar^2}g^{\mu\nu}\phi_\mu\phi_\nu\right)\, ,
\end{align}
The vector field equation reads 
\begin{subequations}
\begin{align}\label{vector_field_equation}
    \partial_\mu \phi^{\mu\nu} - \frac{m^2_U c^2}{\hbar^2}\phi^\nu = 0 \, ,
\end{align}
where we assumed an expanding Universe, with $H \ll m_U c^2/\hbar$, such that the Hubble constant is neglected.
Taking the divergence of Eq.~\eqref{vector_field_equation} and using the antisymmetric property of $\phi^{\mu\nu}$, we find the continuity equation of $\phi^\mu$ i.e
\begin{align}\label{continuity_eq_vector}
    \partial_\mu \phi^\mu &= 0 \, .
\end{align}
\end{subequations}
While a similar equation can be found in electromagnetism by choosing the Lorenz gauge, Eq.~\eqref{continuity_eq_vector} is genuinely a field equation, which means that we are still free to choose a gauge.
As in the scalar field case, the stress-energy tensor of the vector field is defined as 
\begin{align}
T^{\mu\nu} &= \frac{2}{\sqrt{-g}}\frac{\delta (\sqrt{-g}\mathcal{L}_{\phi^\mu})}{\delta g_{\mu\nu}}=\frac{-g^{\mu\nu}\left(\frac{1}{4}\phi_{\alpha\beta}\phi^{\alpha\beta}+\frac{m^2_U c^2}{2\hbar^2}\phi_\alpha \phi^\alpha\right)+\phi^{\alpha\mu}\phi_{\alpha}^{\ \nu} + \frac{m^2_U c^2}{\hbar^2}\phi^\mu\phi^\nu}{\mu_0}\, \label{stress_vector} ,
\end{align}
where we used again the antisymmetricity property of $\phi^{\mu\nu}$. We now assume an equivalent form of the plane wave solution Eq.~\eqref{vector_sol_KG}, $\phi^\mu = Y^\mu \mathcal{R}[e^{ik_\mu x^\mu}]$ (forgetting about the irrelevant phase), such that each field term of Eq.~\eqref{stress_vector} can be written as
\begin{subequations}
\begin{align}
    \phi_{\alpha\beta}\phi^{\alpha\beta} &= 2\left[\left(ik_\alpha Y_\beta\mathcal{R}[e^{ik_a x^a}]\right)\left(ik^\alpha Y^\beta\mathcal{R}[e^{ik_a x^a}]\right)-\left(ik_\beta Y_\alpha\mathcal{R}[e^{ik_a x^a}]\right)\left(ik^\alpha Y^\beta\mathcal{R}[e^{ik_a x^a}]\right)\right]\,\\
    \phi_\alpha \phi^\alpha &= \left(Y_\alpha \mathcal{R}[e^{ik_a x^a}]\right)\left(Y^\alpha \mathcal{R}[e^{ik_a x^a}]\right) \,\\
    \phi^{\alpha\mu}\phi_{\alpha}^{\ \nu} &= \left(ik^\alpha Y^\mu\mathcal{R}[e^{ik_a x^a}]-ik^\mu Y^\alpha\mathcal{R}[e^{ik_a x^a}]\right)\left(ik_\alpha Y^\nu\mathcal{R}[e^{ik_a x^a}]-ik^\nu Y_\alpha\mathcal{R}[e^{ik_a x^a}]\right)\,\\
    \phi^\mu\phi^\nu &= \left(Y^\mu \mathcal{R}[e^{ik_a x^a}]\right)\left(Y^\nu \mathcal{R}[e^{ik_a x^a}]\right) \, .
\end{align}
\end{subequations}
Considering
\begin{align}
    \langle \mathcal{R}[a^\mu e^{ik_a x^a}] \mathcal{R}[b^\nu e^{ik_a x^a}] \rangle &= \frac{1}{2}\mathcal{R}[a^\mu b_*^\nu]
\end{align}
where the star denotes the complex conjugate, the average over several oscillations of the stress-energy tensor of the vector field can be written as
\begin{align}
  \langle T^{\mu\nu} \rangle &= \frac{-g^{\mu\nu}\left(\frac{1}{4}k_\alpha k^\alpha Y_\beta Y^\beta + \frac{m^2_Uc^2}{4\hbar^2}Y_\alpha Y^\alpha\right)+\frac{1}{2}k_\alpha k^\alpha Y^\mu Y^\nu + \frac{1}{2}k^\mu k^\nu Y_\alpha Y^\alpha+\frac{m^2_Uc^2}{2\hbar^2}Y^\mu Y^\nu}{\mu_0}
\end{align}
where we used $k_\mu Y^\mu = 0$ from Eq.~\eqref{continuity_eq_vector}. Using the same identification with perfect fluid as before, we find that the average pressure is
\begin{subequations}
    \begin{align}
        \langle P \rangle &= -\frac{1}{\mu_0}\left\langle\frac{1}{4}k_\alpha k^\alpha Y_\beta Y^\beta + \frac{m^2_Uc^2}{4\hbar^2}Y_\alpha Y^\alpha\right\rangle \equiv 0
    \end{align}
since $k^\mu k_\mu = -(\omega_U/c)^2+|\vec k_U|^2$ and using Eq.~\eqref{dispersion_relation_mass_scalar}. The average energy density is
\begin{align}
    \langle\rho \rangle &= \frac{1}{\mu_0 }\left\langle\frac{1}{4}k_\alpha k^\alpha Y_\beta Y^\beta + \frac{m^2_Uc^2}{4\hbar^2}Y_\alpha Y^\alpha+\frac{1}{2}k_\alpha k^\alpha (Y_0)^2 + \frac{1}{2}\left(\frac{\omega_U}{c}\right)^2 Y_\alpha Y^\alpha+\frac{m^2_Uc^2}{2\hbar^2}(Y_0)^2\right\rangle \,\nonumber\\
    &\equiv \frac{\omega^2_U}{2\mu_0 c^2}Y_\alpha Y^\alpha \, .
\end{align}
\end{subequations}
Then, the amplitude of the total vector field is simply
\begin{subequations}
\begin{align}
    Y_\alpha Y^\alpha &= \frac{2\mu_0 \rho_\mathrm{DM}c^2}{\omega^2_U} \, .
\end{align}
Due to the small galactic velocity in the DM halo, the temporal contribution of the dot product in the previous equation can be neglected. Indeed, parametrizing $Y^\mu = (\phi/c, \vec Y)$, we have
\begin{align}
    Y_\alpha Y^\alpha &= -\left(\frac{\phi}{c}\right)^2 + |\vec Y|^2 = - \left(\frac{c}{\omega_U}\right)^2 \left(\vec k_U \cdot \vec Y\right)^2 + |\vec Y|^2 = \frac{v^2_\mathrm{DM}}{c^2} \left(\hat e_v \cdot \vec Y\right)^2 + |\vec Y|^2 \approx |\vec Y|^2 \, ,
\end{align}
\end{subequations}
where we used the continuity equation $k_\mu Y^\mu = 0$ to set $\phi/c = (\vec k_U \cdot \vec Y) c/\omega_U$ and $\vec k_U = -\omega_U \vec v_\mathrm{DM}/c^2 =- \omega_U v_\mathrm{DM} \hat e_v/c^2$. This implies 
\begin{align}\label{energy_density_vector}
  |\vec Y| &= \frac{\sqrt{2\mu_0 \rho_\mathrm{DM}}c}{\omega_U}  \, .
\end{align}
Eq.~\eqref{energy_density_vector} is a good approximation for experiments where the propagation of the field is negligible, since we assume $\vec k_U = 0$, such that $Y^0 =0$ (from $k_\mu Y^\mu = 0$).
Note that Eqs.~\eqref{energy_density_scalar_DM} and \eqref{energy_density_vector} are related by the change $\mu_0 \rightarrow \kappa$, as expected from the respective Lagrangian densities.

\section{Expressing the dark matter field in various reference frames}

\subsection{Galactic frame}

In the galactic frame, which is assumed to be the local DM rest frame\footnote{Note that we can find a rest frame of DM because the field is massive, as opposed to the electromagnetic field.}, the expression of the (scalar) field Eq.~\eqref{scalar_sol_KG} reduces to 
\begin{align}\label{field_galactic}
    \phi = \phi_0 \cos(\omega_\phi t +\Phi)
\end{align}
with $\hbar \omega_\phi = m_\phi c^2$, since $|\vec k_\phi|= 0$ on average in this frame. However, for experiments lasting longer than the typical coherence time of the field Eq.~\eqref{coherence_DM}, one must take into account the velocity distribution introduced in Eq.~\eqref{DM_vel_distrib}. In this case, the mean velocity of the distribution is $v_\odot \rightarrow v^\mathrm{RF}_\mathrm{DM} = 0$, while the frequency dispersion is given by Eq.~\eqref{eq:freq_DM_std} (with $v_\mathrm{DM} = 0$) $\delta f^\mathrm{RF}_\phi \sim 5 \times 10^{-7} f_\phi$.  

The form of the field Eq.~\eqref{field_galactic} will be used in our study of atom interferometry probes of ULDM in Chapter ~\eqref{chap:AI}. As mentioned previously, we will neglect the frequency broadening due to the velocity distribution Eq.~\eqref{DM_vel_distrib}, such that we will consider DM as a pure monochromatic field.

\subsection{Laboratory frame}

Eq.~\eqref{scalar_sol_KG} is the exact solution of the ULDM field in any other reference frame, in particular laboratory frames on Earth. In such cases, the field is propagating with velocity $|\vec v_\mathrm{DM}| \sim 10^{-3}c$ (see Table ~\ref{Table_DM_parameters}, and neglecting the Earth velocity around the Sun, $|\vec v_\mathrm{Earth}| \sim 3 \times 10^4$ m/s, which averages to $0$ over the year.). 

We would like to know if it is still possible to neglect the propagation term $\vec k_\phi \cdot \vec x$ for the sensitivity estimates of the various experiments under consideration in the following of this thesis. To answer this question, the first point is to know whether the experiment couples to the field itself or to its gradient. In the latter case, it is clear that the full form of the field Eq.~\eqref{scalar_sol_KG} must be used, otherwise its gradient becomes $0$.

In the former case, in order to neglect the propagation, and therefore considering the field as a pure standing wave, the main requirement is that the de Broglie wavelength of the field $\lambda^\mathrm{dB}_\mathrm{DM}$ must be much larger than the length scale of the experiment $\ell_\mathrm{exp}$, i.e, in terms of DM mass
\begin{align}\label{eq:mass_size_relation_lab_frame}
    \ell_\mathrm{exp} \ll 10^{-3} \left(\frac{1\: \mathrm{eV}}{m_\phi c^2}\right) \: \mathrm{m} \, .
\end{align}
As we shall see in the corresponding chapters, this requirement is fulfilled for the experiments detailed in Chapters ~\ref{optical_exp_axion}, \ref{chap:Rydberg_exp_DP}, \ref{chap:SHUKET_exp} and \ref{chap:Classical_tests_UFF}, therefore we will assume the field to be homogeneous over the full size of experiment with the form
\begin{subequations}\label{field_lab_frame}
\begin{align}
    \phi = \phi_0 \cos(\omega_\phi t +\Phi)
\end{align}
with 
\begin{align}\label{eq:compton_freq_mass_corresp_DM}
    \omega_\phi = \sqrt{\left|\vec k_\phi\right|^2 c^2 +\frac{m^2_\phi c^4}{\hbar^2}} = \frac{m_\phi c^2}{\hbar}\sqrt{1+\left(\frac{v_\mathrm{DM}}{c}\right)^2} \approx \frac{m_\phi c^2}{\hbar}
\end{align}
\end{subequations}
where we used $\left|\vec k_\phi\right| = 2\pi /\lambda^\mathrm{dB}_\mathrm{DM}$ and where we neglected the $(v_\mathrm{DM}/c)^2 \sim 10^{-6}$ correction. One can easily notice that this form is equivalent (up to $(v_\mathrm{DM}/c)^2$ corrections) to the one in the galactic frame Eq.~\eqref{field_galactic}.

In Chapter ~\ref{chap:LISA_DM}, we will be interested in the direct detection of ULDM fields by a space-based experiment, \textit{LISA}, which couples to the gradient of the field. and therefore, we will consider the exact form of the field Eq.~\eqref{scalar_sol_KG}.

\chapter{\label{DM_pheno}Couplings to Standard Model fields}

\section{\label{dilaton_pheno}Scalar field}

\subsection{Theoretical motivations}

A scalar particle is the simplest model of ULDM. Light scalar fields appear in various theoretical models, such as dilaton and moduli fields in string theory \cite{Damour94}, or the relaxion field to understand the electroweak hierarchy problem, i.e the large energy difference between the electroweak scale (set by the Higgs field vacuum expectation value $v \sim 246$ GeV) and the gravity scale (set by the Planck mass $M_P c^2 \sim 10^{19}$ GeV) \cite{Graham15}. 
As we shall see in the following, a coupling between such scalar field with SM fields naturally leads to violation of the equivalence principle, which is expected to happen at some scale, as discussed in Section ~\ref{GR_limits}.

\subsection{Rest mass and transition frequency oscillations}

We define the linear interaction Lagrangian of a dimensionless scalar field $\phi$ appearing in Eq.~\eqref{scalar_field_action} as \cite{hees18}
\begin{equation}\label{dilaton_lagrangian}
\mathcal{L}_\mathrm{int} =\frac{\phi}{\sqrt{2}}\left(\frac{d_e}{4\mu_0}F_{\mu\nu}F^{\mu\nu} - \frac{d_g\beta_3}{2g_3}G^a_{\mu\nu}G^{a\mu\nu} - \sum_{i=e,u,d}(d_{m_i} +\gamma_{m_i}d_g)m_i \bar{\psi_i}\psi_i\right) \, ,
\end{equation}
where as stated previously, our convention in the scalar field normalization implies an additional $1/\sqrt{2}$ factor compared to \cite{damour:2010zr,hees18}. $F^{\mu \nu}, G^{a\mu \nu}$ represent the electromagnetic and gluonic strength tensors respectively, $\mu_0$ is the vacuum magnetic permeability, $g_3, \beta_3$ are the dimensionless QCD coupling constant and QCD beta function for the running of $g_3$, $m_i$ the mass of the fermions fields in units of energy, with spinors $\psi_i$ and $\gamma_{m_i}$ the anomalous dimension giving the energy running of the masses of the QCD coupled fermions. The $d_i$ represent the dimensionless coupling constants between the scalar dilaton and the different matter fields. 

Since the energy scale we are interested in is very low (compared to second generation leptons' masses), this low energy effective Lagrangian involves only the low mass fermions, i.e the electron, up and down quarks, as can be noticed in the last term in Eq.~\eqref{dilaton_lagrangian}.

These interactions between the SM particles and the dilaton lead to variation of several fundamental constants of Nature in SM, namely the QCD energy scale $\Lambda_\mathrm{QCD}$, the EM fine structure constant $\alpha$, the electron mass $m_e$ and the light quark masses $m_i$, with $i=\{u,d\}$. Specifically, assuming the usual EM kinetic term in addition to the coupling of the dilaton to the EM sector (which is the first term of Eq.~\eqref{dilaton_lagrangian}), one has
\begin{align}
    \mathcal{L} &\supset -\frac{1}{4\mu_0}\left(1-\frac{ d_e \phi}{\sqrt{2}}\right)F_{\mu\nu}F^{\mu\nu} \approx -\frac{1}{4\mu_0(1+\frac{ d_e\phi}{\sqrt{2}})}F_{\mu\nu}F^{\mu\nu} \, ,
\end{align}
where we used the Taylor expansion of the term in brackets as $d_e \phi \ll 1$. Since $\mu_0 \propto \alpha$, this Lagrangian leads to a small dependency of the fine structure constant on the scalar field as
\begin{subequations}\label{const_variations}
\begin{align}
    \alpha(\phi) &= \left(1+\frac{d_e\phi}{\sqrt{2}}\right)\alpha \, ,
\end{align}
where $\alpha$ is the bare value of the fine structure constant, i.e without the dilaton-EM coupling. The other constants of Nature evolve as \cite{damour:2010zr}
\begin{align}
\Lambda_3(\phi) &= \left(1+\frac{d_g\phi}{\sqrt{2}}\right)\Lambda_3 \, , \\
m_e(\phi) &= \left(1+\frac{d_{m_e}\phi}{\sqrt{2}}\right)m_e \, , \\
m_i(\Lambda_\mathrm{QCD})(\phi) &= \left(1+\frac{d_{m_i}\phi}{\sqrt{2}}\right)m_i(\Lambda_\mathrm{QCD}) \, .
\end{align}
If one of these couplings is non-zero, the corresponding variable would not be constant at two different spacetime positions where the field $\phi$ takes different values, hence the violation of the local position invariance, or more generally of the equivalence principle.  
Following the work of \cite{damour:2010zr}, we introduce the mean light quark mass $\hat{m} = (m_u+m_d)/2$ and the difference of light quark masses $\delta m = m_d-m_u$. Following Eq.~\eqref{const_variations}, the numerical value of these new parameters will change as
\begin{align}
\hat{m}(\phi) &= \left(1+\frac{d_{\hat{m}}\phi}{\sqrt{2}}\right)\hat{m} \, , \\
\delta m(\phi) &= \left(1+\frac{d_{\delta m}\phi}{\sqrt{2}}\right)\delta m \, ,
\end{align}
with 
\begin{align}
d_{\hat{m}} &= \frac{m_u d_{m_u}+m_d d_{m_d}}{m_u+m_d},\,\\
d_{\delta m} &= \frac{m_d d_{m_d}-m_u d_{m_u}}{m_d-m_u}.
\end{align}
\end{subequations} 
Following \cite{damour:2010zr}, we can define dimensionless dilatonic mass and frequency charges for the atom A respectively given by
\begin{subequations}\label{dilatonic_coupling_func}
\begin{align}
[Q^A_M]_d &= \frac{\partial \ln m_A(\phi)}{\partial \phi} \left(\equiv \alpha_A/\sqrt{2}\right) \,  \label{dilatonic_mass_charge} , \\
[Q^A_\omega]_d &= \frac{\partial \ln \omega_A(\phi)}{\partial \phi} \, ,
\end{align}
\end{subequations}
which encode the spacetime variation of the mass and of the transition frequency of the atom A. In Eq.~\eqref{dilatonic_mass_charge}, we show in parenthesis the relation between our definition of the mass charge with the previously defined coupling function in \cite{damour:2010zr,hees18}. 
In a general frame (i.e where the gradient of the DM field is not neglected), using Eqs.~\eqref{scalar_sol_KG} and \eqref{dilatonic_coupling_func}, the rest mass and transition frequency of an atom A oscillate as 
\begin{subequations}\label{mass_freq_osc_dilaton}
\begin{align}
    m_A(t,\vec x)&=m^0_A\left(1+\frac{\sqrt{16 \pi G \rho_\mathrm{DM}}[Q^A_M]_d}{\omega_\phi c}\cos(\omega_\phi t - \vec k_\phi \cdot \vec x +\Phi)\right) \, , \\
    \omega_A(t,\vec x)&=\omega^0_A\left(1+\frac{\sqrt{16 \pi G \rho_\mathrm{DM}}[Q^A_\omega]_d}{\omega_\phi c}\cos(\omega_\phi t - \vec k_\phi \cdot \vec x +\Phi)\right) \, ,
\end{align}
\end{subequations}
where $m^0_A,\omega^0_A$ are respectively the unperturbed rest mass and transition frequency of A.
\begin{table}[t!]
\centering
\begin{tblr}{
    vlines,
    colspec={cccccccc}
}
\hline
Atomic & $Q_{M,e}$ & $Q_{M,m_e}$  & $Q_{M,\hat m}$ & $Q_{M,\delta m}$ & $Q_{\omega, e}$  & $Q_{\omega, \hat m}$ & $Q_{\omega, \delta m} $\\
& [$\times 10^{-3}$] & [$\times 10^{-3}$] & [$\times 10^{-3}$]& [$\times 10^{-3}$] &&[$\times 10^{-3}$]&[$\times 10^{-3}$]
 \\
\hline\hline
$^{195}$Pt \cite{Microscope18} & 4.278 & 0.220 & 85.25 & 0.340 & $-$ &$-$ &$-$\\
$^{48}$Ti \cite{Microscope18} & 2.282 & 0.253 & 82.58 & 0.138 & $-$& $-$ & $-$\\
$^{87}$Rb & 2.869 & 0.234 & 83.95 & 0.254 & 2.34 \cite{Guena12} & -67 \cite{Guena12} & -17.3 \cite{Guena12} \\
$^{85}$Rb & 2.961 & 0.239 & 83.98 & 0.220 & $-$ & $-$ & $-$ \\
$^{40}$Ca & 2.409 & 0.275 & 82.08 & 0 & 2.02 \cite{Angstmann04} & 0.0007 & 0 \\
$^{44}$Ca & 2.116 & 0.250 & 82.29 & 0.155 & 2.02 & 0.00065 & $\mathcal{O}(10^{-7})$ \\
$^{86}$Sr & 3.074 & 0.243 & 84.06 & 0.198 & 2.06 & 0.00030 & $\mathcal{O}(10^{-7})$\\
$^{87}$Sr & 3.027 & 0.240 & 84.05 & 0.215 & 2.06 \cite{Bloom14} & 0.00030 & $\mathcal{O}(10^{-7})$\\
$^{88}$Sr & 2.980 & 0.238 & 84.03 & 0.232 & 2.06 & 0.00030 & $\mathcal{O}(10^{-7})$ \\
$^{171}$Yb & 4.114 & 0.225 & 85.14 & 0.308 & 2.31 \cite{Hinkley13} & 0.00015 & $\mathcal{O}(10^{-7})$ \\
$^{176}$Yb & 3.957 & 0.219 & 85.05 & 0.348 & 2.31 & 0.00020 & $\mathcal{O}(10^{-7})$\\
$^{196}$Hg & 4.469 & 0.224 & 85.35 & 0.312 &  2.81 \cite{Angstmann04} & 0.00013 & $\mathcal{O}(10^{-7})$ \\
$^{202}$Hg & 4.291 & 0.218 & 85.25 & 0.353 &  2.81 & 0.00010 & $\mathcal{O}(10^{-7})$ \\
SiO$_2$ & 1.607 & 0.275 & 79.62 & 0.003 & $-$ & $-$ & $-$ \\
73\%Au-27\%Pt & 2.204 & 0.432 & 85.27 & 0.337 & $-$ & $-$ & $-$ \\ \hline
\end{tblr}
\caption{Dilatonic charges for some species of atoms. The transition is hyperfine for $^{87}$Rb and optical for the rest. The charges are derived from Eqs.~\eqref{dilatonic_mass_charge}, \eqref{dilatonic_hyp_freq_charge} and \eqref{dilatonic_freq_charge}. The value of $Q_{\omega,m_e}$ is universal for all atomic transitions, see Eqs.~\eqref{dilatonic_hyp_freq_charge} and \eqref{dilatonic_freq_charge}, so it is not provided. As we shall see in Chapter ~\ref{chap:AI} and Chapter ~\ref{chap:LISA_DM}, we will not consider the atomic transition of $^{195}$Pt, $^{48}$Ti, $^{87}$Rb, Au-Pt and SiO$_2$, reason why their frequency charges are not provided.}
\label{dilatonic_charge_table}
\end{table}

The rest mass depending differently on all the varying constants of Nature, we can define partial dilatonic mass charges $Q_{M,i}$ such that \cite{damour:2010zr}
\begin{subequations}
\begin{align}
    \label{partial_dil_mass_charge}
    [Q^\mathrm{atom}_M]_d &= \frac{1}{\sqrt{2}}\left(Q^\mathrm{atom}_{M,m_e} (d_{m_e}-d_g)+ Q^\mathrm{atom}_{M,e} d_e + Q^\mathrm{atom}_{M,\hat m}(d_{\hat m}-d_g)+Q^\mathrm{atom}_{M,\delta m}(d_{\delta m}-d_g)\right)\, ,
\end{align}
with \cite{damour:2010zr} 
\begin{align}
Q_{M,\hat{m}} &= 0.093 - \frac{0.036}{A^{1/3}} - 0.02\frac{(A-2Z)^2}{A^2} -1.4 \times 10^{-4} \frac{Z(Z-1)}{A^{4/3}} \, , \\
Q_{M,\delta m} &= 0.0017\frac{A-2Z}{A} \, , \\
Q_{M,m_e} &= 5.5\times 10^{-4} \frac{Z}{A}\, , \\
Q_{M,e} &= \left(-1.4 + 8.2 \frac{Z}{A} + 7.7\frac{Z(Z-1)}{A^{4/3}}\right)\times 10^{-4} \, ,
\label{dilatonic_partial_mass_charge}
\end{align}
\end{subequations}
where A and Z are respectively the mass and charge numbers of the atom and where the couplings $d_i$ are the ones defined in Eq.~\eqref{const_variations}.

The frequency charge depends on dilatonic charges $Q_{\omega,X}$ and on the dilaton/matter coupling coefficients $d_X$ (see e.g. \cite{hees18})
\begin{align}
\label{partial_dil_freq_charge}
    [Q^\mathrm{atom}_\omega]_d &= \frac{1}{\sqrt{2}}\left(Q^\mathrm{atom}_{\omega, m_e}\left(d_{m_e}-d_g\right)+Q^\mathrm{atom}_{\omega, e} d_e + Q^\mathrm{atom}_{\omega,\hat m}\left(d_{\hat m}-d_g\right)+Q^\mathrm{atom}_{\omega,\delta m}\left(d_{\delta m}-d_g\right)\right)\, .
\end{align}
In case of hyperfine transitions, we have $f^\mathrm{hyp}_\mathrm{atom} \propto \alpha^{k_\alpha}(m_e/m_p)(m_q/\Lambda_\mathrm{QCD})^{k_q}$ \cite{Flambaum06}, where $k_\alpha,k_q$ represent respectively the sensitivity coefficients of the hyperfine transition to the fine structure constant and to the ratio of the light quark masses $m_q$ to the QCD mass scale $\Lambda_\mathrm{QCD}$ ratio. Subsequently, the corresponding dilatonic frequency charges are 
\begin{subequations}\label{dilatonic_hyp_freq_charge}
\begin{align}
Q^\mathrm{hyp}_{\omega, m_e} &= 1 \, , \\
Q^\mathrm{hyp}_{\omega,e} &=  k_\alpha \, , \\
Q^\mathrm{hyp}_{\omega, \hat m} &=-0.048+k_q \, , \\
Q^\mathrm{hyp}_{\omega, \delta m} &= 0.0017+k_q \, ,
\end{align}
\end{subequations}
where we used the dependency of the proton mass to the light quark masses $\partial \ln m_p/\partial \ln \hat m = 0.048$, $\partial \ln m_p/\partial \ln \delta m = -0.0017$ \cite{damour:2010zr}.

The optical transition frequencies depend mainly on the electron mass and on the fine structure constant $f^\mathrm{opt}_\mathrm{atom} \propto (m_e m_N) \alpha^{2+\epsilon_\mathrm{atom}}/(m_e+m_N)$ \cite{Arvanitaki15,kim:2022aa} (where $m_N$ is the total nucleus mass and $\epsilon_\mathrm{atom}$ accounts for the relativistic correction that determines the frequency dependence on $\alpha$). Therefore, the frequency dilatonic charges write
\begin{subequations}\label{dilatonic_freq_charge}
\begin{align}
Q^\mathrm{opt}_{\omega, m_e} &= 1 \, , \\
Q^\mathrm{opt}_{\omega,e} &= 2+\epsilon_\mathrm{atom} \, , \\
Q^\mathrm{opt}_{\omega, \hat m} &\approx \frac{2.6 \times 10^{-5}}{A} \, , \\
Q^\mathrm{opt}_{\omega, \delta m} &\approx 9.0 \times 10^{-7}\frac{A-2Z}{A^2} \, ,
\end{align}
\end{subequations}
where for the third charge $Q^\mathrm{opt}_{\omega, \hat m}$, we assumed $m_n=m_p$.
In Table ~\ref{dilatonic_charge_table}, we show the dilatonic mass and frequency charges, for some atoms and corresponding atomic transitions (in particular, for each optical transition, we consider the intercombination line (ICL), i.e the transition $^1S_0 \rightarrow ^3P_1$, see Table ~\ref{tab:alk_isotope_freq}).
\begin{table}\centering
\begin{tblr}{ccc}
\hline
\hline
Species & Transition & Frequency (rad/s)\\
\hline
\hline
$^{88}$Sr & $5s^2\:^1S_0 \rightarrow 5s5p\:^3P_1$ & $2.73 \times 10^{15}$ \\
$^{40}$Ca & $4s4s\:^1S_0 \rightarrow 4s4p\:^3P_1$ & $2.87 \times 10^{15}$ \\
$^{171}$Yb& $6s^2\:^1S_0 \rightarrow 6s6p\:^3P_1$ & $3.39 \times 10^{15}$ \\
$^{196}$Hg& $6s^2\:^1S_0 \rightarrow 6s6p\:^3P_1$ & $7.45 \times 10^{15}$ \\
\hline
\end{tblr}
\caption{Some optical transition frequencies of interest.}
\label{tab:alk_isotope_freq}
\end{table}

\subsection{\label{sec:acceleration_dilaton_general}UFF violating acceleration of a test mass}

The point mass action \cite{damour:2010zr}
\begin{align}
    S_\mathrm{mat}[\Psi_i] = -\sum_A\int_A d\tau (m_A(t, \vec x)c^2+E^\mathrm{int}_A(t, \vec x)) \, ,
    \label{macro_action}
\end{align}
describes the motion of an ensemble of particles with rest mass energy $m_Ac^2$ and internal energy $E^\mathrm{int} = \hbar \omega_A$, with $\omega_A$ the transition frequency of the body under consideration\footnote{This contribution will only be relevant for atoms, see Chapter ~\ref{chap:AI}.}. $\Psi_i$ are the different SM fields, $d\tau$ is the proper time interval defined as $c^2d\tau^2=-g_{\alpha\beta} dx^\alpha dx^\beta$, where $g_{\mu\nu}$ is the spacetime metric. The body A has its own composition therefore its own coupling with the dilaton field. The equivalent Lagrangian description reads
\begin{subequations}
\begin{align}
    \mathcal{L}_A &= - \left(m_A(t, \vec x)c + \frac{\hbar \omega_A(t, \vec x)}{c}\right)\sqrt{-g_{\mu\nu}\frac{dx^\mu}{dt}\frac{dx^\nu}{dt}} \, ,
\end{align}
which, for non relativistic velocities and considering flat spacetime, i.e $\eta_{\mu\nu}=\mathrm{diag}(-1,1,1,1)$, can be approximated by
\begin{equation}
    \mathcal{L}_A = -\left(m_A(t, \vec x)c^2+\hbar \omega_A(t, \vec x)\right)\left(1-\frac{v_A^2}{2c^2}\right) \, ,
    \label{macro_lagrangian}
\end{equation}
\end{subequations}
to first order in $(v_A/c)^2$ (where $v_A$ is the coordinate velocity of the atom).
Using Eq.~\eqref{mass_freq_osc_dilaton}, a simple Euler-Lagrange derivation of Eq.~\eqref{macro_lagrangian} gives
\begin{align}
    \vec a_A(t, \vec x) &= \left[\omega_\phi \vec v_A-\vec k_\phi c^2\right]\frac{\sqrt{16 \pi G \rho_\mathrm{DM}}}{\omega_\phi c}\left([Q^A_M]_d+[Q^A_\omega]_d\frac{\hbar \omega^0_A}{m^0_Ac^2}\right)\sin(\omega_\phi t - \vec k_\phi \cdot \vec x +\Phi) \, ,
    \label{EP_viol_acc_dil}
\end{align} 
at lowest order in $v_A/c$, where the superscript $0$ indicates the bare value of the quantity, at zeroth order in the field $\phi$. In addition, we considered the lowest order expansion in factors of $\hbar \omega^0_A/m^0_Ac^2$, as in general $m^0_Ac^2 \gg \hbar \omega^0_A$. As we have seen previously, the various $[Q^A_M]_d,[Q^A_\omega]_d$ charges are atom-dependent, implying that the acceleration Eq.~\eqref{EP_viol_acc_dil} leads to a violation of the UFF.

\section{\label{axion_pheno}Pseudo-scalar field}

\subsection{Theoretical motivations}

The QCD Lagrangian in SM reads
\begin{align}
    \mathcal{L}_\mathrm{QCD} &= -\frac{1}{4}G^a_{\mu\nu}G^{a,\mu\nu} + \bar \Psi_q\left(i\slashed D-m_qe^{i\theta_q}\right)\Psi_q \, ,
\end{align}
where the first term represents the kinetic energy of the non abelian gluon gauge field, through the strength tensors $G^a_{\mu\nu}$ with $\sqrt{J/m^3}$ units, the second term represents the interaction between gluons and quarks $\Psi_q$ ($\slashed D = \gamma^\mu D_\mu$, where $\gamma^\mu$ are the dimensionless gamma matrices\footnote{The gamma matrices are a set of $4\times 4$ matrices which allow one to construct Lorentz scalar with fermionic spinors. These are the 4-dimensional generalization of Pauli matrices.})  and the third term is the quarks mass term. One could add to this Lagrangian a total derivative term (which would not change the fields equations of motion) of the form 
\begin{align}\label{CP_QCD_lagrangian}
   \mathcal{L}_\theta &= \theta \frac{g^2_3}{32\pi^2} G^a_{\mu\nu}\tilde G^{a,\mu\nu}\, ,
\end{align}
where $\tilde G^{a,\mu\nu}=\epsilon^{\mu\nu\rho\sigma}G^a_{\rho\sigma}/2$ is the dual gluon field strength tensor, with superscript $a$ running over the 8 gluon fields. Despite leaving the equations of motion invariant, the Lagrangian Eq.~\eqref{CP_QCD_lagrangian} breaks CP (Charge-Parity) symmetry if the $\theta$ parameter, being an angle, is different from 0 or $\pi$ \cite{Sikivie21}. To test CP symmetry of QCD, one can measure the neutron electric dipole moment $d_n$ whose best upper limit constraint is $|d_n| < 3 \times 10^{-26}$ e.cm (90\% confidence level) \cite{Pendlebury15}, which translates into an upper limit of the $\theta$ parameter to be $\theta < 10^{-10}$ \cite{Sikivie21}. This extremely small value is puzzling because we should expect $\theta \sim \mathcal{O}(1)$\footnote{In reality, $\theta = \theta_\mathrm
{QCD} + \theta_q$ where $\theta_\mathrm
{QCD}$ is a contribution from QCD and where $\theta_q$ is a contribution from electroweak sector (coming from the quark masses), which are two very different sectors, therefore one should not expect those two contributions to cancel each other exactly.}, and makes the so-called strong CP problem. 
In the 1970s, Peccei and Quinn \cite{Peccei_Quinn77} proposed a solution to solve this problem : promote the previously fixed $\theta$ parameter to a dynamical massive field with a potential with a zero minimum, such that $\theta$ relaxes naturally to 0 \cite{Weinberg78, Wilczek78, Vafa84}. 
More precisely, a new $U(1)_\mathrm{PQ}$ symmetry is introduced which is spontaneously broken at a given energy scale $f_a$. The QCD axion is the Goldstone boson of this broken symmetry, which acquires a mass $m_a$ via its interaction with pions. See \cite{DiLuzio20} for a complete review of the Peccei-Quinn mechanism. 
One can show that the QCD axion mass is inversely proportional to $f_a$, and that it also couples to the photon, nucleon, electron fields and nucleon electric dipole moment with respective couplings $g_{a\gamma}, g_{aN}/m_Nc^2, g_{ae}/m_ec^2,g_d$ (with $m_N,m_e$ the nucleon and electron mass). All these couplings are proportional to $1/f_a$ such that the first three have units of 1/energy and the last coupling $g_d$ has units of $C.m/J$ (because it is also proportional to $d_n$)\cite{Marsh16,Graham13_SN}. Therefore, there is only one free parameter of the theory, which is the QCD axion mass. One can define a more generic pseudo-scalar field, the Axion-Like-Particle (ALP) $a$ which couples to the same fields and has the same mass as the QCD axion, but where the relation of proportionality between all the parameters disappears, i.e all couplings and mass are free parameters of the theory. Therefore, ALP does not solve the strong CP problem, but is still a DM candidate, as shown in Section ~\ref{Osc_field_DM}. The dimensionless ALP field is defined as
\begin{align}
    \theta = \sqrt{\frac{\hbar c^5}{8\pi G}}\frac{a}{f_a} \, ,
    \label{theta_axion}
\end{align}
where $\sqrt{\hbar c^5/8\pi G} \sim 2\times 10^{18} \: \mathrm{GeV} \equiv E_P$ is the reduced Planck energy (which is usually defined in terms of the inverse of the $\kappa$ parameter, reason for the $8\pi$ factor \cite{damour:2010zr}) and its full interaction Lagrangian can be written as \cite{Marsh16}
\begin{align}\label{axion_full_int_lagrangian}
    \mathcal{L}_\mathrm{int} &= E_P\left(\frac{g^2_3}{32\pi^2}\frac{a}{f_a} G^a_{\mu\nu}\tilde G^{a,\mu\nu}- \frac{g_{a\gamma}}{4\mu_0}aF_{\mu\nu}\tilde F^{\mu\nu}+\frac{g_{aN}\hbar c}{2m_N c^2}\partial_\mu a \bar \Psi_N\gamma^\mu \gamma_5 \Psi_N +\right.\,\nonumber \\
    &\left.\frac{g_{ae}\hbar c}{2m_e c^2}\partial_\mu a \bar \Psi_e\gamma^\mu \gamma_5 \Psi_e+\frac{c}{2i}g_d a\bar \Psi_N \sigma_{\mu\nu}\gamma_5 \Psi_N F^{\mu\nu}\right)\, ,
\end{align}
where $\Psi_N,\Psi_e$ are respectively the nucleon and electron fermion fields, where $\gamma_5=i\gamma^0\gamma^1\gamma^2\gamma^3$ the fifth gamma matrix and $\sigma^{\mu\nu} = [\gamma^{\mu},\gamma^\nu]/2$. 

\subsection{\label{axion_QED}Axion electrodynamics}
Following Eqs.~\eqref{axion_field_action} and \eqref{axion_full_int_lagrangian} in flat spacetime ($\sqrt{-g}=1)$ and using space positive metric $g_{\mu\nu}=\mathrm{det}(-1,1,1,1)$, we write the Lagrangian describing the interaction between the classical dimensionless axion pseudo scalar field $a$ with mass $m_a$ and EM through the coupling $g_{a\gamma}$ \cite{Sikivie21}
\begin{align}
\mathcal{L} &= -\frac{1}{4\mu_0}F_{\mu\nu} F^{\mu\nu} + j_\mu A^\mu  - \frac{1}{2\kappa}\partial_\mu a \partial^\mu a- \frac{1}{2\kappa}\frac{m^2_ac^2}{\hbar^2}a^2-E_P\frac{g_{a\gamma}}{4\mu_0}aF_{\mu\nu}\tilde F^{\mu\nu}\, , 
\end{align}
where the EM gauge field $A^\alpha$ has usual units of V.s/m and where $\tilde F^{\mu\nu}$ is the dual electromagnetic strength tensor defined as
\begin{subequations}
\begin{align}
    \tilde F^{\mu\nu} &= \frac{1}{2}\epsilon^{\mu\nu\rho\sigma}F_{\rho\sigma} \equiv \epsilon^{\mu\nu\rho\sigma} \partial_\rho A_\sigma \, ,
\end{align}
with $\epsilon^{\mu\nu\rho\sigma}$ the antisymmetric 4-dimensional Levi-Civita tensor, which has the following properties 
\begin{align}
    \epsilon^{\mu\nu\rho\sigma} &= g^{\mu\alpha}g^{\nu\beta}g^{\rho\delta}g^{\sigma\delta}\epsilon_{\alpha\beta\delta\gamma}\, , \\
    \epsilon_{0ijk} &= \epsilon_{ijk} \, ,
\end{align}
\end{subequations}
where $\epsilon_{ijk}$ is the usual 3-dimensional Levi-Civita tensor. Using these properties, one can show that 
\begin{align}
    F_{\mu\nu}\tilde F^{\mu\nu} &= 4\frac{\vec E \cdot \vec B}{c}\, ,
\end{align}
where we defined $\partial_\mu = (\partial_t/c, \vec \nabla)$, $j^\mu = (\rho c, \vec j)$ and $A^\mu=(\phi/c, \vec A)$ with $\phi, \vec A$ respectively the scalar and vector potentials and introduce the electric and magnetic fields defined as 
\begin{subequations}\label{E_B_potentials}
    \begin{align}
        \vec E &= - \vec \nabla \phi - \partial_t \vec A \, , \\
        \vec B &= \vec \nabla \times \vec A \, .
    \end{align}
\end{subequations}
The equation of motion of the axion and photon fields are derived from the usual Euler-Lagrange equations and reads \cite{Sikivie83, Sikivie84}
\begin{subequations}
\begin{align}\label{EoM_axion_photon}
    \left(\square -\frac{m^2_ac^2}{\hbar^2}\right)a = \frac{\kappa E_P}{\mu_0 c}g_{a\gamma} \vec E \cdot \vec B \, ,
\end{align}
for the axion field and 
\begin{align}
\label{EOM_EM}
\square{A^\nu} - \partial^\nu \partial_\mu A^\mu+ E_Pg_{a\gamma}(\partial_\mu a)\epsilon^{\mu\nu\rho\sigma} \partial_\rho A_\sigma &= -\mu_0 j^\nu \, ,
\end{align}
\end{subequations}
for the photon field.
We can treat separately the temporal and spatial part of Eq.~\eqref{EOM_EM}, i.e 
\begin{subequations}
\begin{align}
\square{\phi} + \partial_t \partial_\mu A^\mu + c E_P g_{a\gamma}\epsilon_{i j k} (\partial_i a)\partial_j A_k &= -\mu_0 \rho c^2 \label{EOM_A0}\, , \\
\square{A^i} - \partial^i \partial_\mu A^\mu +E_P g_{a\gamma}\epsilon^{\mu i\rho\sigma} (\partial_\mu a)\partial_\rho A_\sigma &= - \mu_0 j^i \label{EOM_Ai}\, .
\end{align}
\end{subequations}
Using the temporal equation Eq.~\eqref{EOM_A0}, the definition of the electric and magnetic fields Eq.~\eqref{E_B_potentials} and the zero divergence of the magnetic field, we obtain 
\begin{align}
    \vec \nabla \cdot \left(\vec E - c E_P g_{a\gamma}a \vec B\right) &= \frac{\rho}{\epsilon_0} \, ,
    \label{Gauss_law_modif}
\end{align}
which makes the modified Gauss's law.
From the spatial equation of motion, Eq.~\eqref{EOM_Ai}, we obtain
\begin{align}
\vec{\nabla} \times \vec{B} = \mu_0\left(\vec{j} +\epsilon_0\frac{\partial \vec E}{\partial t}\right)- \frac{E_P}{c} g_{a\gamma}(\dot{a}\vec{B} + \vec\nabla a \times \vec{E})\, ,
\label{Ampere_law_modif}
\end{align}
where the dot represents derivative with respect to time and which corresponds to the modified Amp\`ere's law.

Eqs.\eqref{Gauss_law_modif} and \eqref{Ampere_law_modif} together with the two unchanged Maxwell's equations form what is more commonly known as axion electrodynamics \cite{Sikivie83, Sikivie84}
\begin{subequations}\label{eq:axion_electrodynamics}
\begin{align}
\vec \nabla \cdot \left(\vec E - c E_P g_{a\gamma}a \vec B\right) &= \frac{\rho}{\epsilon_0} \, ,\\
\vec{\nabla }\times \vec{E} + \frac{\partial \vec B}{\partial t} &= 0 \, ,\\
\mu_0\left(\vec{j} +\epsilon_0\frac{\partial \vec E}{\partial t}\right)- \frac{E_P}{c} g_{a\gamma}(\dot{a}\vec{B} + \vec\nabla a \times \vec{E}) &= \vec{\nabla} \times \vec{B} \, ,\\
\vec{\nabla}.\vec{B} &= 0 \, .
\end{align}
\end{subequations}
The various new terms of Eq.~\eqref{eq:axion_electrodynamics} break the electric-magnetic duality which is present in the usual Maxwell's equations : in vacuum, these equations are not invariant under the change $(\vec E, \vec B) \rightarrow (c \vec B, -\vec E/c)$.

\subsection{\label{axion_photon_coupling}Vacuum birefringence and dichroism}

We now focus on the spatial equation of motion of the EM field Eq.~\eqref{EOM_Ai}. We assume propagation in vacuum $\vec j=0$, both temporal and Coulomb gauges for the EM field $A_0=0$ and $\vec{\nabla} \cdot \vec{A} = 0$\footnote{Out of the four degrees of freedom encoded in the 4-potential $A^\mu$, only 2, representing the photon polarization states, are physical, reason why we can fix two gauges.} and the axion field is spatially homogeneous at the scale of the experiment\footnote{As shown in Eq.~\eqref{eq:mass_size_relation_lab_frame}, the constraint on the axion mass depends on the size of experiment considered.} such that the form Eq.~\eqref{field_lab_frame} is valid. Then, Eq.~\eqref{EOM_Ai} becomes \cite{Obata18}
\begin{align}
\ddot{A_i} - c^2 \nabla^2 A_i + cE_P g_{a\gamma}\dot{a}\epsilon_{ijk}\partial_j A_k = 0 \, .
\end{align}

We can decompose $A_i$ into the left and right circular polarization modes of the photon with wave number $\vec k$ as
\begin{align}
A_i(t,x) = \int \frac{d^3k}{(2\pi)^3}\left(A^+_k(t) e^+_i + A^-_k(t) e^-_i\right)e^{i\vec{k}.\vec{x}}\, .
\end{align}
Without loss of generality, if we assume the propagation direction to be along the $\hat z$ direction, $\vec k = (0,0,k)$, the normalized right/left helicities of the photon can be expressed as $e^+ = e^R = (1,-i,0)/\sqrt{2}$ and $e^-=e^L=(1,i,0)/\sqrt{2}$ such that 
\begin{align}
    \epsilon_{ijk}k_j e^\pm_k = \pm i ke^\pm_i  \, .
\end{align}
Then, we can write two separate equations of motion for both polarization modes \cite{Obata18} 
\begin{subequations}\label{eq:EoM_left_right_polar}
\begin{align}
\ddot{A^+_i} + k^2 c^2\left(1 + \frac{E_P}{c^2} \frac{\sqrt{16\pi G \rho_\mathrm{DM}}g_{a\gamma}}{k} \sin(\omega_a t +\Phi)\right)A^+_i= 0 \, \\
\ddot{A^-_i} + k^2 c^2 \left(1 - \frac{E_P}{c^2} \frac{\sqrt{16\pi G \rho_\mathrm{DM}}g_{a\gamma}}{k} \sin(\omega_a t +\Phi)\right)A^-_i= 0 \, ,
\end{align}
\end{subequations}
where we used Eq.~\eqref{energy_density_scalar_DM}. Considering the plane wave ansatz $A^\pm_i(t)= \zeta^\pm_i \exp(-i(\omega_\pm t - \vec k \cdot \vec x))$, Eq.~\eqref{eq:EoM_left_right_polar} leads to 
\begin{align}\label{eq:dispersion_left_right_polar}
    \omega_\pm &= kc\sqrt{1 \pm \frac{E_P}{c^2} \frac{\sqrt{16\pi G \rho_\mathrm{DM}}g_{a\gamma}}{k} \sin(\omega_a t +\Phi)} \, ,
\end{align}
or in other words, the dispersion relation of the right/left circular polarizations is modified such that they travel with different phase velocities $c_\pm = \omega_\pm/k$, respectively 
\begin{align}
\label{delta_c_axion_photon}
c_\pm = c\sqrt{1 \pm \frac{E_P}{c^2} \frac{\sqrt{16\pi G \rho_\mathrm{DM}}g_{a\gamma}}{k} \sin(\omega_a t +\Phi)} \, .
\end{align}
Note that Eq.~\eqref{eq:dispersion_left_right_polar} leads also to a modification of the group velocity of the left and right polarization of light 
\begin{align}
v_g &= \frac{\partial \omega_\pm}{\partial k} \approx c\left(1+\frac{16\pi G \rho_\mathrm{DM}E^2_P g^2_{a\gamma}}{8k^2c^4}\sin^2\left(\omega_a t+\Phi\right)\right)+\mathcal{O}\left(\left(\frac{E_P}{c^2} \frac{\sqrt{16\pi G \rho_\mathrm{DM}}g_{a\gamma}}{k}\right)^3\right) \, .
\end{align}
This modification of the group velocity does not break special relativity since we now assume that light propagates inside a medium full of axions, which is not pure vacuum.

In other words, vacuum becomes birefringent in presence of an axion background. Experiments such as \textit{DANCE} \cite{Michimura20}, \textit{LIDA} \cite{Heinze24} or \textit{ADBC} \cite{Pandey24} attempt to detect axion-photon coupling through this effect.

We now study dichroism effects from axion-photon coupling. To do so, we focus specifically on the axion-photon interaction Lagrangian, written in terms of electric and magnetic fields i.e
\begin{align}
    \mathcal{L}_{a\gamma} &= -E_P\frac{g_{a\gamma}}{\mu_0 c}a\vec E\cdot \vec B \, .
\end{align}
In a medium of permittivity $\epsilon$ and permeability $\mu$ and where external electric and magnetic fields $\vec E_0$ and $\vec B_0$ are applied, the modified Maxwell's equations Eqs.~\eqref{Gauss_law_modif} and \eqref{Ampere_law_modif} show that the axion field is a source of electric charge $\rho_a$ and current density $\vec j_a$ \cite{Sikivie21}
\begin{subequations}
    \begin{align}
        \rho_a &= \epsilon c E_P g_{a\gamma}\vec \nabla a \cdot \vec B_0 \,\\
        \vec j_a &= \frac{\mu E_P}{c}g_{a\gamma}\left(\dot{a}\vec B_0 + \vec \nabla a \times \vec E_0\right)\, ,
    \end{align}
\end{subequations}
implying a conversion between axions and photons.

As derived in \cite{Sikivie21}, if we assume light propagating in the $z$ direction whose polarization lies in the $x-y$ plane and a static and homogeneous magnetic field polarized along the $x$ axis $\vec B_0 \equiv B_0 \hat e_x$, with no electric field, the photon-axion conversion will happen in the same direction as the B polarization direction. This means that the polarization component of light along the $x$-direction is depleted, implying a rotation of the plane of polarization. This is what is called dichroism. The conversion probability is given by \cite{Sikivie21}
\begin{subequations}
\begin{align}
    P(\gamma \rightarrow a, L) &= \sqrt{\frac{\mu}{\epsilon}}\sqrt{\frac{\epsilon_0}{\mu_0}}\frac{\hbar}{\mu_0}\frac{g^2_{a\gamma}B^2_0}{\beta_a q^2}\sin^2\left(\frac{qL}{2}\right) \, ,
\end{align}
where $L$ is the travelling distance, q is the difference in wavevector between axions and photons, i.e $q=|\vec k_a - \vec k_\gamma|$ and $\beta_a = |\vec k_a|/\omega_a \equiv v_a/c^2$ with $v_a$ the speed of incident axions. Note that, by energy conservation, $\omega_a = \omega_\gamma$, since the static magnetic field does not transfer energy.

In addition, the relative phase between the polarizations parallel and perpendicular to the magnetic field changes, and light acquires ellipticity \cite{Sikivie21}. If we assume plane wave solutions for axions and photons, one can show that the $\gamma \rightarrow a$ conversion leads to a modification of the dispersion relation of photons of $x$ polarization, which depends on the amplitude of the magnetic field in this direction. This is equivalent to a relative phase added to the $x$-polarization compared to the $y$-one that reads \cite{Sikivie21}
\begin{align}\label{eq:dichroism_phase_general}
    \phi(L) &=\frac{\hbar c}{\mu_0}\frac{g^2_{a\gamma}B^2_0}{4 q^2}\left(qL-\sin(qL)\right) \, .
\end{align}
\end{subequations}

\subsection{\label{axion_gluon_pheno}UFF violation through the coupling with gluons}

We now consider the coupling between the ALP and gluons which reads, from Eq.~\eqref{axion_full_int_lagrangian}, 
\begin{equation}\label{eq:int_axion_gluon}
	\mathcal L_\mathrm{int} = E_P\frac{g^2_3}{32 \pi^2} \frac{a}{f_a} G^a_{\mu\nu} \tilde G^{a,\mu\nu} \equiv \frac{g^2_3}{32 \pi^2} \theta G^a_{\mu\nu} \tilde G^{a,\mu\nu}\, .
\end{equation}
In \cite{kim:2022aa}, it is shown that the interaction Lagrangian from Eq.~\eqref{eq:int_axion_gluon} induces a dependency of the mass of pions to the axion field, which implies a dependency of the mass of nucleons and atomic binding energy on the axion field. More precisely,
\begin{equation}
	\frac{\partial \ln m_N}{\partial \left(\theta^2\right)} = \frac{\partial \ln m_N}{\partial \ln \left(m_\pi^2\right)} \frac{\partial \ln \left(m_\pi^2\right)}{\partial \left(\theta^2\right)}\, ,
 \label{mass_dep_axions}
\end{equation}
where $m_N$ is the rest mass of the nucleon N. As a consequence, the mass of any atom will also depend on the axion field and its coupling strength with gluons. In this sense, we can define the dimensionless axionic mass and frequency charges of an atom A as 
\begin{subequations}
\begin{align}
    [Q^\mathrm{atom}_M]_a &= \frac{\partial \ln m_\mathrm{atom}}{\partial \left(\theta^2\right)} \, \\
    [Q^\mathrm{atom}_\omega]_a &= \frac{\partial \ln \omega_\mathrm{atom}}{\partial \left(\theta^2\right)} \, ,
    \label{axionic_charge}
\end{align}
\end{subequations}
in a similar way as in Eq.~\eqref{dilatonic_coupling_func}. Therefore, a similar oscillation on both rest mass and transition frequency of an atom A arise
\begin{subequations}
\begin{align}
m_A(t,\vec x)&=m^0_A\left(1+\frac{8 \pi G \rho_\mathrm{DM} E^2_P [Q^A_M]_a}{f^2_a \omega^2_a c^2}\cos(2(\omega_a t - \vec k_a \cdot \vec x +\Phi))\right) \, , \\
\omega_A(t,\vec x)&=\omega^0_A\left(1+\frac{8 \pi G \rho_\mathrm{DM} E^2_P [Q^A_\omega]_a}{f^2_a \omega^2_a c^2}\cos(2(\omega_a t - \vec k_a \cdot \vec x +\Phi))\right) \, .
\end{align}
\end{subequations}
We made a redefinition of the unperturbed mass as $m^0_A \rightarrow m^0_A(1+8 \pi G \rho_\mathrm{DM} E^2_P [Q^A_\omega]_a/f^2_a \omega^2_a c^2)$, and similarly for the frequency. Then, the axion-gluon coupling would induce a UFF violating acceleration on the atom A of the form 
\begin{align}\label{EP_viol_acc_axion}
  \vec a_A(t, \vec x) = \left[\omega_a \vec v_A - \vec k_a c^2\right]\frac{16\pi G \rho_\mathrm{DM}E^2_P}{f^2_a \omega^2_a c^2}\left([Q^A_M]_a + [Q^A_\omega]_a\frac{\hbar \omega^0_A}{m^0_Ac^2}\right)\sin\left(2\omega_a t - 2\vec k_a \cdot \vec x + 2\Phi\right)  \, ,
\end{align}
where we did not include contributions from the axion spin-induced fifth force (see e.g. \cite{Capolupo21}), because we are only interested in the oscillating solution.
Now, we wish to get an analytical expression for both axionic charges.

The rest mass of an atom, with charge number Z and neutron number N, can be parameterized as 
\begin{align}
m_{\mathrm{atom}}c^2 &= m_{\mathrm{const.}}c^2+E_\mathrm{bind} \equiv Z(m_p+m_e)c^2+Nm_n c^2+E_\mathrm{bind}\, ,
\end{align}
where $m_p, m_e, m_n$ are respectively the rest masses of the proton, the electron and the neutron, and where $m_{\mathrm{const.}}, E_\mathrm{bind}$ represent respectively the rest mass of the particle constituents of the atom (proton, neutron, electron) and the nuclear binding energy.

\subsubsection{Nucleons rest mass}

In \cite{kim:2022aa}, it is shown the pion mass is $\theta$ dependent and influence the nucleon mass $m_\mathrm{N}$ through  
\begin{subequations}\label{pions_axions_dep}
	\begin{align}
		\frac{\partial \ln m_N}{\partial \ln m_\pi^2} &\approx 0.06 \, ,\\
		 \frac{\partial \ln m_\pi^2}{\partial \left(\theta^2\right)} & = -\frac{m_u m_d}{2(m_u+m_d)^2}=-0.109 \, \\
   \Rightarrow \frac{\partial \ln m_N}{\partial \left(\theta^2\right)} &\approx -0.065 \, ,
	\end{align}
\end{subequations}
such that, for an atom made of (N+Z) nucleons, the contribution of the nucleons rest mass to the atom rest mass to the axionic charge is given by 
\begin{equation}
	[Q^\mathrm{atom}_M]_a\Big|_\mathrm{const.} = \frac{\partial \ln m_N}{\partial \left(\theta^2\right)} \approx -0.065 \, ,
\label{axionic_charge_rest}
\end{equation}
meaning that it is independent of the number of nucleons inside the atom, or in other words of the atomic species.

\subsubsection{Binding energy of the nuclei}

Let us now focus on the contribution of the binding energy of the nuclei to the axionic mass charge and show that it is composition dependent.

As computed in \cite{damour:2010zr}, the binding energy of the nuclei depends to first order on the mass of the pions and therefore, following Eq.~\eqref{pions_axions_dep}, on the ALP.  We will now use the results from Section IV of \cite{damour:2010zr} to infer the analytical expression of the dependency of the binding energy to the ALP and to the mass number $A$ and the charge number $Z$. 

Four different interactions contribute to the binding energy \cite{damour:2010zr}: the central force $E_{\mathrm{central}}$ coming from the isospin symmetric central nuclear force, the asymmetry energy $E_{\mathrm{asym}}$, i.e the residual energy from the asymmetry between neutrons and protons inside the nucleus, the Coulomb force $E_{\mathrm{Coulomb}}$ depending on how tightly the nucleons are packed together and the pairing energy $E_{\mathrm{pairing}}$ leading to
\begin{align}
E_\mathrm{bind} &= E_{\mathrm{central}}+E_{\mathrm{asym}}\frac{(A-2Z)^2}{A}+E_{\mathrm{Coulomb}}\frac{Z(Z-1)}{A^{1/3}} -\delta \frac{E_{\mathrm{pairing}}}{A^{1/2}} \, ,
\end{align} 
where $\delta = \pm 1$ respectively for even-even and odd-odd number of protons-neutrons in the nuclei, and $\delta = 0$ otherwise.

\paragraph{\textbf{Central force}}\mbox{}
This interaction comes from the isospin symmetric central nuclear force, which is the dominant contribution in the binding of heavy nuclei \cite{damour:2010zr}. 
The dominant interactions are an attractive scalar $\eta_S$ and a repulsive vector $\eta_V$, and \cite{damour:2010zr} shows that the former is more sensitive to the pion mass, implying in our case
\begin{subequations}
\begin{align}
\frac{\partial E_\mathrm{Central}}{\partial (\theta^2)}\approx \frac{\partial E_\mathrm{Central}}{\partial \eta_S} \frac{\partial \eta_S}{\partial \ln m^2_\pi}\frac{\partial \ln m^2_\pi}{\partial (\theta^2)} \, ,
\end{align}
with \cite{Introductory_Nucl_Phys, damour:2010zr}
    \begin{align}
        E_\mathrm{Central} &\approx -(120A-97A^{2/3})\eta_S\, ,\\
        \frac{\partial \eta_S}{\partial \ln m^2_\pi} &= -0.35 \mathrm{ \ MeV} \, .
    \end{align}
\end{subequations}

\paragraph{\textbf{Asymmetry energy}}\mbox{}
The residual energy from the asymmetry between neutrons and protons inside the nucleus contains two components : 1) from the Pauli exclusion principle, requiring that when there are more neutrons than protons, the extra neutrons are in the higher energy states than the protons; and 2) from the nuclear force, which is more attractive for a neutron and a proton than with a pair of neutrons or a pair of protons. 
The asymmetry energy depends mainly on the scalar coupling strength between the nucleons $G_S$ \cite{damour:2010zr} implying 
\begin{subequations}
\begin{align}
\frac{\partial  E_{\mathrm{Asym}}}{\partial (\theta^2)} =\frac{\partial E_{\mathrm{Asym}}}{\partial G_S}\frac{\partial G_S}{\partial \ln m^2_\pi}\frac{\partial \ln m^2_\pi}{\partial (\theta^2)} \, ,
\end{align}
with \cite{damour:2010zr, Introductory_Nucl_Phys}
\begin{align}
    \frac{\partial E_{\mathrm{Asym}}}{\partial G_S}\frac{\partial G_S}{\partial \ln m^2_\pi} &= -19 \mathrm{ \ MeV} \, .
\end{align}
\end{subequations}

\paragraph{\textbf{Coulomb force}}\mbox{}
The Coulomb energy has a dependency on the strong interaction coupling terms since it depends on how tightly the nucleons are packed together. It can be shown that this contribution depends on the same scalar coupling as the asymmetry energy, hence
\begin{subequations}
\begin{align}
\frac{\partial E_{\mathrm{Coulomb}}}{\partial (\theta^2)} =\frac{\partial E_{\mathrm{Coulomb}}}{\partial G_S}\frac{\partial G_S}{\partial \ln m^2_\pi}\frac{\partial \ln m^2_\pi}{\partial (\theta^2)} \, ,
\end{align}
with \cite{damour:2010zr,Introductory_Nucl_Phys}
\begin{align}
    \frac{\partial E_{\mathrm{Coulomb}}}{\partial G_S}\frac{\partial G_S}{\partial \ln m^2_\pi} &=-0.13 \mathrm{ \ MeV} \, .
\end{align}
\end{subequations}

\paragraph{\textbf{Pairing energy}}\mbox{}
The pairing interaction contributes to binding energy and its numerical value is \cite{Introductory_Nucl_Phys}
\begin{equation}
    E_{\mathrm{Pairing}} = 12 \mathrm{ \ MeV} \, .
\end{equation}
\cite{damour:2010zr} shows that this contribution is subdominant for all atoms, due to its dependency to mass number A, compared to the other interactions, hence we will not consider the pairing energy in the calculation of the axionic charges of atoms.

Adding all contributions together, we get the analytic expression for the axionic mass charge 
\begin{align}
&[Q^\mathrm{atom}_M]_a=\frac{1}{m_\mathrm{atom}}\frac{\partial m_\mathrm{atom}}{\partial (\theta^2)}\,\nonumber\\
    &= \frac{1}{m_\mathrm{atom}c^2}\left(\frac{\partial (m_\mathrm{rest \ mass} c^2)}{\partial (\theta^2)}+\frac{\partial E_\mathrm{bind}}{\partial (\theta^2)}\right)\,\nonumber\\
&\approx -0.065 \frac{m_\mathrm{rest \ mass}}{m_\mathrm{atom}}+\frac{1 \mathrm{\: MeV}}{m_\mathrm{atom}c^2}\left(-4.578 A +3.701 A^{2/3} + 2.071 \frac{(A-2Z)^2}{A} +0.014 \frac{Z(Z-1)}{A^{1/3}}\right)\,\nonumber\\
&\approx -0.065+F_A\Big(-4.92+\frac{3.98}{A^{1/3}}+2.22\frac{(A-2Z)^2}{A^2}+1.50\frac{Z(Z-1)}{A^{4/3}}\times 10^{-2}\Big)\times 10^{-3} \label{axionic_mass_charge}\, .
\end{align}
At the last line, we considered $m_\mathrm{rest \ mass}/m_\mathrm{atom} \approx 1 + \langle E_\mathrm{bind}\rangle/m_\mathrm{amu} c^2$ with $\langle E_\mathrm{bind}\rangle \sim 8$ MeV, the average binding energy per nucleon and $m_\mathrm{amu}$ the atomic mass unit \cite{damour:2010zr}. Also, we factorized $F_A=Am_\mathrm{amu}/m_\mathrm{atom}$ which is of order unity at first order for all species of atoms (the relative error is $\mathcal{O}(10^{-3})$)\cite{damour:2010zr}.

Now, we focus on the axionic frequency charge.
We will first consider hyperfine atomic transitions ($\omega^\mathrm{hyp}_\mathrm{atom}$) which are impacted by the axion-gluon coupling from Eq.~\eqref{eq:int_axion_gluon} as \cite{kim:2022aa}
\begin{align}\label{hyp_freq_dep_g}
    \frac{\partial \ln \omega^\mathrm{hyp}_\mathrm{atom}}{\partial (\theta)^2} &= \left(\frac{\partial \ln g}{\partial \ln m^2_\pi}-\frac{\partial \ln m_p}{\partial \ln m^2_\pi}\right)\frac{\partial \ln m^2_\pi}{\partial (\theta)^2} \, ,
\end{align}
where $g$ is the nucleon g-factor whose dependence on pion mass is given by \cite{kim:2022aa}
\begin{subequations}\label{g_dep_pion}
    \begin{align}
        \frac{\partial \ln g}{\partial \ln m^2_\pi} &= K_n \frac{\partial \ln g_n}{\partial \ln m^2_\pi} + K_p \frac{\partial \ln g_p}{\partial \ln m^2_\pi} -0.17 K_b \, ,
    \end{align}
with 
    \begin{align}
        \frac{\partial \ln g_n}{\partial \ln m^2_\pi} &\approx -0.25\, , \\
        \frac{\partial \ln g_p}{\partial \ln m^2_\pi} &\approx -0.17 \, ,
    \end{align}
\end{subequations}
where $g_{n}$ ($g_{p}$) respectively the neutron (proton) gyromagnetic factors and $K_n$, $K_p$ and $K_b$, coefficients computed from chiral perturbation theory (nuclear shell model). These coefficients are not measurable, but they are related to observable parameters, namely, the sensitivity coefficient of the nuclear magnetic moment to light quarks masses $\kappa_q$, to strange quarks masses $\kappa_s$ and to light quarks masses over QCD energy scale $\kappa$ through \cite{Flambaum06}
\begin{subequations}\label{kappa_K_relation}
    \begin{align}
        \kappa_q &= -0.118 K_n -0.087 K_p \, , \\
        \kappa_s &= 0.0013 K_n -0.013 K_p \, , \\
        \kappa &=-0.12 K_n -0.10 K_p -0.11 K_b \, .
    \end{align}
\end{subequations}
Using Eqs.~\eqref{pions_axions_dep}, \eqref{hyp_freq_dep_g}, \eqref{g_dep_pion} and \eqref{kappa_K_relation}, we define the dimensionless axionic frequency charge of the atom as
\begin{align}
    [Q^\mathrm{atom}_\omega]_a  &\approx \left(-16.8 \kappa -5.69 \kappa_q +25.1\kappa_s + 0.65\right) \times 10^{-2} \, .
    \label{axionic_freq_charge}
\end{align}
In particular, Eq.~\eqref{axionic_freq_charge} is relevant  for $^{87}$Rb hyperfine transition, whose associated $\kappa$ parameters are $\kappa = -0.016, \kappa_q = -0.046, \kappa_s = -0.010$ \cite{Flambaum06}.\\
\noindent
\begin{minipage}{\textwidth}
\begin{minipage}[b]{0.49\textwidth}
    \centering
    \begin{tblr}{
        vlines,
        colspec={ccc}
    }
    \hline
    Species & $Q_M $ [$\times 10^{-3}$] & $Q_\omega$ [$\times 10^{-3}$] \\
    \hline\hline
    $^{195}$Pt \cite{Microscope18} & -69.065 & $-$ \\
    $^{48}$Ti \cite{Microscope18} & -68.770 & $-$\\
    $^{87}$Rb & -68.920 & 9.30 \cite{Flambaum06} \\
    $^{85}$Rb & -68.924 & $-$ \\
    $^{40}$Ca & -68.715 & -0.00188 \\
    $^{44}$Ca & -68.738 & -0.182 \\
    $^{86}$Sr & -68.933 & -0.554 \\
    $^{87}$Sr & -68.932 & -0.552 \\
    $^{88}$Sr & -68.932 & -0.00550 \\
    $^{171}$Yb & -69.054 & -0.0121 \\
    $^{176}$Yb & -69.043 & -0.0119 \\
    $^{196}$Hg & -69.077 & -0.00684 \\
    $^{202}$Hg & -69.066 & -0.00677 \\
    SiO$_2$ & -68.442 & $-$ \\ 
    73\%Au-27\%Pt & -69.067 & $-$ \\ \hline
    \end{tblr}
    \captionof{table}{Axionic charges for some species of atoms. The transition is hyperfine for $^{87}$Rb and optical for the rest (as in Table ~\ref{dilatonic_charge_table}, all of the optical transition are the $^1S_0 \rightarrow ^3P_1$ ICL transition, see Table ~\ref{tab:alk_isotope_freq}). The charges are derived from Eqs.~\eqref{axionic_mass_charge}, \eqref{axionic_freq_charge}, \eqref{axionic_freq_charge_2nd_order_1} and \eqref{axionic_freq_charge_2nd_order_2}. As we shall see in Chapters ~\ref{chap:AI} and \ref{chap:LISA_DM}, we will not consider the atomic transition of $^{195}$Pt, $^{48}$Ti, SiO$_2$ and Au-Pt, reason why their frequency charge is not provided.}\label{axionic_charge_table}
  \end{minipage}
  \hfill%
  \begin{minipage}[b]{0.49\textwidth}
   Let us now consider optical transitions whose frequency does depend neither on the nucleon g-factor nor on the proton mass at lowest order. This implies that the axionic frequency charge at lowest order is 0. Higher order contributions would lead to non-zero axionic frequency charge, in particular if one considers its dependence on the fine structure constant $\alpha$ which arises at loop level \cite{Beadle23}, or to the nuclear charge radius \cite{Zhao24} and are therefore highly suppressed. We now compute the axionic optical frequency charge of an atom A, considering both of these effects for some atomic optical transition. Using the dependence of optical transition on the fine structure constant $\alpha$, the gluon-photon coupling at loop level leads to a charge 
    \begin{subequations}\label{axionic_freq_charge_2nd_order_1}
    \begin{align}
    [Q^A_\omega]_a &= c_{F_2}\frac{(2+\epsilon_A) \alpha}{4\pi^2} \, ,
    \end{align}
    where $c_{F_2}$ is the parameter encoding the explicit symmetry-breaking, generating the one-loop coupling. While for QCD axion, this parameter is $\mathcal{O}(1)$, it is much smaller in the case of ALP (due to the introduction of an additional dark sector, e.g a dark photon, see \cite{Beadle23}). It is nevertheless possible to have $c_{F_2} \sim 10^{-2}$ \cite{Beadle23}. Therefore, the corresponding axionic charge is, e.g. for $^{171}$Yb,
    \begin{align}
    [Q^{Yb}_\omega]_a & \sim 1.2 \times 10^{-8} \, .
    \end{align}
    \end{subequations}
    \end{minipage}
\end{minipage}
\noindent
The second contribution, arising from the coupling of the ALP to the nuclear charge radius has the form \cite{Zhao24, Seltzer69, Nörtershäuser20}
\begin{subequations}\label{axionic_freq_charge_2nd_order_2}
\begin{align}
     [Q^A_\omega]_a &= \frac{\partial \ln \omega_A}{\partial \ln \langle r^2_N \rangle}\frac{\partial \ln \langle r^2_N \rangle}{\partial \ln m^2_\pi}\frac{\partial \ln m^2_\pi}{\partial (\theta)^2} = \frac{\beta F}{\omega_A} \frac{\partial \ln m^2_\pi}{\partial (\theta)^2}\left(\langle r^2_N  \rangle + \frac{C_2}{C_1} \langle r^4_N  \rangle\right) \, ,
\end{align}
where $\langle r^2_N \rangle, \langle r^4_N \rangle$ are respectively the mean squared and quartic charge nuclear radius, $F$ is the difference in field shift factor between excited and ground state, $\beta = \partial \ln \langle r^2_N \rangle/\partial \ln m^2_\pi \equiv -0.2$ \cite{Zhao24} and $C_1,C_2$ are Seltzer coefficients, to account for higher order radial moments in the computation of the field shift (which are relevant for heavy atoms) \cite{Nörtershäuser20} . For $^{171}$Yb optical transition, $F = -2\pi \times 10.955$ GHz/fm$^2$ \cite{Schelfhout21}, $r_N = 5.2906$ fm \cite{Angeli13}, $C_1= 6.23 \times 10^{-4}$ and $C_2 =-0.579$ \cite{Seltzer69}, which leads to
\begin{align}
    [Q^{Yb}_\omega]_a &\sim -1.21 \times 10^{-5} \, .
\end{align}
\end{subequations}
For other atomic transition, we will use the fact that the field shift $F\langle r^2_N \rangle$ approximately scales as $Z^2/A^{1/3}$ \cite{Nörtershäuser20}. 
Note that, as expected, those two contributions are at least two orders of magnitude smaller than the axionic charges of hyperfine transitions (see Table ~\ref{axionic_charge_table}).

\section{\label{DP_pheno}Vector field}

\subsection{Theoretical motivations}

In addition to the current SM gauge group $SU(3)\times SU(2)\times U(1)$, many theories predict an additional gauge group which would resolve various particle physics and/or cosmology problems, e.g. the $\nu$MSM model solves DM, neutrino masses and baryogenesis by incorporating three right-handed $SU(2)_R \times U(1)$ neutrinos \cite{Asaka05}. The minimal gauged extension would be an additional $U(1)$ symmetry, carried by a new vector field, coupled to a local gauge symmetry current, in particular a linear combination of the baryon number $B$, the lepton number $L$ and the hypercharge $Y$ \cite{Fayet89, Fayet90}. Supersymmetry predicts a coupling to the $B-L$ current \cite{Fayet89, Fayet90}. This new vector field, sometimes referred as $U$-boson, heavy photon or dark photon (DP), would induce a new force, proportional to the current to which it couples. 

\subsection{\label{DP_photons_coupling}Oscillating electric field through the coupling with photons}

We now are interested with the interaction of the DP vector field $\phi^\mu$ of mass $m_{U}$ with EM. Using Eq.~\eqref{vector_field_action}, the Lagrangian describing such interaction is given by \cite{Horns}
\begin{align}\label{Int_lagrangian_vector}
  \mathcal L =& - \frac{1}{4\mu_0}F^{\mu\nu}F_{\mu\nu}+j^\mu A_\mu  - \frac{1}{4\mu_0}\phi^{\mu\nu}\phi_{\mu\nu} - \frac{m^2_U c^2}{2\mu_0\hbar^2}\phi^\mu\phi_\mu -\frac{\chi}{2\mu_0}F_{\mu\nu}\phi^{\mu\nu}\, ,
\end{align}
where $\chi$ is the dimensionless kinetic mixing coupling parameter which characterizes the coupling between the DP and the EM field.

In order to ease the calculations, we make a non-unitary transformation of the fields, i.e \cite{nelson:2011tv}
\begin{equation}\label{eq:barA}
	\bar A^\mu=A^\mu+\chi \phi^\mu
\end{equation}  
while $\phi^\mu$ remains unchanged at first order in $\chi$. This allows us to redefine the fields in terms of mass eigenstates called massless and massive photons, which couples to the standard EM current. Using this change of variable, Eq.~\eqref{Int_lagrangian_vector} becomes, at first order in $\chi$,
\begin{align}
	\mathcal L =& - \frac{1}{4\mu_0}\bar F^{\mu\nu} \bar  F_{\mu\nu}+j^\mu \left(\bar A_\mu -\chi \phi_\mu\right)  - \frac{1}{4\mu_0}\phi^{\mu\nu}\phi_{\mu\nu} - \frac{m^2_Uc^2}{2\mu_0\hbar^2}\phi^\mu\phi_\mu \, ,
\end{align}
with $ F_{\mu\nu} = \bar F_{\mu\nu}-\chi \phi_{\mu\nu}$.  The field equations read
\begin{subequations}\label{eq:field_eq}
  \begin{align}
    \partial_\alpha \bar F^{\beta\alpha} &= \mu_0 j^\beta \, , \\
    \partial_\alpha \phi^{\beta\alpha} &= -\chi \mu_0 j^\beta - \frac{m^2_Uc^2}{\hbar^2} \phi^\beta \, .
  \end{align}
The antisymmetry of both strength tensors leads to the conservation of the electromagnetic 4-current $\partial_\mu j^\mu = 0$, and the continuity equation for the DP field $\partial_\mu \phi^\mu = 0$, as shown in Section ~\ref{vector_field_DM}. 
\end{subequations}
Using the Lorenz gauge for EM, $\partial^\mu \bar A_\mu = 0$, Eqs.~(\ref{eq:field_eq}) become 
  \begin{subequations}\label{eq:field_mass}
    \begin{align}
      \Box \bar A^\beta &= -\mu_0 j^\beta \, , \\
      \Box \phi^\beta  &= \frac{m^2_Uc^2}{\hbar^2} \phi^\beta + \chi \mu_0 j^\beta \, , \label{eq:k2m}
    \end{align} 
  \end{subequations}    
where $\Box = \eta^{\mu\nu}\partial_\mu \partial_\nu \equiv -\frac{1}{c^2}\partial^2_t + \nabla^2$. These equations
admit two classes of solutions in vacuum (where the current vanishes): a massless vector field (standard EM) and a massive one,  the latter being characterized by solutions : 
\begin{subequations}
\begin{align}
	\phi^\beta &=Y^\beta e^{i k_\mu x^\mu + \Phi}\,   \label{eq:osc_DP} , \\
	\bar A^\alpha &= 0 \, ,
\end{align}
\end{subequations} 
where $k_\mu k^\mu = -(\omega_U/c)^2 + \left|\vec k_U\right|^2 = -m^2_Uc^2/\hbar^2$, as derived in Eq.~\eqref{dispersion_relation_mass_scalar}, and where Eq.~\eqref{eq:osc_DP} is equivalent to Eq.~\eqref{vector_sol_KG}.
As can be noticed from Eq.~(\ref{eq:barA}), this solution induces an ordinary electromagnetic field in a vacuum (see also \cite{Horns})
\begin{equation}\label{eq:A}
	A^\beta = -\chi Y^\beta e^{i k_\mu x^\mu + \Phi}\, . 
\end{equation}
Due to the coupling between the DP and the EM field (coupling characterized by the mixing parameter $\chi$), the DP field will induce a small electromagnetic field, whose strength is proportional to $\chi$ and to the DP field amplitude, see Eq.~(\ref{eq:A}). Then the induced electromagnetic field consists in an oscillating electric and magnetic fields of the form \cite{Horns}
\begin{subequations}
\begin{align}\label{eqs:EDM_BDM_general}
	 E^j_\mathrm{DM} &= -\frac{\partial A^0}{\partial x_j} - \frac{\partial  A^j}{\partial t} = \chi \left(k^j_U Y^0 - \omega_U  Y^j\right) \sin\left(\omega_U t - \vec k_U \cdot \vec x + \Phi\right)\, \\
     B^j_\mathrm{DM} &= \epsilon_{ijk}\partial^j A^k  = -\chi \epsilon_{ijk} k^j_U  Y^k \sin\left(\omega_U t - \vec k_U \cdot \vec x + \Phi\right)\, ,
\end{align}
The magnetic component is suppressed by a factor $v_\mathrm{DM}/c \sim 10^{-3}$, therefore only the electric field component will be considered in Chapters ~\ref{chap:Rydberg_exp_DP} and \ref{chap:SHUKET_exp}. In addition, if one considers that $\vec k_U=0$, we have $Y^0 = 0$ using the continuity equation $k_\mu Y^\mu =0$, and the electric field does not propagate, i.e 
\begin{align}\label{eqs:EDM}
    E^j_\mathrm{DM} &= -\chi \omega_U  Y^j \sin\left(\omega_U t + \Phi\right)
\end{align}
where its amplitude is directly related to the local DM density (from Eq.~\eqref{energy_density_vector})
\begin{equation}
	\left|\vec E_\mathrm{DM}\right| = \chi c\sqrt{2\mu_0 \rho_{\mathrm{DM}}}\, .
 \label{amp_E_field}
\end{equation}
\end{subequations}
The idea of several experiments searching for DP is to focus this small electromagnetic field in order to enhance it and hopefully make it detectable \cite{SHUKET,Tokyo1,Tokyo2,Tokyo3,Tokyo4,FUNK}.

\subsection{UFF violation through the coupling with the $B - L$ current}

We now study the interaction between the vector field and the $B-L$ current $j^\mu_\mathrm{B-L}$ through the coupling $\epsilon$, which reads \cite{Fayet19, Fayet18}
\begin{align}\label{B_L_vector_lagrangian}
  \mathcal L =& - \frac{1}{4\mu_0}\phi^{\mu\nu}\phi_{\mu\nu} - \frac{m^2_U c^2}{2\mu_0\hbar^2}\phi^\mu\phi_\mu + \epsilon e j^\mu_\mathrm{B-L}\phi_\mu\, ,
\end{align}
where the coupling is parameterized with respect to the usual EM charge $e$. 
Varying Eq.~\eqref{B_L_vector_lagrangian} with respect to $\phi^\mu$ leads to its field equation
\begin{align}
    \partial_\nu \phi^{\mu\nu} &= -\frac{m^2_U c^2}{\hbar^2} \phi^\mu + \mu_0 \epsilon e j^\mu_\mathrm{B-L} \, .
\end{align}
The conservation of the $B-L$ current leads to the continuity equation of the DP field $\partial_\mu \phi^\mu = 0$. Therefore, one can simplify the field equation to 
\begin{align}\label{eqs:field_eq_B-L}
    \square \phi^\mu &= \frac{m^2_U c^2}{\hbar^2} \phi^\mu - \mu_0 \epsilon e j^\mu_\mathrm{B-L} \, .
\end{align}
Using Eq.~\eqref{macro_lagrangian}, one finds the equation of motion for a test particle A in vacuum of mass $m_A$, velocity $\vec v_A$ and acceleration $\vec a_A$
\begin{subequations}
\begin{align}\label{eq:acc_B_L_gen}
    m_A \vec a_A = \epsilon e [Q^A_\mathrm{B-L}]\left(\vec E_\mathrm{DP} + \vec v_A \times \vec B_\mathrm{DP}\right) \, ,
\end{align}
where we have defined the $B-L$ current associated to A as $j^\mu_\mathrm{B-L} = (c,\vec v_A)[Q^A_\mathrm{B-L}]\delta^\mathrm{(3)}\left(\vec x - \vec x_A\right)$, with $[Q^A_\mathrm{B-L}]$, the $B-L$ charge of A, and where
\begin{align}
    \vec E_\mathrm{DP} &= -\frac{\partial \vec \phi}{\partial t} - \vec \nabla \phi^0= \left(-\vec k_U Y^0 + \omega_U  \vec Y\right) \sin\left(\omega_U t - \vec k_U \cdot \vec x + \Phi\right)\,\\
    \vec B_\mathrm{DP} &= \vec \nabla \times \vec \phi = \left(\vec k_U \times \vec Y\right) \sin\left(\omega_U t - \vec k_U \cdot \vec x + \Phi\right) \, ,
\end{align}
\end{subequations}
using Eqs.~\eqref{eq:osc_DP} and \eqref{eqs:EDM_BDM_general} ($\vec E_\mathrm{DP}, \vec B_\mathrm{DP} = \vec E_\mathrm{DM}/\chi, \vec B_\mathrm{DM}/\chi$), as we are considering the vacuum solution of Eq.~\eqref{eqs:field_eq_B-L}. Note that a static solution of such equation exists and is associated with a Yukawa-type force sourced by a massive body.

The small galactic velocity implies that 1) $|\vec k_U| c \ll \omega_U$, and together with the continuity equation, 2) $\vec k_U Y^0 \ll \omega_U \vec Y$, such that Eq.~\eqref{eq:acc_B_L_gen} simplifies to \cite{Yu23, Pierce18, Michimura20,Morisaki21}
\begin{align}\label{eq:acc_B_L}
   \vec a_A (t, \vec x) = \epsilon e \omega_U \vec Y \frac{[Q^A_\mathrm{B-L}]}{m_A}\sin(\omega_U t - \vec k_U \cdot \vec x_A + \Phi) \, ,
\end{align}
which, in the same way as Eq.~\eqref{EP_viol_acc_dil}, violates UFF.

\clearpage
\pagestyle{plain}
\printbibliography[heading=none]
\clearpage
\pagestyle{fancy}

\part{Electromagnetic cavities probes}
\chapter{\label{optical_exp_axion}Search for axions with an optical cavity and an optical fiber}

Usually, cavities can be used to search for axion-photon coupling $g_{a\gamma}$ through the axion-photon conversion under a strong magnetic field. This is the case for \textit{ADMX} \cite{ADMX_2002,ADMX:2010,ADMX:2018_1,ADMX:2019,ADMX:2021} or \textit{CAPP} \cite{Capp:2020_1, Capp:2020_2, CAPP:2020_3} experiments. In practice, these experiments aim at detecting photons sourced by this strong magnetic field and the DM axion field through the inverse Primakoff effect. Then, the photons would resonate inside the cavity in order to increase drastically the signal. In this case, the frequency of the produced photons correspond to the axion mass, and therefore these experiments are sensitive to a very narrow mass interval (which is essentially the width of the resonant frequency of the cavity).

Cavities can also search for this $g_{a\gamma}$ coupling through the effect derived in Section \ref{axion_photon_coupling}, i.e the phase velocity difference between the left and right polarization of light. In the following, we will consider an optical cavity to compute the phase observable. Then, in Chapter ~\ref{chap:sens_experiments}, we will make some estimation on the sensitivity of the optical cavity used in the \textit{DAMNED} experiment \cite{Savalle21}. We will show that unfortunately, such experiments are not competitive for the search of the axion-photon coupling.

\section{\label{axion_photon_phase_vel}Birefringence and oscillation of the length of the cavity}

We consider a setup where two optical cavities are used : one in which a linear polarized light is used, i.e the phase velocity is exactly the speed of light $c$, and another where we send a right polarized light. This could be done, e.g, by inserting a quarter wave plate in front of one of the cavities which would change a linear polarization to a circular one. Then, we measure the phase difference between the output of the two cavities. 

From Eq.~\eqref{delta_c_axion_photon}, the left and right circular polarized light respectively evolve with phase velocity
\begin{subequations}
\begin{align}
c \left(1 - \frac{E_P}{c^2} \frac{\sqrt{4\pi G \rho_\mathrm{DM}}g_{a\gamma}}{k}\sin(\omega_a t + \Phi)\right) \equiv c - \delta c \sin(\omega_a t+\Phi)\, \\
c \left(1 + \frac{E_P}{c^2} \frac{\sqrt{4\pi G \rho_\mathrm{DM}}g_{a\gamma}}{k}\sin(\omega_a t + \Phi)\right) \equiv c + \delta c \sin(\omega_a t+\Phi)\, ,
\end{align}
with $\omega_a$ the oscillation frequency of the axion field, $E_P$ the reduced Planck energy and
\begin{align}\label{delta_c_axion_photon_right}
    \delta c = \frac{\sqrt{4\pi G \rho_\mathrm{DM}} E_P g_{a\gamma}}{\omega_0} \, ,
\end{align}
\end{subequations}
with $\omega_0 = k c$, the unperturbed light angular frequency.
In this model, the cavity length $\ell$ is unchanged, but is directly related to the light travel time $\tau_\leftrightarrow(t)$ inside the cavity as
\begin{align}
2 \ell &= \int_0^{\tau_\leftrightarrow(t)}c(t')dt' \, ,
\end{align}
where $c(t)$ is the time dependent phase velocity of light whose exact form depends on the light polarization Eq.~\eqref{delta_c_axion_photon}.
Assuming a right handed photon starting its propagation at the left mirror at time $t-\tau_\leftrightarrow(t)$, it will come back to the same spatial point at time t, where the travel time is given by
\begin{subequations}
\begin{align}\label{time delay}
\tau_\leftrightarrow(t) = \frac{2\ell}{c} - \int_{t-\tau_\leftarrow}^t \frac{\delta c_\leftarrow(t')}{c}dt' - \int_{t-\tau_\leftarrow-\tau_\rightarrow}^{t-\tau_\leftarrow} \frac{\delta c_\rightarrow(t')}{c}dt' \,,
\end{align}
where $\tau_\rightarrow$ and $\tau_\leftarrow$ are the time intervals the photon takes on the trip there and back respectively and
\begin{align}
\delta c_\rightarrow(t) &= +\delta c \sin(\omega_at+\Phi) \\
\delta c_\leftarrow(t) &= -\delta c \sin(\omega_at+\Phi) \, ,
\end{align} 
\end{subequations}
respectively for right and left handed polarizations, since the photon has opposite circular polarization on the way there and back due to the mirror reflection. Indeed, the mirror reverses one of the linear polarizations while keeping the other unchanged, which implies that a circular polarization changes other (up to an irrelevant phase)
We can then interpret this oscillation of phase velocity of light as an oscillation of the cavity length itself, i.e we can write 
\begin{subequations}
\begin{align}\label{eq:tau_ell}
    \tau_\leftrightarrow(t) &= \frac{2(\ell + \delta\ell(t))}{c} \, ,
\end{align}
where
\begin{align}\label{eq:delta-ell}
\delta\ell(t) = \delta c f(t) \,,
\end{align}
with 
\begin{align}\label{eq:f(t)}
f(t) = \frac{1}{2}\left[\int_{t-\tau_\leftarrow}^t \sin(\omega_at'+\Phi) dt' - \int_{t-\tau_\leftarrow-\tau_\rightarrow}^{t-\tau_\leftarrow} \sin(\omega_at'+\Phi)dt' \right] \,.
\end{align}
\end{subequations}
As we are interested in the first order solution of the phase shift in the perturbation $g_{a\gamma}$, and since $\delta c = \mathcal{O}(g_{a\gamma})$, we can solve explicitly $f(t)$ at zeroth order in the perturbation, i.e using $\tau_\leftarrow = \tau_\rightarrow = \tau_0 \equiv \ell/c$ in the integral bounds
\begin{align}
f(t) &= \frac{2}{\omega_a}\sin^2\left(\frac{\omega_a \ell}{2c}\right)\cos\left(\omega_a\left(t-\frac{\ell}{c}\right)+\Phi\right) \,.
\end{align}
such that the variation of length of the cavity becomes
\begin{align}
    \delta \ell(t) = \frac{\sqrt{16\pi G \rho_\mathrm{DM}} E_P g_{a\gamma}}{\omega_0\omega_a}\sin^2\left(\frac{\omega_a \ell}{2c}\right)\cos\left(\omega_a\left(t-\frac{\ell}{c}\right)+\Phi\right)
    \label{delta_l}
\end{align}

\section{\label{phase_optical_cavity_axion}Phase shift at the output of the cavity}

In order to obtain the phase shift of the electric field resulting from the length oscillation Eq.~\eqref{delta_l}, we will sum over a large number $N$ of field contributions that are reflected back and forth in the cavity (i.e $N \rightarrow \infty$).

The calculation is explicitly presented in Appendix ~\ref{ap:phase_shift_cavity}.  Here, we immediately show the amplitude of the phase shift (oscillating at the axion frequency $\omega_a$) between the transmitted field and the input field as 
\begin{align}
|\Delta \phi(t)| &\approx \frac{2\sqrt{16\pi G \rho_\mathrm{DM}} E_P g_{a\gamma}}{\omega_a c\sqrt{1-2r^2\cos\left(\frac{2\omega_a \ell}{c}\right)+r^4}}\sin^2\left(\frac{\omega_a \ell}{2c}\right) \label{eq:phase_final_DAMNED}\, ,
\end{align}
where $r$ is the mirror reflectivity of the cavity. This means the phase observable is independent of the laser frequency. When the axion frequency corresponds to an even mode of the cavity, $\omega_a \ell/c = (2n +1) \pi$, and considering that $r = 1- \epsilon$, $\epsilon \ll 1$, the signal scales in $\epsilon^{-1}$, and therefore increases drastically. This phase shift expression has not been derived in the literature explicitly, in the case of the axion-photon coupling. In Chapter ~\ref{chap:sens_experiments}, we will use it to derive the sensitivity of \textit{DAMNED} to $g_{a\gamma}$. 

\section{\label{fibers_phase_shift}Phase shift along an optical fiber}

We now consider a very simple experiment where we use two fibers of same length L, in which we send a left polarized light A in the first one and a right polarized light B in the other one. At the output, we measure the phase shift between both light signals. Mathematically, this situation is equivalent to having only one single fiber and sending both signals in the two opposite directions of the fiber. Both signals have the same angular frequency $\omega_0$ (at zeroth order in the perturbation), and after having traveled through the fiber, the electric fields have the form
\begin{subequations}
\begin{align}
E_A(t) &= E^0_A e^{-i\omega_0(t-t_\rightarrow)} \,\\
E_B(t) &= E^0_B e^{-i\omega_0(t-t_\leftarrow)}\,
\end{align}
\end{subequations}
with $t_\leftarrow$ and $t_\rightarrow$ being respectively the time for the left and right polarized signals to cross the full fiber.
At the output of the fiber, the phase shift between the two electric fields is just given by 
\begin{subequations}
\begin{align}
   \Delta \phi(t) = \omega_0 (t_\leftarrow - t_\rightarrow)
\end{align}
where 
\begin{align}
    t_\leftarrow &= \frac{L}{c} - \int^{L/c}_0 \frac{\delta c_\leftarrow }{c}\sin(\omega_a t + \Phi) \, \\
    t_\rightarrow &= \frac{L}{c} - \int^{L/c}_0 \frac{\delta c_\rightarrow }{c}\sin(\omega_a t + \Phi) \, ,
\end{align}
\end{subequations}
where the amplitudes $\delta c_\rightarrow = - \delta c_\leftarrow$ are given in Eq.~\eqref{delta_c_axion_photon_right}. Then, it is straightforward to find the amplitude of the phase shift as 
\begin{align}\label{eq:phase_fiber_axion_photon}
|\Delta \phi(t)| &= \frac{2\sqrt{16\pi G \rho_\mathrm{DM}} E_P g_{a\gamma}}{\omega_a c} \left|\sin\left(\frac{\omega_a L}{2c}\right)\right| \, .
\end{align}
Similarly as in the previous section, we will use this result to derive the sensitivity of an optical fiber to $g_{a\gamma}$ in Chapter ~\ref{chap:sens_experiments}.

\chapter{\label{chap:Rydberg_exp_DP}Search for dark photons in a microwave cavity with Rydberg atoms}

As mentioned in Section ~\ref{DP_photons_coupling}, an oscillating DP coupled to the standard electromagnetic field will induce a small oscillating electric field in a vacuum. One very peculiar feature of this electric field is that it does not propagate, i.e. its wave-vector vanishes $\vec k_U=0$ (to first order in $v_\mathrm{DM}/c$). This is due to the fact that this electric field is induced by a massive vector field and therefore its dispersion relation is given by Eq.~\eqref{dispersion_relation_mass_scalar}. In this chapter, we will show how an electromagnetic cavity can be used to search for the electric field induced by a DP. In addition, we will show how to use atoms as a tool to detect this electric field through the Stark effect, i.e. the displacement of the energy levels of an atom under a perturbation by a static electric field (or by an electric field whose frequency is much lower than the transition frequency of the atom). This experimental proposal to search for DP using atoms inside a microwave cavity is the object of a published article \cite{Gue23}. 

There are mainly two reasons to consider a cavity as an experiment to search for DP. First of all, as for other DP experiments using resonators, the mirrors of the cavity will enhance the electric field induced by the DP. Indeed, the electric field parallel to the surface of a perfect conductor has to vanish. Therefore, because of the presence of the oscillating DP-induced electric field, the mirror will generate a standard electromagnetic field that will propagate perpendicularly to the mirror and whose amplitude is such that it will cancel the DP-induced field parallel to the surface. Physically, the DP-induced electric field will induce an oscillation of the electrons within the mirror which will create a standard electromagnetic field. Since a cavity consists in two mirrors, this boundary condition can, under some conditions, produce resonances that will significantly enhance the small DP-induced electric field. 

The second reason to consider a cavity is related to the use of atoms to measure the electric field inside the cavity through the Stark effect, which is sensitive to the square of the electric field. If one applies a standard electromagnetic wave inside the cavity (whose electric field will be denoted by $\vec E_A$) the DP contribution to the square of the electric field inside the cavity is $\sim \vec E_\mathrm{DM}\cdot \vec E_A$ (to first order in $\chi$, the kinetic mixing between EM and DP), which can also be enhanced by a resonant $\vec E_A$. In addition to enhancing the amplitude of the signal to be measured, applying an external field is also important to produce a signal at low, but non-zero, frequency, where the Stark effect can be realistically measured. More precisely, for a cavity whose length is of the order of a few cm, we will be interested in searching for DP oscillating at a frequency $f_U=\omega_U/2\pi$ of the order of a few GHz. 
The difficulty is that such a rapid oscillation of the atomic transition frequency will be very hard to measure as the interrogation cycle of the atoms is much longer. But, if one applies en external field at a angular frequency $\omega_A$ which is close to $\omega_U$, then the cross term between the DP electric field and the applied electric field will have a component oscillating at the low angular frequency $\left|\omega_U-\omega_A\right|$, which can be measured by the atoms.

As in the previous chapter, we will compute the expected signal, leaving the experimental parameters free. Afterwards, in Chapter ~\ref{chap:exp_summary}, we will discuss the sources of noise of such experiment, and in Chapter ~\ref{chap:sens_experiments}, we will derive the expected sensitivity of the experiment to the $\chi$ coupling, by considering specific experimental parameters.

\section{Contributions from dark photons and applied electric field inside the cavity}

In this section, we derive the electric field induced by both the DP and the applied field at the center of the cavity. We summarize the methodology, the derivation and discuss the main results. 

The finesse $\mathcal{F}$ of a cavity is a measure of the narrowness of resonances $\delta f$ compared to their frequency. More precisely, it is defined as the ratio between $\delta f$ and the frequency separation of resonances $\Delta f$
\begin{align}
    \mathcal{F} &= \frac{\Delta f}{\delta f} \, ,
\end{align}
which is frequency independent.
Considering a cavity consisting of two flat mirrors of reflectivity $r$ separated by a distance $L$\footnote{For simplicity, we assume that the transverse size of the mirrors is $\gg L,\lambda$, where $\lambda$ is the wavelength of the fields of interest.}, we can relate the reflectivity and the finesse of a cavity through ($\mathcal{F} \gg$ 1)\cite{Suter14}
\begin{align}
	r &\approx 1 - \frac{\pi}{2\mathcal{F}} \, .
\end{align}
While the finesse is usually associated to an optical cavity, it is the quality factor $Q$ that is a more common parameter for microwave cavities. The quality factor is usually defined as the ratio between the energy stored inside the cavity with the energy dissipated. It then becomes obvious that finesse and quality factor are two quantities profoundly connected. The quality factor can also be defined as \cite{Suter14}
\begin{align}
    Q &= \frac{f}{\delta f} \, ,
\end{align}
Although being frequency dependent, one can notice that the quality factor of the first resonance of the cavity corresponds exactly to the finesse. In the following of this chapter, we will assume only the first resonances of our microwave cavity, such that $\mathcal{F} \sim Q$. 

The principle of calculation is similar for both the DP and the applied electric fields and has already been used in Chapter ~\ref{phase_optical_cavity_axion}, inspired from \cite{savallePhD}. The idea is to propagate the electric field an infinite number of round trips inside the cavity and to sum these infinite number of contributions at a given location. To perform cavity round trips, the field is propagated along one direction and when it reaches a mirror, its amplitude is multiplied by $-r$ and its wave-vector is flipped. For $r<1$, the infinite sum converges and can be calculated explicitly.

\begin{figure}[h!]
\centering
\includegraphics[width=\textwidth]{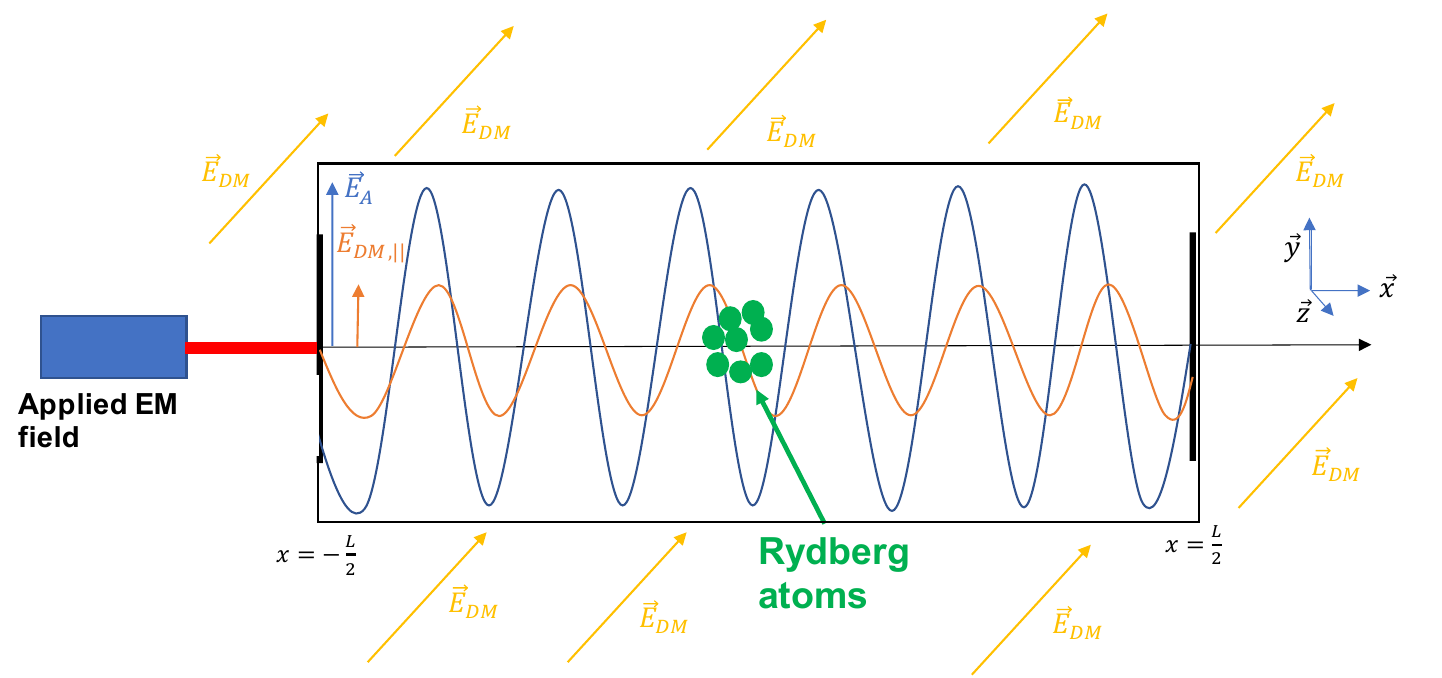}
\caption{Experimental scheme that we propose for the search of the DP-photons coupling : an external field (in blue) is applied at the cavity edge. The standing DP electric field (in yellow) generates a propagating electric field inside the cavity (in orange). At the center of the cavity, the transition frequency of Rydberg atoms is impacted by $|E|^2$ through the Stark effect (see text.).}
\label{DP_cavity}
\end{figure}
First, let us apply this procedure to the applied external field. We assume that the external electric field is applied on the left edge of the cavity, see Fig.~\ref{DP_cavity}. Ideally, the applied field can be parameterized as $\vec E_{A}=\vec X_A \Re \left[  e^{-i(\omega_At - k_A \left(x+\frac{L}{2}\right)+\phi_A})\right]$, with its amplitude $X_A$, angular frequency $\omega_A$ and phase $\phi_A$. Still assuming a transmission coefficient of the mirror $\sqrt{1-r^2}$, the first contribution at the center of the cavity reads
\begin{align}
\vec E^0_A(x=0,t) &= \Re \left[\sqrt{1-r^2}\vec{X}_A e^{-i(\omega_At-k_A\frac{L}{2}+\phi_A)}\right]\, .
\end{align}
This contribution propagates until the other cavity boundaries, gets reflected once with coefficient $-r$ such that boundary conditions are respected, then comes back to the center, implying that the second contribution reads 
\begin{align}
\vec E^1_A(x=0,t) &= -r\vec E^0_A\left(x=0,t+\frac{L}{c}\right) = \Re\left[-r\sqrt{1-r^2}e^{ik_AL}\vec{X}_A e^{-i(\omega_At-k_A\frac{L}{2}+\phi_A)}\right]\, , 
\end{align}
the additional phase $e^{ik_AL}$ shows the time delay of $E^1$ compared to $E^0$ after half a round-trip. This occurs several times and after an infinite number of round trips $N$, the full contribution of the external applied field inside the cavity is
\begin{subequations}
\begin{align}
\vec E^\mathrm{tot}_A(x=0,t) &= \sum_{n=0}^{N=+\infty} \vec E^{n}_A(x=0,t) \,\\
&=\sqrt{1-r^2}\vec X_A\Re \left[e^{-i(\omega_At+\phi_A)}\frac{e^{i\frac{k_AL}{2}}}{1+re^{ik_AL}}\right] \, \\
&\equiv \vec A(\omega_A)\cos(\omega_At+\phi_A)+\vec B(\omega_A)\sin(\omega_At+\phi_A) \, ,\label{full_field_applied_simplified}
\end{align}
\end{subequations}
where we assumed $r < 1$ such that $r^N \rightarrow 0$ and with 
\begin{subequations}\label{eq:A_B}
\begin{align}
    \vec A(\omega_A) &= \frac{\vec X_A \left(1+r\right)\sqrt{1-r^2}\cos\left(\frac{\omega_AL}{2c}\right)}{1+2r\cos(\frac{\omega_AL}{c})+r^2}\, , \\
    \vec B(\omega_A) &= \frac{\vec X_A \left(1-r\right)\sqrt{1-r^2}\sin\left(\frac{\omega_AL}{2c}\right)}{1+2r\cos(\frac{\omega_AL}{c})+r^2}\, .
\end{align}
\end{subequations}
One can notice a resonance at the center of the cavity for even modes of the cavity, as expected.

Let us now focus on the contribution from the DP field. From Eqs.~(\ref{eqs:EDM}), one can, without loss of generality, write the expression of the electric field related to the DP as $\vec E_\mathrm{DM} = \vec X_\mathrm{DM} \Re \left[ e^{-i\omega_U t}\right]= X_\mathrm{DM} \hat e_\mathrm{DM} \Re \left[ e^{-i\omega_U t}\right]$ where $\hat e_\mathrm{DM}$ is a unit vector characterizing the polarization of the DP field and $X_\mathrm{DM}=\chi c\sqrt{2\mu_0\rho_\mathrm{DM}}$ (from Eq.~\eqref{amp_E_field}), and where the phase of the field $\Phi$ disappeared because it can be reabsorbed in $\phi_A$ when computing the interference between the DP and applied fields. Because of this electric field, both mirrors will generate a propagating standard electromagnetic field such that the total component of the electric field parallel of the mirrors' surface vanishes. 
The same procedure as above can be realized to know the DM electric field amplitude at the center of the cavity. The subtleties of this calculation are that : 1) the field is emitted by the mirrors towards the center of the cavity, therefore the transmission coefficient $\sqrt{1-r^2}$ factor is not present ; 2) only the DM polarization transverse to the mirror is re-emitted, noted $\vec X_\mathrm{DM,\parallel}$; and 3) there are two different contributions, in phase, each being emitted from one of the edges of the cavity. The total DM contribution at the center is then
\begin{subequations}
\begin{align}
\vec E^\mathrm{tot}_\mathrm{DM}(x=0,t) &= \Re \left[\vec X_\mathrm{DM}e^{-i\omega_U t} + 2 \vec{X}_\mathrm{DM,\parallel}e^{-i(\omega_U t-\frac{kL}{2})}\frac{1}{1+re^{ikL}}\right]\, \label{full_field_DM} \\
&\equiv \vec C(\omega_U)\cos(\omega_U t)+\vec D(\omega_U)\sin(\omega_U t) \, , \label{full_field_DM_simplified}
\end{align}
\end{subequations}
with
\begin{subequations}\label{eq:C_D}
\begin{align}
    \vec C(\omega_U) &= \vec X_\mathrm{DM} +\frac{2\vec X_\mathrm{DM,\parallel} (1+r)\cos(\frac{\omega_U L}{2c})}{1+2r\cos(\frac{\omega_U L}{c})+r^2}\, , \\
    \vec D(\omega_U) &= \frac{2\vec X_\mathrm{DM,\parallel}(1-r)\sin(\frac{\omega_U L}{2c})}{1+2r\cos(\frac{\omega_U L}{c})+r^2}\, .
\end{align}
\end{subequations}
The first term of Eq.~\eqref{full_field_DM} corresponds to the background oscillating DM field at the center, which is always present, even without the cavity. The second term is the DM contribution from the cavity, which is almost equivalent, in its form, to the total contribution of the applied field, with an additional factor two, due to the emission of a field from both edges of the cavity (instead of only one for the applied field). 

The calculations presented in this section have been carried out to leading order in $\mathcal O(\chi)$. In particular, at each interaction between the EM waves and the mirrors a small quantity of EM energy will be transformed into DP. The amplitude of such a process is proportional to $\chi$ and therefore neglected as it contributes terms of order $\mathcal O(\chi^2)$. Furthermore the corresponding energy loss is much smaller than the one coming from the finite reflection coefficient $r$. 

\section{\label{Total_E_field_DP}Total electric field squared}

As mentioned in the beginning of this chapter, the main idea of the experiment proposed is to detect the hypothetical electric field induced by the DP field by using atoms to measure it through the quadratic Stark effect. The Stark effect consists in a shift in the energy levels of an atom under the perturbation of a static (or slowly evolving\footnote{As long as the angular frequency of oscillation is much smaller than the atomic transition angular frequency from state k to i, $\Delta \omega \ll \omega_{ik}$.}) electric field, and is given by \cite{cohen-tannoudji:1986aa}
\begin{equation}\label{stark}
\Delta \nu = -\frac{\Delta \alpha}{2h} \left| \vec E \right|^2  \, ,
\end{equation}
at quadratic order in the electric field, thus the quadratic Stark effect, where $h$ is the Planck constant, $\Delta \nu$ is the frequency shift induced by the slowly evolving electric field $\vec E$ and $\Delta \alpha$ is the differential polarizability of the atomic transition considered. 
Taking into account both contributions from the applied electric field and the DP field computed previously, the total electric field power at the center of the cavity is
\begin{align}
&|\vec E(\omega_U,\omega_A)|^2 =  \left|\vec E^\mathrm{tot}_A +\vec E^\mathrm{tot}_\mathrm{DM}\right|^2 =\left(\vec A(\omega_A)\cdot \vec C(\omega_U)+\vec B(\omega_A)\cdot \vec D(\omega_U)\right)\cos(\Delta\omega t+\phi_A)+\, \label{general_E_power} \\
&\left(\vec B(\omega_A)\cdot \vec C(\omega_U)-\vec A(\omega_A)\cdot \vec D(\omega_U)\right)\sin(\Delta\omega t+\phi_A) + \mathrm{constant \ and \ fast \ oscillating \ terms} \, \nonumber , 
\end{align}
with $\Delta \omega = \omega_A - \omega_U$. In the following, we will not consider the constant terms. Indeed, in the experimental scheme proposed here, we will be interested in the oscillatory behavior of the atomic frequencies. Moreover, we discarded the fast oscillating terms whose angular frequencies are $2\omega_A$, $2\omega_U$ or $\omega_A+\omega_U$, with periods ($\mathcal{O}(10^{-9})$ s) much shorter than the atom interrogation time, such that on average, their impact vanishes.

From Eq.~\eqref{general_E_power}, the signal amplitude can be written as (see Appendix ~\ref{ap:amp_field_cavity} for more details)
\begin{subequations}
\begin{align}
    \label{eq:Etot2}
    \frac{\sqrt{1-r^2}X_AX_\mathrm{DM}\beta}{\sqrt{1+2r\cos(\frac{\omega_A L}{c})+r^2}}\sqrt{1+4\frac{1+(1+r)\cos(\frac{\omega_U L}{2c})}{1+2r\cos(\frac{\omega_U L}{c})+r^2}} \equiv \chi S(\omega_U,\omega_A;\rho_\mathrm{DM}, X_A;L,r)\, ,
\end{align}
where 
\begin{align}
\label{beta_DP}
    \beta &= \hat e_\mathrm{DM} \cdot \frac{\vec X_A}{X_A} \, ,
\end{align}
\end{subequations}
and $X_\mathrm{DM} \propto \chi$, following Eq.~\eqref{amp_E_field}.

If the polarization is fixed and does not change each coherence time, $\beta = \cos\theta$, with $\theta$ the angle between $\vec X_A$ and $\vec X_\mathrm{DM}$. If the DP field is isotropically distributed $ \beta=1/\sqrt{3}$. To avoid any orthogonality between the two polarizations, i.e $\beta =0$, one must run the experiment for a significant time, i.e at least several days (see discussion in \cite{Caputo}).
\begin{figure}[ht!]
    \centering
    \begin{subfigure}{.5\textwidth}
        \centering
        \includegraphics[width=.8\linewidth]
        {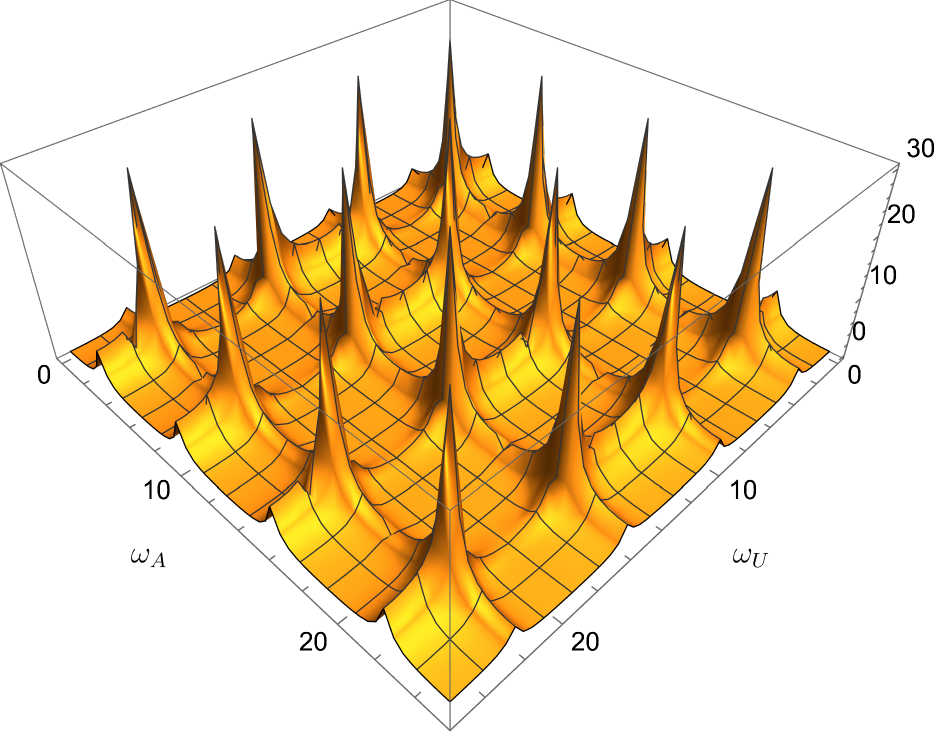}
    \end{subfigure}\hfill
    \begin{subfigure}{.5\textwidth}
        \centering
        \includegraphics[width=.65\linewidth]{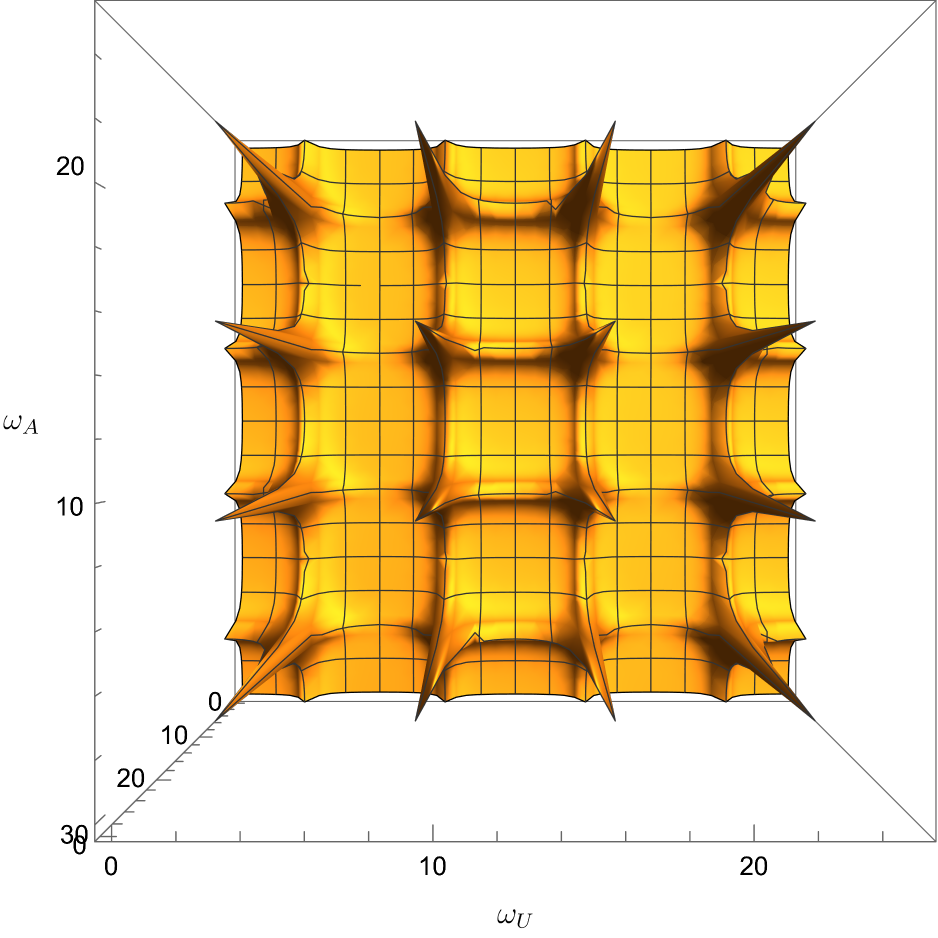}
    \end{subfigure}
    \caption{Signal contribution Eq.~\eqref{eq:Etot2} (arb. units) as function of $\omega_U, \omega_A$, both presented in units of $c/L$. The resonance peaks appear clearly when both frequencies correspond to an odd mode of the cavity.}
    \label{3D_signal}
\end{figure}
In Fig.~\ref{3D_signal} are shown 3D plots of the signal contribution Eq.~\eqref{eq:Etot2} as function of the angular frequencies of the DM field $\omega_U$ and the applied field $\omega_A$. One can notice the various resonance peaks when both frequencies correspond to an odd mode of the cavity. In the experiment scheme presented here, we require the two frequencies to be close, hence we are only interested in regions where $\omega_U \sim \omega_A$. This corresponds to the diagonal of the plot seen from the top. However, in other types of experiment where both frequencies could be separated by some hundreds of MHz or more, other regions of this plot could be studied. 

The amplitude of the signal is linear in $\chi$ and depends non-linearly on the local DM density, on the amplitude and frequency of the applied external field and on the mass ($m_Uc^2=\hbar \omega_U$) of the DP field. It depends also on some geometric factor of order unity that characterizes how the detector couples to the DP field polarization.

For DM and applied frequencies close to odd resonances of the cavity, i.e $\omega_U L/c \approx \omega_A L/c \approx \pi + 2n\pi$, with $n \in \mathbb{N}$, the amplitude Eq.~\eqref{eq:Etot2} becomes proportional to the quality factor $Q$, as expected. Additionally, as it is also proportional to the injected amplitude $X_A$, we directly see the interesting feature of looking at the field squared amplitude, the applied field acting as an amplification for the DM field.

Finally, the signal we are interested in consists in atomic frequency oscillations
with respect to an unperturbed reference 
\begin{subequations}
\begin{align}\label{signal_stark}
    \nu(t) = \nu_0 + \Delta\nu \cos\left(\Delta \omega t +\phi\right) \, ,
\end{align}
whose amplitude is given by
\begin{align}\label{signal}
\Delta \nu = - \chi \frac{\Delta\alpha S(\omega_U,\omega_A;\rho_\mathrm{DM},  X_A;L,r)}{2h} \, ,
\end{align}
\end{subequations}
which depends on both the applied external field and the DP field.

\section{Measurement with atoms through the quadratic Stark effect}

\subsection{\label{sec:Rydberg_presentation}Rydberg atoms}

The transition frequency measurement can be performed using a regular atomic clock, which allows very good uncertainty on the frequency measurement, but is not very sensitive to the Stark effect e.g. for the $5s^2\,^1S_0 \to 5s5p\,\,^3P_0$ clock transition in Sr the differential polarizability $\Delta \alpha/2h \approx 3.1\times 10^{-6}$~Hz/(V/m)$^2$ \cite{Middelmann}, thus requiring a strong applied field.

To overcome this lack of sensitivity to $E^2$, ``regular'' atoms can be replaced by Rydberg atoms, which are in a quantum state with high principal quantum number $n$ \cite{Gallagher_1988}. The electrons are much further from the nucleus thus the atom has much higher polarizability. For large $n$ the corresponding polarizabilities in Sr can reach $\Delta\alpha/2h \approx 10^{5}$~Hz/(V/m)$^2$ ($n>60$) \cite{Millen2011}.

Typically, $Sr$ Rydberg measurements use laser cooled $Sr$ atoms that are excited to Rydberg states using two photon transitions \cite{Millen2011,Bridge2016} and direct spectroscopy of e.g. the $5s5p\,^3P_1 \to 5sns\,^3S_1$ transition is performed, with $n$ up to 81 \cite{Millen2011,Bridge2016}.
We give a more explicit summary of the measurement process done in \cite{Bridge2016}. The atoms are initially in a given atomic state $|i\rangle$ and they transition to a Rydberg state $|r\rangle$ using a UV beam whose frequency (of the order of some hundreds of THz) is continuously controlled to match the transition frequency, that can be perturbed by, e.g., environmental noise. More explicitly, this UV beam is locked on a reference cavity, whose length is modified using a piezo-tuning. This spectroscopic measurement allows one to determine the Rydberg transition frequency with or without the effect of the DP-EM coupling. 
Now, let us add the effect of such coupling. The quadratic electric field induces a Stark shift on the Rydberg state, implying that at a given time $t$, the transition frequency corresponds to $\nu(t)$ from Eq.~\eqref{signal_stark}. Therefore, in order to maximize the transition probability, the UV light will continuously follow $\nu(t)$, and therefore it will also slowly oscillate at $\Delta \omega$.  Afterwards, when the electrons have transitioned to the Rydberg state, another laser pulse is used to ionize them and a small electric field is used to drift the ions to a metallic plate and count them. By counting the ions, one can infer the number of atoms which made the transition to the Rydberg state and therefore measure their transition frequency from the UV beam.

\subsection{\label{bandwidth_exp}Bandwidth of detection}

In order to see the oscillations at $\Delta\omega$ in the total field power Eq.~\eqref{general_E_power}, we require the DM and the applied fields to have different frequencies. We also require the Nyquist frequency of the apparatus to be higher than the angular frequency of the signal $\Delta \omega < \pi f_s$ to be able to detect any oscillatory behavior in the transition frequency of the atoms.

For a given applied angular frequency $\omega_A$, sampling frequency $f_s$, Rydberg atoms perform the measurements of the electric field squared during $T_\mathrm{obs}$ at an angular frequency $\Delta \omega$ (more precisely, $T_\mathrm{obs} \times f_s$ measurements will be taken for each $\omega_A$). The time $T_\mathrm{obs}$ is arbitrary; if it is longer than the measurement process comprising excitation and ionisation, one has to prepare again the atoms to their Rydberg state after de-excitation accordingly. As detailed above, the experiment is sensitive to any $\Delta \omega$ such that $2\pi/T_\mathrm{obs} \leq |\Delta \omega| \leq \pi f_s$, which, in terms of DM angular frequency, is equivalent to
\begin{align}\label{eq:DP_Rydberg_bandwidth}
\omega_U  \in [\omega_A - \pi f_s;\omega_A - \frac{2\pi}{T_\mathrm{obs}}] \cup [\omega_A + \frac{2\pi}{T_\mathrm{obs}};\omega_A + \pi f_s] \, .
\end{align}
This consists in one measurement sequence which will probe the range of DP masses corresponding to Eq.~\eqref{eq:DP_Rydberg_bandwidth}. The logic of the experiment is then to shift the frequency of the applied electric field to scan another part of the mass range. By applying this procedure for various well chosen $\omega_A$, one can scan a large part of the DP mass range.

This scheme can be repeated N times, as much as time allows. At the end of this loop, the total experimental time is simply $T_{\mathrm{tot}} = NT_\mathrm{obs}$. The corresponding total DM frequencies band scanned is $f_{\mathrm{tot}} = Nf_s = T_{\mathrm{tot}}f_s/T_\mathrm{obs}$. The larger the total experimental time, the larger the band of scanned DM frequencies. Additionally, the blind spots at exact $\omega_A$ can be avoided, and sensitivity can be optimised (see below) by shifting $\omega_A$ by a little less than $2\pi f_s$, at the expense of increasing the overall experimental duration. 

For DM frequencies in the GHz range, the coherence time of the DM field Eq.~\eqref{coherence_DM} $\tau(\omega_U/2\pi) \sim 10^{-3}$ s, i.e it will be considered much smaller than $T_{\mathrm{obs}}$, of the order of several seconds. 

Here, we have presented the experimental scheme of an experiment involving Rydberg atoms inside a microwave cavity for the detection of the EM-DP coupling $\chi$. In Chapter ~\ref{chap:exp_summary}, we will discuss the sources of noises of the apparatus, which will allow us to derive an expected sensitivity of the experiment in Chapter ~\ref{chap:sens_experiments}.

\clearpage
\pagestyle{plain}
\printbibliography[heading=none]
\clearpage
\pagestyle{fancy}

\part{Dish antennas detectors}
\chapter{\label{chap:SHUKET_exp}Search for dark photons using dish antennas}

As a reminder, DP will induce an electric field everywhere in space due to its kinetic mixing with EM. In Chapter ~\ref{chap:Rydberg_exp_DP}, we focused on an experiment to enhance this electric field with a cavity. In this chapter, we will focus on a different way to enhance this electric field, using a spherical dish antenna. Indeed, one can use a dish antenna to focus the small electric field in order to enhance it inside a detector located at the curvature center of the mirror \cite{SHUKET,Tokyo1,Tokyo2,Tokyo3,Tokyo4,Qualiphide, Kotaka23, Knirck23, Bajjali23}. One of the main assumptions used in existing studies is that the power generated by the dish is entirely focused on the curvature center, hence can be detectable by the detector \cite{Horns}. 

In this chapter, we will show that this assumption is too optimistic and we will derive an improved modeling for such experiments. More precisely, we will derive analytically the EM field measured by the antenna detector which is highly dependent on the DP Compton frequency. This is done in two steps: (i) we compute the electric field at the location of the antenna and (ii) we determine how much of this electric power is measured by the antenna. This study is the object of an article \cite{Gue:2024gws} published in \textit{Physical Review D}.

\section{Electromagnetic emission from the dish}

\begin{figure}[h]
    \centering
    \includegraphics[scale=0.4]{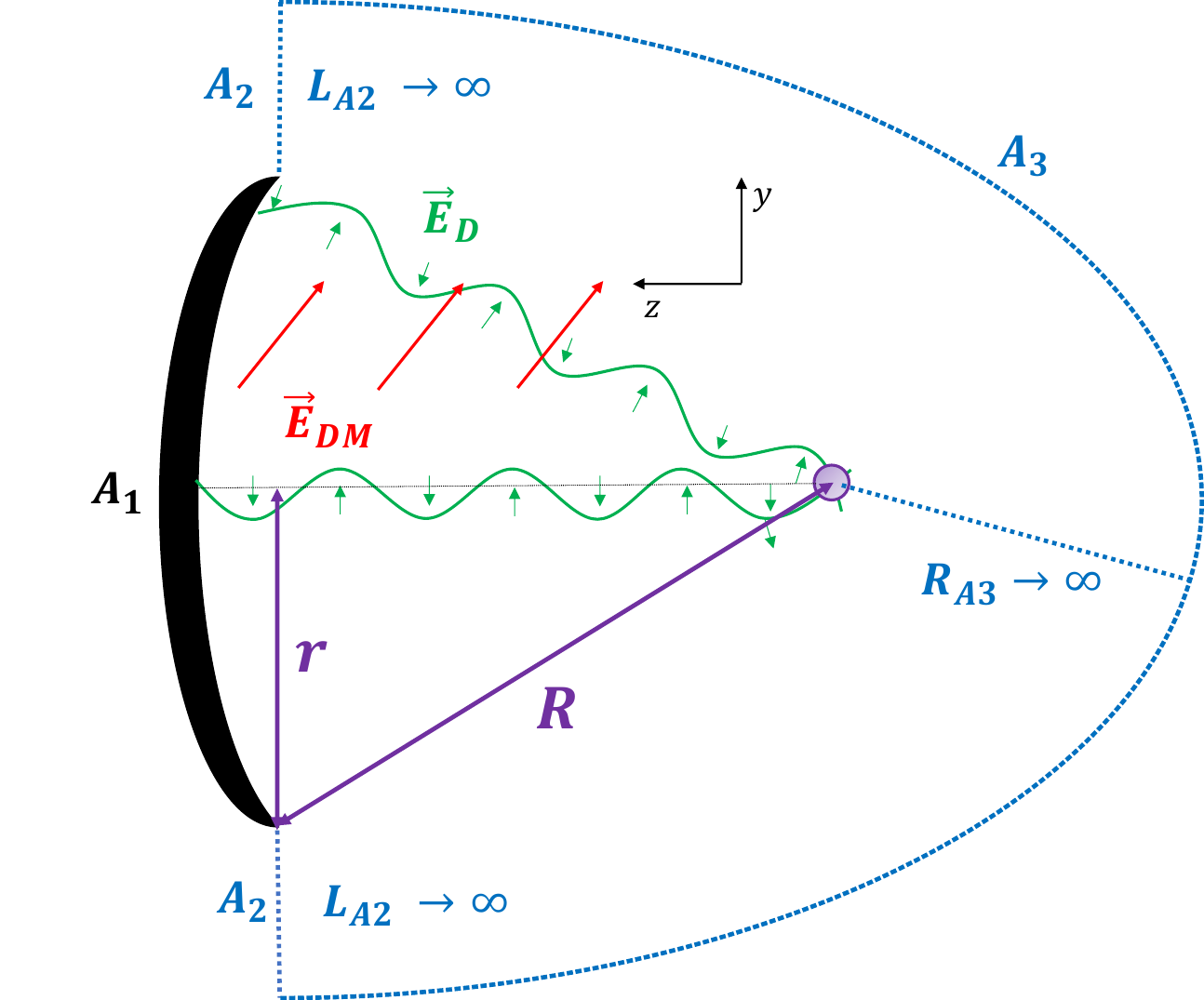}
    \caption{The oscillating standing DP electric field $\vec E_\mathrm{DM}$ is shown with red arrows. The dish emitter in black, with curvature center shown in purple, acts as reflector and emits a classic propagating electric field $\vec E_\mathrm{D}$ in green. The Kirchhoff integral boundary surfaces $A$ considered for the computation of the electric field Eq.~\eqref{E_Dirichlet} are shown in blue (in addition to the dish, surface $A_1$). It consists of the junction of a semi-sphere centered on the curvature center of the dish with infinite radius (surface $A_3$), and one additional vertical infinite plane surfaces to close the boundary surface (surface $A_2$). This closed surface is chosen such that the contributions from surface $A_2$ and $A_3$ are zero at the curvature center of the dish.}
    \label{fig:SHUKET_Kirchhoff}
\end{figure}

For our analysis, we consider a dish of radius $r$ and curvature radius $R$, as depicted in Fig.~\ref{fig:SHUKET_Kirchhoff}, which works as an EM reflector of the oscillating electric field Eq.~\eqref{eqs:EDM}. Indeed, boundary conditions $\vec E_\mathrm{tot,\parallel} = 0$ at the edge of the dish requires the emission of a regular EM wave, at every point of the dish, directed according to the normal vector of the dish, and with same frequency as the background DM oscillating field, i.e it has the form
\begin{equation}
  \vec E_{D}(\theta,\varphi)=i\chi\omega_U \vec Y_{\parallel,D}(\theta, \varphi) e^{-i\omega_U t} \, ,
  \label{eq:E_out}
\end{equation}
with $\vec Y_{\parallel,D}(\theta,\varphi)$ the component of the DP polarization vector $\vec Y$ defined in Eq.~\eqref{eq:osc_DP}, parallel to the radial vector at cylindrical coordinates $(\theta, \varphi)$ such that the electric field parallel to the dish vanishes at the mirror's surface. In addition, we choose a coordinate system whose origin is at the curvature center of the dish.  

In the literature (see e.g \cite{Horns}), one of the main assumptions used in this kind of experiment is that provided that the diameter of the dish $d_\mathrm{dish} \gg \lambda$, the wavelength of the emitted field (i.e the Compton wavelength of the DP field), diffraction effects are negligible and the power generated by the dish is entirely focused on the curvature center of the mirror \cite{Horns}. If we further assume that the field is homogeneous over the surface of the spherical mirror $A_\mathrm{dish}$ (i.e the DM field coherence length $\ell_\mathrm{DM} = \hbar/m_\mathrm{DM}v_\mathrm{DM} \gg d_\mathrm{dish}$), this total power emitted is \cite{Horns}
\begin{align}\label{power_emit_dish}
    P_\mathrm{emit} &= A_\mathrm{dish} \langle |E_D(\theta,\phi)|^2\rangle = \langle\beta^2\rangle A_\mathrm{dish} \chi^2 \rho_\mathrm{DM} c 
\end{align}
where $\beta$ is the projection of the DP field on the surface of the dish, similarly defined as in Eq.~\eqref{beta_DP} and $\langle \rangle$ represents the average value over the surface. Assuming random polarization scenario, $\beta = 2/3$, as we consider a planar emitter. 

While all previous analysis were based on these assumptions, we will show in this chapter that they are over-optimistic. Indeed, we will provide a more detailed modeling of such experiments by deriving properly the propagation of this emitted field from the dish using Kirchhoff integral. We will show that the amount of power received at the curvature center does depend on a various number of parameters, mainly the Compton frequency of the DP field, and that in some cases, reaches only a few percent of the emitted power, even when the radius of the dish is larger than the wavelength by more than one order of magnitude, i.e when diffraction effects appear negligible.

\section{\label{sec:propag_field}Propagation of the field from the dish to an antenna}
In this section, we will derive the expression of the total electric field induced by the reflection of the DP field by the dish at a given location. In Sec.~\ref{sec:Kirchhoff}, we will give a brief overview of the Kirchhoff method and show that it cannot be directly used to get an analytical expression of the electric field. Instead, we will decompose the problem into two sub-problems. First, we will propagate the field from the dish to the plane closing the dish (displayed in orange in Fig.~\ref{fig:Dish_plane_Vinet}) using an approximation of the Kirchhoff method valid for thin optical element as presented in Sec.~\ref{sec:dish_plane}. Subsequently, in Sec.~\ref{sec:plane2receptor}, we will propagate the electric field from the plane to the position of the detector using exactly the Kirchhoff method.

\subsection{Kirchhoff integral}\label{sec:Kirchhoff}

The Kirchhoff theorem allows one to solve the wave equation at a given position $\vec x$ by computing an integral over a closed surface around $\vec x$ \cite{Jackson}. In practice, this means that one can derive the electric field at position $\vec x$ as function of the field at all points on this closed surface.  

In this formalism, the temporal dependence of the field is separated from its spatial dependence. Therefore, we decompose the emitted electric field at the dish's surface  as 
\begin{align}
\vec E_D(\vec x',t) &=  \Re[\vec U_D(\vec x')e^{-i\omega t}]\, , \label{eq:Uout}
\end{align}
where the complex function $\vec U_D(\vec x')$ denotes the spatial part of the field and $\vec x'$ is a point on the dish's surface.

We now consider a closed surface, denoted A, which encloses the point $\vec x$ where the value of the field is calculated.
From Kirchhoff integral theorem, the general expression of $\vec U_D$ is \cite{Jackson},
\begin{align}\label{vector_Kirchhoff}
\vec{U}_D(\vec{x}) &= \int_{A} \mathrm{dS'} \left(\vec{U}_D(\vec{x'})(\hat{n}'\cdot \vec \nabla G(\vec{x},\vec{x}')) - G(\vec{x},\vec{x}')(\hat{n}' \cdot \vec \nabla) \vec{U}_D(\vec{x'})\right)\, ,
\end{align}
where $\hat n'$ is a unit vector normal to the surface A directed inwardly, the derivatives are with respect to the emission coordinates $x'$ and $G$ is a Green function, appropriately defined for the situation.

In the situation depicted in the previous section, from Eq.~\eqref{eq:E_out}, we know exactly the components of the electric field at the surface of the emitting dish. Therefore, an appropriate Green function is the Dirichlet Green function $G_D$ \cite{Jackson}
\begin{align}
G_D(\vec{x},\vec{x}')= 0 \quad \forall \: \vec x' \in A \, .  
\end{align}
Then, Eq.~\eqref{vector_Kirchhoff} becomes
\begin{equation}\label{E_Dirichlet}
\vec U_D(\vec{x}) = \int_{A} \mathrm{dS} \left(\vec U_D(\vec{x'}) (\vec \nabla G_D(\vec{x},\vec{x}')\cdot \hat{n}')\right)\, .
\end{equation}

In the following of the chapter, unless otherwise specified, whenever we mention electric fields, we refer to $\vec U_D$, i.e its spatial part, following Eq.~\eqref{E_Dirichlet}.

In the present situation, the closed surface A is appropriately defined by the junction of the dish, a surface with radius $R\rightarrow \infty$ and one additional plane surface along $(x,y)$ plane to close the surface, such that only the field emitted by the dish will contribute. The Kirchhoff integral surface boundary is shown in blue in Fig.~\ref{fig:SHUKET_Kirchhoff}.

Our system is therefore composed of a spherical dish and Eq.~\eqref{E_Dirichlet} requires the derivation of a Dirichlet Green function for a portion of sphere. However, no analytical expression has been found for such geometry, which makes it impossible to find an exact analytical solution for our problem. Instead, we will decompose the problem into two sub-problems: using an approximation, we will propagate the electric field from the dish to the plane closing the dish (see Fig.~\ref{fig:Dish_plane_Vinet}) and then, using the Kirchhoff integral, from the plane up to the detector.

\subsection{\label{sec:dish_plane}Propagation of the electric field from the dish to the plane}

\begin{figure}
    \centering
    \includegraphics[scale=0.22]{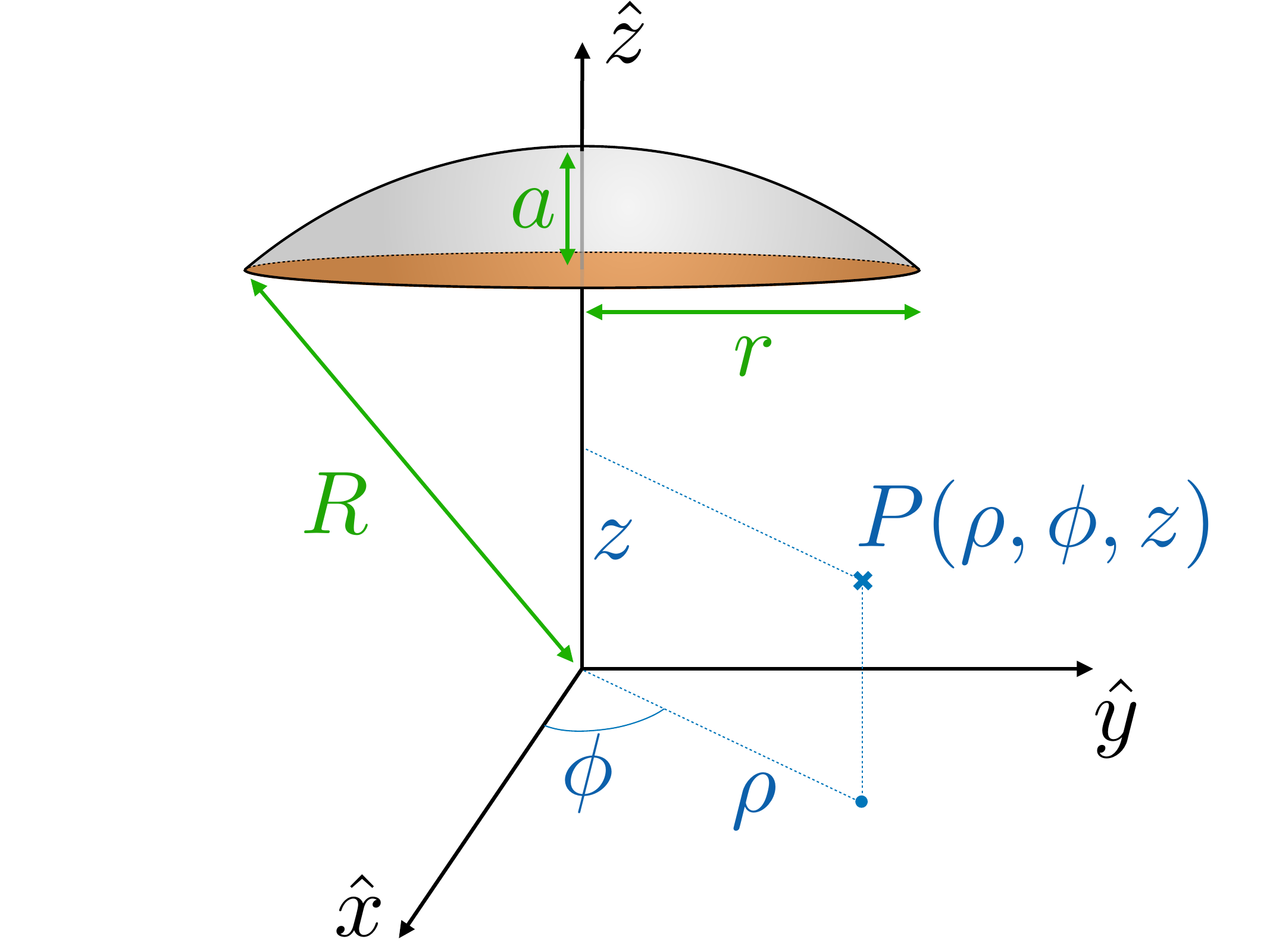}
    \caption{To compute the electric field induced by the dish at a location $P$ at cylindrical coordinates $(\rho,\phi,z)$, we first propagate the  electric field emitted by the spherical dish (in light grey) into the fictional plane (in orange) following Eq.~\eqref{Vinet}. Then, we propagate this electric field to the receiver using the Kirchhoff integral. This procedure is valid only for a spherical dish with low curvature, i.e whose radius $r$ is much smaller than its curvature radius $R$.}
    \label{fig:Dish_plane_Vinet}
\end{figure}

This situation is depicted in Fig.~\ref{fig:Dish_plane_Vinet}, where the distance between the fictional plane (orange on the figure) and the dish at coordinates $x=y=0$ is noted $a$.

As mentioned above, it is not possible to find an analytical solution to Eq.~(\ref{E_Dirichlet}) for this particular geometry. Instead, we use the thin optical element approximation presented in details in Sec. 2.2.7 from \cite{Vinet} which allows one to find an analytical expression for the field $\vec U(\vec x)$ of Eq.~\eqref{E_Dirichlet} on this fictional plane. This approximation is valid if the two following conditions are fulfilled:
\begin{itemize}
    \item The transverse propagation modes $p,q$ (which can be viewed as the coordinates of the wavevector in the $x-y$ plane \cite{Vinet}) need to be much smaller than the longitudinal one $k$. In our case, the transverse modes are determined by the spatial extent of the dish antenna in the $x-y$ plane $\Delta x, \Delta y$, with $p \sim q \sim 1/\Delta x \sim 1 \: \mathrm{m}^{-1} \ll \omega_U/c = k \sim 10^3 \: \mathrm{m}^{-1}$, in the GHz range. In addition, the  galactic velocity distribution in the DM halo induces a transverse contribution \cite{An23}, of the order of $\omega_U v_\mathrm{DM}/c^2 \sim 3 \times 10^{-3} \: \mathrm{m}^{-1} \ll \omega_U/c\sim k$. Therefore, this condition is fulfilled.
    \item The dish needs to have a low curvature, or in other words, the radius is much smaller than the curvature radius $r \ll R$. Therefore, in the following, we will restrict ourselves to low curvature dish emitter. 
\end{itemize}
A straightforward but lengthy calculation presented in details in \cite{Vinet} shows that, under the conditions detailed above, the Kirchhoff integral Eq.~\eqref{E_Dirichlet} simply reduces to 
\begin{align}
\vec U_P\left(x,y,z_\mathrm{plane}\right) &= -e^{ikf(x,y)}\vec U_D\left(x,y,z_\mathrm{dish}\left(x,y\right)\right)\, ,
\label{Vinet}
\end{align}
where $\vec U_D$ is the electric field emitted by the dish and $\vec U_P$ is the electric field at the location of the closing plane. Furthermore, the function  $f(x,y)=z_\mathrm{dish}\left(x,y\right)-z_\mathrm{plane}$ is the surface equation of the dish. In other words, within the approximation of thin optical element, the curvature of the dish translates into simple phase factors determined by the distance between the closing plane and the dish \cite{Vinet}. 

We now introduce a cylindrical coordinate system ($\rho,\phi,z$), see Fig.~\ref{fig:Dish_plane_Vinet}. 
Because of the cylindrical symmetry, the surface equation that writes
\begin{align}
    f(x,y)&=f(\rho) = \sqrt{R^2-\rho^2}-R+a \approx \frac{r^2 -\rho^2}{2R} \, ,
    \label{eq:f}
\end{align}
depends only on the cylindrical coordinate $\rho$, where we used $a \approx r^2/2R$, since the dish is a portion of sphere, with small curvature. 

Inserting the spatial part of the field on the dish from Eq.~\eqref{eq:E_out} into Eq.~(\ref{Vinet}), we can now estimate the expression of the field in this fictional plane as
\begin{equation}\label{eq:E_plane}
\vec U_P(\rho,\phi,z_\mathrm{plane})=-i\chi \omega_U e^{ikf(\rho)}\vec Y_{\parallel, D}(\rho,\phi,z_\mathrm{plane}+f(\rho))\, ,
\end{equation} 
where $\rho\leq r$ and $z_\mathrm{plane}=R-a$ is the $z$-coordinate of the plane. This expression is only valid in the thin optical element approximation, whose conditions are detailed at the beginning of this section.

\subsection{\label{sec:plane2receptor}Propagation of the electric field from the plane to the receiver}

We can now focus on the propagation from an emitting plane, a subject that has been vastly treated in the literature, e.g in \cite{Jackson, Vinet}. The idea is to use the Kirchhoff integral from Eq.~(\ref{vector_Kirchhoff}) using an appropriate Dirichlet Green function that vanishes on the plane. In the case of a plane emitter located at $z_\mathrm{plane}=R-a$, such a Dirichlet Green function is given by
\begin{align}
G_D(\vec x,\vec x')=\frac{e^{ik L'}}{4\pi L'} -\frac{e^{ik L''}}{4\pi L''}\, ,
\label{Dirichlet}
\end{align}
where \cite{Jackson,Vinet}
\begin{subequations}\label{eq:Green_distances}
\begin{align}
L' &= |\vec x - \vec x'| = \sqrt{\rho^2+\rho'^2-2\rho\rho'\cos(\phi-\phi')+(R-a+\Delta z-z')^2}\,\\
L'' &= |\vec x - \vec x''| = \sqrt{\rho^2+\rho'^2-2\rho\rho'\cos(\phi-\phi')+(R-a-\Delta z-z')^2}\, ,
    \end{align}
\end{subequations}
with $\vec x''$ the point symmetrical to $\vec x'$ with respect to the plane. In Eq.~\eqref{eq:Green_distances}, we defined
\begin{align}
    \Delta z &= z- (R-a) \, ,
\end{align}
the difference of $z$ coordinates between the receiving point $z < z'$ and the emitter on the plane. One can easily show that this property leads to $L'=L''$ if $x'$ is on the plane and therefore the required condition $G_D(\vec x, \vec x_\mathrm{plane}) = 0$ is satisfied. To compute Eq.~\eqref{E_Dirichlet}, we use the directional derivative since $\hat n' = \hat z'$
\begin{subequations}
\begin{align}
&\vec \nabla G_D(\vec{x},\vec{x}')\cdot \hat{n}' = \frac{\partial G_D}{\partial z'}\,\\
&=-\frac{(ik L'-1)(R-a+\Delta z-z')e^{ik L'}}{4\pi L'^3}+\frac{(ik L''-1)(R-a-\Delta z-z')e^{ik L''}}{4\pi L''^3}\, .
\end{align}
\end{subequations}
Now, plugging $z'= z_\mathrm{plane}$ and considering reception at $\vec x=(\rho,\phi,z)$, we get
\begin{align}
\frac{\partial G_D}{\partial z'}\Big|_\mathrm{z' \ \in \ plane}&= -\frac{(ik L-1)\Delta z}{2\pi L^3}e^{ik L} \, ,
\label{Green_plane}
\end{align}
with $L = \sqrt{\rho^2+\rho'^2-2\rho\rho'\cos(\phi-\phi')+(\Delta z)^2}$.

Then, the electric field on the fictional plane propagating towards $z<R-a$ reads
\begin{align}
\vec U_\mathrm{dish}(\rho,\phi, z) &\approx -\frac{i\chi \omega_U \Delta z}{2\pi}\int_0^r  d\rho' \rho'e^{ikf(\rho')} \int_0^{2\pi}d\phi' \frac{ikL-1}{L^3}e^{ikL}\vec Y_{\parallel, D}(\rho',\phi',f(\rho')+R-a) \, ,
\label{E_field_complete}
\end{align}
This result provides the expression of the electric field induced by the DP field reflected by the dish anywhere in space under the approximation that the dish is a thin optical element. In particular, this formula includes diffraction effects that were implicitly neglected in previous studies \cite{Horns}. This integral is generally not solvable analytically and some approximations might be made to simplify it.

\section{\label{sec:detection_field}Detection of the electric field with a horn antenna}

While the previous section was devoted to the emission and propagation of the electric field induced by the dark photon and enhanced by the dish, we will now focus on the detection of this electric field.  In this section, we consider the detection system to be a horn antenna of long side $A$ and small side $B$.
At the output of the antenna is located a resistance $R_0$ such that when an oscillating electric field is applied to the horn antenna, it is translated into a measurable voltage.

We will compare two different approaches to estimate this measurement: the first one is based on an analytical calculation which computes the overlap integral between the electric field at the location of the antenna and the antenna mode ; the second one consists in using the antenna gain factor provided by the antenna's manufacturer.

\subsection{Computation using the modes overlap}

Let us first consider an  antenna of internal resistance $R_0$ as an emitter by applying  an oscillating voltage $V(t) = V_0 \cos (\omega_U t)$ to its terminals. The reciprocity theorem ensures that this will be equivalent to considering the antenna as a receiver. As a result, the antenna will emit an electric field predominantly in the TE$_{10}$ mode, i.e. polarized parallel to the small side of the rectangular horn, which is the most widely used fundamental mode for pyramidal horn antennas \cite{Balanis05}. \\
\newline
\begin{figure}[h!]
\begin{minipage}{\textwidth}
  \begin{minipage}{0.42\textwidth}
    \centering
    \includegraphics[width=\textwidth]{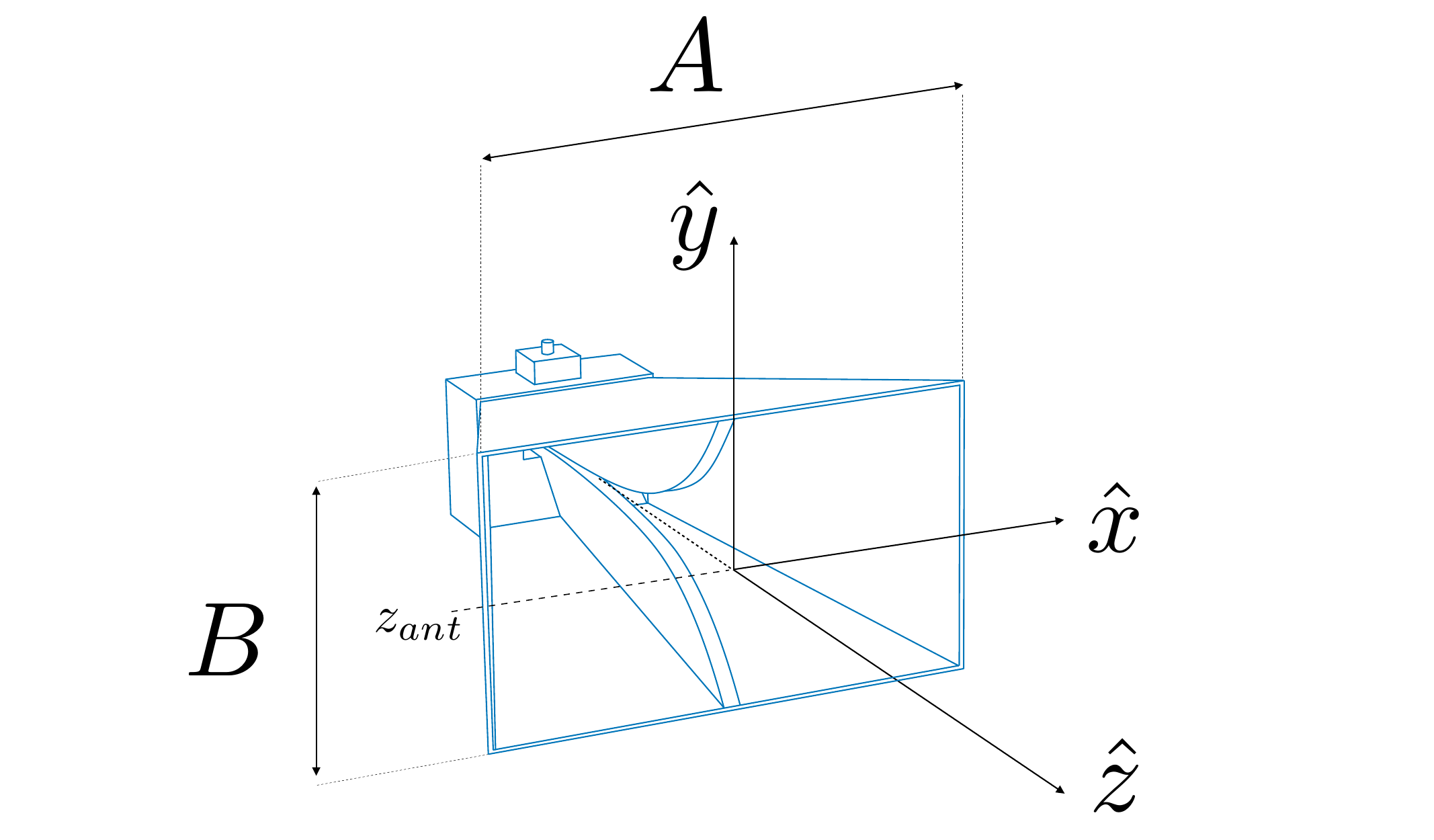}
    \caption{Rectangular horn antenna and definition of its physical surface aperture.}
    \label{fig:antenna}
    \end{minipage}
    \hfill
    \begin{minipage}{0.55\textwidth}
    Therefore, in a coordinate system with the $z$-axis perpendicular to the surface aperture of the antenna, as shown in Fig.~\ref{fig:antenna}, the electric field generated by the antenna can be written as 
    \begin{align}\label{eq:E_mode_ant}
    \vec E_\mathrm{ant}(t,x,y,z) &= \Re\left[V_\mathrm{ant} \vec M_\mathrm{ant}(x,y) e^{i(kz-\omega_U t+\phi)}\right] \, ,
    \end{align}
    where $\phi$ is the phase of the electric field and $V_\mathrm{ant}$ characterizes the amplitude of this field (Energy conservation implies that $V_0^2/2R_0=V_\mathrm{ant}^2/2Z_0$, where $Z_0=376.7 \: \Omega$ is the vacuum impedance.). \\
    \newline
  \end{minipage}
\end{minipage}
\end{figure}
\newline
In general, horn antennas make an axial detection of the field, i.e. only on one single axis, the small axis of the rectangular horn, the $\hat y$ axis. The real antenna mode can be written as
\begin{subequations}
\begin{align}\label{eq:mode_ant}
\vec M_\mathrm{ant}(x,y) &= m_{TE_{10}}\hat{y}\cos\left( \frac{\pi x}{A_\mathrm{eff}} \right) \,,
\end{align}
where $A_\mathrm{eff}$ is the effective width and $m_{TE_{10}}$ (dimension L$^{-1}$) is a normalisation factor that ensures the mode is normalized $\int dS_\mathrm{eff} |\vec M_\mathrm{ant}(x,y)|^2 = 1$ on the effective aperture of the antenna, i.e.
\begin{align}\label{eq:m_value}
    m_{TE_{10}} = \sqrt{\frac{2}{S_\mathrm{eff}}} \,.
\end{align}
\end{subequations}
The effective aperture of the antenna $S_\mathrm{eff}$ depends on the frequency $f_U=\omega_U/2\pi$ of the field and in general differs from the physical aperture of the antenna. Indeed, the effective area of the horn, which can be expressed as the product of an effective width $A_\mathrm{eff}$ with an effective height $B_\mathrm{eff}$, depends on the frequency as \cite{Balanis05}
\begin{align}
S_\mathrm{eff}(\omega_U) &= A_\mathrm{eff}(\omega_U)B_\mathrm{eff}(\omega_U)=\frac{e_r\pi G(\omega_U)c^2}{\omega^2_U} \, ,
\label{eq:effective_surface}
\end{align}
where $e_r$ is the realistic efficiency of the antenna, which represents reflection losses inside the antenna, and where $G(\omega)$ is the frequency dependent gain of the antenna \cite{Balanis05}. Note that according to the IEEE Standards, the gain does not include reflection and polarization losses \cite{Balanis05}, this is why we separate $e_r$ from $G(\omega)$. The time-averaged electromagnetic power generated by the antenna is given by
\begin{equation}\label{eq:Pout}
    P_\mathrm{out} = \frac{1}{2Z_0} \int dS_\mathrm{eff} \left|\vec E_\mathrm{ant}\right|^2 =\frac{V_\mathrm{ant}^2}{2Z_0} \, .
\end{equation} 
We will now invert this reasoning and consider the antenna as a receiver that will output a voltage in response to an electric field 
\begin{align}\label{eq:E_def}
    \vec E(t,x,y,z)=\Re[\vec U(x,y,z)e^{-i\omega_U t}]= \left|\vec U\right|\hat e_U\cos(\omega_U t +\varphi) \, ,
\end{align} 
where $\varphi$ is an irrelevant phase, $\hat e_U$ is the unit polarization vector of $\vec U$, $\left|\vec U\right|=\left(\vec U\cdot \vec U^\dagger \right)^{1/2}$ denotes the complex modulus. The antenna will  measure electric fields that are propagating into the $\hat z$ direction and matching its mode $\vec M_\mathrm{ant}(x,y)$. More precisely, by taking the dot product of Eq.~(\ref{eq:E_mode_ant}) with $\vec M_\mathrm{ant}$, one gets $V_\mathrm{ant}\cos \left(\omega t -kz -\phi\right)=\int dS_\mathrm{eff} \vec M_\mathrm{ant}(x,y)\cdot \vec E_\mathrm{ant}(t,x,y,z)$. Using the reciprocity properties of electromagnetism, one can replace $\vec E_\mathrm{ant}$ by the electric field we are trying to measure Eq.~(\ref{eq:E_def}). This reasoning shows that the antenna will output a voltage proportional to
\begin{align}\label{eq:antenna_output_gen}
    V_\mathrm{ant}(z)\cos(\omega_U t+\phi'(z))=\cos(\omega_U t+\varphi) \int dS_\mathrm{eff}(\omega_U) \left|\vec U(x,y,z)\right| \hat e_U \cdot \vec M_\mathrm{ant}(x,y) \, ,
\end{align} 
where $\phi'(z)=-kz-\phi$ and the 2D integral is carried out in the $x-y$ plane at the $z$-coordinate of the antenna.

Let us factorize the electric field generated by the dish at the location of the antenna provided by Eq.~(\ref{E_field_complete}) in a voltage factor $V_\mathrm{dish}$, a real mode factor $\vec M_\mathrm{dish}$ and a plane wave factor, such that it takes the following form in the antenna aperture plane $(x,y)$
\begin{subequations}
\begin{align}\label{eq:E_mode_dish}
\vec {E}_\mathrm{dish}(t,x,y,z) &= V_\mathrm{dish} \vec{M}_\mathrm{dish}(x,y,z) \cos(\omega t+\phi'(z)) \equiv \left|\vec U_\mathrm{dish}(x,y,z)\right|\hat e_U\cos(\omega t+\varphi) \, ,
\end{align}
where $\vec{M}_\mathrm{dish}(x,y,z)$ is normalized over the infinite surface of the antenna plane $S_\infty$ at $z=z_\mathrm{ant}$ i.e
\begin{align}
    \int dS_\infty |\vec{M}_\mathrm{dish}(x,y,z_\mathrm{ant})|^2 &= 1 \, .
    \label{mode_dish_norm}
\end{align}
\end{subequations}
The constant amplitude $V_\mathrm{dish}$ in Eq.~\eqref{eq:E_mode_dish} can be computed through energy conservation: the total energy generated by the dish should equal the energy received on the infinite antenna plane $S_\infty $. Then, using Eqs.~\eqref{power_emit_dish} and \eqref{mode_dish_norm}, we have
\begin{align}
 P_\mathrm{dish} &= \int dS_\infty \frac{\left|\vec U_\mathrm{dish}\right|^2}{2Z_0} =\frac{V_\mathrm{dish}^2}{2Z_0} \Rightarrow V_\mathrm{dish} = \sqrt{\frac{4Z_0A_\mathrm{dish}\rho_\mathrm{DM}c}{3}}\chi \label{eq:Vdish} \, .
\end{align}

Then, in our experimental scheme, the time averaged power generated by the horn antenna is obtained by combining Eqs.~\eqref{eq:Pout}, ~\eqref{eq:antenna_output_gen} and \eqref{eq:E_mode_dish} and is provided by
\begin{equation}\label{eq:overlap_int}
P_\mathrm{rec}(z_\mathrm{ant},\omega_U) = \frac{V_\mathrm{dish}^2}{2Z_0}\left(\int dS_\mathrm{eff}(\omega_U) \vec{M}_\mathrm{dish}(x,y,z_\mathrm{ant})\cdot \vec{M}_\mathrm{ant}(x,y)\,\right)^2 \, ,
\end{equation}
where $\vec M_\mathrm{dish}$ depends explicitly on $z_\mathrm{ant}$ and where the integral is performed at every position $(x,y)$ on the effective surface of the horn antenna. 

Then, the ratio between the time averaged power measured by the horn antenna and the total power emitted by the dish is simply given by
\begin{align}
\gamma(z_\mathrm{ant},\omega_U)_\mathrm{Overlap} &= \left(\int dS_\mathrm{eff}(\omega_U) \vec{M}_\mathrm{dish}(x,y,z_\mathrm{ant})\cdot \vec{M}_\mathrm{ant}(x,y)\right)^2 \, ,
    \label{ratio_powers}
\end{align}
where $\vec{M}_\mathrm{dish}$ directly depends on the Kirchhoff integral Eq.~\eqref{E_field_complete}.

\subsection{Computation using the antenna factor\label{sec:antenna_factor_gen}}

Another approach to estimate the output of the horn antenna is to consider the antenna factor (AF), a characteristic of the antenna provided by the manufacturer. 

The AF is defined by considering an incident plane wave, or in other words, an incoming electromagnetic field whose mode is constant over the aperture of the antenna. It is defined by (see e.g. \cite{Mclean02})
\begin{align}
\mathrm{AF}(\omega_U) &= \frac{\left|\vec U_\mathrm{dish}(\omega_U)\right|}{V_0} = \frac{V_\mathrm{dish} M_\mathrm{dish}}{V_0}  \, , 
\end{align}
where $V_\mathrm{0}$ is the generated tension, $\vec U_\mathrm{dish}$ the amplitude of the incoming electric field, $\vec M_\mathrm{dish} = M_\mathrm{dish} \hat e_U$ from Eq.~\eqref{eq:E_mode_dish}, with ${M}_\mathrm{dish}$ the value of the constant mode of the electric field on the effective antenna area. The AF depends on the frequency of the incident field. 
It is measured experimentally and therefore takes into account any loss inside the antenna, represented by the $e_r$ parameter. 

Using energy conservation, we immediately get the expression of time-averaged generated power of the antenna
\begin{align}
P_\mathrm{rec}(z_\mathrm{ant},\omega_U) = \frac{V_0^2}{2R_0}=\frac{V^2_\mathrm{dish}M^2_\mathrm{dish}(z_\mathrm{ant},\omega_U)}{2 R_0 \mathrm{AF}(\omega_U)^2} \, .
\end{align}
However, the definition of the antenna factor assumes perfectly aligned polarization modes \cite{Mclean02}, which is not necessarily true in our case. Indeed, we assume a antenna polarization mode along the $\hat y$ axis while the electric field emitted by the dish has its polarization lying in the $(x,y)$ plane. 

If we assume the DP polarization to be isotropically distributed, the polarization of the electric field emitted by the dish has a cylindrical symmetry and a linearly polarized antenna will be sensitive only to half of the power from the electric field.
 Taking into account this additional factor, the ratio $\gamma$ becomes
\begin{align}
\gamma(z_\mathrm{ant},\omega_U)_\mathrm{AF} &= \frac{Z_0 M^2_\mathrm{dish}(z_\mathrm{ant},\omega_U)}{2 R_0 \mathrm{AF}(\omega_U)^2} \, .
\label{ratio_AF}
\end{align}

From there, we have all the necessary ingredients to estimate the power received by the antenna as function of its position, as long as $z_\mathrm{ant} < z_\mathrm{plane}$. Indeed, Eqs.~\eqref{ratio_powers} and \eqref{ratio_AF} give us what portion of the emitted power by the dish Eq.~\eqref{power_emit_dish} is actually received in the antenna wires. Therefore, from a given constraint of an experiment which assumes 100\% of power emitted is received by the antenna, one can infer how to correct this constraint using this ratio. In particular, in Chapter ~\ref{chap:sens_experiments}, we will use these results to derive the expected power received by the antenna in \textit{SHUKET} \cite{SHUKET}, an experiment using a dish and a horn antennas to detect the small electric field Eq.~\eqref{eq:E_out}. In \cite{SHUKET}, it is assumed that $\gamma=1$, i.e all the power generated by the dish is transmitted to the antenna. Using Eqs.~\eqref{ratio_powers} and \eqref{ratio_AF}, we will show that this is an over optimistic result and that the power received by the antenna is multiple orders of magnitude lower.

\clearpage
\pagestyle{plain}
\printbibliography[heading=none]
\clearpage
\pagestyle{fancy}

\part{Search through tests of the equivalence principle}
\chapter{\label{chap:Classical_tests_UFF}Classical tests of the equivalence principle}

\section{\label{sec:exp_test_UFF}Experiments testing the equivalence principle}

As described in Section ~\ref{EP_general}, EP is not a fundamental symmetry of the universe, and many Beyond the Standard Model (BSM) theories predict its violation \cite{Uzan11, Damour12,Damour94,hees18,Bertone18,Hui17}, in particular by the spacetime variation of fundamental constants such as the fine structure constant $\alpha$ or the electron mass $m_e$.


In order to quantify such violation, one can measure the acceleration of two test bodies A and B in an external gravitational field and compare them through the so-called E\"otv\"os parameter defined as
\begin{align}
    \label{Eotvos_param}
    \eta &= 2\frac{(\vec a_A - \vec a_B)}{|\vec a_A + \vec a_B|}\cdot \hat e_\mathrm{meas}
\end{align}
where $\hat e_\mathrm{meas}$ is the sensitive axis of the apparatus considered.

\subsection{\textit{MICROSCOPE}}

The \textit{MICROSCOPE} mission consists in a micro-satellite launched in circular orbit around Earth at a 710 km altitude \cite{Microscope17}. It is the first mission aiming at measuring a violation of EP in space, taking advantage of a very quiet environment and long durations of free fall.  It was launched on 25 april 2016 for a total duration of 120 orbits of the satellite around Earth \cite{Berge23}, which as we shall see in Chapter ~\ref{chap:sens_experiments} corresponds approximately to $8.1 \times 10^6$ s integration time. The satellite measured the difference of acceleration between a Platinum and a Titanium based test masses and it is sensitive to both Eqs.~\eqref{EP_viol_acc_dil} and \eqref{EP_viol_acc_axion}. Their final constraint on $\eta$ is \cite{Microscope22}
\begin{equation}\label{eq:eta_MICROSCOPE}
\eta= (-1.5 \pm 2.8)\times 10^{-15}\, ,
\end{equation}
at a 1$\sigma$ confidence-level. 

\subsection{E\"otv\"os torsion balances}

Torsion balance is an apparatus able to measure very weak forces, once used by Cavendish to measure the gravitational constant $G$, and now used for the measurement of the weak EP. In usual torsion balances experiments, a bar is suspended by a thin fiber, which acts as a torsion spring. If a force is applied to one of the ends of the bar with direction perpendicular to it, the bar rotates and twists the fiber. This way, the magnitude of this unknown force can be derived by measuring the rotation angle of the bar. By setting two different test masses with two different compositions at each end of the bar (but with the same mass), one can measure the differential gravitational acceleration that act on them and therefore a violation of the equivalence principle. The current best constraint on the E\"otv\"os parameter from Torsion balances experiments is \cite{Schlamminger08}
\begin{align}
    \eta &= (0.3 \pm 1.8) \times 10^{-13} \, ,
\end{align}
at a 1$\sigma$ confidence-level. 

\subsection{Lunar Laser Ranging}

Thanks to retro-reflectors array installation on the surface of the Moon by several american and russian space missions, starting from 1969, it is now possible to measure the Earth-Moon distance by sending light signals to this array and measure the round trip time of photons. This is what is called Lunar Laser Ranging (LLR). 

Both the Earth and the Moon have different compositions and are in the solar gravitational potential, a violation of the EP would imply the two systems fall at different rates towards the Sun. This would distort the lunar orbit, which could be measuring using LLR \cite{Williams09}.

Since it would be unsatisfactory to model planetary and stellar bodies as point-like particles, we must include the self gravitation energy of each body, which is covered by the strong EP. In other words, LLR measurements provide us a constrain on a combination of strong and weak EP, but not on one of them individually.

\section{Acceleration difference induced by oscillating rest mass}

As derived in Sections ~\ref{dilaton_pheno} and \ref{axion_gluon_pheno}, axion and dilaton fields can produce oscillations of the rest mass and transition frequencies of atoms from their different couplings with SM fields. 

We make the calculations in the laboratory frame, where the gradient of the DM field is not neglected. 
The rest mass and transition frequency oscillations of an atom A are shown in Eq.~\eqref{mass_freq_osc_dilaton}, which leads to an UFF violating acceleration Eq.~\eqref{EP_viol_acc_dil}
\begin{subequations}
\begin{align}
    \vec a_A(t, \vec x) &= \left[\omega_\phi \vec v_A-\vec k_\phi c^2\right]\frac{\sqrt{16 \pi G \rho_\mathrm{DM}}[Q^A_M]_d}{\omega_\phi c}\sin(\omega_\phi t - \vec k_\phi \cdot \vec x +\Phi) \, \\
    &\approx \frac{\sqrt{16 \pi G \rho_\mathrm{DM}}\vec v_\mathrm{DM}[Q^A_M]_d}{c}\sin(\omega_\phi t - \vec k_\phi \cdot \vec x +\Phi) \label{eq:a_UFF_lab_frame}
\end{align}
\end{subequations}
where, as explained below Eq.~\eqref{EP_viol_acc_dil}, we immediately got rid off the transition frequency contribution to the acceleration, as it is suppressed by a factor $\hbar \omega^0_A/m^0_A c^2 \sim 10^{-10}$. At the second line, we used the fact that $-\vec k_\phi c^2 = \omega_\phi \vec v_\mathrm{DM}$ and $v_\mathrm{DM} \gg v_A$.
It follows a differential acceleration between two macroscopic bodies of different composition $A$ and $B$ with the same initial velocities and position given by
\begin{subequations}\label{delta_a_UFF_osc}
\begin{align}
    \Delta \vec a(t,\vec x)_d &\approx \frac{\sqrt{16 \pi G \rho_\mathrm{DM}}\vec v_\mathrm{DM}}{c}\left([Q^A_M]_d-[Q^B_M]_d\right)\sin(\omega_\phi t - \vec k_\phi \cdot \vec x +\Phi) \, . \label{eq:delta_a_dil_lab_frame}
\end{align}
Similarly, a differential acceleration between two bodies of different composition arises through the axion-gluon coupling, as derived in Section ~\ref{axion_gluon_pheno} of the form
\begin{align}
    \Delta \vec a(t,\vec x)_a &\approx \frac{16\pi G \rho_\mathrm{DM} \vec v_\mathrm{DM}E^2_P}{f^2_a \omega_a c^2}\left([Q^A_M]_a - [Q^B_M]_a\right)  \sin\left(2\omega_a t - 2\vec k_a \cdot \vec x + 2\Phi\right) \,\label{eq:delta_a_axion_lab_frame}.
\end{align}
\end{subequations}
As we have derived it in Sections ~\ref{dilaton_pheno} and \ref{axion_gluon_pheno}, the charges $[Q_M]_d, [Q_M]_a$ are atom dependent, as they depend on the mass and charge number of the atom. Therefore, the differential accelerations Eq.~\eqref{delta_a_UFF_osc} are non zero, and lead to a violation of the UFF, i.e $\eta \neq 0$ in Eq.~\eqref{Eotvos_param}. The various experiments listed in Sec.~\ref{sec:exp_test_UFF} are sensitive to such non-zero differential acceleration, therefore they can put constraint on dilaton and axion couplings to SM fields. 
In Chapter ~\ref{chap:sens_experiments}, we will derive such sensitivities for \textit{MICROSCOPE}. In this case, the two test-masses are not precisely at the same position but are approximately $\sim 20$ $\mu$m apart \cite{Microscope17}, which sets the typical length scale of the experiment. As we shall see in Chapter ~\ref{chap:exp_summary}, \textit{MICROSCOPE} is sensitive to oscillations up to $0.3$ Hz. Following Eq.~\eqref{eq:mass_size_relation_lab_frame}, this means that the propagation term of the plane wave in Eq.~\eqref{delta_a_UFF_osc} can be neglected, and that we can approximate that both test-masses are at the same position. 

In the next chapter, we will be interested in atom interferometry, another experimental tool which could detect such UFF violating accelerations.

\chapter{\label{chap:AI}Atom interferometry as a quantum test of the equivalence principle}

As derived in the previous chapter, the time-dependent mass and internal frequency produces a differential acceleration between two atoms. Standard UFF tests search for such differential acceleration of macroscopic test masses. Atom interferometry (AI) constitutes the quantum equivalent of such classical experiments. Therefore, as we shall see in this chapter, AI experiments are also sensitive to oscillations from Eq.~\eqref{delta_a_UFF_osc}.

In this chapter, we will derive the observable signatures produced by oscillating atomic rest mass and transition energy on various AI configurations. In the most generic case, AI consists of generating interference between atomic wavepackets that are split and then recombined using atom-laser interactions. We classify the various AI schemes into two classes: (i) AI schemes using two photon transitions and (ii) AI schemes using single photon transitions. In the former, the transition from a stable momentum state to another stable momentum state is done using two photons, while in the latter, only one photon is needed for such a transition. Those results are based on an article \cite{Gue:2024onx}, published in \textit{Physical Review D}.

\section{Two-photon transitions : $\pi/2-\pi-\pi/2$ setups}

\subsection{General principle}

In this section, we present the basics of atom interferometry and we will focus on the most common AI setups : two-photon transition interferometers. 

\begin{figure}[h!]
\centering
\includegraphics[scale=0.5]{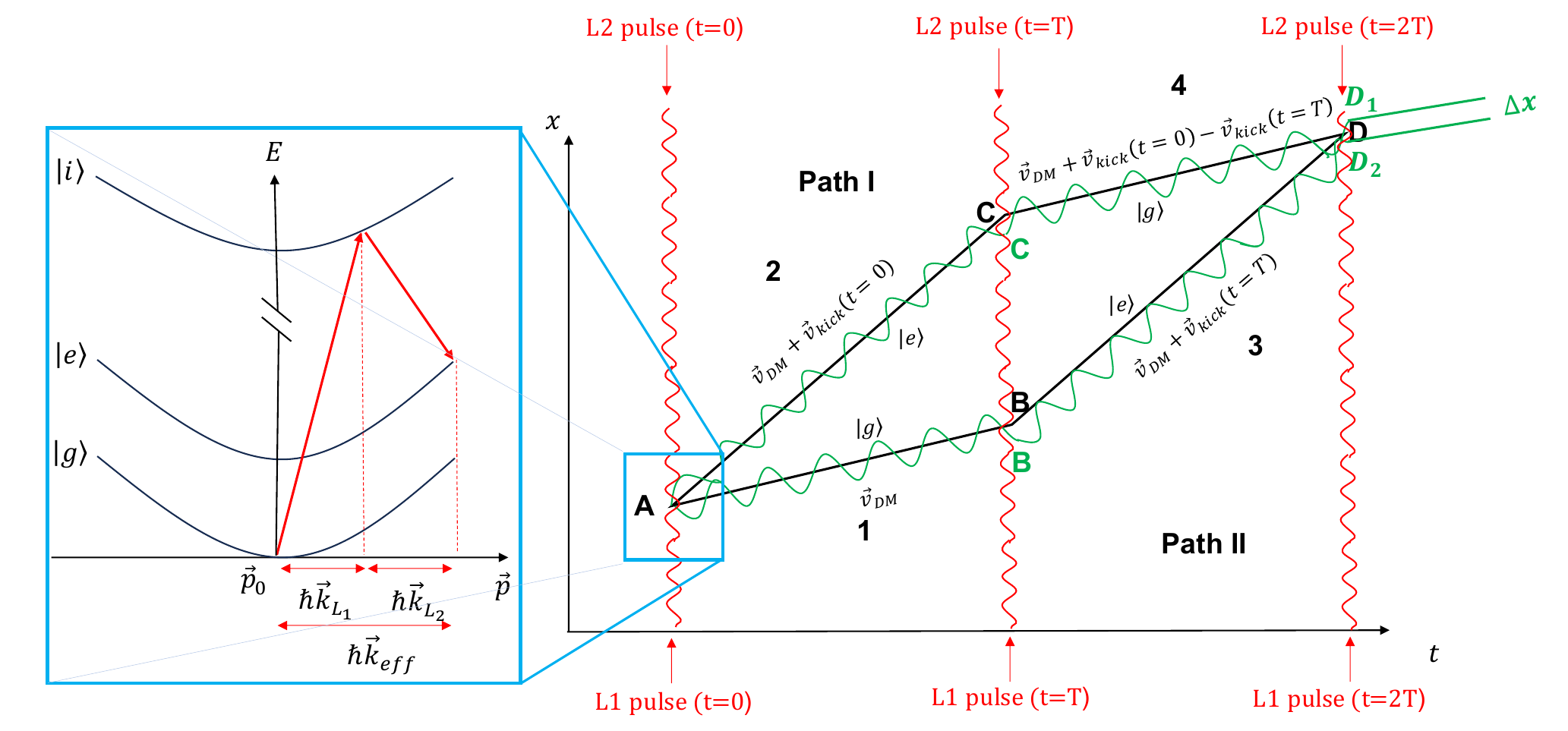}
\caption{On the left, scheme of the two photon Raman transition considered in the present section. The atom starts in the ground state $|g\rangle$, absorbs a photon which makes it transition to a high energetic state $|i\rangle$,  then emits a photon to transition to the excited state of the hyperfine transition $|e\rangle$. On the right, full Raman interferometric sequence, based on such two-photon transitions. At each atom-laser interaction, the laser pulses change both internal energy state and external momentum state of the wavepackets. The Bragg-type equivalent has the same spacetime diagram, but the atom wavepackets do not change their internal energy state. In black are shown the atomic paths without any perturbation, i.e straight lines (the gravitational field is neglected here). In green is shown the perturbed motion of the atoms induced by Eq.~\eqref{EoM_atom_AI}, with exaggerated amplitude of oscillation.}
\label{Mach-Zehnder_perturbed}
\end{figure}

The first interferometric scheme considered is known as two-photon transition Raman interferometry. Its sequence is depicted in Fig.~\ref{Mach-Zehnder_perturbed}. In this setup, two-level free falling atoms $A$ enter the interferometer with an initial momentum $\hbar k$ and in their internal energetic ground state, noted $|g\rangle$ (i.e. their initial state is defined as $|g,\hbar k\rangle$). After entering the interferometer, they interact with a pair of laser waves $L_1$ and $L_2$ with respective frequencies $\omega_\mathrm{L_1}, \omega_\mathrm{L_2}$, whose energy difference is resonant with the transition of the $|g\rangle \rightarrow |e\rangle$ where $|e\rangle $ is the excited state of the two-levels atom, i.e $\omega^0_A = \omega_\mathrm{L_1}-\omega_\mathrm{L_2}$. This means that the atom first absorbs a photon from $L_1$, which excites it to an intermediate state with energy $\hbar \omega_i \gg \hbar \omega^0_A$, and afterwards, the atom stimulately emits a photon in $L_2$. This process splits the atoms into two spatially distant wavepackets, which means the state of each atom becomes the superposition of the two internal states: (i) the ground state that remains unchanged $|g,\hbar k\rangle$ and (ii) the excited state that has changed its momentum because of its interaction with the two lasers $|e,\hbar(k +k_\mathrm{eff})\rangle$, where $\hbar \vec k_\mathrm{eff} = \hbar (\vec k_\mathrm{L_1}-\vec k_\mathrm{L_2})$ denotes the effective momentum transfer, considering both lasers.

This whole experimental manoeuvre is called a $\pi/2$ laser pulse since the transition amplitude from ground state to excited state corresponds to a probability $1/2$, i.e the state of the atoms is now a half-half superposition of the two different states.

The atom freely propagates inside the interferometer and at time $t=T$, both wavepackets undergo a $\pi$ pulse, which will invert the state of all atoms. In other words, depending on the internal state of the atom prior to the $\pi$ pulse, two state transitions happen : $|e,\hbar(k +k_\mathrm{eff})\rangle \rightarrow |g,\hbar k\rangle$ and $|g,\hbar k\rangle \rightarrow  |e,\hbar(k +k_\mathrm{eff})\rangle$. 

A final laser-atom interaction happens at time $t=2T$, where a second $\pi/2$ pulse divides the two incoming wavepackets into four different ones: two of them are in the state $|g,\hbar k\rangle$ and the remaining two are in the state $|e,\hbar(k +k_\mathrm{eff})\rangle$. The study of interference pattern between the wavepackets in the same state allows one to measure a phase shift difference. The full interferometric sequence is shown in Fig.~\ref{Mach-Zehnder_perturbed}.

The second two-photon transfer AI scheme considered is the two-photons transition Bragg-type interferometer. This scheme is similar to the Raman interferometry presented above except that the atoms remain in the same energy state during all the inteferometric path, i.e, the laser pulses only change the external state of the atom (their momentum).

For both interferometers described above, the effective wavevector $k_\mathrm{eff}$ depends on the setup of the experiment. If counterpropagating lasers are used, the atom absorbs a photon in one direction as a result of the interaction with the first laser, and emits another photon in the opposite direction during the interaction with the second laser, implying $k_\mathrm{eff}=k_\mathrm{L_1}+k_\mathrm{L_2}$. In general, this effective wave vector is multiple orders of magnitude larger than the transition frequency of the atom, i.e $k_\mathrm{eff} \gg \omega^0_A/c$, because both laser are usually operating at hundreds of THz frequencies (this is why the intermediate state is much higher in energy, see previous paragraphs and Fig.~\ref{Mach-Zehnder_perturbed}). On the opposite, if  co-propagating laser waves are used in the experimental setup, the absorption and emission directions are the same, implying $k_\mathrm{eff}=k_{L_1}-k_{L_2}$. In that case, $k_\mathrm{eff} = \omega^0_A/c$.

\subsection{\label{phase_contrib_MZ}Phase contributions}

As in the previous chapter, we will make the detailed calculations considering a coupling between dilatons and SM fields, and we will generalize the main results to the axion case. 

The calculation of the observable phase shift at the output of the interferometer follows closely the ones presented in \cite{Storey}. This method relies on Feynman path integrals and can be used for Lagrangians which are at most quadratic in the position and velocity \cite{Storey} and considering atomic plane waves at initial time $t=t_0$. This suggests that the calculations of the phase shift in our case can formally only be done in the galactic frame, where there is no  $\cos(\omega t - \vec k \cdot \vec x)$ terms\footnote{As we shall see in the following, this framework can also work when the propagation is negligibly small, i.e when $kL \ll 1$, where $L$ is the typical scale of the experiment.}.
Therefore, we will make all calculations in the galactic frame, where the field simply reads $\phi =\phi_0 \cos(\omega_\phi t + \Phi)$. In the next section, we will make a discussion on the possibility of making the calculations in the laboratory frame. 
Therefore, we consider an atom A whose nominal rest mass and transition frequency $m^0_A,\omega^0_A$ are perturbed such that they oscillate in phase as in Eq.~\eqref{mass_freq_osc_dilaton} but without the propagation term $\vec k_\phi \cdot \vec x$ in the phase. 
\begin{subequations}\label{mass_freq_acc_DM_rest_frame}
\begin{align}
    m_A(t)&=m^0_A\left(1+\frac{\sqrt{16 \pi G \rho_\mathrm{DM}}[Q^A_M]_d}{\omega_\phi c}\cos(\omega_\phi t +\Phi)\right) \label{mass_DM_rest_frame}\, \\
    \omega_A(t)&=\omega^0_A\left(1+\frac{\sqrt{16 \pi G \rho_\mathrm{DM}}[Q^A_\omega]_d}{\omega_\phi c}\cos(\omega_\phi t + \Phi)\right) \label{freq_DM_rest_frame}\, ,
\end{align}
\end{subequations}
By the same procedure as in Section ~\ref{sec:acceleration_dilaton_general}, the macroscopic Lagrangian is
\begin{subequations}
\begin{align}\label{eq:macro_lagrangian_DM_rest_frame}
    \mathcal{L}_A = -\left(m_A(t)c^2+\hbar \omega_A(t)\right)\left(1-\frac{v_A^2}{2c^2}\right) \, ,
\end{align}
which induces an acceleration on the atom A
\begin{align}
\vec a_A(t) &= \frac{\sqrt{16 \pi G \rho_\mathrm{DM}}\vec v_A [Q^A_M]_d}{c}\sin(\omega_\phi t +\Phi) \label{acc_DM_rest_frame}\, ,
\end{align}
\end{subequations}
where $\vec v_A$ is the velocity of the atom in the galactic frame, i.e $\vec v_A \approx \vec v_\mathrm{DM}$ at leading order. One can already notice the similarity with Eq.~\eqref{eq:delta_a_dil_lab_frame}. Indeed, for situations where the gradient of the field is almost constant on the size of experiment $L$ considered, i.e  $\lambda^\mathrm{dB}_\mathrm{DM} \gg L$, one can neglect the propagation term $\vec k_\phi \cdot \vec x$ in Eq.~\eqref{eq:delta_a_dil_lab_frame} and the two accelerations are identical.

In practice, we will assume that the AI experiment is performed in a laboratory frame which has a velocity $\vec v_\mathrm{DM}$ with respect to the DM rest frame where the Lagrangian takes the form given by Eqs.~(\ref{mass_freq_acc_DM_rest_frame}) and (\ref{eq:macro_lagrangian_DM_rest_frame}), such that $\vec v_A = \vec v_\mathrm{DM}+\vec {\tilde v}_0$ where the second term is the initial velocity of the atoms with respect to the laboratory reference frame. $\tilde v_0$ is the velocity impacting systematic effects such as gravity gradients or second-order Doppler shift of the atoms when interacting with light beams (see e.g \cite{Dimopoulos07} for a comprehensive list of such effects).

\begin{figure}
\centering
\includegraphics[scale=0.5]{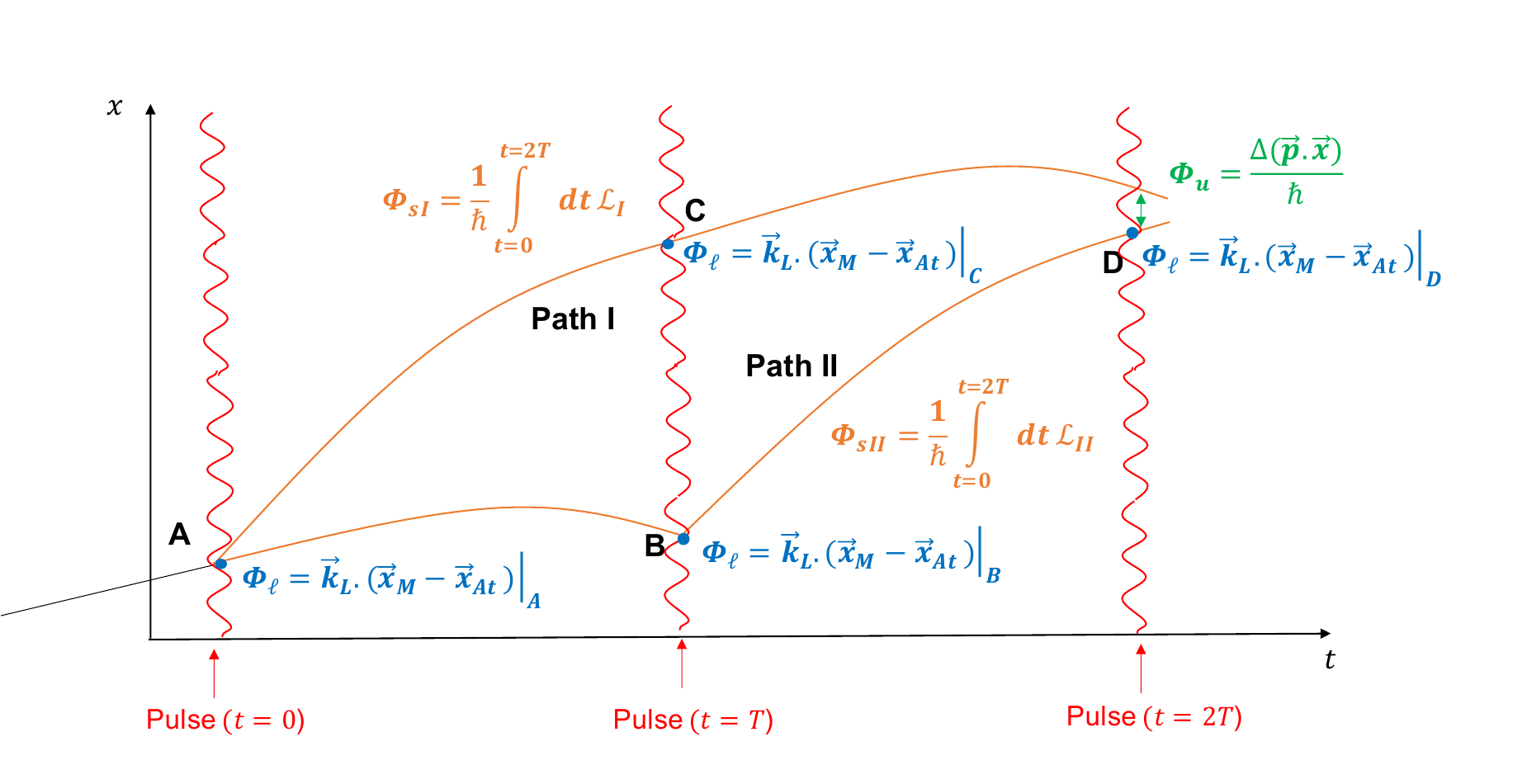}
\caption{The various phase contributions along the AI path : in orange are the propagation phase shifts calculated along the wavepackets paths ; in blue are the laser phase shifts, calculated at light-matter interaction vertices ; and in green is the separation phase shift, calculated at the output of the interferometer.}
\label{Mach-Zehnder_phase}
\end{figure}

As derived in e.g. \cite{Wolf, Wolf04,Storey}, there are three independent contributions to the total phase shift in an AI: (i) the separation phase noted $\Phi_\mathrm{u}$ which corresponds to a spatial incoincidence between the two output wavepackets, (ii) the laser phase $\Phi_\mathrm{\ell}$ which gathers the additional phase factors of the laser, due to displacement of the light-matter interaction vertices and (iii) the propagation phase shift denoted $\Phi_\mathrm{s}$ which is essentially the phase accumulated by the atom wavepacket over the full interferometric path. All of these contributions must be calculated accurately to predict the phase shift at the output of the interferometer. Those contributions are summarized in Fig.~\ref{Mach-Zehnder_phase}.

Finally, we will make all calculations assuming a Raman interferometer, such that the transition frequency of the atom is relevant. When relevant, we will show how the various Bragg interferometer phases differ from the Raman AI. We also make all calculations at first order in the perturbations.

We describe the quantum state of the different atom wavepackets by a wavefunction $\Psi$, which we break down into two different wavefunctions $\Psi_\mathrm{I}$ and $\Psi_\mathrm{II}$, where I,II respectively the up/down paths in Fig.~\ref{Mach-Zehnder_perturbed}, depending on the classical path the atom followed. 
Considering free particles at the input of the interferometer, the atomic plane waves have the form \cite{Storey}
\begin{align}
\Psi_\mathrm{init}(t_0,\vec x_0) = \Psi_0 e^{i\Phi(t_0,\vec x_0)} \, ,
\label{wavefunction_init}
\end{align}
with the amplitude of the wavefunction $\Psi_0$, $\Phi(t_0,\vec x_0) = \vec k \cdot \vec x_0 - \omega t_0 - \varphi$ with $\omega, \vec k, \varphi$ respectively its angular frequency, wavevector and constant phase.
At the output of the interferometer, at a time $T_f \geq 2T$, considering all the different phase contributions listed previously, the atomic plane wave along the trajectory j read
\begin{align}
\Psi_\mathrm{j}(T_f,\vec x_f) = \Psi_\mathrm{init}(T_f,\vec x_f) \times  e^{i\Phi_\mathrm{j}} \, ,
\label{wavefunction_fin}
\end{align}
where $\Phi_\mathrm{j}=\Phi_\mathrm{sj}+\Phi_\mathrm{\ell j}+\Phi_\mathrm{uj}$ represents the trajectory dependent phase factor.
Then, at some detection time $T_\mathrm{d} \geq T_\mathrm{f}$, a detector measures by fluorescence the number of atoms on each quantum state, which is essentially the measurement of the probability that the two wavepackets are in the same quantum state, i.e
\begin{align}
  \int \Big|\Psi_\mathrm{I}(T_\mathrm{d},\vec x_\mathrm{d}) + \Psi_\mathrm{II}(T_\mathrm{d},\vec x_\mathrm{d}) \Big|^2 dS \, ,
  \label{overlap}
\end{align}
where the integral is taken over the detector area S.  Note that for Raman schemes, two output quantum states can differ by their internal energy and momentum state, while in Bragg, they only differ by their momentum. Plugging Eq.~\eqref{wavefunction_fin} into Eq.~\eqref{overlap} and neglecting loss of contrast due to decoherence, we find that the measurement result is proportional to (1 + $\cos \Delta \Phi$) where $\Delta \Phi = \Phi_\mathrm{I} - \Phi_\mathrm{II}$ is the phase difference between the two wave functions at $T_\mathrm{d}$.
For simplicity, we consider $T_\mathrm{d} \equiv 2T$, because any additional phase shifts cumulated during $T_d-T_f$ is negligible (as in general $T_d-T_f \ll 2T$, and because the two wavefunctions on each output port have almost the same position, momentum and energy.).

\subsection{\label{sec:calc_phase_shift}Calculation of the phase shift observable}

We first derive the perturbed equations of motion of the atom following the perturbation to the acceleration Eq.~\eqref{acc_DM_rest_frame}, to get the motion followed by the different wavepackets along the trajectories presented in Fig.~\ref{Mach-Zehnder_perturbed}. To simplify the reading of the next set of calculations, we set $\sqrt{16 \pi G \rho_\mathrm{DM}}/\omega_\phi c \equiv X_\mathrm{DM}$ (which is dimensionless). 
These equations of motion read 
\begin{subequations}\label{EoM_atom_AI}
\begin{align}
&\vec v_A(t,t_0) \approx \vec v_\mathrm{DM} \left(1-X_\mathrm{DM}[Q^A_M]_d\left(\cos(\omega_\phi t+\Phi)-\cos(\omega_\phi t_0+\Phi)\right)\right) \ \, , \\
&\vec x_A(t,t_0) \approx  \vec x_0 + \vec v_\mathrm{DM}(t-t_0) -\frac{X_\mathrm{DM}\vec v_\mathrm{DM} [Q^A_M]_d }{\omega_\phi}\left[\sin(\omega_\phi t+\Phi)-\sin(\omega_\phi t_0+\Phi)\right.\,\\
&\left.-\omega_\phi(t-t_0)\cos(\omega_\phi t_0+\Phi)\right] \, \nonumber,
\end{align}
\end{subequations}    
at first order in the perturbation $[Q^A_M]_d$, with $\vec x_0 = \vec x(t_0)$ the initial position of the atom when entering the interferometer. In the following, we set $x_0 = t_0 = 0$ and $\vec v_\mathrm{DM} = v_\mathrm{DM}\hat e_v$. Note that in Eq.~\eqref{EoM_atom_AI}, one of the terms arising in the position of the atom is $\propto \omega_\phi t$, which induces a temporal linear drift of the position of the atom, due to its initial inertia. 

As already discussed in \cite{Geraci16}, the velocity kick experienced by the atom during its interaction with the laser beams is perturbed by the effective mass of the atom at the time of the kick. In addition to that effect, the beams being locked to a given frequency reference, their own frequency oscillates as $\omega_L(t) = \omega^0_L(1+\left(\sqrt{16 \pi G \rho_\mathrm{DM}}[Q^L_\omega]_d /\omega_\phi c \right)\cos(\omega_\phi t + \Phi))$, neglecting the additional phase coming from the propagation of photons to the atom\footnote{In the following sections, the travelling distance of the photon to reach the freely falling atoms will be of the order of $100$ m maximum, which would induce a significant phase for oscillation frequencies $\omega_\phi \gtrapprox 10^5$ rad/s, way above the DM frequencies of interest for this study, see Chapter ~\ref{chap:sens_experiments}.} .
Therefore, if the atom interacts with the laser beam at time $t$, still considering $[Q^A_M]_d, [Q^L_\omega]_d \ll 1$,
\begin{align}
v^A_\mathrm{kick}(t) = \frac{\hbar k_\mathrm{eff}(t)}{m_A(t)} &\approx \frac{\hbar k_\mathrm{eff}}{m^0_A}\left(1+X_\mathrm{DM}([Q^L_\omega]_d-[Q^A_M]_d)\cos(\omega_\phi t +\Phi)\right)\equiv v^A_\mathrm{kick, \ 0}+\delta v^A_\mathrm{kick}(t)\, ,
\label{kick_modif}
\end{align}
where $v^A_\mathrm{kick, \ 0} = \hbar k_\mathrm{eff}/m^0_A$ is the nominal kick velocity, i.e without perturbation, and $\delta v^A_\mathrm{kick}(t)$ is the perturbed contribution to the total velocity kick imparted to the atom.
In Appendix ~\ref{ap:pos_vel_atoms_AI}, we compute explicitly the position and velocity of the wavepackets along the various interferometric paths, which will be useful for the computation of the various phase contributions.

\paragraph{Propagation phase contribution}
The first component of phase shift is the one coming from the phase accumulated by atoms throughout the whole interferometric paths taking into account modified equations of motion and perturbed kicks. In the special case of quadratic Lagrangian in the position and velocity of the atom at maximum, this phase is by the principle of least action the integral of the Lagrangian over the path from initial point $i$ to final point $f$ $(t_i,x_i)\rightarrow (t_f,x_f)$
\begin{align}\label{eq:prop_phase}
    \Phi_\mathrm{s} = \frac{1}{\hbar}\int_{t_i}^{t_f} L(x,\dot{x}) dt \, ,
\end{align}
where the Lagrangian is defined in Eq.~\eqref{macro_lagrangian}. The internal energy term of the Lagrangian being associated with the oscillation of the transition energy, it only contributes when the atom is on the excited state, i.e on paths 2 and 3 in Fig.~\ref{Mach-Zehnder_perturbed}.
Then, the phase accumulated by the atoms on the path I of the interferometer is 
\begin{subequations}
\begin{align}
\Phi_\mathrm{sI} &=-\omega^0_A\int_0^{T}dt (1+X_\mathrm{DM}[Q^A_\omega]_d \cos(\omega_\phi t +\Phi))\left(1-\frac{|\vec v^{(2)}_A(t)|^2}{2c^2}\right)-\,\nonumber\\
&\frac{m^0_Ac^2}{\hbar}\left[\int_0^T dt \left(1+X_\mathrm{DM}[Q^A_M]_d\cos(\omega_\phi t +\Phi)\right)\left(1-\frac{|\vec v^{(2)}_A(t)|^2}{2c^2}\right)+\right.\,\nonumber\\
&\left.\int_T^{2T} dt\left(1+[Q^A_M]_d\cos(\omega_\phi t +\Phi)\right)\left(1-\frac{|\vec v^{(4)}_A(t)|^2}{2c^2}\right)\right] \, ,
\end{align}
while the atom wavepacket on the path II accumulates a phase 
\begin{align}
\Phi_\mathrm{sII} &=-\omega^0_A\int_T^{2T}dt (1+X_\mathrm{DM}[Q^A_\omega]_d \cos(\omega_\phi t +\Phi))\left(1-\frac{|\vec v^{(3)}_A(t)|^2}{2c^2}\right)-\,\nonumber\\
&\frac{m^0_Ac^2}{\hbar} \left[\int_0^{T} dt \left(1+X_\mathrm{DM}[Q^A_M]_d\cos(\omega_\phi t +\Phi)\right)\left(1-\frac{|\vec v^{(1)}_A(t)|^2}{2c^2}\right)+\right.\,\nonumber\\
&\left.\int_T^{2T} dt\left(1+X_\mathrm{DM}[Q^A_M]_d\cos(\omega_\phi t +\Phi)\right)\left(1-\frac{|\vec v^{(3)}_A(t)|^2}{2c^2}\right)\right] \, ,
\end{align}
\end{subequations}
where $v^{(1)},v^{(2)},v^{(3)},v^{(4)}$ are respectively the atom velocities along portions 1, 2, 3 and 4 in Fig.~\ref{Mach-Zehnder_perturbed}, which can all be recovered explicitly using the set of calculations described in Appendix ~\ref{ap:pos_vel_atoms_AI}.

Then, the propagation phase shift between the two perturbed trajectories is 
\begin{subequations}
\begin{align}
&\Phi_\mathrm{s} = \Phi_\mathrm{sI}-\Phi_\mathrm{sII}\,\\
&=-\frac{4X_\mathrm{DM}}{\omega_\phi}\left[k_\mathrm{eff}\left(v_\mathrm{DM}\hat e_v \cdot \hat e_\mathrm{kick}+\frac{\hbar k_\mathrm{eff}}{2m^0_A}\right)[Q^A_M]_d+\omega^0_A[Q^A_\omega]_d\right]\sin^2\left(\frac{\omega_\phi T}{2}\right)\sin(\omega_\phi T + \Phi) +\, \nonumber \\
&4k_\mathrm{eff} T X_\mathrm{DM} [Q^L_\omega]_d\left(v_\mathrm{DM}\hat e_v \cdot \hat e_\mathrm{kick} +\frac{\hbar k_\mathrm{eff}}{2m^0_A}\right)\sin\left(\frac{\omega_\phi T}{2}\right)\sin\left(\frac{\omega_\phi T}{2} + \Phi\right) + \mathcal{O}\left(\left(\frac{v_\mathrm{DM}}{c}\right)^2\right)\, ,
\label{action_phase}
\end{align}
\end{subequations}
where we used $m^0_Av_\mathrm{kick}=\hbar k_\mathrm{eff}$ at zeroth order in the perturbation, following Eq.~\eqref{kick_modif} and where we defined $\vec v_\mathrm{kick} = v_\mathrm{kick}\hat e_\mathrm{kick}$. The terms $\propto (v^2_\mathrm{DM}/c^2)$ arise from the small contribution of the atoms velocity to the internal state kinetic energy.
The Bragg propagation phase can be simply recovered by remembering that in this configuration, one effectively manipulates one-level atom, i.e $\omega^0_A=0$.

\paragraph{Laser phase contribution}

We now consider the phase from light-matter interaction between the laser and the atoms.
In both Bragg and Raman schemes, we assume that the atoms are freely falling inside a "vacuum tower" where the lasers are on ground, located on a mount at initial position $x_G(0)=0$, while a retro-reflective mirror at initial position $x_M(0)=L$ is used to reflect the beams in order to create the counter-propagating scheme. In the Bragg case, only one laser beam is used and retro-reflected, while for the Raman case, two beams $L_1$, $L_2$ with respective frequencies $\omega_{L_1},\omega_{L_2}$ (such that $\omega_{L_1}-\omega_{L_2}=\omega^0_A$) are retro-reflected and the atomic wavepackets interact only with $L_1$ going up and $L_2$ going down. Considering that the whole tower on Earth is freely falling during the entire interferometric process, its own mass composition is affected by the oscillating mass behavior Eq.~\eqref{mass_freq_acc_DM_rest_frame}, i.e $m_M = m^0_M\left(1+Q^M_M\cos(\omega_\phi t +\Phi)\right)$, with $m^0_M, Q^M_M$ respectively its unperturbed mass and mass charge, implying it follows the same perturbed equations of motion as the atom Eq.\eqref{EoM_atom_AI}. Therefore, both retro-reflective mirror and mount on which laser rest upon are oscillating together\footnote{At next-to-leading order,  $L$ is also oscillating as the interatomic bonds oscillate too.}.
At each spacetime points of light-matter interaction is associated a phase, which contributes to the laser phase shift. These points are denoted $A, B, C, D_2$ in Fig.~\ref{Mach-Zehnder_perturbed}, since we assume the measurement of the interference pattern between the two wavepackets ending up on the internal ground state $|g\rangle$.

Keeping in mind that $\Phi_L(t,\vec x) = \vec k_L \cdot \vec x - \omega_L t - \varphi$, with $\omega_L, \vec k_L$ the light angular frequency and wavevector respectively, and that $\vec k_L \cdot \vec x - \omega_L t = 0$ along a photon geodesic in a flat spacetime (which is the case at the surface of the Earth, see \cite{Badurina22}), the total light phase felt by the atom along a path $X$ is given by 
\begin{align}
    \Phi_{\ell X} &=- \sum_{j=0}^n  \varphi_\mathrm{j}(t_i) =- \sum_{j=0}^n s_j\left(\varphi_\mathrm{j}(t^\mathrm{down}_i) -\varphi_\mathrm{j}(t^\mathrm{up}_i)\right) \, ,
\end{align}
where we sum on the total number of interaction points of path X (occurring at times $\{0, T, 2T\}$), and where $t_i$ corresponds respectively to the time of laser at emission of the photon (the superscripts $up$ and $down$ refer to a photon coming from the up or down laser). At the second line, we took into account the fact that we have two-photon interactions, where the atom takes a photon from one of the laser and emits in the second one, and the $s_j = \pm 1$ parameter depends on the transition of the wavepacket at interaction j (it is $+1$ for a $|g\rangle \rightarrow |e\rangle $ transition and $-1$ for a $|g\rangle \rightarrow |e\rangle $ transition). Therefore, only the initial phase of the laser at the time of photon emission is needed. Since the laser frequency is locked on an atom ensemble whose frequency oscillates through Eq.~\eqref{mass_freq_acc_DM_rest_frame}, the phase, as the integral of the frequency, is also time dependent, i.e 
\begin{align}\label{eq:init_phase_laser}
    \varphi_\mathrm{j}(t) &= \int^t_0 dt' \omega_L(t') \equiv \int^t_0 dt' \omega^0_L\left(1+X_\mathrm{DM}Q^L_\omega \cos(\omega_\phi t' + \Phi)\right) \, .
\end{align}
\begin{figure}[h!]
\begin{minipage}{\textwidth}
  \begin{minipage}{0.45\textwidth}
    \centering
    \includegraphics[width=\textwidth]{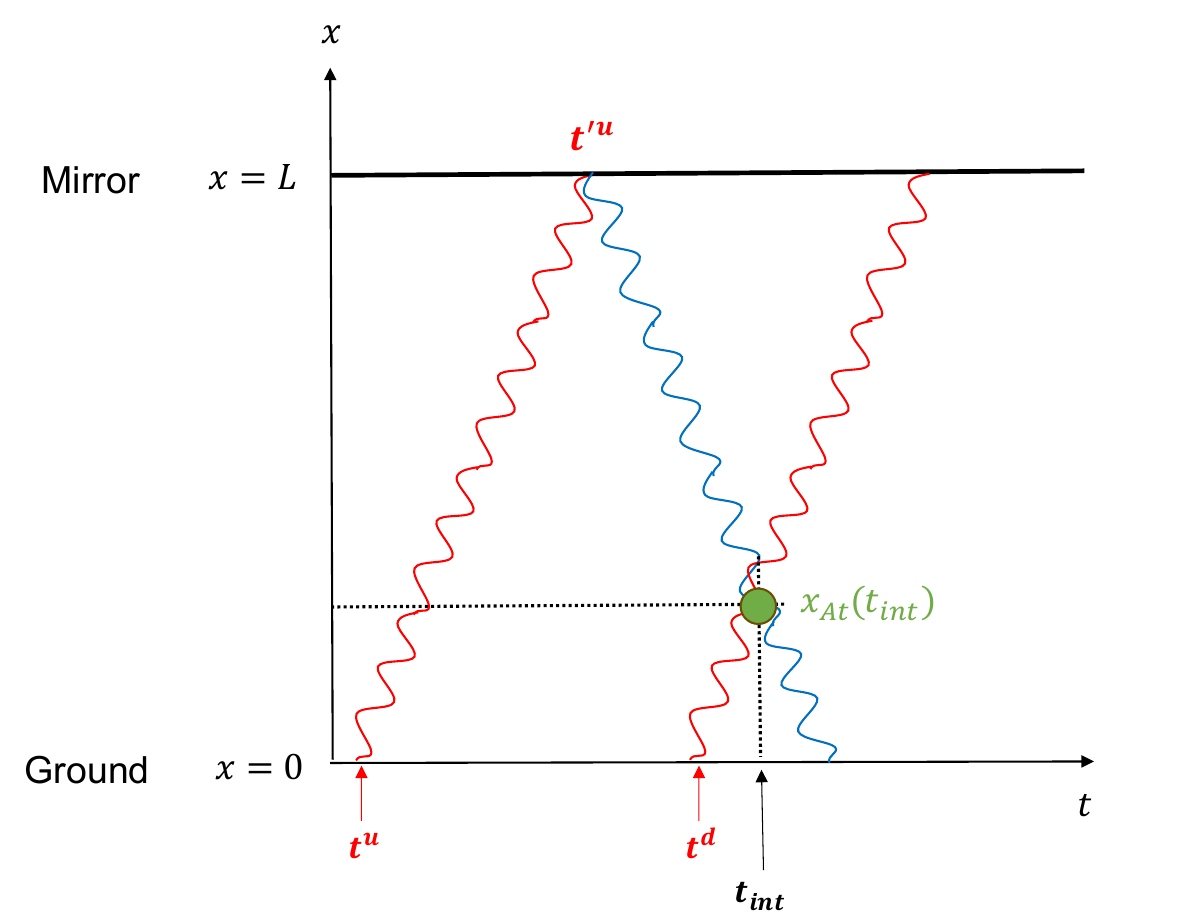}
    \caption{Simple scheme on the computation of the time of emission of photons from both lasers.}
    \label{fig:laser_phase}
    \end{minipage}
    \hfill
    \begin{minipage}{0.53\textwidth}
    To compute the laser phase, we need to know the different times $t_i$ of emission of photons. For a given interaction time $t_\mathrm{int}$, they are given by (see Fig.~\ref{fig:laser_phase}).
    \begin{subequations}\label{eq:emission_times_laser}
    \begin{align}
        t^u &= t_\mathrm{int}-\frac{2x_M(t_\mathrm{int}-t^{'u})-x_\mathrm{At}(t_\mathrm{int})-x_G(t^{'u}-t^u)}{c} \, , \\
        t^d &= t_\mathrm{int} - \frac{x_\mathrm{At}(t_\mathrm{int})-x_G(t_\mathrm{int}-t^d)}{c} \, ,
    \end{align}
    \end{subequations}
    respectively for the beam retro-reflected (up) and not retro-reflected (down), where $x_\mathrm{M}$ is the mirror coordinate, $x_\mathrm{G}$ is the mount coordinate and $x_\mathrm{At}$ is the perturbed vertex of the atomic path. In order to solve Eq.~\eqref{eq:emission_times_laser}, the various $t^u,t^{'u},t^d$ on the right side are treated as unperturbed times.
  \end{minipage}
\end{minipage}
\end{figure}

Then, for Raman AI, the total laser phase is
\begin{subequations}
\begin{align}\label{eq:laser_phase_Raman}
   &\Phi_\mathrm{\ell} = \Phi_\mathrm{\ell I}-\Phi_\mathrm{\ell II}\,\\ &=\frac{4X_\mathrm{DM}}{\omega_\phi}\left(k_\mathrm{eff}v_\mathrm{DM}\left([Q^A_M]_d-[Q^M_M]_d\right)\hat e_v\cdot \hat e_\mathrm{kick}+\frac{\hbar k^2_\mathrm{eff}}{m^0_A}[Q^A_M]_d+\omega^0_A [Q^L_\omega]_d\right)\sin^2\left(\frac{\omega_\phi T}{2}\right)\sin(\omega_\phi T + \Phi) +\, \nonumber \\
   &4[Q^L_\omega]_d X_\mathrm{DM}\left(L\left(k_\mathrm{eff} -\frac{\omega^0_A}{c}\right)\sin^2\left(\frac{\omega_\phi T}{2}\right)\cos(\omega_\phi T + \Phi)-\frac{\hbar k^2_\mathrm{eff} T}{2m^0_A}\sin\left(\frac{\omega_\phi T}{2}\right)\sin\left(\frac{\omega_\phi T}{2}+\Phi\right)\right) \,  . 
\end{align}
\end{subequations}
The Bragg laser phase is obtained by setting $\omega^0_A = 0$.

\paragraph{Separation phase contribution}

Following Eq.~\eqref{overlap} and considering that, at the end of the classical paths at $t=T_d$, the wave packet of the path $i$=$\{$I,II$\}$ can be expressed as \cite{Wolf04}
\begin{align}
    \Psi_i = \Psi_0 e^{i\left(\Phi_{si}+\Phi_{\ell i}+\frac{\vec p_i}{\hbar}(\vec x-\vec x_i)\right)} \, ,
\end{align}
where $\Phi_s,\Phi_\ell$ are respectively the propagation and laser phase contributions and where $\vec p_i,\vec x_i$ are respectively the momentum and position of the wavepacket just after the last $\pi/2$ pulse at $t=T_d$. Then, it can be shown easily that for small difference in momentum between the two wavepackets $\Delta \vec p = \vec p_\mathrm{I} - \vec p_\mathrm{II}$ (i.e $|\Delta \vec p| L_\mathrm{det}/\hbar \ll 1$, with $L_\mathrm{det}$, the size of the detector, which we show in Appendix ~\ref{ap:pos_vel_atoms_AI}), the separation phase shift can be expressed as 
\begin{align}
\label{unclosed_path_gen}
    \Phi_u &= \frac{\Delta \vec p \cdot \vec x_\mathrm{det., COM}(2T)}{\hbar}-\frac{\Delta (\vec p \cdot \vec x)(2T)}{\hbar} \, ,
\end{align}
where we compute the difference of momenta $\vec p = \hbar \vec k$ and position $\vec x$ between the two wavepackets in the same energy state, i.e $|g\rangle$ in our calculation. $\vec x_\mathrm{det., COM}$ represents the detector center of mass position at time $t=T_d=2T$, i.e in our case it is simply $\vec x_\mathrm{det., COM}(2T) = 2 \vec v_\mathrm{DM} T$, since its equation of motion reads $\vec x_\mathrm{det., COM}(t) = \vec v_\mathrm{DM}t$ (its own oscillation would induce a second order effect).

As we have shown in Appendix ~\ref{ap:pos_vel_atoms_AI}, both positions and velocities of the wavepackets at the end of paths $3$ and $4$ in Fig.~\ref{Mach-Zehnder_perturbed} are different in our framework, therefore, the separation phase is a function of both differences of positions and velocities at the end of the two paths 
\begin{subequations}
\begin{align}
    \Delta \vec x(2T) &= 4\vec v_\mathrm{kick} X_\mathrm{DM}\left([Q^L_\omega]_d T-\frac{[Q^A_M]_d}{\omega_\phi}\sin\left(\frac{\omega_\phi T}{2}\right)\right)\sin\left(\frac{\omega_\phi T}{2}\right)\sin(\omega_\phi T +\Phi) \, \\
    \Delta \vec v(2T) &= -4\vec v_\mathrm{kick}X_\mathrm{DM}[Q^L_\omega]_d\sin^2\left(\frac{\omega_\phi T}{2}\right)\cos(\omega_\phi T +\Phi) \, ,
\end{align}
\end{subequations}
implying that the difference in velocities is small (of order $[Q^L_\omega]_d$), justifying the form of the separation phase Eq.~\eqref{unclosed_path_gen}.
Then, the separation phase shift is 
\begin{subequations}
\begin{align}
    &\Phi_u = \frac{m^0_A}{\hbar}\left(\Delta \vec v(2T)\cdot\vec x_\mathrm{det., COM}(2T) -\Delta \vec v(2T)\cdot \vec x(2T) -\vec v(2T) \cdot \Delta \vec x(2T)\right)\, \\
    &=\frac{4k_\mathrm{eff}X_\mathrm{DM}}{\omega_\phi}\left(v_\mathrm{DM} \hat e_v \cdot \hat e_\mathrm{kick}\left[[Q^A_M]_d\sin\left(\frac{\omega_\phi T}{2}\right)\sin(\omega_\phi T + \Phi)-\omega_\phi T [Q^L_\omega]_d \sin\left(\frac{\omega_\phi T}{2} + \Phi\right)\right]\right.+\, \nonumber\\
    &\left.\frac{\hbar k_\mathrm{eff} \omega_\phi T}{m^0_A}[Q^L_\omega]_d\sin\left(\frac{\omega_\phi T}{2}\right)\cos(\omega_\phi T +\Phi)\right)\sin\left(\frac{\omega_\phi T}{2}\right) \, ,
    \label{unclosed_path_shift}
\end{align}
\end{subequations}
where $\vec v(2T), \vec x(2T)$ respectively correspond to the unperturbed velocity and position of the wavepackets at $t=2T$.

\subsection{Total phase shift}

Adding all contributions of phase shift Eqs.\eqref{action_phase}, ~\eqref{eq:laser_phase_Raman} and \eqref{unclosed_path_shift}, the total phase of the Raman AI is 
\begin{subequations}\label{eq:MZ_full_phase_dil}
\begin{align}
    &[\Phi^\mathrm{Raman}_A]_d = \frac{4\sqrt{16 \pi G \rho_\mathrm{DM}}}{\omega^2_\phi c}\left[\left(k_\mathrm{eff}v_\mathrm{DM}\Delta [Q_M]_d\hat e_v \cdot \hat e_\mathrm{kick}+\omega^0_A\Delta [Q_\omega]_d\right)\sin^2\left(\frac{\omega_\phi T}{2}\right)\sin(\omega_\phi T + \Phi) \right.+ \, \nonumber \\
    &\left.\left(\frac{\hbar k^2_\mathrm{eff}[Q^A_M]_d}{2 m^0_A}\sin(\omega_\phi T + \Phi)-k_\mathrm{eff}\omega_\phi[Q^L_\omega]_d \left(L\left(1-\frac{\omega^0_A}{k_\mathrm{eff}c}\right)+\frac{\hbar k_\mathrm{eff}T}{m^0_A}\right)\cos\left(\omega_\phi T + \Phi\right)\right)\sin^2\left(\frac{\omega_\phi T}{2}\right)\right]\,  ,
\end{align}
where $\Delta [Q_M]_d=[Q^A_M]_d-[Q^M_M]_d$ and $\Delta [Q_\omega]_d=[Q^L_\omega]_d-[Q^A_\omega]_d$.
For Bragg AI, one simply needs to set $\omega^0_A=0$ and the phase reads
\begin{align}
    &[\Phi^\mathrm{Bragg}_A]_d = \frac{4\sqrt{16 \pi G \rho_\mathrm{DM}}}{\omega^2_\phi c}\left[k_\mathrm{eff}v_\mathrm{DM}\Delta [Q_M]_d\hat e_v \cdot \hat e_\mathrm{kick}\sin^2\left(\frac{\omega_\phi T}{2}\right)\sin(\omega_\phi T + \Phi) + \right.\, \\
    &\left.\left(\frac{\hbar k^2_\mathrm{eff}[Q^A_M]_d}{2m^0_A}\sin(\omega_\phi T + \Phi)-k_\mathrm{eff}\omega_\phi[Q^L_\omega]_d \left(L+\frac{\hbar k_\mathrm{eff}T}{m^0_A}\right)\cos\left(\omega_\phi T + \Phi\right)\right)\sin^2\left(\frac{\omega_\phi T}{2}\right)\right]\, \nonumber ,
\end{align}
\end{subequations}
at lowest order in $v_\mathrm{DM}/c$. Similarly, using Eq.~\eqref{EP_viol_acc_axion}, we find the expected signal resulting from the axion-gluon coupling in a Raman AI, which reads
\begin{subequations}\label{eq:MZ_full_phase_axion}
\begin{align}
    &[\Phi^\mathrm{Raman}_A]_a = \frac{16\pi G \rho_\mathrm{DM}E^2_P}{f^2_a \omega^3_a c^2}\left[\left(k_\mathrm{eff}v_\mathrm{DM}\Delta [Q_M]_a\hat e_v \cdot \hat e_\mathrm{kick}+\omega^0_A\Delta [Q_\omega]_a\right)\sin^2\left(\omega_a T\right)\sin(2\omega_a T + \Phi)+\right. \, \nonumber \\
    &\left.\left(\frac{\hbar k^2_\mathrm{eff}[Q^A_M]_a}{2 m^0_A}\sin(2\omega_a T + \Phi)-k_\mathrm{eff}\omega_a[Q^L_\omega]_a \left(L\left(1-\frac{\omega^0_A}{k_\mathrm{eff}c}\right)+\frac{\hbar k_\mathrm{eff}T}{m^0_A}\right)\cos\left(2\omega_a T + \Phi\right)\right)\sin^2\left(\omega_a T\right)\right] \, ,
\end{align}
where $\Delta [Q_M]_a=[Q^A_M]_a-[Q^M_M]_a$ and $\Delta [Q_\omega]_a=[Q^L_\omega]_a-[Q^A_\omega]_a$. Again, for Bragg AI, one simply needs to set $\omega^0_A=0$ and the phase reads
\begin{align}
    &[\Phi^\mathrm{Bragg}_A]_a = \frac{16\pi G \rho_\mathrm{DM}E^2_P}{f^2_a \omega^3_a c^2}\left[k_\mathrm{eff}v_\mathrm{DM}\Delta [Q_M]_a\hat e_v \cdot \hat e_\mathrm{kick}\sin^2\left(\omega_a T\right)\sin(2\omega_a T + \Phi)+\right. \,  \\
    &\left.\left(\frac{\hbar k^2_\mathrm{eff}[Q^A_M]_a}{2 m^0_A}\sin(2\omega_a T + \Phi)-k_\mathrm{eff}\omega_a[Q^L_\omega]_a \left(L+\frac{\hbar k_\mathrm{eff}T}{m^0_A}\right)\cos\left(2\omega_a T + \Phi\right)\right)\sin^2\left(\omega_a T\right)\right] \, \nonumber .
\end{align}
\end{subequations}

As we derived it in Chapter ~\ref{DM_pheno}, the mass and frequency charges in both dilaton and axion models are composition-dependent, therefore the terms appearing in the first line in Eqs.~\eqref{eq:MZ_full_phase_dil} and \eqref{eq:MZ_full_phase_axion} are a signature of the violation of the Einstein EP. In the following, the terms proportional to $v_\mathrm{DM}$, which are present in both Bragg and Raman configurations, will be denoted as the mass terms, while the terms proportional to $\omega^0_A$, only present in Raman configuration will be denoted as the frequency terms. Depending on the setup, the $Q^L_\omega$ and $Q^A_\omega$ charges can be the same, cancelling this term completely. 

All the terms appearing at the second lines of the same equations, i.e the terms quadratic in the effective wavevector and respectively proportional to the mass charge of the atom A and the frequency charge of the laser and the terms proportional to the mirror coordinate $L$ still remain, even in the universal charges case\footnote{At this point, the interpretation of such terms in terms of the Einstein Equivalence Principle starts to be subject to interpretation. One can argue that these would be a sign of violation of the UFF as they would still induce a non-zero phase shift when measuring a differential phase shift between two different atomic species. Another interpretation would be a violation of the Lorentz Position Invariance as those terms appear due to the variation of the laser phase, which now depends on the spacetime position of the interaction.}. In fact, the terms quadratic in the effective wavevector can be understood as a non-local measurement, whose macroscopic counterpart would be to compare the free fall of two test masses with different velocities. In the following, we will neglect these term, since $v_\mathrm{DM} \sim 10^5 \: \mathrm{m/s} \gg 10^{-2} \: \mathrm{m/s} \sim \hbar k_\mathrm{eff}/m^0_A$. The argument is the same for the term $\propto L$. We will also drop this term in the following, as it is also much smaller than the leading order terms, in particular when computing a differential phase shift (see next paragraph). Therefore, in the Bragg case, the mass term is the dominant term.

Two-photon transition AI are commonly operating using hyperfine transitions of alkaline-Earth atoms (e.g the hyperfine transition of Rb, Cs, K atoms \cite{STE-QUEST, Gauguet08, Gillot16}). For both co-propagating and counter-propagating configurations, the laser beams are locked onto the optical transition of an atomic ensemble of the same species as the atoms in free fall inside the interferometer, and the frequency difference $\omega_\mathrm{L_1}-\omega_\mathrm{L_2} = \omega^0_A$ is provided by a radio-frequency source (in the GHz range). Then $Q^L_\omega$ is the charge of that source, which, depending on the experimental configuration, may or may not be the same as $Q^A_\omega$ the charge of the hyperfine transition of the atoms in the AI. As discussed previously, in the case of co-propagating lasers, the effective wavevector corresponds to the frequency transition of the atom, i.e $k_\mathrm{eff}=\omega^0_A/c$. Therefore, the mass term in Eq.~\eqref{eq:MZ_full_phase_dil} and Eq.~\eqref{eq:MZ_full_phase_axion} is suppressed by a factor $v_\mathrm{DM}/c$ compared to the frequency term. For counter-propagating laser beams, the effective wavevector $k_\mathrm{eff} \gg \omega^0_A/c$, implying that the mass term in these equations is much bigger than in the co-propagating case. In that case, both mass and frequency terms are relevant and need to be taken into account. The mass term has already been derived in \cite{Geraci16}, in the case of oscillating mass coming from a coupling between matter and a classical oscillating dark matter field. However, the calculation in \cite{Geraci16} was performed in the lab frame but only considering the velocity of the atoms in this frame, i.e $v_A \sim 10$ m/s, while we argue that another component, due to the galactic velocity $v_\mathrm{DM} \sim 10^5 \: \mathrm{m/s} \gg v_A$ should be taken into account, as we demonstrated it in the beginning of this section. Adding this contribution would improve the expected sensitivity to the charges $Q^A_M$ by several orders of magnitude. In addition, our calculations take into account the contribution from Earth oscillation (through $Q^M_M$) which was not the case in \cite{Geraci16}. 

Note that in double diffraction interferometers, i.e when two pairs of laser beams transfer opposite momentum to the atom \cite{Giese13}, such that the spatial separation between the two coherent wavepackets is twice as large, the total phase shift is the same with the change $k_\mathrm{eff} \rightarrow 2 k_\mathrm{eff}$, as expected. In addition, in our calculation, the retro-reflective mirror is used to create the counter propagating waves, i.e we implicitely assumed a counter propagating scheme. For the co-propagating situation, one simply needs to set $L=0$, as there is no mirror. 

Dual atom interferometers using two atomic species $A$ and $B$ with different mass and frequency charges will measure the difference of the interferometric phases, whose amplitude is given by
\begin{subequations}\label{eq:MZ_delta_phase_shift}
\begin{align}
    &\left|\Delta \Phi^\mathrm{Raman}_\mathrm{AB}\right|_d = \frac{4\sqrt{16 \pi G \rho_\mathrm{DM}}}{\omega^2_\phi c}\left(v_\mathrm{DM}\hat e_v \cdot \hat e_\mathrm{kick}\left(k^A_\mathrm{eff}([Q^A_M]_d-[Q^M_M]_d)-k^B_\mathrm{eff}([Q^B_M]_d-[Q^M_M]_d)\right)+\right.\,\nonumber\\
    &\left.\omega^0_A\left([Q^{L,A}_\omega]_d -[Q^A_\omega]_d\right)-\omega^0_B\left([Q^{L,B}_\omega]_d -[Q^B_\omega]_d\right)\right)\sin^2\left(\frac{\omega_\phi T}{2}\right)\, \\
    &\left|\Delta \Phi^\mathrm{Raman}_\mathrm{AB}\right|_a = \frac{16\pi G \rho_\mathrm{DM}E^2_P}{f^2_a \omega^3_a c^2}\left(v_\mathrm{DM}\hat e_v \cdot \hat e_\mathrm{kick}\left(k^A_\mathrm{eff}([Q^A_M]_a-[Q^M_M]_a)-k^B_\mathrm{eff}([Q^B_M]_a-[Q^M_M]_a)\right)+\right.\,\nonumber\\
    &\left.\omega^0_A\left([Q^{L,A}_\omega]_a -[Q^A_\omega]_a\right)-\omega^0_B\left([Q^{L,B}_\omega]_a -[Q^B_\omega]_a\right)\right)\sin^2\left(\omega_a T\right)\, ,
\end{align}
for Raman interferometers and 
\begin{align}
    &\left|\Delta \Phi^\mathrm{Bragg}_\mathrm{AB}\right|_d = \frac{4\sqrt{16 \pi G \rho_\mathrm{DM}}v_\mathrm{DM}\hat e_v \cdot \hat e_\mathrm{kick}}{\omega^2_\phi c}\left(k^A_\mathrm{eff}([Q^A_M]_d-[Q^M_M]_d)-k^B_\mathrm{eff}([Q^B_M]_d-[Q^M_M]_d)\right)\sin^2\left(\frac{\omega_\phi T}{2}\right)\, \\
    &\left|\Delta \Phi^\mathrm{Bragg}_\mathrm{AB}\right|_a = \frac{16\pi G \rho_\mathrm{DM}E^2_P v_\mathrm{DM}\hat e_v \cdot \hat e_\mathrm{kick}}{f^2_a \omega^3_a c^2}\left(k^A_\mathrm{eff}([Q^A_M]_a-[Q^M_M]_a)-k^B_\mathrm{eff}([Q^B_M]_a-[Q^M_M]_a)\right)\sin^2\left(\omega_a T\right)\, ,
\end{align}
\end{subequations}
for Bragg interferometers, where we used Eqs.~\eqref{eq:MZ_full_phase_dil} and \eqref{eq:MZ_full_phase_axion}, neglected the subdominant terms $\propto k^2_\mathrm{eff}, L$ and where $Q^{L,A}_\omega, Q^{L,B}_\omega$ are respectively the frequency charge of the beams used for the transition of the atomic species A and B in the Raman case. The mass terms of Eq.~\eqref{eq:MZ_delta_phase_shift} are the quantum equivalent of the classical calculation Eq.~\eqref{delta_a_UFF_osc}. Indeed, for small $\omega_\phi T, \omega_a T$ and assuming $k^A_\mathrm{eff}=k^B_\mathrm{eff} \equiv k_\mathrm{eff}$, the Taylor expansion of the mass terms of Eq.~\eqref{eq:MZ_delta_phase_shift} is related to Eq.~\eqref{delta_a_UFF_osc} through $|\Delta \Phi_\mathrm{AB}| = k_\mathrm{eff}T^2|\Delta \vec a_\mathrm{AB}|$, which is the usual relation between phase shift and acceleration in atom interferometers \cite{Storey}. The second terms of Eq.~\eqref{eq:MZ_delta_phase_shift} of Raman phase, proportional to the frequency charges $Q_\omega$ have no classical counterpart. Assuming mass and frequency charges of same order of magnitude, the frequency terms of Eq.~\eqref{eq:MZ_delta_phase_shift} will dominate in co-propagating laser Raman interferometers, while for counter-propagating laser Raman interferometers, both terms contribute to the phase shift. 

\section{Are the calculations doable in the laboratory frame ?}

In the previous section, the observable phase shift is computed in the galactic reference frame (the DM rest frame) where the perturbation to the rest mass/atomic frequency is proportional to $\cos \left(\omega_\phi t + \Phi\right)$ (in the dilatonic case, but the argument is the same for the axion). As it was shown in Chapter ~\ref{Cosmo_evolution} (but this can be recovered by a simple Lorentz transformation), the field (and by extension the perturbation to the rest mass or atomic frequency) is proportional to $\cos(\omega_\ell t_\ell - \vec k_\ell \cdot \vec x_\ell + \Phi)$ in a lab-centric frame, moving at velocity $\vec v_\mathrm{DM}$ with respect to the Galactocentric frame (see Eq.~\eqref{eq:lab_frame_w_k} for their explicit expression). The $\ell$ subscript denotes quantities expressed in the lab-centric frame. The reason why we performed the calculations in the galactocentric frame relies on the fact that the method presented in \cite{Storey}  is valid only for Lagrangians at most quadratic in the position and velocity, which is the case in the galactocentric case but not in the lab-centric one. For this reason, the quantities appearing in Eqs.~\eqref{eq:MZ_full_phase_dil}, \eqref{eq:MZ_full_phase_axion} and \eqref{eq:MZ_delta_phase_shift} (and as we shall see in the next sections, this works also for Eqs.~\eqref{eq:phase_gradio} and \eqref{eq:phase_new_prop}) are quantities evaluated in the galactocentric frame. The first goal of this section is to justify why one can safely replace the values of $\omega^0_A$ , $k_\mathrm{eff}$ and $L$ appearing in these equations by their lab-centric counterpart. The result from Eqs.~\eqref{eq:MZ_full_phase_dil} and \eqref{eq:MZ_full_phase_axion} differs from the one obtained in \cite{Geraci16} by the fact that in \cite{Geraci16}, the mass terms are proportional to the lab-centric initial velocity of the atoms while in the previous section, it corresponds to its galactic counterpart (i.e the galactic velocity, which is multiple orders of magnitude larger). 

To strengthen our point, we will explicitly derive the classical equations of motion of the atoms in the lab-centric frame and show that the galactic velocity is indeed expected to appear in the phase shift. Finally, we will explain how a solution can be derived in the lab-centric frame (at first order in $v_\mathrm{DM}/c$) using the formalism from \cite{Storey} and show that this is also consistent with Eqs.~(\ref{eq:MZ_full_phase_dil}).

First, the two reference frames are related to each other by a Lorentz transformation. This means that the angular frequency $\omega$ and wave vector $\vec k$ of the laser beam transform following
\begin{subequations}\label{eq:lab_frame_w_k}
    \begin{align}
        \omega_\ell &= \gamma \left(\omega - \vec v_\mathrm{DM} \cdot \vec k\right) = \omega \left(1 + \mathcal O\left(\frac{v_\mathrm{DM}}{c}\right) \right) \, , \\
        \vec k_\ell & = \vec k + \frac{1}{v_\mathrm{DM}^2}\left(\gamma -1 \right)\left(\vec v_\mathrm{DM} \cdot \vec k\right)\vec v_\mathrm{DM} - \frac{1}{c^2}\gamma\omega \vec v_\mathrm{DM} = \vec k\left(1  + \mathcal O\left(\frac{v_\mathrm{DM}}{c}\right)\right) \, ,
    \end{align}
\end{subequations}
where $\gamma=(1-v_\mathrm{DM}^2/c^2)^{-1/2}$ and $v_\mathrm{DM}/c\sim 10^{-3}$. For this reason, although Eqs.~\eqref{eq:MZ_full_phase_dil} and \eqref{eq:MZ_full_phase_axion} are expressed in the galactocentric reference frame, one can safely replace $\vec k_\mathrm{eff}$, $\omega_A^0$ by their lab-centric counterpart. This would lead to a correction three orders of magnitude smaller than the leading order term. A similar argument applies for the other quantities such as $L$, $T$ the interrogation time, etc. This demonstrates that one can safely use lab-centric quantities in Eqs.~\eqref{eq:MZ_full_phase_dil} and \eqref{eq:MZ_full_phase_axion} (and this applies to all the equations following them, including the ones derived in the next two sections).

Let us now convince ourselves that a derivation directly performed in the lab-centric frame would also lead to a phase shift whose main term is also proportional to the lab velocity with respect to the galactic reference frame, in agreement with Eqs.~\eqref{eq:MZ_full_phase_dil} and \eqref{eq:MZ_full_phase_axion}. 
In Chapter ~\ref{chap:Classical_tests_UFF}, we derived the acceleration felt by an atom A in the laboratory frame Eq.~\eqref{eq:a_UFF_lab_frame}. Therefore, in the lab frame, the acceleration is proportional to the galactic velocity $v_\mathrm{DM}$, at leading order.
In addition, the AI schemes considered in Eqs.~\eqref{eq:MZ_full_phase_dil} and \eqref{eq:MZ_full_phase_axion} act as accelerometers that provide, to first order, a phase shift $\Delta \Phi = k_\mathrm{eff} a T^2$ where $a$ is the local acceleration. For this reason, as an output to the perturbative acceleration form Eq.~(\ref{eq:a_UFF_lab_frame}), one expects that the leading order term for $\Delta \Phi$ is $\propto k_\mathrm{eff} v_\mathrm{DM} [Q^A_M]_d T^2$, consistent for $\omega T\ll 1$ with the first term from Eqs.~\eqref{eq:MZ_full_phase_dil} and \eqref{eq:MZ_full_phase_axion}. This reasoning provides an argument showing that the velocity of the laboratory with respect to the galactocentric reference frame also appears when reasoning directly in the lab frame.
 
Formally, the method from \cite{Storey} cannot be used to compute the AI phase shift from the lab frame Lagrangian provided by Eq.~(\ref{macro_lagrangian}) since it is not at most quadratic in the position. Nevertheless, it is possible to perform an approximate calculation in the lab frame in the case where the de Broglie wavelength of the DM field (which for demonstration we assume is the dilaton field) is much larger than the typical size of the experiment, i.e. if $\vec k_\phi \cdot \vec x_\ell \ll 1$ (which is the case for the AI experiments considered in this thesis (see Chapter ~\ref{chap:exp_summary}) characterized by  $\vec k_\phi \cdot \vec x_\ell \lessapprox 5 \times 10^{-8} \ll 1$\footnote{We consider DM frequencies characterized by $m_\phi c^2 \leq 10^{-13}$ eV and size of experiments $L \leq 100$ m).}). In such a case, an expansion of the Lagrangian Eq.~\eqref{macro_lagrangian} leads to 
\begin{align}
    \mathcal{L}_A \approx& -m^0_Ac^2\left(1-\frac{v^2_{A,\ell}}{2c^2}\right)\times \left(1-\frac{\sqrt{16\pi G \rho_\mathrm{DM}}[Q^A_M]_d}{\omega_\phi c}\left(\cos(\omega_{\phi} t_\ell + \Phi)+\vec k_\phi \cdot \vec x_\ell \sin(\omega_\phi t_\ell+\Phi)\right)\right) \, .
\end{align}
From this Lagrangian, one can derive the UFF violating acceleration in the laboratory frame, which reads
\begin{subequations}
\begin{align}
    \vec a_A &\approx \left(\omega_\phi \vec v_{A,\ell} - \vec k_\phi c^2 \right)\frac{\sqrt{16\pi G \rho_\mathrm{DM}}[Q^A_M]_d}{\omega_\phi c}\sin(\omega_\phi t_\ell + \Phi) \,\\
    &= \frac{\left(\vec v_{A,\ell} +\vec v_\mathrm{DM}\right)}{c}\sqrt{16\pi G \rho_\mathrm{DM}}[Q^A_M]_d\sin(\omega_\phi t_\ell + \Phi) \, \label{a_vUFF_lab} .
\end{align}
\end{subequations}
This Lagrangian is now linear in the position and can be used to compute the phase shift using the method from \cite{Storey}. Note that there is an important difference compared to the calculation performed in \cite{Geraci16} where the second term has implicitly been neglected.

The equations of motion deriving from this Lagrangian consists in the ones from Eq.~(\ref{eq:a_UFF_lab_frame}) where one neglects the $\vec k_\ell \cdot \vec x_\ell$ within the sine function. The calculation of the phase shift directly in the lab frame follows exactly the equations presented in the previous section. There are mainly two differences in the derived equations. First, the perturbed trajectory and velocity of the atom in the lab frame now read (taking $t_0 = 0$ immediately and $t_\ell \rightarrow t$)
\begin{subequations}
\begin{align}
&\vec v_{A,\ell}(t) \approx \vec v_{\ell,0} - \left(\vec v_{\ell,0}+\vec v_\mathrm{DM}\right)X_\mathrm{DM}[Q^A_M]_d\left(\cos(\omega_\phi  t+\Phi)-\cos(\Phi)\right) \, , \\
&\vec x_{A,\ell}(t) \approx  \vec x_{\ell,0} + \vec v_{\ell,0} t -\frac{\vec v_{\ell,0}+\vec v_\mathrm{DM}}{\omega_\phi}X_\mathrm{DM}[Q^A_M]_d\left(\sin(\omega_\phi t+\Phi)-\sin(\Phi) - \omega_\phi t \cos(\Phi)\right) \, ,
\end{align}
\end{subequations}
where $v_{\ell,0}$ is the initial velocity of the atom in the lab frame, which corresponds to the launch velocity. These equations replace Eqs.~(\ref{EoM_atom_AI}). Secondly, one needs to keep the $\vec k_\phi \cdot \vec x_\ell$ term in the Lagrangian when computing the propagation phase using Eq.~(\ref{eq:prop_phase}). The full calculation following the method from the last section leads to a result that is consistent with Eqs.~\eqref{eq:MZ_full_phase_dil} and \eqref{eq:MZ_full_phase_axion} to first order in $v_\mathrm{DM}/c$, in Bragg and Raman AI respectively, which demonstrates the equivalence between the two frames.

\section{Gradiometers}

We will now focus on interferometric setup that involves single photon transitions.  The first setup considered is known as a gradiometer, i.e a setup where two atom interferometers are stacked at different altitudes, for the study of e.g gravity gradients, and which has already been studied in \cite{Graham13,Badurina22}.

\begin{figure}
    \centering
    \includegraphics[width=0.7\textwidth]{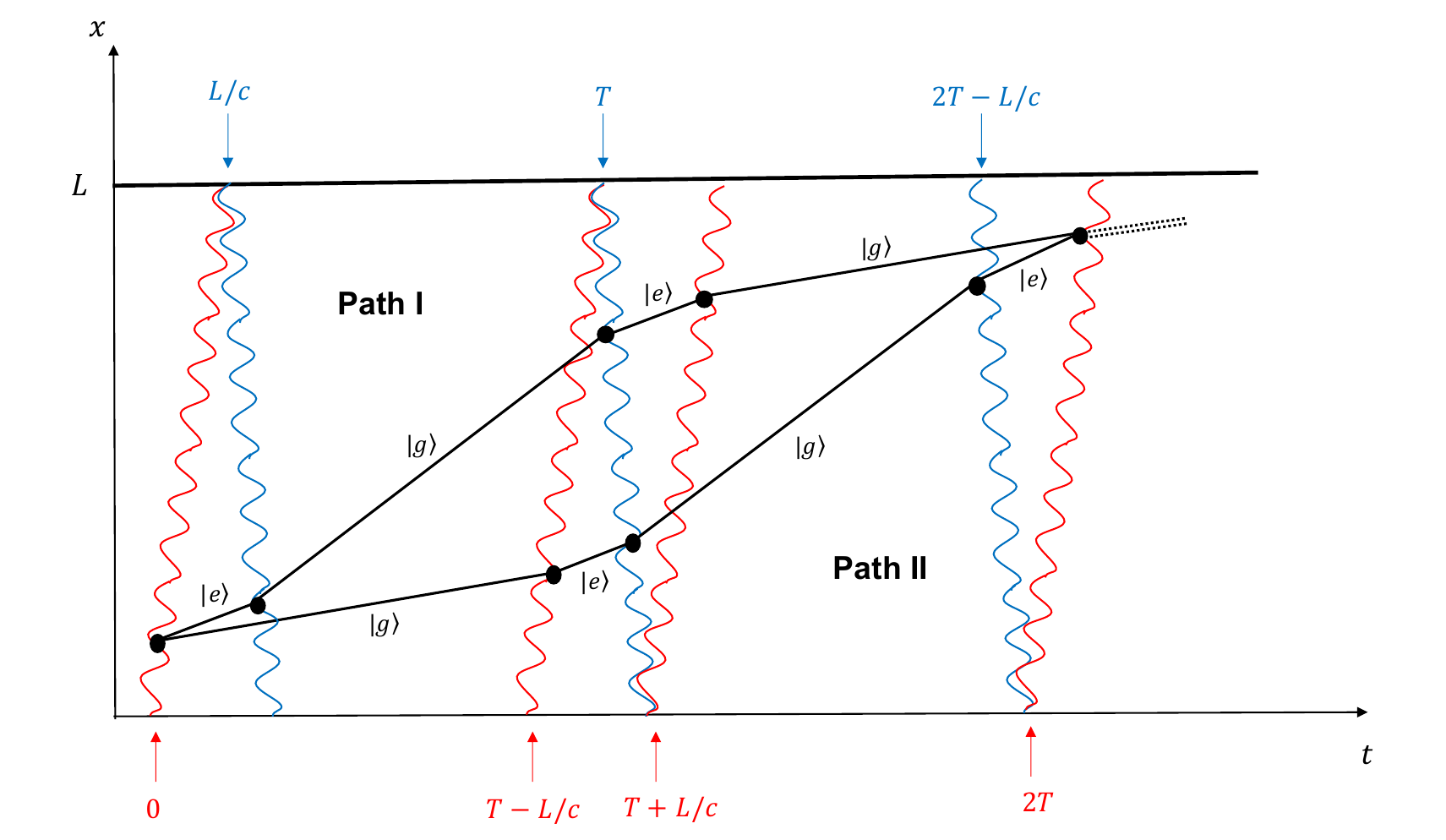}
    \caption{Spacetime diagram of a single photon transition, as proposed in \cite{Graham13} with $n=2$ (using the convention of \cite{Arvanitaki18}, see text.) Two lasers are used, one located at $x=0$, with emission (in red) towards the second laser located at $x=L$ (with blue emission). The frequencies of the lasers are selected carefully to interact only with the wavepacket of interest (see text). The black dots indicate the location of matter-wave interactions.}
    \label{fig:gradio}
\end{figure}
We are interested in the setup initially proposed by \cite{Graham13} and then studied in e.g \cite{Arvanitaki18,Badurina22}. Practically, we consider two ensembles of atoms $A$ (one for each interferometer) located respectively at $x_1$ and $x_2$, all initially in the state $|g,\hbar \vec k\rangle$ and two lasers, one at coordinate $x=0$ with  wavevector $\vec k_1$ and the other at coordinate $x=L$ with  wavevector $\vec k_2$.  In those single photon interactions configurations, $k_1 \approx k_2 = k = \omega^0_A/c$. 

At an initial time $t_0$, the first laser sends a beam which interacts with both atom ensembles and which corresponds to a $\pi/2$ pulse. Then, the second laser beam sends a $\pi$ Doppler shifted laser pulse in order to interact only with the excited state wavepackets, whose motion induces a change in transition frequency, and to convert entirely this wavepacket to the ground state. After this sequence, all wavepackets are in the ground state, however one of them has gained momentum  $2\hbar k$. Following the convention of \cite{Arvanitaki18, Badurina22}, at the end of the sequence, the fast wavepackets have received a Large Momentum Transfer (LMT) photon kick of order 2\footnote{Note that in the original proposal by \cite{Graham13}, the LMT has a slightly different definition.}. If another pair of $\pi$ pulses (with the first one from the bottom laser and the second one from the top laser in the same way as before) is sent to the faster half of the atom, one makes a large momentum transfer (LMT) beam splitter of order 4. More generally, if $m$ pairs of $\pi$ pulses are sent, the order of the LMT beam splitter is $2(m+1)$ and the faster wavepacket has gained total momentum $2(m+1)\hbar k$. In other words, the order $n$ of the LMT is defined as $n=2(m+1)$.

Later at time $t=T$, a  sequence of state inversion similar to the one used in two-photon transitions interferometers is performed, but this time, with three different $\pi$ pulses, the first one coming from the bottom laser, the second one from the top laser and the last one from the bottom laser again. However, in order to slow down the faster wavepacket, $m$ pairs of $\pi$ pulses are added before this sequence, such that it loses $2(m+1)\hbar k$ momentum. Symmetrically, $m$ other pairs of LMT pulses are added after the state inversion to accelerate the other wavepacket, such that it gains $2(m+1)\hbar k$ momentum.

Finally at time $t=2T$, a sequence of pulses opposite to the one sent at the initial $t=0$ is sent to the wavepackets, i.e $m$ LMT pairs of $\pi$ pulses are sent to the wavepackets before the final $\pi-\pi/2$ pulses used for recombination. This whole sequence is depicted in Fig.~\ref{fig:gradio} with $n=2$ (i.e $m=0$) and only one interferometer.

\begin{figure}
    \centering
    \includegraphics[width=0.7\textwidth]{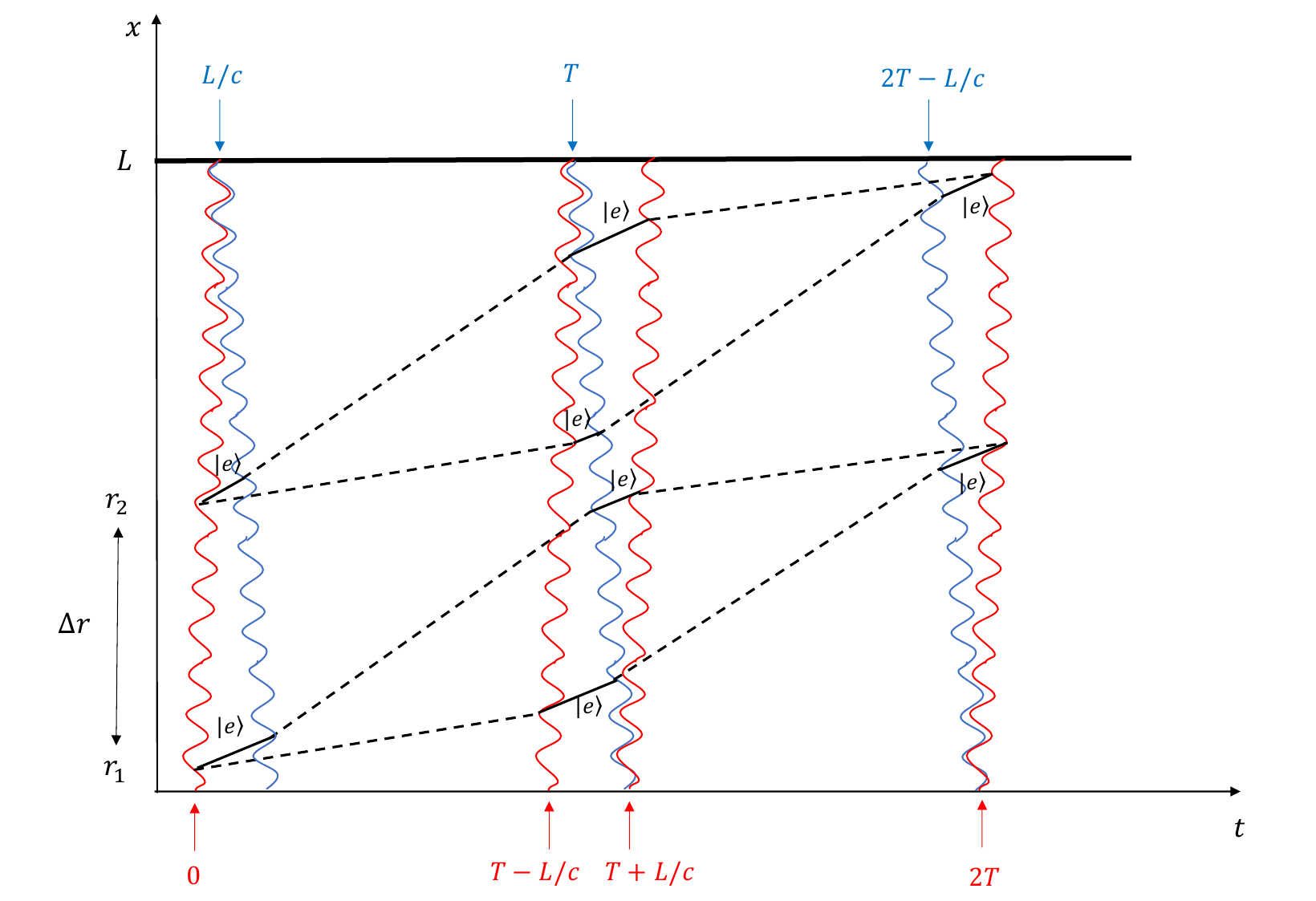}
    \caption{Spacetime diagram of a gradiometer, where two interferometers as shown in Fig.~\ref{fig:gradio} are stacked at different altitudes. In this figure, trajectories of the atom in its ground stat are represented by dashed lines, while full lines represent paths where the atom is in its excited state. }
    \label{fig:gradio_multiple}
\end{figure}

In gradiometers, two such interferometers are stacked at different altitudes, separated by a distance $\Delta r$, and the same laser beam is used for the laser-atom interaction in both interferometers, which is depicted in Fig.~\ref{fig:gradio_multiple}.
The main advantage of this setup is that laser phase noise is entirely cancelled when measuring the differential phase shift between the two interferometers.

Assuming $\omega_\phi L/c,\omega_a L/c \ll 1$ and $n L/c \ll T$ where the LMT kick is of order $n$, and $L$ is the baseline separation between the two lasers, which we assume for simplicity to be the distance between the two interferometers, the differential phase shift can be computed following the exact same methodology as in Section~\ref{sec:calc_phase_shift} and reads
\begin{subequations}\label{eq:phase_gradio}
\begin{align}
|\Delta \Phi^\mathrm{Grad}_\mathrm{A}|_d &\approx \frac{4\sqrt{16\pi G \rho_\mathrm{DM}}n\omega^0_A \Delta r [Q^A_\omega]_d}{\omega_\phi c^2}\sin^2\left(\frac{\omega_\phi T}{2}\right) \, \\
|\Delta \Phi^\mathrm{Grad}_\mathrm{A}|_a &\approx \frac{32\pi G \rho_\mathrm{DM}E^2_P n\omega^0_A \Delta r [Q^A_\omega]_a}{f^2_a \omega^2_a c^3}\sin^2\left(\omega_a T\right) \, .
\end{align}
\end{subequations}
We recover the leading order phase shift amplitude already derived in \cite{Graham13,Badurina22}.

\section{\label{sec:SPID_prop_pres}Single Photon Isotope Differential interferometer}

We study another setup employing single photon transitions, that we name \textit{SPID} in the following, for Single Photon Isotope Differential AI, which is a variation of the interferometric sequence presented in the previous section. We will show that this experimental setup is more sensitive to oscillations in atom rest mass and transition frequency compared to regular gradiometers. This type of setup has already been proposed, without any detail, for ULDM detection in \textit{MAGIS-100} \cite{Abe21} (see end of this section for a discussion). The goal of this section is to show the expected signals of such a setup, in order to compare it directly with gradiometers, as the one expected to be used in \textit{AION-10} \cite{Badurina22}.

\begin{figure}
    \centering
    \includegraphics[width=0.8\textwidth]{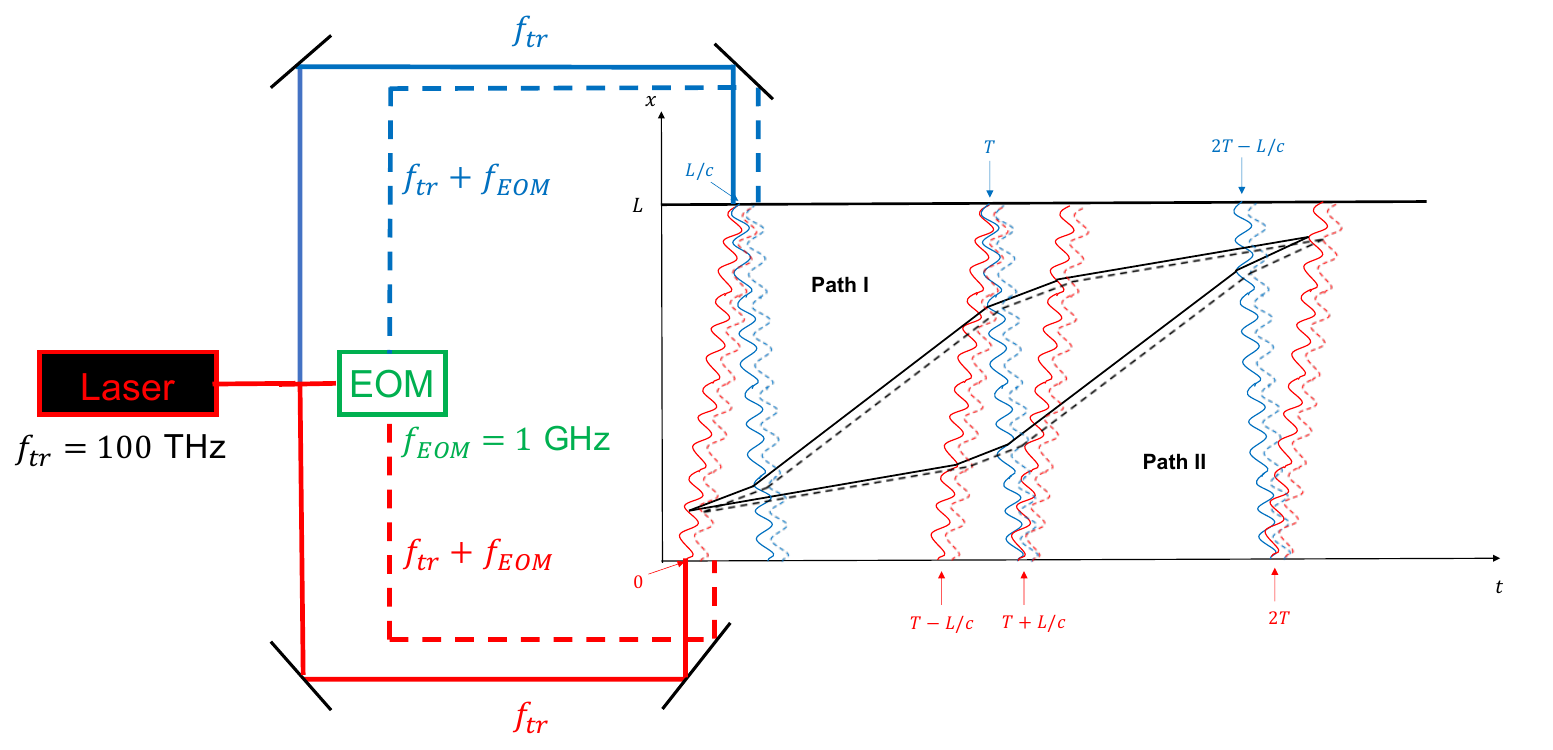}
    \caption{Simplified setup for the \textit{SPID} experiment. For the sake of simplicity, we assume two different isotopes with respective transition frequencies $f_1=f_\mathrm{tr}=10^{14}$ Hz, $f_2 =f_\mathrm{tr}+f_\mathrm{EOM}=(1+10^{-5})\times10^{14}$ Hz. The laser is locked on the transition frequency of isotope 1 and split in three: the first two outputs are used for the isotope 1 AI (in red). The third output enters an electro-optic modulator (EOM) to shift its frequency by $f_\mathrm{EOM}$ to account for the isotope shift in order to be used with the second isotope. As for the other experimental AI schemes, we do not show the various acousto-optic modulators (AOM) that shift the frequency of one of the laser input of each interferometer, to account for the Doppler shift of the freely falling atoms.}
    \label{fig:isotope_AI}
\end{figure}

Contrary to usual gradiometers which only use one single species of atom, we consider two different atom isotopes, each of them undergoing individually the interferometric path described earlier in the previous section. The setup is presented in Fig.~\ref{fig:isotope_AI}. The two interferometers overlap at the same elevation, so the experiment will test the universality of free fall between the two isotopes. This setup employs single photon transition (meaning we will consider only optical transitions) and measures the differential acceleration between two isotopes, so we will refer to this setup as Single Photon transition Isotope Differential (\textit{SPID}).

Typically, an optical transition has a frequency of the order of $\sim$10$^{14}$~Hz while the typical frequency shift between two isotopes is of order $\sim$10$^{9}$~Hz (i.e. typically 5-6 orders of magnitude smaller than the nominal transition frequency, see \cite{Takano17} for Sr). 
Then, a unique laser source can be used in this setup and separated in two different beams: one which is directly used in the laser-atom interactions inside the interferometer of the first isotope; and the other one whose frequency is shifted, e.g. using an electro-optic modulator (EOM), in order to interact with the second isotope. 

The calculation of the differential phase shift for such a setup can be computed following the exact same methodology as in Section~\ref{sec:calc_phase_shift}. If we assume the optical transition frequencies to be close, i.e $\omega^0_A \approx \omega^0_B \equiv \omega_0$ and the initial velocity ($\propto v_\mathrm{DM}$) to be much larger than the velocity kick $\hbar k_\mathrm{eff}/m^0$, the differential phase shift between the two interferometers reads
\begin{subequations}\label{eq:phase_new_prop}
\begin{align}
    |\Delta \Phi^\mathrm{SPID}_\mathrm{AB}|_d &\approx \frac{4\sqrt{16 \pi G \rho_\mathrm{DM}}v_\mathrm{DM}n\omega_0}{\omega^2_\phi c^2}\left|([Q^A_M]_d-[Q^B_M]_d)\hat e_v \cdot \hat e_\mathrm{kick}\right|\sin^2\left(\frac{\omega_\phi T}{2}\right)\, \\
    |\Delta \Phi^\mathrm{SPID}_\mathrm{AB}|_a &\approx \frac{16 \pi G \rho_\mathrm{DM} E^2_P v_\mathrm{DM}n\omega_0}{f^2_a\omega^3_a c^3}\left|([Q^A_M]_a-[Q^B_M]_a)\hat e_v \cdot \hat e_\mathrm{kick}\right|\sin^2\left(\omega_a T\right)\, ,
\end{align}
\end{subequations}
where we kept only lowest order terms in $\omega L/c$. In this result, we did not take into account the effect of an oscillating EOM frequency through $Q^\mathrm{EOM}_\omega \neq 0$. Nonetheless, its effect is suppressed by a factor $\omega L/v_\mathrm{DM} \times \Delta \omega^0/\omega^0 \leq 10^{-7}$ compared to the leading term, where $\Delta \omega^0$ is the isotope shift. In such a case (and similarly for $L=0$), we recover the leading order Bragg phase shift derived in Eq.~\eqref{eq:MZ_delta_phase_shift} with $n\omega_0/c=k_\mathrm{eff}$, as expected.

The advantage of this setup compared to usual gradiometers can be immediately visualized by comparing Eq.~\eqref{eq:phase_new_prop} and Eq.~\eqref{eq:phase_gradio}: the \textit{SPID} setup does not suffer from a small factor $\omega \Delta r/v_\mathrm{DM}$. More precisely, the ratio of amplitude of signals between this variation and usual gradiometers is roughly $v_\mathrm{DM}/(\omega\Delta r) \times (Q^A_M-Q^B_M)/Q^A_\omega \sim \left(1 \mathrm{\: rad.s^{-1}}/\omega\right)$ for the following values: $\Delta r \sim 5$ m, $v_\mathrm{DM} \sim 10^{-3} c$, $\hat e_v \cdot \hat e_\mathrm{kick} \sim \mathcal{O}(1)$ and mass and frequency charges presented in Tables ~\ref{dilatonic_charge_table} and \ref{axionic_charge_table}.). This implies that, at low angular frequency ($\omega < 1$ rad/s), the signal of the single photon transition isotope differential AI will be larger than the one in a gradiometer. The simple reason for this difference in signal amplitudes is that the gradiometer leading order phase shift in Eq.~\eqref{eq:phase_gradio} is proportional to the frequency charge $Q_\omega$, or in other words, it is proportional to the time in which the wavepackets are in their excited state. As can be noticed from Fig.~\ref{fig:gradio}, this happens for a limited time, of the order of $nL/c$. Conversely, the signal amplitude of the \textit{SPID} variation is proportional to mass charges $Q_M$ whose effect is imprinted in the phase shift, whatever the internal state, i.e for a time $\sim T$.

As it was pointed out in the beginning of this section, \textit{MAGIS-100} \cite{Abe21} will operate a similar mode as \textit{SPID} for the search of ULDM and we want to derive the expected sensitivity of such a large scale experiment to axion and dilaton signals. More specifically, \textit{MAGIS-100} will run a Bragg interferometer (i.e with two photons transitions instead of single-photon transition, as in \textit{SPID}). However, as mentioned in the text after Eq.~\eqref{eq:phase_new_prop}, the leading order signal of both setup is the same, provided that the beams are locked on the same optical transition in both cases. The main difference between the two setup is that for \textit{SPID}, the wavepackets spend some time in their excited state (which they do not in Bragg configuration), but as shown in Eq.~\eqref{eq:phase_new_prop}, the additional phase shift is next-to-leading order, and thus negligible. In addition, as we shall see in Chapter ~\ref{chap:sens_experiments}, the phase noise levels will be equivalent in both setups. Therefore, the optimal choice between the two setup is a matter of practicality. In particular, we will be deriving the sensitivity of such setup considering optical transitions summarized in Table ~\ref{tab:alk_isotope_freq}. In these choices of isotopes and transitions, we ignore the small lifetime of the various excited states (of hundreds of $ns$ to hundreds of $\mu s$), which could limit the number of atoms detected at the end of the sequence due to spontaneous emission\footnote{This limitation could be overcome for optical transitions due to the large Rabi frequency, implying a very short $\pi$ pulse duration \cite{Rudolph20}.}

\clearpage
\pagestyle{plain}
\printbibliography[heading=none]
\clearpage
\pagestyle{fancy}

\part{Effects on gravitational waves detectors}
\chapter{Gravitational waves and \textit{LISA}}

\section{Gravitational waves physics basics}

As quickly mentioned in the introduction of this thesis, one of the main prediction of GR is the production of GW by massive objects. More precisely, only systems with time dependent quadrupolar moment (or higher order moment) produce such waves. 

GW are solutions of the linearized Einstein equations. Assuming a static Minkowski background $\eta_{\mu\nu}$ perturbed by $h_{\mu\nu} \ll \eta_{\mu\nu}$ such that the spacetime metric is $g_{\mu\nu} = \eta_{\mu\nu} + h_{\mu\nu}$, around a system energy momentum tensor $T_{\mu\nu}$, one can show that \cite{Flanagan05}
\begin{subequations} 
\begin{align}\label{GW_wave_equation}
    \square \bar h_{\mu\nu} = \frac{-16 \pi G}{c^4}T_{\mu\nu} \, ,
\end{align}
where $\square \equiv \eta^{\alpha\beta}\partial_\alpha \partial_\beta$ and 
\begin{align}
 \bar h_{\mu\nu} = h_{\mu\nu}-\frac{1}{2}\eta_{\mu\nu}h^{\alpha}_{\:\alpha} \, .
\end{align}
\end{subequations}
In presence of matter-energy, Eq.~\eqref{GW_wave_equation} describes the emission of the GW sourced by $T_{\mu\nu}$, while in vacuum where $T_{\mu\nu}=0$, it describes the propagation of the GW. The energy-momentum tensor being symmetric, it consists of $10$ independent components, therefore  Eq.~\eqref{GW_wave_equation} is a set of $10$ independent equations. Imposing Lorenz gauge $\partial_\mu \bar h^{\mu\nu} =0$ and the so-called $TT$ (traceless-transverse) gauge $\bar h^\mu_{\: \mu} = 0, \partial^i \bar h_{ij}=0$ leaves only two independent radiative degrees of freedom, which are the two GW polarizations. Therefore, in the $TT$ gauge, we have $h^\mathrm{TT}_{\mu\nu}=\bar h^\mathrm{TT}_{\mu\nu}$. Assuming a GW wave with frequency $f=\omega/2\pi$ propagating in the $z$-direction, we have $h^\mathrm{TT}_{\mu\nu} = \Re[A^\mathrm{TT}_{\mu\nu}e^{-i(\omega t - kz)}]$ with $A^\mathrm{TT}_{11} = A_+$ and $A^\mathrm{TT}_{12}= A_\times$, and the amplitude of the wave is
\begin{align}
    A^\mathrm{TT}_{\mu\nu} = \begin{pmatrix}
        0 & 0 & 0 & 0 \\
        0 & A_+ & A_\times & 0 \\
        0 & A_\times & -A_+ & 0 \\
        0 & 0 & 0 & 0
    \end{pmatrix} &= A_+ \begin{pmatrix}
        0 & 0 & 0 & 0 \\
        0 & 1 & 0 & 0 \\
        0 & 0 & -1 & 0 \\
        0 & 0 & 0 & 0
    \end{pmatrix} + A_\times\begin{pmatrix}
        0 & 0 & 0 & 0 \\
        0 & 0 & 1 & 0 \\
        0 & 1 & 0 & 0 \\
        0 & 0 & 0 & 0
    \end{pmatrix} \,\nonumber \\
    &\equiv A_+ \epsilon^+_{\mu\nu} + A_\times \epsilon^\times_{\mu\nu}
\end{align}
The amplitudes $A_+, A_\times$ can be found by solving Eq.~\eqref{GW_wave_equation}. Using Green's functions and the conservation of the energy-momentum tensor $\partial_\mu T^{\mu\nu}=0$, one can show that the spatial solutions of Eq.~\eqref{GW_wave_equation} is the perturbed spacetime at coordinates $(t,\vec x)$ induced by a GW emitted by a source located at spatial coordinates $\vec y$, and has the form \cite{Flanagan05} 
\begin{subequations}
\begin{align}
    h^\mathrm{TT}_{ij}(t, \vec x) = \frac{2G}{c^4 R}\mathcal{P}_{ijab}\frac{\partial^2}{\partial t^2}\int d^3 \vec y \left(y_a y_b - \frac{1}{3} \delta_{ab} |\vec y|^2 \right) \: T_\mathrm{00}\left(t-\frac{|\vec x-\vec y|}{c}, \vec y\right)\, , 
\end{align}
where the argument of the integral is the quadrupolar moment tensor of the source and where we introduced the projection operators
\begin{align}
    \mathcal{P}_{ijab}&=\mathcal{P}_{ia}\mathcal{P}_{jb}-\frac{1}{2}\mathcal{P}_{ij}\mathcal{P}_{ab} \,\\
    \mathcal{P}_{ij} &=\delta_i \delta_j - \hat k_i \hat k_j \, ,
\end{align}
\end{subequations}
that ensures that $h^\mathrm{TT}_{ij}$ is traceless and transverse to the propagation vector $\hat k$. This result is derived assuming that the distance $|\vec x -\vec y| \gg |\vec y|$, the size of the source.

The effect of the GW is the modification of physical distances between test masses in the plane of polarization. Considering the same GW as above, assuming two freely falling bodies located at $z=0$, separated on the $x$-axis by a coordinate distance $L_0$. When the GW passes by, this coordinate distance is unchanged, because the coordinates move with the GW \cite{Flanagan05}. The relevant and physical quantity is the proper distance, as the distance measured by a light signal and a clock, which is shown to change as \cite{Flanagan05}
\begin{align}\label{eq:GW_strain}
    L(t) &= \int_0^{L_0} dx \sqrt{g_{xx}} \approx L_0\left(1+\frac{1}{2}h^{TT}_{xx}(t,z=0)\right) \, .
\end{align}
More generally, if we assume a ring of particles in the $x-y$ plane, the $+$ polarization of the GW makes the ring oscillating in the $x-y$ directions while the $\times$ polarization acts on the ring along the $x'-y'$ directions which are rotated by $45\degree$ with respect to $x-y$ coordinates (i.e $\hat e_{x'} = \sqrt{2}(\hat e_x + \hat e_y)/2$ and $\hat e_{y'} = \sqrt{2}(-\hat e_x + \hat e_y)/2$). This is the reason for the $+,\times$ polarization names, as they respectively distort a circular ring of particles to make a $+$ and $\times$ shape. 

Since the GW strain (i.e the relative displacement $(L(t)-L_0)/L_0 \equiv \delta L/L_0$ in Eq.~\eqref{eq:GW_strain}) is expected to be extremely small, of the order of $10^{-21}$ or less, one must use the most precise method to measure distances, i.e optical interferometers. For example, the Earth-based GW detectors \textit{LIGO/VIRGO} are operating a Michelson optical interferometer in a $L$-shape in order to maximize the effect of the GW, i.e stretch one arm and compress the other. In such a case, a photon of wavelength $\lambda$ that travels back and forth on the arm whose distance has changed acquires a phase shift $\Delta \phi$.
Note that even if light is stretched by the passage of the GW (i.e its frequency changes), one can still use it as a "ruler" to detect any change in physical distance between two freely falling objects. This is because light is not used as a ruler but as a clock, i.e one does not measure the total number of crests that fits in a given distance (e.g. between reflecting mirror and beam splitter in an usual GW detector) but rather measures the number of crests of light that arrives at the beam splitter per unit of time. As the GW stretches (or compresses light), the crests will arrive at the beam splitter with a time delay compared to situations with no GW, and this creates an observable phase shift \cite{Saulson97}. 

\section{\textit{LISA}, the first european space-based gravitational waves detector}

After the success of the ground-based GW detectors such as \textit{LIGO} and \textit{VIRGO}, one of the next important mission aiming at detecting GW is \textit{LISA} (Laser Interferometer Space Antenna), see \cite{Amaro17} for a complete review of the mission and its scientific goals. In this manuscript, we will only describe the important features of the mission that will be needed for its understanding. 
\begin{figure}
    \centering
    \includegraphics[width=0.6\textwidth]{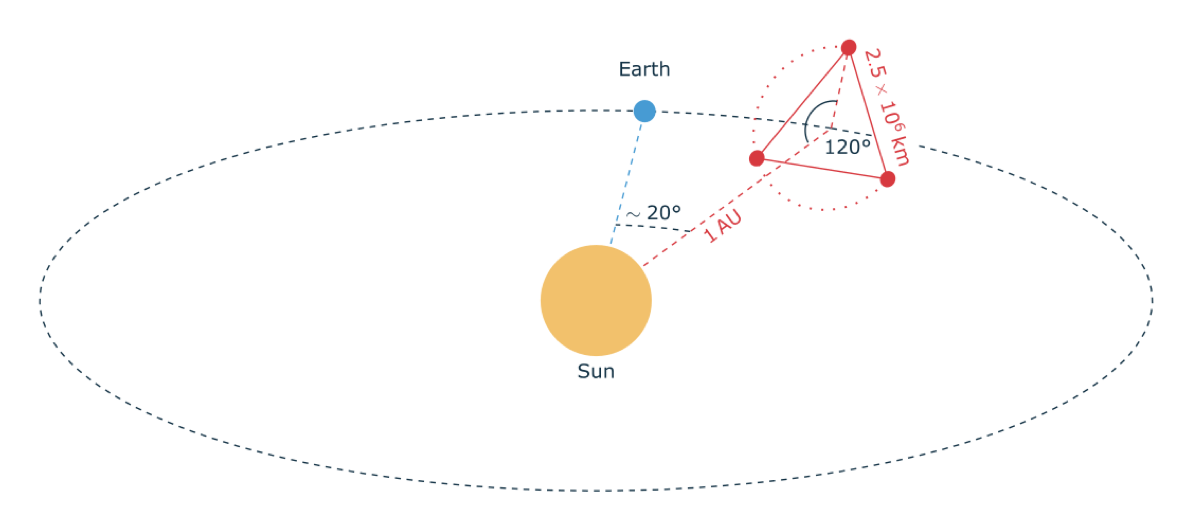}
    \caption{\textit{LISA} constellation heliocentric orbit (from \cite{Colpi24}).}
    \label{fig:orbit_LISA}
\end{figure}
As pictured in Fig.~\ref{fig:orbit_LISA}, \textit{LISA} will consist of a constellation of three spacecrafts, in the following noted $1, 2, 3$, which will follow Earth on an heliocentric orbit. In practice, \textit{LISA} will be able to detect GW of frequencies in the $10^{-4}-1$ Hz range, which is complementary to \textit{LIGO/VIRGO} sensitivity bands (around $10-10^4$ Hz). However, the mean spatial separation between the spacecrafts being $L = 2.5 \times 10^9$ m, at frequencies $f> c/L \sim 0.1$ Hz, the sensitivity decreases due to the arm length penalty \cite{Robson19}, i.e the GW wavelength is shorter than the arm length, which induces averaging of the signal. 
All spacecrafts will continuously produce two local laser signals through two local oscillators and exchange it with the other spacecrafts, and GW signals will be detected by interferometric measurement between those optical signals. Indeed, at the passage of a GW, the distance between spacecrafts changes and therefore, the observed phase of the laser signal that travelled in vacuum changes as well. More precisely, inside one given spacecraft (e.g. $1$), interferometric measurements will be made between the local oscillator of $1$ and the laser received from spacecraft $2$ using a phasemeter.  A very specific optical bench inside the spacecrafts allows to recombine the two optical signals and make the interference. A more precise description of such bench will be done in Chapter ~\ref{chap:axion_photon_LISA}. Overall, in order to make interferometric measurements between all the optical signals, each spacecraft contains two optical benches.
In addition, inside each of the spacecrafts is located a freely-falling test mass surrounded by metallic shields in order to protect from any exterior perturbations, e.g. radiation pressure from the Sun. This way, the test mass is insensitive to non-gravitational forces which allows it to stay on its geodesic. By measuring the distance between the test mass and the spacecraft around it, one knows when the spacecraft has left its geodesic, due to perturbations, and can replace the spacecraft on its geodesy using mini-rockets on board. Combining all these measurements in all spacecrafts, one can reconstruct the distance between the test masses.

Finally, let us discuss sources of noise in the apparatus, in particular laser noise. It is admitted that the laser noise amplitude spectral density (ASD) is $\sqrt{S_f} \sim 30$ Hz/$\sqrt{\mathrm{Hz}}$ \cite{Petiteau08}, which in terms of the nominal frequency of the laser $\nu_0 \sim 2.82 \times 10^{14}$ Hz \cite{Petiteau08}, corresponds to $\sqrt{S_f}/\nu_0 \sim 10^{-13}$ Hz$^{-1/2}$. In the same unit, the GW signal ASD is $\sim 10^{-21}$ Hz$^{-1/2}$, i.e at this level, the laser noise completely dominates the signal. Fortunately, there exists a method to effectively lower drastically the laser noise by combining signals from the different spacecrafts in such a way that the resulting observable is independent of laser noise, but still contains the GW signal. This is called \textit{Time Delay Interferometry}, or TDI and were first introduced in \cite{Armstrong99, Tinto99, Dhurandhar02}. The principle relies on the fact that each laser noise is measured several times. 

It can be shown mathematically (for details, we refer to \cite{Petiteau08}) that some time retarded signals combinations between the three different arms completely reduces the laser noise to zero, thus the name of the method. Here, we will only present some of the TDI combinations of interest. 

First, we will label the optical benches with two indices $ij$, where the spacecraft $i$ receives light signal (host optical bench) from emitting spacecraft $j$. Defining the laser phase of the laser on spacecraft $i$ by $\Phi_i$, the beatnote phase between $\Phi_i$ and $\Phi_j$ at time of reception $t_r$ is
\begin{subequations}
\begin{align}
    \Phi_{ij}(t_r) &= \Phi_j\left(t_r-\frac{L_{ij}(t_r)}{c}\right)-\Phi_i(t_r) \, ,
\end{align}
where $L_{ij}$ is the arm length between the two spacecrafts and where we assume that the measurements inside different spacecrafts are synchronized. We define the relative phase fluctuation
\begin{align}
    \delta \Phi_{ij}(t_r) &= -\frac{\omega_j \delta L_{ij}}{c} \equiv \frac{\omega_j L_{ij} \gamma_{ij}}{c}\, ,
\end{align}
where $\delta L_{ij}$ is the arm length fluctuation induced by the GW, $\omega_j$ is the local frequency of spacecraft $j$ and where the interferometric measurement is expressed in terms of relative arm length fluctuation induced by the GW
\begin{align}
    \gamma_{ij}(t_r) &= -\frac{\delta L_{ij}}{L_{ij}} \, .
\end{align} 
\end{subequations}
One can also define relative frequency fluctuation signals (i.e Doppler shifts) as 
\begin{align}\label{eq:Doppler_phase_shift_link}
    y_{ij}(t_r) &= \frac{1}{\omega_j}\frac{d \delta \Phi_{ij}(t_r)}{dt_r} = -\frac{1}{c}\frac{d\delta L_{ij}}{dt} \, . 
\end{align}
Using the sign convention of \cite{Bayle21}, we define the first (and second) generation Michelson $X_1$ ($X_2$) combination as \cite{Nam23}
\begin{subequations}
\begin{align}\label{eq:TDI_X_gen}
    X_1(t) &= y_{13}+D_{13}y_{31}+D_{131}y_{12}+D_{1312}y_{21}-\left(y_{12}+D_{12}y_{21}+D_{121}y_{13}+D_{1213}y_{31}\right)\, \\
    X_2(t) &= (1-D_{12131})\left(y_{13}+D_{13}y_{31}+D_{131}y_{12}+D_{1312}y_{21}\right)-\,\nonumber\\
    &(1-D_{13121})\left(y_{12}+D_{12}y_{21}+D_{121}y_{13}+D_{1213}y_{31}\right) \, ,
\end{align}
\end{subequations}
where we used the compacted writing of TDI where $D_{ij}y_{jk}(t) = y_{jk}(t-L_{ij}/c)$ and $D_{ijk}y_{kl}(t) = y_{kl}(t-L_{ij}/c-L_{jk}(t-L_{ij}/c))$. 
It can be shown geometrically that the circulation of the laser beams in that case is very similar to an usual optical Michelson interferometers, thus the name of the combination. The first generation combinations are not sensitive to the laser noise in the constant armlength approximation while the second generation takes into account effects of rotation of the detector (Sagnac effect) and of the time variation of armlengths. The residuals of laser noise in the second generation TDI is proportional to the difference in spacecraft velocities squared \cite{Hartwig22}.
In the constant armlength approximation and considering that the Doppler shifts are harmonic functions of the frequency $\omega$, the second generation TDI X combination  Eq.~\eqref{eq:TDI_X_gen} can be rewritten as
\begin{subequations}
\begin{align}
    X_2 &= \left(1-e^{-\frac{4i\omega L}{c}}\right)\left(y_{13}+e^{-\frac{i\omega L}{c}}y_{31}+e^{-\frac{2i\omega L}{c}}y_{12}+e^{-\frac{3i\omega L}{c}}y_{21}\right)-\,\nonumber \\
    &\left(1-e^{-\frac{4i\omega L}{c}}\right)\left(y_{12}+e^{-\frac{i\omega L}{c}}y_{21}+e^{-\frac{2i\omega L}{c}}y_{13}+e^{-\frac{3i\omega L}{c}}y_{31}\right) \, \\
    &=-4\sin \left(\frac{\omega L}{c}\right)\sin \left(\frac{2\omega L}{c}\right) \left(e^{\frac{-3i\omega L}{c}} \left(y_{13} -y_{12} + e^{\frac{-i\omega L}{c}}\left(y_{31}-y_{21}\right)\right)\right) \label{eq:TDI_X2_approx}\, ,
\end{align}
\end{subequations}
where we used the fact that one delay operator corresponds to a factor $\exp(-i\omega L/c)$.
The other two Michelson $Y,Z$ combinations can be obtained from Eq.~\eqref{eq:TDI_X_gen} by a cyclic permutation of the indices $1 \rightarrow 2 \rightarrow 3 \rightarrow 1$. Other important combinations are the $A,E,T$ combinations which are linear combinations of $X,Y,Z$ \cite{Prince02}
\begin{subequations}\label{eq:AET_TDI}
    \begin{align}
        A(t) &= \frac{1}{\sqrt{2}}\left(Z(t) - X(t)\right) \, \\
        E(t) &= \frac{1}{\sqrt{6}}\left(X(t)-2Y(t)+Z(t)\right)\,\\
        T(t) &= \frac{1}{\sqrt{3}}\left(X(t)+Y(t)+Z(t)\right) \, ,
    \end{align}
\end{subequations}
which we will be using for the analysis. These combinations are useful because their cross noise power spectral density are vanishing in the equal armlength approximation \cite{Nam23}. For convenience, we also introduce the Sagnac combination $\alpha$ defined as \cite{Hartwig23}
\begin{subequations}\label{eq:Sagnac_combination}
\begin{align}
    \alpha_1(t) &= -\left(y_{13} + D_{13}y_{32} + D_{132}y_{21} -\left(y_{12}+D_{12}y_{23}+D_{123}y_{31} \right)\right) \, \\
    \alpha_2(t) &=\left(1-D_{1231}\right)\alpha_1(t) \, ,
\end{align}
\end{subequations}
respectively for the first and second generations.

\section{\textit{LISA}'s expected performance}

As mentioned in the previous section, the Earth-based GW detectors  \textit{LIGO/VIRGO} are sensitive to GW in the $ \sim 10-10^4$ Hz frequency range. Out of the $\sim 90$ various signals detected by the collaboration, all of them are transients signals \cite{Abbott23}, which are GW signals that last for a few seconds, and which correspond to the last periods of inspirals between the two bodies and their merge into a new body. 

On the contrary, \textit{LISA} is expected to be sensitive to various different signals \cite{Petiteau08}. In particular, it will be able to detect the inspiral phase of binary systems much before the merge. In such cases, the two bodies rotate around each other at an almost constant frequency, up to corrections (see below), for a time which exceeds largely the time of observation of the mission. Therefore, GW emitted by such sources are quasi-monochromatic. As we shall see in the following, this is an important point as DM induces also a monochromatic oscillation on the test masses. 

In terms of noise, we will be interested in the single-link optical metrology noise and single test mass acceleration noise which respectively read \cite{Robson19,Babak21}
\begin{subequations}\label{eq:noise_PSD_LISA}
\begin{align}
S_\mathrm{oms} &= \left(1.5 \times 10^{-11} \: \mathrm{m}\right)^2\left(1+\left(\frac{2 \: \mathrm{mHz}}{f}\right)^4\right) \: \mathrm{Hz}^{-1} \, \\
S_\mathrm{acc} &= \left(3 \times 10^{-15}\: \mathrm{m.s}^{-2}\right)^2\left(1+\left(\frac{0.4 \: \mathrm{mHz}}{f}\right)^2\right)\left(1+\left(\frac{f}{8 \: \mathrm{mHz}}\right)^4\right)\: \mathrm{Hz}^{-1} \, ,
\end{align}
\end{subequations}
corresponding to the instrument performance requirements \cite{Colpi24}. 

\chapter{\label{chap:LISA_DM}Search for oscillations of rest mass in \textit{LISA}}

Many of the various couplings that we have seen so far violate the UFF, by inducing an oscillating acceleration on atoms that is element-dependent. Even though the different test masses used in \textit{LISA} have the same composition, those oscillations could still be seen in gravitational waves detectors as a Doppler effect when photons are exchanged between one spacecraft and another.

\section{Signatures of dark matter and gravitational waves on \textit{LISA} arms}

In this section, we derive the Doppler on photons exchanged on one arm of \textit{LISA} induced by both ULDM fields and GW\footnote{As we shall see in the following, we discuss the Doppler induced by the dynamical motion of the test masses induced by ULDM. As it is pointed out in \cite{Grote19}, a change in the length of optical elements (like beam splitters) arises also through the variation of fundamental constants of Nature. While this effect is dominant in ground-based GW detectors (and was used for search for ULDM signals in \textit{LIGO}, see e.g \cite{Gottel24}), it is sub-dominant in space-based GW detectors.}. We make the calculations in the barycentric reference frame, where the spacecrafts positions are defined \cite{Petiteau08}. In this frame, the directions of DM wind and GW sources are defined with the ecliptic latitude $\beta$ and longitude $\lambda$.

\subsection{Oscillating rest mass}

\subsubsection{One-arm Doppler shift through oscillating rest mass}

In this section, we derive the Doppler shift on the \textit{LISA} interferometric measurements induced by the various ULDM candidates.

We start by the effects of scalar fields, i.e the UFF violating acceleration induced by the oscillation of the rest mass of a test mass A. In this part, we first consider only interactions between SM and a pure dilatonic field $\phi$, then the Doppler shift will be generalized to the axion field as well. From Eq.~\eqref{EP_viol_acc_dil}, the acceleration of a test mass $A$ in a generic frame is 
\begin{align}
    \vec a_A(t,\vec x) &= \left[\omega_\phi \vec v_A -\vec k_\phi c^2\right]\frac{\sqrt{16 \pi G \rho_\mathrm{DM}}}{\omega_\phi c} [Q^A_M]_d \sin(\omega_\phi t - \vec k_\phi \cdot \vec x +\Phi) \, , \label{EP_viol_acc_rm}
\end{align}
where $\vec v_A$ is the velocity of the test mass, $\omega_\phi,\vec k_\phi$ are respectively the Compton frequency and the wavenumber of the DM field, and $[Q^A_M]_d$ is the mass charge of the test mass, i.e the amplitude of oscillation of the rest mass of $A$.
In the barycentric frame, the velocity of the test mass is roughly the same as the Earth velocity around the Sun, i.e $v_A \sim 3 \times 10^4$ m/s, which is one order of magnitude smaller than the galactic velocity, therefore the second term $\propto -\vec k_\phi c^2 = \omega_\phi v_\mathrm{DM}$ dominates. One can treat the position $\vec x(t)$ in Eq.~\eqref{EP_viol_acc_rm} as the unperturbed position $\vec x_0(t)$ at leading order and write it as $\vec x(t) = \vec x_0(t) + \delta \vec x_A(t)$, where the latter is $\mathcal{O}([Q^A_M])$. Considering that $\vec x(t) = \vec x_\mathrm{AU} \cos(\omega_E t)$ at leading order, where $|\vec x_\mathrm{AU}| \sim 1.5 \times 10^{11}$m is one astronomical unit distance and $\omega_E$ is the Earth rotation frequency around the Sun, one can integrate twice this expression, which leads to the oscillation of the position of the test mass\footnote{To make this integration, one needs to expand the $\sin(\omega_\phi t - \vec k_\phi \cdot \vec x +\Phi)$ in power of $\vec k_\phi \cdot \vec x_\mathrm{AU} \cos(\omega_E t)\ll 1$, integrate each term of the series and then reconstruct the sine. We also considered $\omega_E \ll \omega_\phi$.}
\begin{align}\label{osc_pos_LISA}
    &\delta \vec x_A(t,t_0,\vec x) = \frac{\vec k_\phi c\sqrt{16 \pi G \rho_\mathrm{DM}}}{\omega^3_\phi} [Q^A_M]_d \left[\sin(\omega_\phi t - \vec k_\phi \cdot \vec x_\mathrm{AU}\cos(\omega_E t) +\Phi)-\right.\, \\
    &\left.\sin(\omega_\phi t_0 - \vec k_\phi \cdot \vec x_\mathrm{AU}\cos(\omega_E t_0) +\Phi)-\omega_\phi(t-t_0)\cos\left(\omega_\phi t_0 - \vec k_\phi \cdot \vec x_\mathrm{AU}\cos(\omega_E t_0) +\Phi\right)\right] \, \nonumber .
\end{align}
The three last terms are in general not taken into account in various analysis. Indeed, two of them have no time dependence. However, note that one of them $\propto \omega_\phi t$ induces a temporal drift of the test mass inside the spacecraft, which is due to its initial inertia. In the following, we will drop those terms and only work with the time oscillating signal, because they are outside the \textit{LISA} band. In addition, we will not keep the explicit expression of $\vec x(t)$, but we will come back to it later. The slow oscillation at $\omega_E$ will produce side bands in the Fourier signal at frequencies $\omega_\phi \pm \omega_E$. However, note that in the following, we will assume a time of integration of one year $T_\mathrm{obs}=2\pi/\omega_E$, such that the size of one Fourier bin is exactly $f_E=\omega_E/2\pi$. Therefore, these side bands might not be visible but will most likely enlarge the signal at $\omega_\phi$.

Assuming that spacecraft $e$ (at position $\vec x_e$) sends a signal to spacecraft $r$ (at position $\vec x_r$, the one-way Doppler shift of light associated to this rest mass oscillation is obtained by the projection onto the corresponding \textit{LISA} arm of the time derivative of the variation of the position $\delta \vec x$ i.e \cite{Yu23}
\begin{subequations}
\begin{align}
    y_{re} &= -\frac{1}{c}\left(\hat n_{re} \cdot \frac{d\delta \vec x}{dt}\right) \, ,
\end{align}
where $\hat n_\mathrm{re}$ is a unit vector pointing from $e$ to $r$. We can now express the Doppler shift induced by the dilaton-SM  interaction Eq.~\eqref{osc_pos_LISA}, but also considering an axion-gluon interaction which induces an acceleration Eq.~\eqref{EP_viol_acc_axion}. We respectively note $y^d_{re}$ and $y^a_{re}$ these one-way Doppler shifts which read
\begin{align}
    y^d_{re} &= \left(\hat n_{re} \cdot \hat e_v\right)\frac{\sqrt{16 \pi G \rho_\mathrm{DM}}v_\mathrm{DM}[Q_M]_d}{\omega_\phi c^2}\Re\left(e^{i\left(\omega_\phi t - \vec k_\phi \cdot \vec x_r + \Phi\right)}- e^{i\left(\omega_\phi t - \frac{\omega_\phi L}{c}-\vec k_\phi \cdot \vec x_e + \Phi\right)}\right) \,\label{eq:Doppler_scalar_DM} , \\
    y^a_{re} &= \left(\hat n_{re} \cdot \hat e_v\right)\frac{8 \pi G \rho_\mathrm{DM} v_\mathrm{DM} E^2_P[Q_M]_a}{f^2_a \omega^2_a c^3}\Re\left(e^{2i\left(\omega_a t - \vec k_a \cdot \vec x_r + \Phi\right)}-  e^{2i\left(\omega_a t - \frac{\omega_a L}{c}-\vec k_a \cdot \vec x_e + \Phi\right)}\right) \, \,\label{eq:Doppler_pseudoscalar_DM},
\end{align}
\end{subequations}
respectively for dilaton and axion interactions, (using Eq.~\eqref{EP_viol_acc_axion}), where all the test masses are made of same alloy, therefore they all share the same mass charge $[Q_M]_d$ (and similarly for the axion mass charge), and $\vec k_\mathrm{DM} = - \omega_\mathrm{DM} v_\mathrm{DM}\hat e_v /c^2$ ($\hat k = - \hat e_v$, where $\hat e_v$ is the unit vector in the direction of the Sun velocity in the galactic halo).

As shown in Chapter ~\ref{DP_pheno}, the light vector field induces an oscillating acceleration on test masses by a small Lorentz force that acts on them. From Eq.~\eqref{eq:acc_B_L}, the leading order contribution to the acceleration on the test mass A is 
\begin{align}
    \vec a_A (t, \vec x) = \epsilon e \omega_U \vec Y \frac{[Q^A_\mathrm{B-L}]}{m_A}\sin(\omega_U t - \vec k_U \cdot \vec x + \Phi) \, .
\end{align}
Using a similar approach as before with Eq.~\eqref{energy_density_vector}, we can derive the one-way Doppler shift associated to this oscillation as \cite{Yu23} 
\begin{align}
    y^U_{re} &= -\left(\hat n_{re} \cdot \hat e_Y \right)\frac{\sqrt{2\mu_0 \rho_\mathrm{DM}}\epsilon e [Q_\mathrm{B-L}]}{\omega_U m_\mathrm{TM}}\Re\left(e^{i\left(\omega_U t - \vec k_U \cdot \vec x_r + \Phi\right)}-  e^{i\left(\omega_U t - \frac{\omega_U L}{c}-\vec k_U \cdot \vec x_e + \Phi \right)}\right) \, ,
\end{align}
where we consider the test masses have all the same mass $m_\mathrm{TM}$, $\vec Y = |\vec Y|\hat e_Y$ and, as in the scalar cases, the test masses have the same $B-L$ charge $[Q_\mathrm{B-L}]$.

\subsubsection{Transfer function of the second generation TDI}

Following \cite{Yu23}, the transfer function of the one-arm Doppler shift induced by the pure DM scalar field is defined as the Fourier transform of $y_{re}(t)$ Eq.~\eqref{eq:Doppler_scalar_DM} normalized by the constant amplitude
\begin{align}\label{eq:TF_DM}
    \mathcal{T}^\mathrm{DM}_{re}(\omega) &=\left(\hat n_{re}\cdot \hat e_v\right)\Re\left(e^{-i\left(\vec k\cdot \vec x_r -\Phi\right)}-e^{-i\left(\frac{\omega L}{c}+\vec k\cdot \vec x_e-\Phi\right)}\right) \, .
\end{align}
Using Eq.~\eqref{eq:TDI_X2_approx}, we find the DM transfer function of the second generation X combination in the constant arm length approximation as
\begin{subequations}
\begin{align}
&\mathcal{T}^\mathrm{DM}_X(\omega) =-4\sin \left(\frac{\omega L}{c}\right)\sin \left(\frac{2\omega L}{c}\right) \Re\left[e^{\frac{-3i\omega L}{c}} \left(\mathcal{T}^\mathrm{DM}_{13} -\mathcal T^\mathrm{DM}_{12} + e^{\frac{-i\omega L}{c}}\left(\mathcal T^\mathrm{DM}_{31}-\mathcal T^\mathrm{DM}_{21}\right)\right)\right]\, \\
&= -4\sin \left(\frac{\omega L}{c}\right)\sin \left(\frac{2\omega L}{c}\right)\Re\left[e^{\frac{-3i\omega L}{c}}e^{-i(\vec k \cdot \vec x_1 - \Phi)}\left(1+e^{\frac{-2i\omega L}{c}}\right)\left(\hat n_{13}-\hat n_{12}\right)\cdot \hat e_v -\right.\,\nonumber \\
&\left.2e^{\frac{-4i\omega L}{c}}\left(\hat n_{13}\cdot \hat e_v e^{-i(\vec k \cdot \vec x_3-\Phi)}-\hat n_{12}\cdot \hat e_v e^{-i(\vec k \cdot \vec x_2-\Phi)}\right)\right]\, \\
&= -8\sin \left(\frac{\omega L}{c}\right)\sin \left(\frac{2\omega L}{c}\right)\Re\left[e^{\frac{-4i\omega L}{c}}\left(\left(\hat n_{13}-\hat n_{12}\right)\cdot \hat e_v \cos\left(\frac{\omega L}{c}\right)e^{-i(\vec k \cdot \vec x_1 - \Phi)}-\right.\right.\,\nonumber \\
&\left.\left.\left(\hat n_{13} e^{-i(\vec k \cdot \vec x_3-\Phi)}-\hat n_{12} e^{-i(\vec k \cdot \vec x_2-\Phi)}\right)\cdot \hat e_v\right)\right] \, \\
&=-8\sin \left(\frac{\omega L}{c}\right)\sin \left(\frac{2\omega L}{c}\right)\Re\left[e^{-i\left(\frac{4\omega L}{c}+\vec k \cdot \vec x_1 - \Phi\right)}\left(\hat n_{13}\cdot \hat e_v\left(\cos\left(\frac{\omega L}{c}\right)-1+ikL \hat n_{13}\cdot \hat e_v\right)-\right.\right.\,\nonumber\\
&\left.\left.\hat n_{12}\cdot \hat e_v\left(\cos\left(\frac{\omega L}{c}\right)-1+ikL \hat n_{12}\cdot \hat e_v\right)\right)\right]\label{eq:TF_X2_DM_full}\,,
\end{align}
\end{subequations}
where we used $\vec x_2 = \vec x_1 - L \hat n_{12}$ and $\vec x_3 = \vec x_1 - L\hat n_{13}$ and we took the first order expansion of the exponent $\propto kL \ll 1$ for all frequencies of interest (for example, at $f=1$ Hz, which is the maximum frequency of the \textit{LISA} band, $kL \sim 0.05$, with the galactic velocity $v_\mathrm{DM} = 10^{-3} \: c$). The amplitude of the transfer function is then
\begin{align}\label{eq:amp_TF_X2}
    &\left|\mathcal{T}^\mathrm{DM}_X(\omega)\right| = 16\sin \left(\frac{\omega L}{c}\right)\sin \left(\frac{2\omega L}{c}\right) \: \times \, \nonumber \\
    &\sqrt{\left(\hat n_{23}\cdot \hat e_v\right)^2\sin^4\left(\frac{\omega L}{2c}\right)+\left((\hat n_{13}\cdot \hat e_v)^2-(\hat n_{12}\cdot \hat e_v)^2\right)^2\left(\frac{\omega L|\vec v_\mathrm{DM}|}{2c^2}\right)^2} \, ,
\end{align}
where we used $\hat n_{13}-\hat n_{12}=\hat n_{23}$. 
As it can be noticed from Eq.~\eqref{eq:amp_TF_X2}, the first term dominates as long as $(\omega L/c)^2 > (|\vec v_\mathrm{DM}|/c)^2$ (and $\overline{(\hat n_{23} \cdot \hat e_v)^2} \neq 0$), i.e for all frequencies $f \geq 5 \times 10^{-5}$ Hz, which fully contains the \textit{LISA} band. Then, we can neglect the second term and the amplitude of the transfer function becomes
\begin{align}\label{eq:TDI_X2_DM_low_freq}
    \left|\mathcal{T}^\mathrm{DM}_X(\omega)\right| &\approx 8\left(\frac{\omega L}{c}\right)^4 \left|\hat n_{23}\cdot \hat e_v\right| \, ,
\end{align}
i.e the transfer function scales as $f^{4}$. We do not recover the scaling of the transfer function of \cite{Yu23}, because Eq.~\eqref{eq:amp_TF_X2} corresponds to the amplitude of the transfer function of the second generation TDI, while \cite{Yu23} used the 1.5 generation for their calculation. However, using Eq.~\eqref{eq:X2_sl}, one can show easily that the transfer function of the first generation TDI is $\left|\mathcal{T}^\mathrm{DM}_X(\omega)\right|/2\sin (2\omega L/c)$, which allows us to recover the results of \cite{Yu23}\footnote{To compute the transfer function, we did not average over all possible DM wind directions $\hat k$ as \cite{Yu23} did, since it is fixed to a single direction when the field is a coherent superposition of plane waves.}.

\subsection{Monochromatic gravitational waves}

\subsubsection{One arm Doppler shift}
We first model the GW with frequency $f_\mathrm{GW}= \omega_\mathrm{GW}/2\pi$ in the source frame propagating along the $\hat k$ direction as \cite{LDC_tech_note}
\begin{subequations}
\begin{align}
    h_{\mu\nu}(\xi) = A_+\cos(\varphi(\xi))\epsilon^+_{\mu\nu} + A_\times\sin(\varphi(\xi))\epsilon^\times_{\mu\nu} \, ,
\end{align}
where $\xi = t - \hat k \cdot \vec x/c$ represents the surfaces of constant phase and   
\begin{align}
    \varphi(\xi) &= \omega_\mathrm{GW} \xi + \frac{1}{2}\dot \omega_\mathrm{GW}\xi^2+\Phi_\mathrm{GW} \, ,
\end{align}
\end{subequations}
where $\Phi_\mathrm{GW}$ is the initial phase of the wave and where we assume a non zero time derivative of the frequency $\dot \omega_\mathrm{GW}$.
In the barycentric reference frame, we define the triad $(\hat u, \hat v, \hat k)$ from the ecliptic latitude $\beta$ and ecliptic longitude $\lambda$ as \cite{LDC_tech_note,Petiteau08}
\begin{align}
    \hat u &= \begin{pmatrix}
         \sin(\lambda)\\
        -\cos(\lambda)\\
        0
    \end{pmatrix}\: , \:
    \hat v = \begin{pmatrix}
        -\sin(\beta)\cos(\lambda)\\
        -\sin(\beta)\sin(\lambda)\\
        \cos(\beta)
    \end{pmatrix}\: , \: 
    \hat k = \hat u \times \hat v = -\begin{pmatrix}
        \cos(\beta)\cos(\lambda)\\
        \cos(\beta)\sin(\lambda)\\
        \sin(\beta)
    \end{pmatrix}\, .
\end{align}
We can further define two functionals $\xi_+(\hat u, \hat v, \hat n)$ and $\xi_\times(\hat u, \hat v, \hat n)$ \cite{Petiteau08}
\begin{subequations}
\begin{align}
    \xi_+(\hat u, \hat v, \hat n) &= \left(\hat u \cdot \hat n\right)^2-\left(\hat v \cdot \hat n\right)^2\, , \\
    \xi_\times(\hat u, \hat v, \hat n) &= 2\left(\hat u \cdot \hat n\right)\left(\hat v \cdot \hat n\right) \, ,
\end{align}
\end{subequations}
such that, in the barycentric frame (superscript SSB), the one-arm Doppler shift induced by the GW can be written as \cite{Cornish03,LDC_tech_note,Petiteau08}
\begin{subequations}
\begin{align}
    y^\mathrm{GW}_\mathrm{re} &= \frac{-\mathcal{A}}{2\left(1-\hat n_{re} \cdot \hat k \right)}\Re\left[\left(\hat h^\mathrm{SSB}_+\xi_+(\hat u, \hat v, \hat n_{re})+\hat h^\mathrm{SSB}_\times\xi_\times(\hat u, \hat v, \hat n_{re})\right)\left(e^{i\varphi(\xi_r)}-e^{i\varphi(\xi_e)}\right)\right] \,\\
    &\equiv \frac{-\mathcal{A}\hat n^i_{re} \hat n^j_{re}}{2\left(1-\hat n_\mathrm{re} \cdot \hat k \right)}\Re\left[\hat h^\mathrm{SSB}_{ij}\left(e^{i\varphi(\xi_r)}-e^{i\varphi(\xi_e)}\right)\right] \label{eq:Doppler_GW} \, ,
\end{align}
\end{subequations}
where 
\begin{subequations}
\begin{align}
    \hat h^\mathrm{SSB}_+ &= \cos(2\Psi)\left(1+\cos^2(\imath)\right)+2i\sin(2\Psi)\cos(\imath) \, , \\
    \hat h^\mathrm{SSB}_\times &= -\sin(2\Psi)\left(1+\cos^2(\imath)\right)+2i\cos(2\Psi)\cos(\imath) \, , \\
    \hat h^\mathrm{SSB}_{ij} &= (u_i u_j-v_iv_j)\hat h^\mathrm{SSB}_++(u_iv_j+u_jv_i)\hat h^\mathrm{SSB}_\times \, ,
\end{align} 
\end{subequations}
where $\Psi$ is the polarization angle, $\imath$ is the inclination of the source and $\mathcal{A}$ is the amplitude of the wave which depends only on the source parameters, and defined in the barycentric frame as
\begin{align}
    A_+ = -\mathcal{A}(1+\cos^2(\imath)) \: , \: A_\times = - 2\mathcal{A}\cos(\imath) \, .
\end{align}

\subsubsection{Transfer function of the second generation TDI}

\begin{figure}
    \centering
    \includegraphics[width=0.6\textwidth]{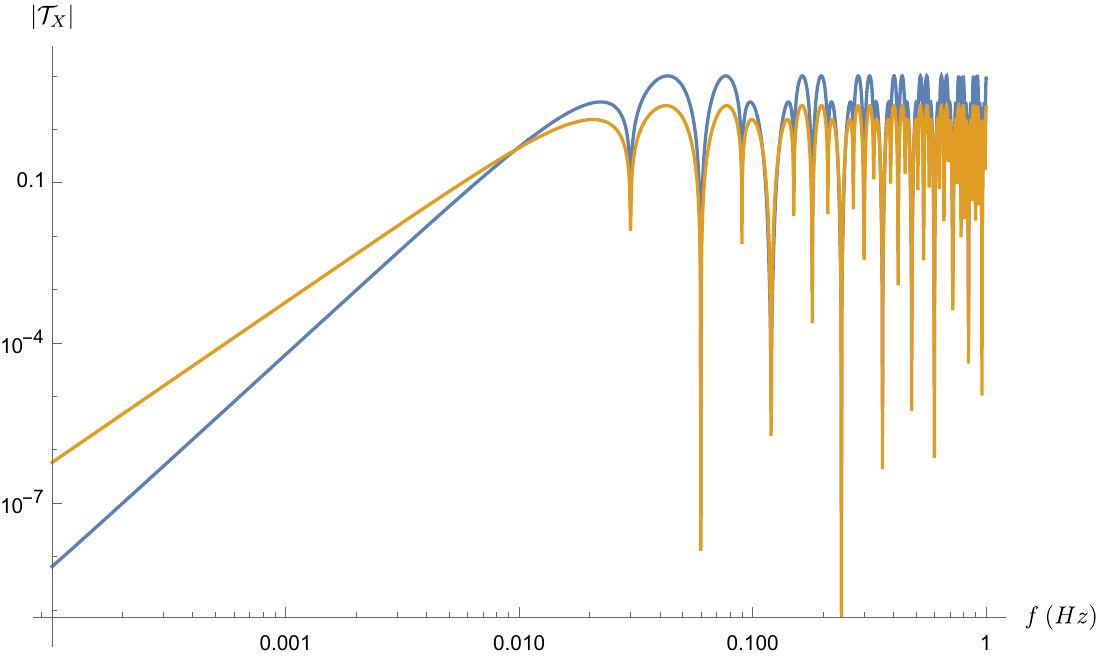}
    \caption{Amplitude of transfer functions of TDI $X$ combinations for both scalar ULDM (in blue) and GW (in orange) from Eqs.~\eqref{eq:amp_TF_X2} and \eqref{eq:TDI2_GW_TF} respectively. For this plot, we neglected the impact of the geometric factors.}
    \label{fig:TF_LISA}
\end{figure}

We now compute the transfer function for a GW (assuming $\dot f = 0$ for simplicity), which reads (using Eqs.~\eqref{eq:Doppler_GW} and \eqref{eq:TDI_X2_approx})
\begin{subequations}
\begin{align}\label{eq:TDI2_GW_TF}
    &\mathcal{T}^\mathrm{GW}_X(\omega) = 2\sin \left(\frac{\omega L}{c}\right)\sin \left(\frac{2\omega L}{c}\right) \Re\left[\hat h^\mathrm{SSB}_{ij}e^{-i\left(\frac{3\omega L}{c}+\vec k \cdot \vec x_1 - \Phi\right)} \: \times \right.\,\\
    &\left.\sum_{\ell=2,3} w_\ell \frac{\hat n^i_{1\ell} \hat n^j_{1\ell}}{1-\left(\hat n_{1\ell} \cdot \hat k  \right)^2}\left((1+\hat n_{1\ell} \cdot \hat k)\left(1-e^{-i\frac{\omega L}{c}(1-\hat n_{1\ell} \cdot \hat k)}\right)-e^{-i\frac{2\omega L}{c}}(1-\hat n_{1\ell} \cdot \hat k)\left(1-e^{i\frac{\omega L}{c}(1+\hat n_{1\ell} \cdot \hat k)}\right)\right)\right]\, \nonumber \\
    &=2\sin \left(\frac{\omega L}{c}\right)\sin \left(\frac{2\omega L}{c}\right) \Re\left[\hat h^\mathrm{SSB}_{ij} e^{-i\left(\frac{5\omega L}{c}+\vec k \cdot \vec x_1 - \Phi\right)} \: \times \right.\,\\
    &\left.\sum_{\ell=2,3} w_\ell \frac{\hat n^i_{1\ell} \hat n^j_{1\ell}}{1-\left(\hat n_{1\ell} \cdot \hat k  \right)^2}\left((\hat n_{1\ell} \cdot \hat k)\left(1-2e^{\frac{i \omega L}{c}(1+\hat n_{1\ell} \cdot \hat k)}+e^{\frac{2i\omega L}{c}}\right)+2ie^{\frac{i\omega L}{c}}\sin\left(\frac{\omega L}{c}\right)\right)\right] \, \nonumber,
\end{align}
where $w_\ell = \pm 1$ respectively for $\ell=3,2$. One can simplify this expression when $\omega L/c \ll 2\pi$ to find the amplitude of the transfer function
\begin{align}\label{eq:TDI_X2_GW_low_freq}
    |\mathcal{T}^\mathrm{GW}_X(\omega)| &= 8\left(\frac{\omega L}{c}\right)^3\left|\hat h^\mathrm{SSB}_{ij} \left(\hat n^i_{13}\hat n^j_{13}-\hat n^i_{12}\hat n^j_{12}\right)\right| \, .
\end{align}
\end{subequations}
Therefore, at small $\omega L/c$, $|\mathcal{T}^\mathrm{GW}_X| \propto f^3$ which means that the DM signal will, on average (i.e neglecting the geometric factor), be suppressed by a factor $\omega L/c$ compared to the GW signal, as it can be noticed in Fig.~\ref{fig:TF_LISA}.

As it can be noticed from Eqs.~\eqref{eq:TDI_X2_DM_low_freq} and \eqref{eq:TDI_X2_GW_low_freq} and in Fig.~\ref{fig:TF_LISA}, at low frequency, the DM and GW transfer function do not scale equally as function of the frequency, in particular it exists a third order term in $\omega L/c$ in the GW transfer function and not in the DM one. 
There are two reasons for this difference. The first one is that the GW travels at speed $c$, while the DM field is non relativistic.
Indeed, one can notice that assuming a "relativistic DM" which travels at speed $c$ instead of $10^{-3}\:c$, the transfer function Eq.~\eqref{eq:TF_X2_DM_full} scales as $(\omega L/c)^3$ at low frequency, as the GW transfer function. This suggests that for a dipolar signal, the velocity of the wave impacts the scaling of the transfer function on frequency.
The second reason is that DM is dipolar while GW is quadrupolar. Indeed, assuming now a non-relativistic GW (i.e neglecting terms $\propto kL \ll 1$), one can notice that the GW transfer function Eq.~\eqref{eq:TDI2_GW_TF} also scales as $(\omega L/c)^3$ at low frequency, as the relativistic GW. This suggests that for a quadrupolar signal, the velocity of the wave does not influence the scaling of the transfer function on the frequency. 
Therefore, it is the fact that the DM produces a non-relativistic \textit{and} dipolar signal which makes a difference in transfer function compared to the GW.
Now, let us give a physical explanation. Assuming a low frequency and non-relativistic wave (i.e $\omega \ll c/L \equiv 1/\tau$, where $\tau$ is the light travel time between spacecrafts ; and with velocity $v \ll c$, such that the wavelength is much larger than the armlength $\lambda \gg L$), the phase of the wave at the location of the emitting and receiving spacecrafts is very close (i.e $\Delta \Phi \ll 2\pi$) and therefore, the one-way Doppler seen by the photon is small as well (from Eq.~\eqref{eq:TF_DM}, it is $\mathcal{O}(\omega L/c)$). If we now assume a full round trip emitter-receiver-emitter, the wave has not propagated much and therefore, the outward effect is, at leading order, the same as the inward effect (up to a sign). Therefore, for DM, since we are essentially looking at this inward-outward effect projected on a dipole, the signal decreases (for example, $\mathcal{T}^\mathrm{DM}_{13}+\exp(-i\omega L/c)\mathcal{T}^\mathrm{DM}_{31}$ is $\mathcal{O}(\omega L/c)^2$). For the quadrupolar GW, the outward effect is still almost the same as the inward effect but they also share the same sign, therefore the signal is still $\mathcal{O}(\omega L/c)$. This is why the polarization makes a difference (for non relativistic waves).
Now, if we assume that both waves are relativistic, but the polarization still differs, the phase of oscillation between the emitting and receiving spacecrafts will be different for both dipolar and quadrupolar waves (even for low frequencies), and therefore, both effects are of the same order of magnitude. 
This is why DM, as a non relativistic wave and with a dipolar effect on \textit{LISA} arm, has not the same transfer function as GW.

\subsection{Summary of the expected one-arm signals}

In summary, if we consider an ULDM or GW background that produces a Doppler shift at angular frequency $\omega$ (i.e $\omega = \omega_\phi = 2 \omega_a = \omega_U = \omega_\mathrm{GW}$), one can write the one-link Doppler shift in a very concise way, similarly as in \cite{Yu23}
\begin{subequations}
\begin{align}\label{eq:Doppler_LISA}
    y_{re} = \Re\left[\mu_{re} \left(e^{i(\omega t - \vec k \cdot \vec x_r + \Phi)}-e^{i(\omega t - \frac{\omega L}{c} - \vec k \cdot \vec x_e + \Phi)}\right)\right] \, ,
\end{align}
where the wavevector $\vec k$ depends on the velocity of the wave i.e $|\vec k|=\omega/c$ for GW and $|\vec k|=\omega v_\mathrm{DM}/c^2$ for DM, and with the unitless $\mu_\mathrm{re}$ are given by
\begin{align}\label{eq:geo_factor_DM_GW}
\mu_{re} =
    \begin{cases}
    \displaystyle   
     \left(\hat n_{re} \cdot \hat e_v\right)\frac{\sqrt{16 \pi G \rho_\mathrm{DM}}v_\mathrm{DM}[Q_M]_d}{\omega c^2} \: \: \: \: &\mathrm{for \:DM \:   pure \: scalar \: field} \, \\
     \displaystyle 
     \left(\hat n_{re} \cdot \hat e_v \right)\frac{32 \pi G \rho_\mathrm{DM} v_\mathrm{DM} E^2_P[Q_M]_a}{f^2_a \omega c^3} \: \: \: \: &\mathrm{for \: DM \:  pseudo \: scalar \: field} \, \\
     \displaystyle 
     -\left(\hat n_{re} \cdot \hat e_Y \right)\frac{\sqrt{2\mu_0 \rho_\mathrm{DM}}\epsilon e [Q_\mathrm{B-L}]}{\omega m_\mathrm{TM}}  \: \: \: \:
     &\mathrm{for \: DM \:  vector \: field} \, \\
     \displaystyle 
     -\left(\frac{\hat n^i_{re}\hat n^j_{re}\hat h^\mathrm{SSB}_{ij}(\imath, \Psi)}{2(1-\hat n_{re} \cdot \hat k)}\right) \mathcal{A} \: \: \: \: &\mathrm{for \:  gravitational \: wave} 
    \end{cases}
\end{align}
\end{subequations}
and where we neglected the frequency derivative in the case of the GW $\dot f = 0$, for simplicity and concise equations purposes. These $\mu_{re}$ factor are the product of an amplitude and a pure geometric factor which shows how the detector couples to DM or GW, and which gives us indication on the nature of the wave. Indeed, ULDM signal being dipolar, the \textit{LISA} arm couples to a single direction (which corresponds to either the galactic velocity in the scalar cases, or to the vector polarization direction in the vector case) while GW signal being quadrupolar, the arm couples to the two orthogonal directions corresponding to the quadrupolar GW polarization.   

On one hand, among other signals, \textit{LISA} will be sensitive to (almost) monochromatic GW, i.e GW emitted from galactic binaries whose inspiral phase (with frequency which can be approximated as a constant) lasts longer than the total mission duration \cite{Petiteau08}.
On the other hand, considering that the coherence time of the ULDM field exceeds the time of the mission, which we will consider to be 1 year in the following\footnote{Following Eq.~\eqref{coherence_DM}, this implies that we need $f \leq 10^{-2}$ Hz.}, the acceleration induced on the test masses would be monochromatic as well. This implies that both monochromatic signals from ULDM and GW Eq.~\eqref{eq:Doppler_LISA} would lead to similar shapes in Fourier space (essentially one Dirac at the wave frequency). This suggests that the two signals might resemble each other, which could lead to degeneracy between them.

In addition, one can find simple detection setup where the quadrupolar nature of the GW signal would be seen by the detector as a dipole, therefore where the ULDM and GW signals would look similar and therefore be degenerate.
For example, let us assume a simple detection setup where the test mass $A$ is static and the test mass $B$ is located at a distance $L$ and is rotating counterclockwise around $A$ in the $x-y$ plane with angular frequency $\omega_R$. Both test masses are sending light signals to each other. We assume that the DM field is a pure scalar field of frequency $\omega$ whose propagation direction is along the $x$-axis in this coordinate system. Let us further assume a monochromatic GW of the same angular frequency $\omega$ which propagates along the $y$-axis, and whose polarization is purely $+$ in the $x-z$ plane. One can easily deduce that the GW polarization in the $z$ direction will not affect the light signals emitted between the two test masses, only the $x$ polarization will be noticed. Therefore, the GW will be seen as a dipole by the detector. 

However, in the case of a more complex detector like \textit{LISA}, we might wonder if this difference in polarization would clearly lead to a different spectral shape of the signal. In addition to this polarization difference, DM and GW propagates at different speed, i.e $|\vec k_\mathrm{GW}| = \omega_\mathrm{GW}/c$ while $|\vec k_\mathrm{DM}| = \omega_\mathrm{DM} v_\mathrm{DM}/c^2 \sim 10^{-3} \omega_\mathrm{DM}/c$ in Eq.~\eqref{eq:Doppler_LISA}, which might also lead to a difference in the spectral signature for both signals. Note that such a difference would not be visible for time scales much smaller than one year, where the geometric factor would simply be constant. 

Therefore, in the following, we will try to answer the following question : using realistic orbits, would \textit{LISA} be able to break the degeneracy between GW emitted from a galactic binary and (scalar) ULDM fields ?

To do so, the first step is to simulate one signal induced by a galactic binary and one signal induced by a (scalar) DM candidate. Then, we will model one galactic binary and one scalar DM candidate using Bayesian analysis on each of these data sets. Finally we will compare the efficiency of these models to see if, for each data set, one is preferred compared to the other.

\section{\label{sec:ULDM_GW_discriminate}Bayesian inference to discriminate between dark matter and gravitational waves}

\subsection{A short introduction on Bayesian inference}

In this section, we will review how Bayesian inference works. It is based one Bayes' theorem, which for two random events $A$ and $B$, allows to write the probability of $A$ under the condition $B$ $P(A|B)$ as
\begin{align}\label{eq:Bayes_theorem}
    P(A|B) &= \frac{P(A)}{P(B)}P(B|A) \, ,
\end{align}
where $P(B|A)$ is the probability of $B$ under the condition $A$ and $P(A), P(B)$ respectively the probabilities of $A$ and $B$.
When considering a vector of data points $\vec d$ depending on a vector of parameters $\vec p$, Eq.~\eqref{eq:Bayes_theorem} becomes
\begin{align}
    \mathcal{P}\left(\vec p|\vec d\right) &= \frac{\mathcal{L}\left(\vec d|\vec p\right)\pi\left(\vec p\right)}{\mathcal{N}\left(\vec d\right)} \, .
\end{align}
$\pi\left(\vec p\right)$ is the prior distribution of the parameters $\vec p$. This represents what we know of the parameters, prior to looking at the data $\vec d$. It can e.g. come from prior measurements on another set of data. For example, the velocity distribution Eq.~\eqref{DM_vel_distrib_scalar} can be used as a prior distribution for the galactic velocity (this is what will be done in the following). On the contrary, if no information on the given parameter is available, one might use a flat uniform prior distribution. Therefore, in the Bayesian framework, the parameters $\vec p$ are treated as random variables and not as fixed constants.

$\mathcal{L}\left(\vec d | \vec p\right)$ is the likelihood density function and represents the agreement between a given set of parameters $\vec p$ with the data $\vec d$. In the most simple case, the data (of length $j$) has white noise with variance $\sigma^2$, and we model the signal as $s(\vec p)$ such that the likelihood reduces to 
\begin{align}\label{eq:gaussian_likelihood}
    \log \mathcal{L}\left(\vec d | \vec p\right) &= -\frac{1}{2}\sum_{n=1}^j \frac{\left(d_n-s_n(\vec p)\right)^2}{\sigma^2} \, .
\end{align}
Maximizing the likelihood means minimizing the sum, which is very similar to $\chi^2$ method. 

$\mathcal{P}\left(\vec p | \vec d\right)$ is the posterior probability density function and represents the updated conditional probability of the parameters $\vec p$ after observing the data $\vec d$. For a given set of parameters $\vec p_w$, $\mathcal{P}\left(\vec p_w | \vec d\right)\rightarrow 0$ means that $\vec p_w$ is not compatible with the data.   
As the Bayes' theorem suggests it, the posterior distribution is the product of the prior distribution and the likelihood (up to $\mathcal{N}(\vec d)$, a normalization constant, independent of the parameters), i.e it is the combination of our knowledge of the parameters before and after analyzing the data.  

There exists two big classes of algorithms to reconstruct the full posterior probability distribution of the parameters $\vec p$ given a dataset $\vec d$.
The first one gathers Markov Chain Monte Carlo (MCMC) algorithms. At each step $i$ of the algorithm, a random sample of parameters $\vec p_i$ is generated from the sample $\vec p_{i-1}$ and the acceptance of this new sample only depends on $\vec p_i$ and $\vec p_{i-1}$. The ratio of posteriors between parameters of $i-1$ and $i$ steps is computed in order to know if newly found parameters explain the data better than the previous one. In general, the distance in parameter space between $\vec p_i$ and $\vec p_{i-1}$ is small (compared to the parameter space volume), and therefore, the algorithm can be trapped in a local maximum, which we want to avoid.
The second method is called Nested sampling \cite{Skilling04} and is a way of avoiding being stuck in a local maximum likelihood. We first need to introduce the prior mass $X$ as the fraction of prior contained within an iso-likelihood contour. Since we expect large likelihood to be contained in a small region of the parameter space, the likelihood $\mathcal{L}(X)$ must be a decreasing function of $X$. The principle of nested sampling is to choose a set of $N$ random points covering uniformly the parameter space, and which have different likelihoods. At the iteration $i$, one deletes the point $N_i$ with lowest likelihood $\mathcal{L}(X_i)$ and finds a new point in the parameter space $N_{i+1}$ such that $\mathcal{L}(X_{i+1}) >\mathcal{L}(X_{i})$. This way, the remaining prior mass decreases and converges to the region with largest likelihood. Since one can store the value of $\mathcal{L}(X_i)$ at each iteration, one has access to the integral of $\mathcal{L}(X)$, which  corresponds to the model evidence. When comparing two models, one uses the Bayes factor, which is the ratio of model evidence. More specifically, one can write Bayes' rule in terms of $\vec p$ which depends on $\vec d$ but also on a specific model $M$, i.e 
\begin{align}
    \mathcal{P}(\vec p | \vec d, M) &=\frac{\mathcal{L}(\vec d | \vec p, M)\pi(\vec p | M)}{p(\vec d | M)} \, ,
\end{align}
where $p(\vec d | M)$ is known as the model evidence, which is the probability of the data $\vec d$ given $M$. This does not depend on the parameters $\vec p$, which have been integrated out (or marginalized out). Considering two given models $M_1$, $M_2$, the Bayes factor 
\begin{align}\label{eq:Bayes_factor}
    \mathcal{B} &= \frac{p(\vec d | M_1)}{p(\vec d | M_2)} \, ,
\end{align}
indicates which model best explain the data. 
Therefore, nested sampling is very convenient to compare different models on the same data.
In the following, we will be using a nested sampler algorithm, called \textit{Nessai} \cite{Nessai} for the parameter estimation of both a galactic binary and a scalar ULDM field. 

\subsection{\label{sec:fastDM}Fast likelihood modeling}

We now focus on the analysis of TDI dataset induced by a galactic binary (GB) on one side and (scalar) ULDM on the other side. 
For a given model, we want to optimize the computation of the likelihood, in particular to make it faster. To do so, we work in the Fourier domain, because it is much more convenient for the signals that we are interested in.  Indeed, for (quasi) monochromatic signals, the idea is to separate the fast oscillation (at the wave frequency) whose Fourier transform can be done analytically, from the slowly oscillating part which can be Fourier transformed numerically \cite{Cornish07}. This speeds up the numerical integration necessary for the computation of the likelihood. As we shall see in the following, the TDI combination are very simple to implement in Fourier space.

In Appendix ~\ref{ap:fastDM_fastGB}, we show explicitly how to express the slow oscillating part of TDI $X_2$ combination for both scalar ULDM and GB signals. The two \textit{Python} softwares that we use which follow this idea, are \textit{FastGB} and \textit{FastDM}, respectively for the modeling of GB and scalar DM. 

\subsection{\label{sec:Generation_signal_LISA}Generation of signals and parameter space}

We now explicitly describe the general process from the data simulation to the data modeling.

We first simulate a dataset which contains either GB or scalar DM signal. To do so, we simulate in time domain the second generation TDI Eq.~\eqref{eq:TDI_X_gen} induced by either a GB or a scalar ULDM signals through Doppler effects derived in the previous sections, see Eqs.~\eqref{eq:Doppler_scalar_DM} and \eqref{eq:Doppler_GW}.  The positions of the spacecrafts are obtained using \textit{LISAOrbits} \cite{Bayle22}, and we use Earth-trailing orbits for their orbits \cite{Martens21},  which we simulate for one full year, i.e $T_\mathrm{obs} = 365 \times 86400$ s.  The positions of the spacecrafts are sampled with frequency $f_s=0.1$ Hz, i.e $N=T_\mathrm{obs}f_s$ measurements are performed. For the generation of data, no noise is added i.e these are "perfect" dataset\footnote{Therefore, the results of the analysis, and its possible bias, will not depend on noise.}.

For the data modeling, we fit the TDI $A,E$ data combinations expressed in Fourier domain. More precisely, writing the TDI data in time domain $d^{A,E}(t)$ as a vector $(d^{A,E}_0,d^{A,E}_1,...,d^{A,E}_{N-1})$ where $d^{A,E}_i = d^{A,E}(i/f_s)$ is the ith sampled element of $d^{A,E}(t)$, we perform the discrete Fourier transform (DFT) $\tilde d^{A,E}$ defined as
\begin{align}
    \tilde d^{A,E}_k &= \sum^{N-1}_{\ell=0} e^{-\frac{2\pi i \ell k}{N}} d^{A,E}_\ell \, ,
\end{align}
and we fit a model $\tilde m^{A,E}$ to such dataset in Fourier space.
We use a Gaussian likelihood expressed in time domain as \cite{Derevianko18}
\begin{subequations}
\begin{align}\label{eq:likelihood_time_domain_general_LISA}
 \log \mathcal{L} &= \sum_{x=A,E}\left(-\frac{1}{2}\log\left(\det (2\pi \mathbf{C}^x)\right) - \frac{1}{2}\left(\vec d^x - \vec m^x\right)^\dag (\mathbf{C}^x)^{-1}\left(\vec d^x - \vec m^x\right)\right)\, ,
\end{align}
where $\vec d,\vec m$ represent respectively the time vector or data and model and where $\mathbf{C}^x$ is the noise covariance matrix of the $x$ TDI combination.
Using the DFT of the data and model, the likelihood can be written as \cite{Savalle21}
\begin{align}\label{eq:likelihood_general_LISA}
 \log \mathcal{L} &= \sum_{x=A,E}\left(-\frac{1}{2}\log\left(2\pi \Pi_{k} \tilde C^x_{kk}\right) - \sum^{N/2}_{j=0}\frac{|\tilde d^x_j- \tilde m^x_j|^2}{\tilde C^x_{jj}}\right)\, ,
\end{align}
where $\tilde C^x_{kk}$ is the diagonal element at Fourier bin $k$ of the two-sided PSD matrix of the $x=A,E$ TDI combinations defined as \cite{Savalle21}
\begin{align}
    \tilde C^x_{kk} &= N f_s \delta_{kl}N_x(f_l) \, ,
\end{align}
\end{subequations}
where $\delta_{kl}$ is the Kronecker delta and therefore $\tilde C^x_{kk}$ is diagonal because the noise is stationary \cite{Derevianko18}, and where the $N_x(f_k)$ is the noise PSD of the second TDI $A,E$ combinations at frequency bin $k$, which are defined as \cite{Nam23}
\begin{align}\label{eq:TDI_A_noise_PSD}
    N_{A,E}(f) &= 32\sin^2\left(\frac{2 \pi f L}{c}\right)\sin^2\left(\frac{4 \pi f L}{c}\right)\left(2\left(3+2\cos\left(\frac{2 \pi f L}{c}\right)+\cos\left(\frac{4 \pi f L}{c}\right)\right)S_\mathrm{acc}(f)+\right.\,\nonumber \\
    &\left.\left(2+\cos\left(\frac{2 \pi f L}{c}\right)\right)S_\mathrm{oms}(f)\right) \, ,
\end{align}
assuming no correlation between noises, and which is obtained by extending the single-link noise PSD Eq.~\eqref{eq:noise_PSD_LISA} on the full TDI combination.
Since the noise PSD are equivalent for $A,E$ combinations, we can rewrite Eq.~\eqref{eq:likelihood_general_LISA} as 
\begin{align}\label{eq:likelihood_final_LISA}
    \log \mathcal{L} &= -\frac{1}{2}\log\left((2\pi)^2 f_s \Pi_{k} N_{A,E}(f_k)\right) - \sum_{x=A,E}\left(\frac{1}{N f_s}\sum^{N/2}_{j=0}\frac{|\tilde d^x_j- \tilde m^x_j|^2}{N_x(f_j)}\right) \, .
\end{align}

The GB parameters that will be fitted (using Eqs.~\eqref{eq:alpha}, \eqref{eq:mathcal_Y} and \eqref{eq:beta}) are the amplitude $\mathcal{A}$, the frequency $f_\mathrm{GW}$, the drift in frequency $\dot f_\mathrm{GW}$, the sky localization ($\beta,\lambda$), the inclination $\imath$, the polarization angle $\Psi$ and the phase $\Phi_\mathrm{GW}$.
There is no specific knowledge on the physics of one particular GB, therefore we will use isotropic prior distributions for each of those parameters.

The scalar DM parameters that will be fitted (using Eqs.~\eqref{eq:alpha_DM}, \eqref{eq:mathcal_Y_DM} and \eqref{eq:beta_DM}) are the mass charge\footnote{From Eq.~\eqref{partial_dil_mass_charge}, we remind that $[Q_M]_d$ is a sum of partial dilatonic mass charges and dilatonic couplings.} $[Q_M]_d\equiv \varepsilon$, the frequency $f_\phi$, the direction of the DM 'wind' $(\beta_\mathrm{DM},\lambda_\mathrm{DM})$, its amplitude, which is given by the galactic velocity amplitude $|\vec v_\mathrm{DM}|$ and the phase $\Phi$. For most of these parameters, we will use isotropic priors, except for the galactic velocity amplitude prior that will be given by the amplitude of the galactic velocity distribution Eq.~\eqref{DM_vel_distrib_scalar}. Even though we are working on time scales lower than the coherence time of the field, such that the DM wave is essentially monochromatic, the prior will help constraining the simulation to consistent regions of the parameter space\footnote{As we shall see in the following, for most frequencies under consideration, the phase shift induced by DM propagation (through $|\vec v_\mathrm{DM}|$) is small, and therefore hard to fit with accuracy.}. Note that, by choosing $\vec v_\mathrm{DM}$ as a random variable and not a fixed parameter, we will derive realistic sensitivity curves for ULDM of \textit{LISA}. By doing so, we simplify the system by considering that the velocity amplitude $|\vec v_\mathrm{DM}|$ is uncorrelated with the DM wind direction, while it is in reality not true as both parameters are constrained together by the same distribution Eq.~\eqref{DM_vel_distrib}\footnote{As mentioned above, we will use isotropic priors for both ecliptic latitude and longitude of the DM wind, while the galactic velocity distribution Eq.~\eqref{DM_vel_distrib} provides prior information on such direction. First, we make this choice because there is no analytical expression for the angular distributions only. Second, by doing the numerical integration of Eq.~\eqref{DM_vel_distrib} over the velocity amplitude, one can show that the width of the angular distributions are much larger than the posterior distributions of these parameters (see below, Section ~\ref{sec:model_DM_signal_DM}), and therefore using isotropic prior will not impact the results.}.

\subsection{Model of galactic binary onto galactic binary signal}

\begin{figure}
\centering
\includegraphics[width=\textwidth]{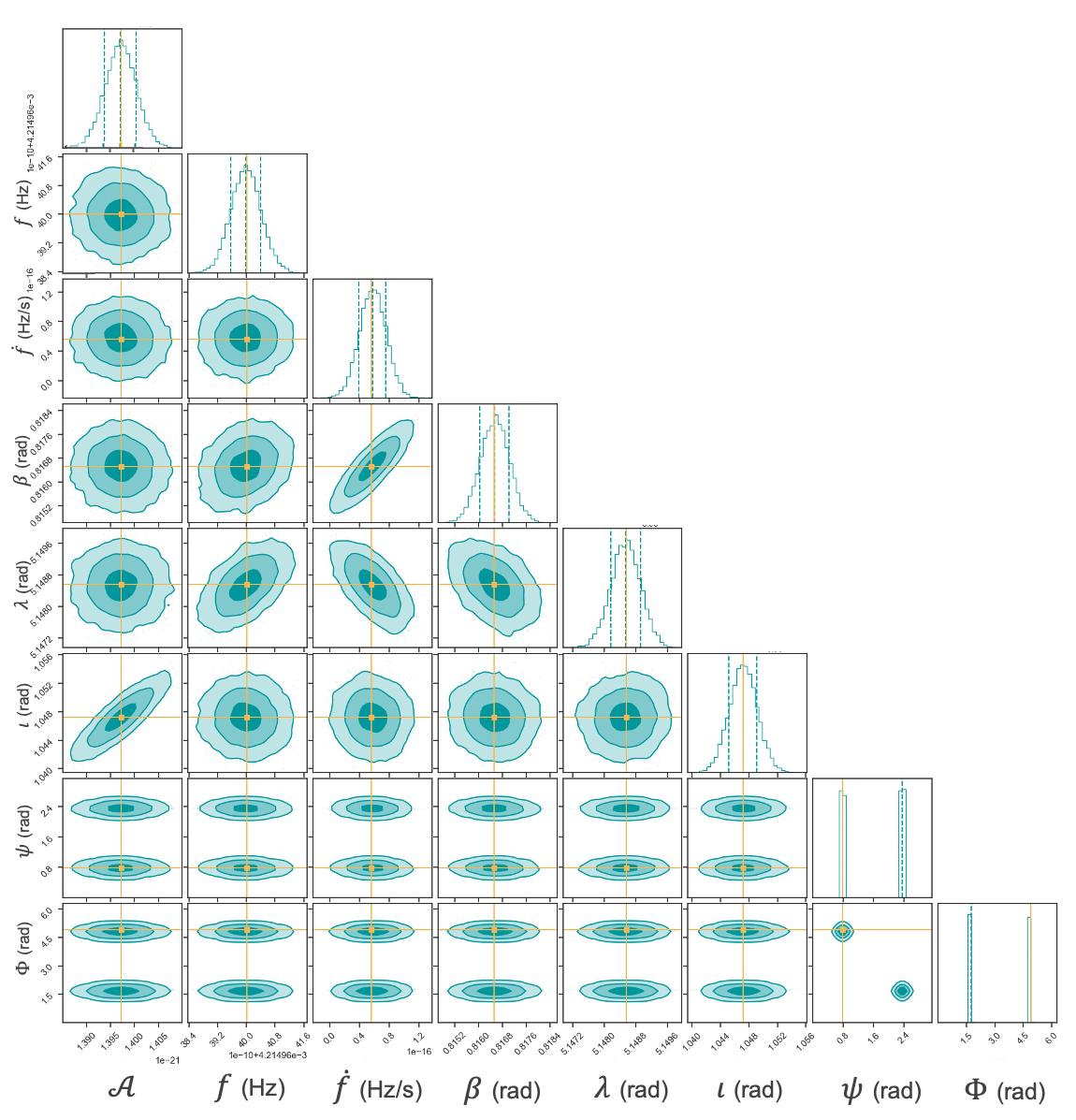}
\caption{Results of the fit of a galactic binary signal by a galactic binary model. The yellow lines indicate the true values of the parameters. The analysis recovers correctly the injected parameters.}
\label{fig:corner_plot_GW_GW}
\end{figure}

\begin{figure}[h!]
\begin{minipage}{\textwidth}
  \begin{minipage}[b]{0.49\textwidth}
    \centering
    \includegraphics[width=\textwidth]{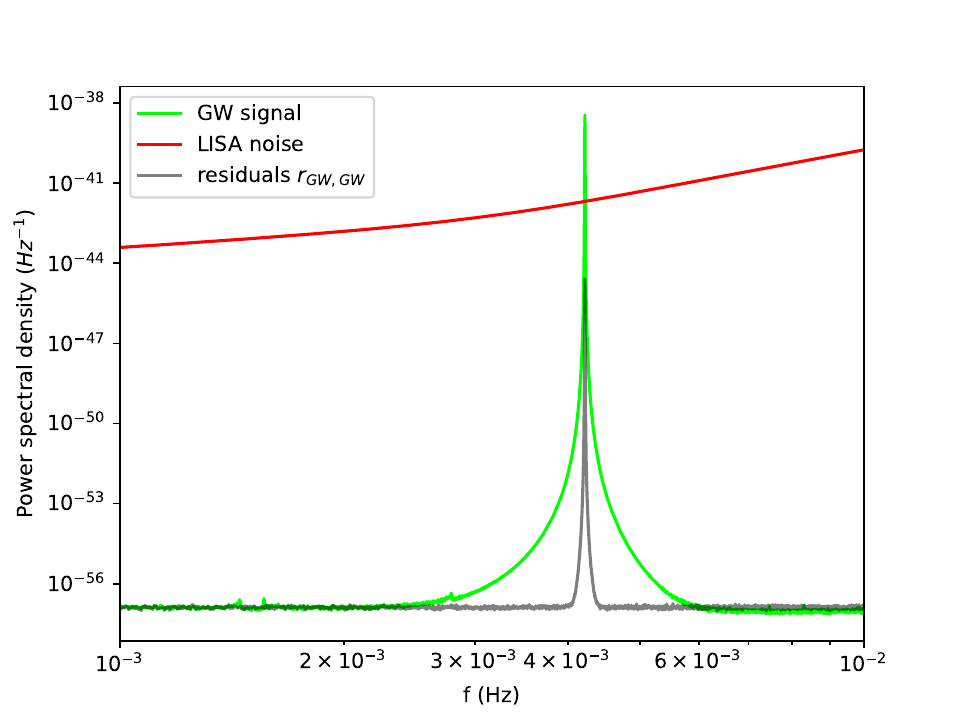}
    \caption{GW signal PSD (in lime) compared to LISA noise PSD of the TDI $A$ combination (in red) in Fourier domain. The signal is well above the noise. We also show the $r_\mathrm{GW,GW}$ residuals power in grey, which are below the noise.}
    \label{fig:GW_signal}
    \end{minipage}%
    \hfill
    \begin{minipage}[b]{0.49\textwidth}
    \centering
    \includegraphics[width=\textwidth]{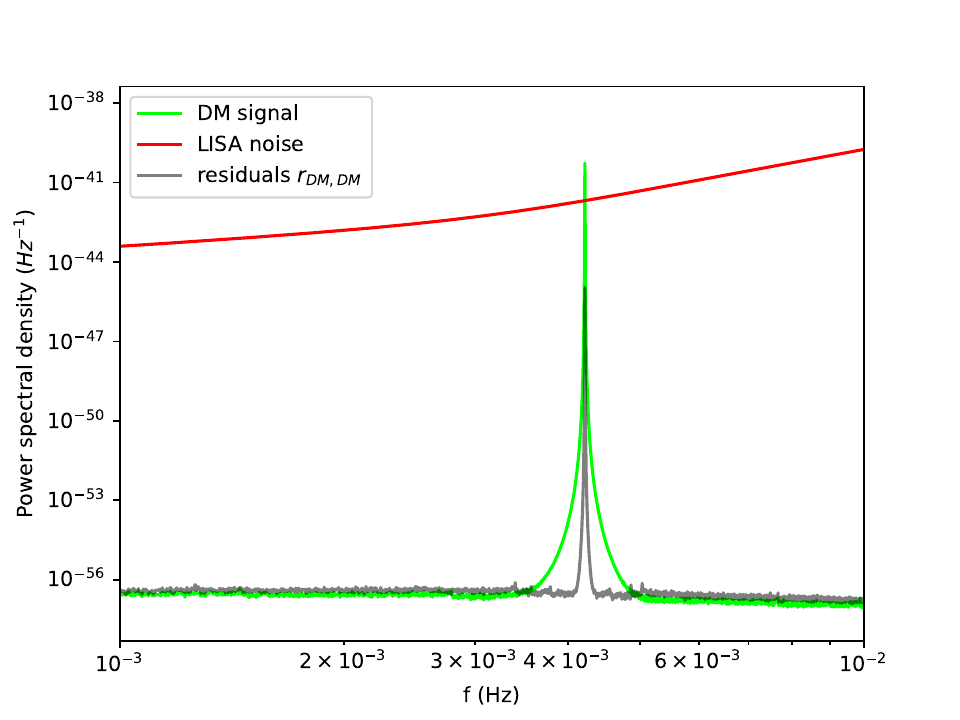}
    \caption{DM signal PSD (in lime) compared to LISA noise PSD of TDI $A$ combination (in red) in Fourier domain. The signal is above the noise. We also show the $r_\mathrm{DM,DM}$ residuals power in grey, which are below the noise.}
    \label{fig:DM_signal}
  \end{minipage}
\end{minipage}
\end{figure}

We first fit a dataset $d_\mathrm{GW}$ from a galactic binary using \textit{FastGB} whose parameters $D_\mathrm{GW}$ are shown on the left of Table ~\ref{tab:GW_GW}, and check for efficiency of the model in reproducing data. The 2D posterior distributions are shown in Fig.~\ref{fig:corner_plot_GW_GW}, with the joint posterior distribution for all parameters. The yellow lines represent the true values of the parameters, i.e the values injected in the simulation. One can notice strong correlations between the amplitude $\mathcal{A}$ and the inclination $\imath$, and between the ecliptic latitude $\beta$ and the drift in frequency $\dot f$. One can also notice that for the polarization angle $\Psi$ and initial phase $\Phi$ parameters, two modes exist which are respectively shifted by $\pi/2$ and $\pi$. From Eq.~\eqref{eq:Doppler_GW}, it is straightforward to show that a change in polarization angle $\Psi \rightarrow \Psi + \pi/2$ is equivalent to a change in phase $\Phi \rightarrow \Phi + \pi$. Overall, the analysis recovers correctly the injected parameters.  \\
\noindent
\begin{minipage}{\textwidth}
  \begin{minipage}[b]{0.49\textwidth}
    The model best fit parameters $M^\mathrm{GW}_\mathrm{GW}$, that we define as the posterior with largest likelihood, in addition to the mean and standard deviation of the distribution of posteriors are shown in Table ~\ref{tab:GW_GW} (only the $\sim 10 \: 000$ last iterations are taken into account by the algorithm).
    Let us now analyze the result of the fit, in particular, its efficiency. First, one can compare the signal to noise levels (i.e compute the signal-to-noise ratio (SNR)) for the GW using \cite{Babak21}\footnotemark
    \begin{align}
    \sqrt{\mathrm{SNR}} = \sqrt{4\Re\left(\int df \frac{\tilde d_\mathrm{GW} \tilde d^\dag_\mathrm{GW}}{N(f)}\right)} \,,
    \end{align}
    where $\tilde d^\dag_\mathrm{GW}$ is the conjugate transpose of $\tilde d_\mathrm{GW}$, the GW data in Fourier domain, and $N(f)$ is the noise PSD. 
    For the GW data from Table ~\ref{tab:GW_GW}, one finds
    \begin{subequations}\label{eq:SNR_GW_data}
    \begin{align}
    \sqrt{\mathrm{SNR}^\mathrm{GW}_A} &\approx 632 \, \\
    \sqrt{\mathrm{SNR}^\mathrm{GW}_E} &\approx 679 \, ,
    \end{align}
    \end{subequations}
    for both $A$ and $E$ TDI combinations. The signal power is well above the noise, as it can be noticed in Fig.~\ref{fig:GW_signal}.
    \end{minipage}%
    \hfill
    \begin{minipage}[b]{0.49\textwidth}
    \centering
    \begin{tblr}{vlines, colspec={cccc}}
        \hline
        & \SetCell[c=1]{} $D_\mathrm{GW}$ & \SetCell[c=1]{} $M^\mathrm{GW}_\mathrm{GW}$ & \SetCell[c=1]{} Mean $\pm$ 1 $\sigma$ \\
        \hline
        \hline
        $\mathcal{A}$ $(10^{-21})$ & $1.397$& $1.398$ &$ 1.397 \pm 0.003 $\\
        \hline
        $f$ (mHz) & $4.215$ & $4.215$ & $4.215 \pm 0.000$\\
        \hline
        $\dot f$ (aHz/s) & $55.91$& $56.26$ &$57.36 \pm 18.20$\\
        \hline
        $\beta$ (rad) & $0.817$ & $0.8164$ & $0.8165 \pm 0.0005$ \\
        \hline
        $\lambda$ (rad) & $5.149$ & $5.1485$ & $5.1485 \pm 0.0004$\\
        \hline
        $\imath$ (rad) & $1.047$ & $1.048$ &$1.047 \pm 0.002$ \\
        \hline
        \SetCell[r=2]{}$\Psi$ (rad) & $0.785$ & $0.785$ & $0.785 \pm 0.002$ \\
        & & & $2.356 \pm 0.002$\\
        \hline
        \SetCell[r=2]{}$\Phi$ (rad) & $4.890$ & $4.888$ & $4.889 \pm 0.007$\\
        & & & $1.748 \pm 0.007$\\
        \hline
    \end{tblr}
    \captionof{table}{On the left column, we show the GW injected parameters in the simulation. On the middle, we show the GW model best fit, i.e the posterior with the largest likelihood. The model correctly recovers the injected parameters with less than 1\% deviation. On the right column, we present the mean and standard deviation of the full distribution of posteriors.}     
    \label{tab:GW_GW}
  \end{minipage}
\end{minipage}
\footnotetext{The definition of SNR is slightly changed compared to \cite{Babak21}, in order to be consistent with its previous definition of ratio of signal to noise power.}

In a second step, in order to estimate the accuracy of the model, we simulate the TDI time series of a galactic binary with parameters given by $M^\mathrm{GW}_\mathrm{GW}$, which we call $m^\mathrm{GW}_\mathrm{GW}$ and we construct the residuals (in time domain)
\begin{align}
        r_\mathrm{GW,GW} &= d_\mathrm{GW} - m^\mathrm{GW}_\mathrm{GW} \, ,
\end{align}
which we compare with the \textit{LISA} noise PSD. 
In Fig.~\ref{fig:GW_signal}, we compare the residuals to LISA noise in Fourier space. One can notice that the residuals are well below the noise, indicating that the model extracted all the necessary information out of the data\footnote{Despite its efficiency, the residuals still present a peak at the GW frequency. This is due to the approximations done in \textit{FastGB} (e.g. neglecting the spacecraft velocity during light travel time).}. We now compute the residuals-to-noise ratio (RNR) between the residual in Fourier domain $\tilde r_\mathrm{GW,GW}$ given by \cite{Babak21}
\begin{align}   
\sqrt{\mathrm{RNR}} = \sqrt{4\Re\left(\int df \frac{\tilde r \tilde r^\dag}{N(f)}\right)} \,,
\end{align}
where $\tilde r^\dag$ is the conjugate transpose of $\tilde r$. For the GW model on GW data, one finds
\begin{subequations}
\begin{align}
\sqrt{\mathrm{RNR}^\mathrm{GW, GW}_A} &\approx 0.54 \, \\
\sqrt{\mathrm{RNR}^\mathrm{GW, GW}_E} &\approx 0.41 \, .
\end{align} 
\end{subequations}
The RNR are well below unity and the SNR values, indicating that the model extracts all the necessary information out of the data, as expected from what is shown in Fig.~\ref{fig:GW_signal}.\\
\noindent
\noindent

\subsection{Model of scalar dark matter onto a scalar dark matter signal}\label{sec:model_DM_signal_DM}

\begin{figure}
    \centering
    \includegraphics[width=\textwidth]{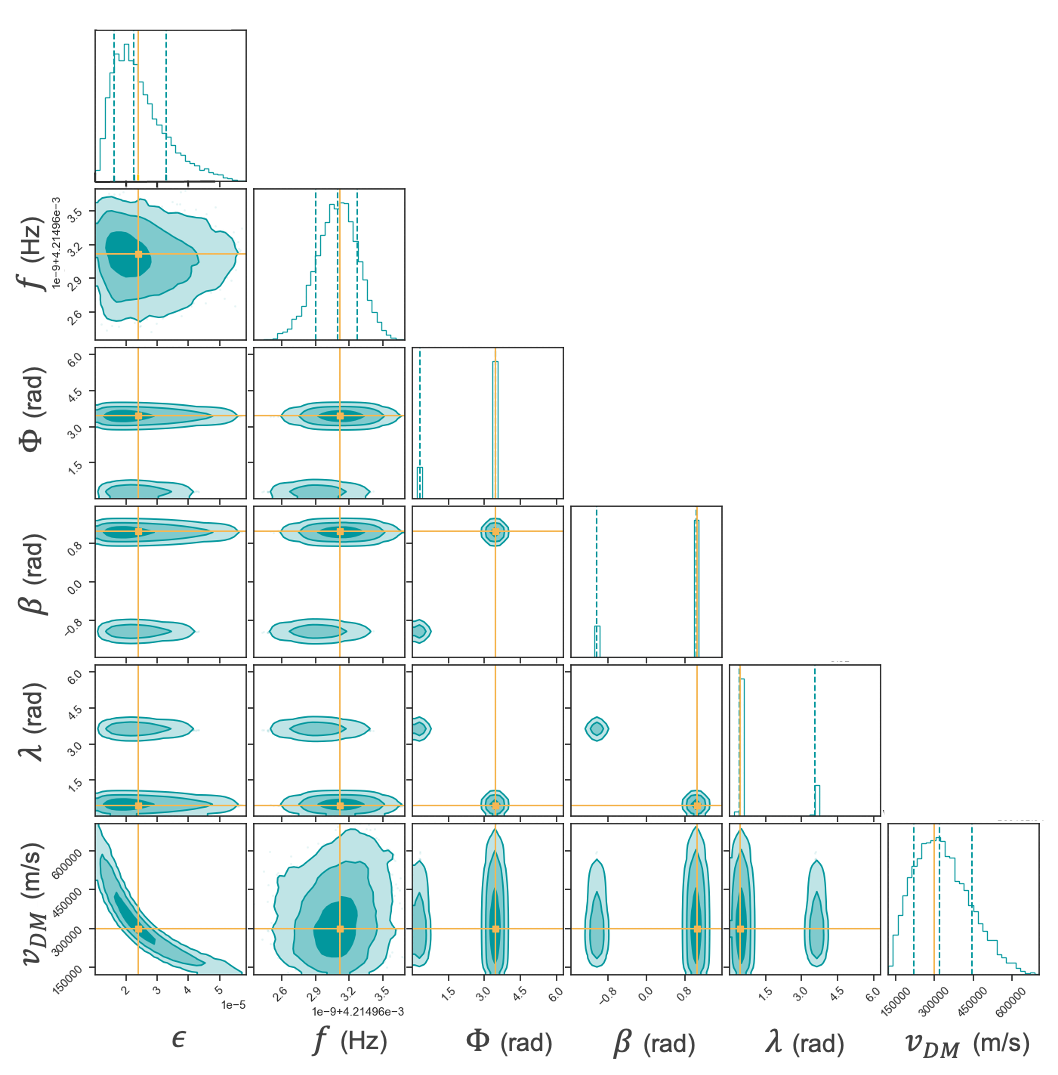}
    \caption{Results of the fit of the DM signal by a DM model. The yellow lines indicate the true values of the parameters. One can notice strong correlations between the coupling $\varepsilon$ and the velocity $v_\mathrm{DM}$.}
    \label{fig:cornerplot_DM_DM}
\end{figure}

Using \textit{FastDM}, we now turn to the analysis of DM signal. For consistency with the galactic binary that was analyzed in the previous section, we consider a DM signal with the same frequency (in order to compare to the same noise level) and intrinsic amplitude of oscillation on the test masses. From Eq.~\eqref{eq:geo_factor_DM_GW}, this means that we choose the coupling $\varepsilon$ such that 
\begin{align}\label{eq:equiv_A_epsilon}
    \varepsilon &= \mathcal{A}\frac{2\pi f c^2}{\sqrt{16 \pi G \rho_\mathrm{DM}}v_\mathrm{DM}} \, ,
\end{align}
where $f, \mathcal{A}$ are given in Table ~\ref{tab:GW_GW}.\\
\noindent
\begin{minipage}{\textwidth}
  \begin{minipage}[b]{0.4\textwidth}
    In addition, we will use the mean velocity direction and amplitude of the velocity distribution Eq.~\eqref{DM_vel_distrib}. The mean direction is well known and it points towards $(\beta_\mathrm{DM},\lambda_\mathrm{DM})=(59.91\degree,24.67\degree)\sim(1.046 \: \mathrm{rad},0.431\: \mathrm{rad})$\footnotemark. This is the DM wind direction we will use. For the amplitude, we use $v_\mathrm{DM}= 3 \times 10^{5}$ m/s for the velocity of the wave, i.e the mean of the galactic velocity distribution Eq.~\eqref{DM_vel_distrib}. The result of the fit of a DM model onto a DM signal is presented in Fig.~\ref{fig:cornerplot_DM_DM}
    \end{minipage}
    \hfill
    \begin{minipage}[b]{0.55\textwidth}
    \centering
    \begin{tblr}{vlines, colspec={cccc}}
        \hline
        & \SetCell[c=1]{} $D_\mathrm{DM}$ & \SetCell[c=1]{} $M^\mathrm{DM}_\mathrm{DM}$ & \SetCell[c=1]{} Mean $\pm$ 1 $\sigma$ \\
        \hline
        \hline
        $\varepsilon$ $(10^{-5})$ & $2.396$& $2.785$ &$2.430 \pm 0.855$\\
        \hline
        $f$ (mHz) & $4.215$ & $4.215$ & $4.215 \pm 0.000$\\
        \hline
        $\beta$ (rad) & $1.046$ & $1.048$ & $1.045 \pm 0.010$ \\
        \hline
        $\lambda$ (rad) & $0.431$ & $0.435$ & $0.430 \pm 0.021$\\
        \hline
        $v_\mathrm{DM}$ $(10^5)$ (m/s) & $2.998$ & $2.580$ & $3.305 \pm 1.079$ \\
        \hline
        $\Phi$ (rad) & $3.444$ & $3.444$ & $3.444 \pm 0.010$\\
        \hline
    \end{tblr}
    \captionof{table}{On the left column, we show the DM injected parameters in the simulation. On the middle, we show the DM model best fit, i.e the posterior with the largest likelihood. On the right column, we present the mean and standard deviation of the full distribution of posteriors.}
    \label{tab:DM_DM}
  \end{minipage}
\end{minipage}
\footnotetext{The Sun velocity in the galactic halo points towards $\alpha$ Cygni, the biggest star of the Cygnus constellation \cite{Miuchi20}, which corresponds to a right ascension $\alpha_\mathrm{DM}=310.36\degree \mathrm{E}$ and declination $\delta_\mathrm{DM}=45.28\degree \mathrm{N}$ \cite{vanLeeuwen07}, in the equatorial frame. By transforming it to the ecliptic coordinate system \cite{Leinert98}, we find $(\beta_\mathrm{DM},\lambda_\mathrm{DM})=(59.91\degree,24.67\degree)$.}
One can notice that there exists a strong correlation between the coupling $\varepsilon$ and the galactic velocity parameters $v_\mathrm{DM}$, as expected from the amplitude of the Doppler Eq.~\eqref{eq:Doppler_scalar_DM}, which is proportional to both parameters. One can notice from Table ~\ref{tab:DM_DM}, that the analysis correctly retrieve the correct value of the true parameter (at the 1\% error level), except for the coupling $\varepsilon$ and galactic velocity $v_\mathrm{DM}$, whose best fit values differ from the true value of $\sim 15\%$, and whose posterior distribution width is much larger than for the rest of the parameters ($\sigma_{v_\mathrm{DM}}/v_\mathrm{DM} \sim \sigma_\varepsilon/\varepsilon \sim 1$ while for the rest of the parameters $\sigma_p/p \ll 1$). The reason is that at this frequency, \textit{LISA} cannot probe accurately the DM wind velocity, since it produces a phase in the Doppler Eq.~\eqref{eq:Doppler_scalar_DM} which is too small to be seen.  As a consequence, we have a large uncertainty on this parameter, which is translated to the coupling due to the large correlation between the two parameters. Indeed, one can notice from Table ~\ref{tab:DM_DM} that the typical width of the phase $\Phi$ is $\sigma_\Phi = 10^{-2}$ rad. As we have used an uniform prior on $\Phi$, the width of the posterior is essentially the width on the likelihood $\mathcal{L}(d_\mathrm{DM}| \Phi)$, which is directly related to the noise PSD. At $f \sim 4\times 10^{-3}$ Hz, the amplitude of the phase induced by the propagation of the field is $\sim \omega v_\mathrm{DM} |\vec x| \cos(\beta)/c^2 \sim 6 \times 10^{-3} < \sigma_\Phi$ (where $\beta$ is the ecliptic latitude of the DM wind direction, considering $|\vec x| \sim 1.5 \times 10^{11} \: \mathrm{m} \equiv 1 \: \mathrm{AU}$, as mentioned previously and that the orbit of the spacecrafts are close to the ecliptic plane). In Section ~\ref{sec:delta_varepsilon}, we will analytically derive the typical width of the coupling $\varepsilon$ parameter to be able to predict its value at any frequency in the \textit{LISA} band.

For now, let us compute the SNR of such DM data
\begin{subequations}\label{eq:SNR_DM_data}
\begin{align}
    \sqrt{\mathrm{SNR}^\mathrm{DM}_A} &\approx 77 \, \\
    \sqrt{\mathrm{SNR}^\mathrm{DM}_E} &\approx 65 \, ,
\end{align}
\end{subequations}
i.e the signal is still higher than the noise, but the SNR is smaller than in the case of the GW data. This difference in $\sqrt{\mathrm{SNR}}$ between DM and GW of a factor $\sim 10$ comes from the difference in transfer functions at low frequency Eqs.~\eqref{eq:TDI_X2_DM_low_freq} and \eqref{eq:TDI_X2_GW_low_freq}\footnote{From those equations and Eq.~\eqref{eq:AET_TDI}, one can easily obtain the low frequency transfer functions of e.g. the TDI $A$ combination. Then, using the true value of DM and GW parameters (i.e $f,\hat e_v$ for DM and $f,\Psi,\imath, \hat k$ for GW), one can compute numerically the various geometric factors $\hat n \cdot \hat e_v$ and $\hat h^\mathrm{SSB}_{ij}\hat n^i\hat n^j$ from the spacecrafts orbits. We find $\mathcal{T}^\mathrm{GW}_A/\mathcal{T}^\mathrm{DM}_A \sim 8.3$ while the ratio of Eqs.~\eqref{eq:SNR_GW_data} and \eqref{eq:SNR_DM_data} gives $\sim 8.2$ for TDI $A$ combination. Note that from Fig.~\ref{fig:TF_LISA}, the ratio of the two $X$ transfer functions is $\sim 3$ at $f\sim 4.215$ mHz, but this is because we did not take into account the geometric factors for this plot.}.

Eq.~\eqref{eq:SNR_DM_data} is consistent with what is shown in Fig.~\ref{fig:DM_signal}, where we notice that the DM signal PSD is well above LISA noise\footnote{The Fourier peak is narrower compared to the GW case, because the DM signal is completely monochromatic (up to small deviations from the orbit), while we implemented a small frequency drift for the GW.}.
Calling $m^\mathrm{DM}_\mathrm{DM}$ the TDI time series of the model with parameters $M^\mathrm{DM}_\mathrm{DM}$, we can define the residuals as 
    \begin{align}
        r_\mathrm{DM,DM} &= d_\mathrm{DM} - m^\mathrm{DM}_\mathrm{DM} \, ,
    \end{align}
    and compare them to the noise level. 
    The RNR of the DM model on the DM data gives 
    \begin{subequations}
    \begin{align}
    \sqrt{\mathrm{RNR}^\mathrm{DM, DM}_A} &\approx 0.35 \, \\
    \sqrt{\mathrm{RNR}^\mathrm{DM, DM}_E} &\approx 0.18 \, ,
    \end{align}
    \end{subequations}
    smaller than 1 and the SNR values Eq.~\eqref{eq:SNR_DM_data}. As in the GW case above, this suggests, that the model extracts all the information out of the data. We can also check the residuals power in Fourier space Fig.~\ref{fig:DM_signal}, and we notice that it is below the noise, as expected.

\subsection{Can scalar dark matter be misinterpreted as a galactic binary ?}

\begin{figure}[h!]
\begin{minipage}{\textwidth}
  \begin{minipage}[b]{0.49\textwidth}
    \centering
    \includegraphics[width=\textwidth]{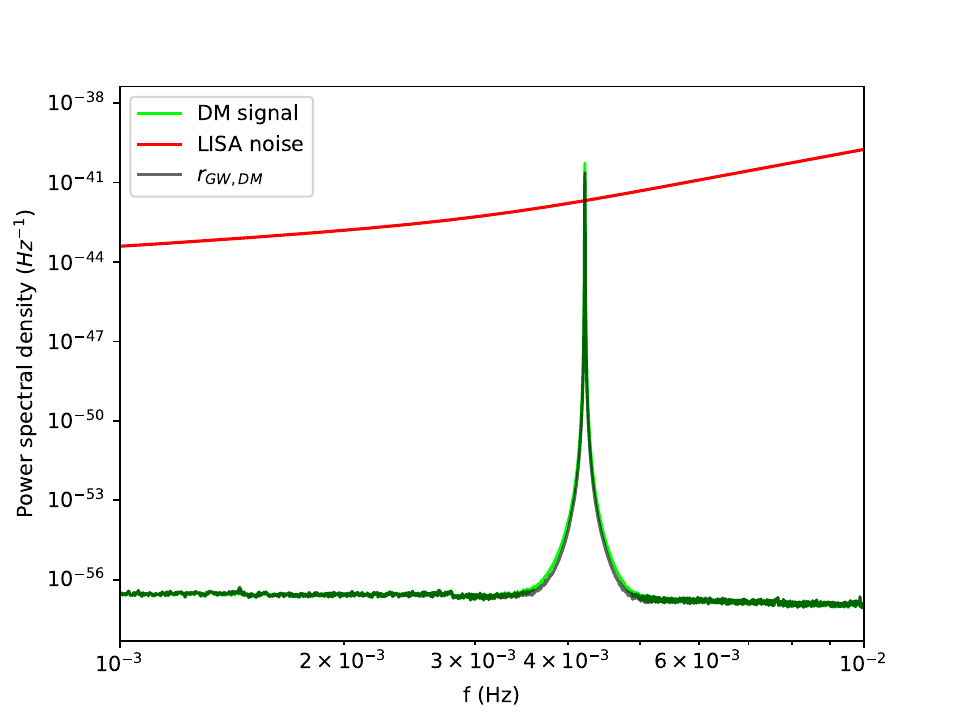}
    \caption{DM signal PSD (in lime) compared to \textit{LISA} noise PSD of TDI $A$ combination (in red) in Fourier domain (as in Fig.~\ref{fig:DM_signal}). This time, the $r_\mathrm{GW,DM}$ residuals power in grey are close to the signal power and above the noise. This indicates the model fits poorly the data.}
    \label{fig:GW_DM_res}
    \end{minipage}
    \hfill
    \begin{minipage}[b]{0.49\textwidth}
    \centering
    \includegraphics[width=\textwidth]{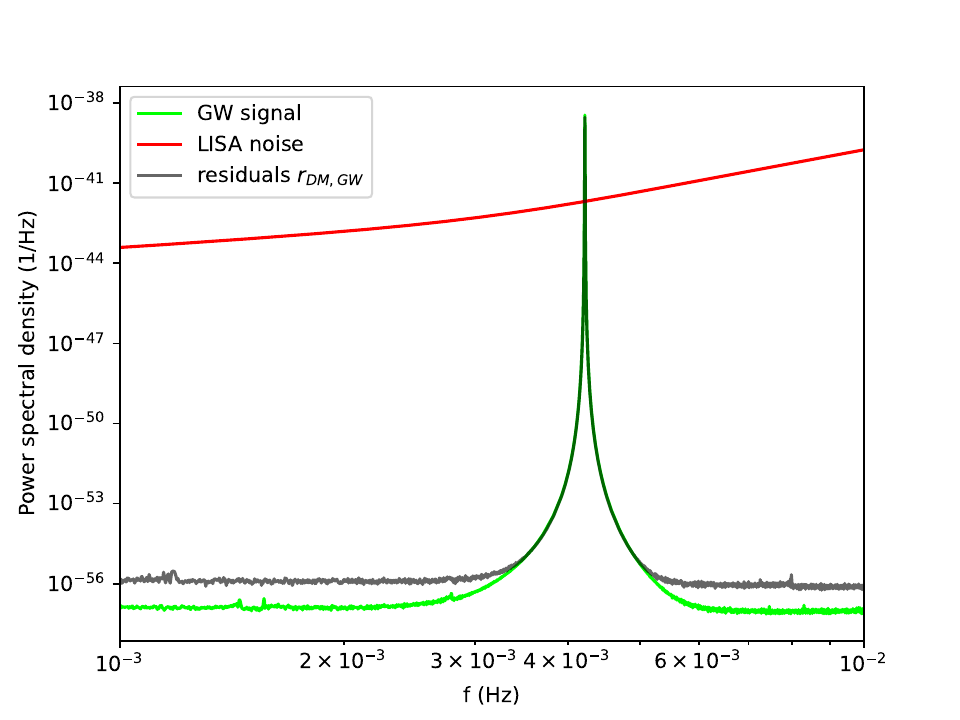}
    \caption{GW signal PSD (in lime) compared to \textit{LISA} noise PSD of TDI $A$ combination (in red) in Fourier domain (as in Fig.~\ref{fig:GW_signal}). This time, the $r_\mathrm{DM,GW}$ residuals power in grey are close to the signal power and above the noise. This indicates the model fits poorly the data.}
    \label{fig:DM_GW_res_signal}
  \end{minipage}
\end{minipage}
\end{figure}

\begin{minipage}{\textwidth}
  \begin{minipage}[b]{0.49\textwidth}
   As it was mentioned in the beginning of this chapter, we now tackle one of the main goals of this study, i.e we try to answer the following question : can a real DM signal be misinterpreted as a GB ? To do so, we simulate the same DM signal as previously (the parameters are shown in Table ~\ref{tab:DM_DM}) and we fit this signal with a GB model using \textit{FastGB}. 
   The fitted parameters are shown in Table ~\ref{tab:GW_DM}. We use the same procedure as before, i.e we compute the residuals between the GW model time series $m^\mathrm{GW}_\mathrm{DM}$ (from the model parameters $M^\mathrm{GW}_\mathrm{DM}$, see Table ~\ref{tab:GW_DM}) and the DM data $d_\mathrm{DM}$ 
    \begin{align}
        r_\mathrm{GW,DM} &= d_\mathrm{DM} - m^\mathrm{GW}_\mathrm{DM} \, ,
    \end{align}
    which allows us to compute the RNR
    \begin{subequations}
    \begin{align}
        \sqrt{\mathrm{RNR}^\mathrm{GW, DM}_A} &\approx 51 \, \\
        \sqrt{\mathrm{RNR}^\mathrm{GW, DM}_E} &\approx 52 \, .
    \end{align}
    \end{subequations}
    Those values must be compared with the SNR values of DM data Eq.~\eqref{eq:SNR_DM_data}.
    \end{minipage}
    \hfill
    \begin{minipage}[b]{0.49\textwidth}
    \centering
    \begin{tblr}{vlines, colspec={cccc}}
        \hline
         & \SetCell[c=1]{}$D_\mathrm{DM}$ & \SetCell[c=1]{} $M^\mathrm{GW}_\mathrm{DM}$& \SetCell[c=1]{} Mean $\pm$ 1 $\sigma$  \\
        \hline
        \hline
        $\mathcal{A}$ $(10^{-21})$ & $1.397 $ & $0.129$ & $ 0.128 \pm 0.004$\\
        \hline
        $f$ (mHz) & $4.215$ & $4.215$ & $4.215 \pm 0.000$\\
        \hline
        $\dot f$ (aHz/s) & $-$ & $-8.040$ & $38.14 \pm 208.6$\\
        \hline
        $\beta$ (rad) & $1.046$ & $-1.520$ &$-1.522 \pm 0.004$\\
        \hline
        $\lambda$ (rad) & $0.431$ &$5.326$ & $5.307 \pm 0.021$\\
        \hline
        $\imath$ (rad) & $-$ & $1.990$ & $1.995 \pm 0.024$\\
        \hline
        \SetCell[r=2]{}$\Psi$ (rad) & $-$ & $3.114$ & $3.079 \pm 0.046$\\
        & & & $1.571 \pm 0.082$\\
        & & & $0.067 \pm 0.053$\\
        \hline
        \SetCell[r=2]{} $\Phi$ (rad) & $3.444$ & $4.740$ & $4.740 \pm 0.066$\\
        & & & $1.597 \pm 0.069$ \\
        \hline
    \end{tblr}
    \captionof{table}{On the left, the GW parameters injected in the simulation equivalent to the DM parameters shown in Table ~\ref{tab:DM_DM} (see Eq.~\eqref{eq:equiv_A_epsilon}). On the middle and on the right, the GW best fit and the mean and standard deviation of the full distribution of posteriors. One can notice different modes for $\Psi$ and $\Phi$ parameters.}\label{tab:GW_DM}
  \end{minipage}
\end{minipage}
\noindent\\
One can notice that the SNR and RNR are of the same order of magnitude, and are larger than 1, meaning that the model does not fit the data (see Fig.~\ref{fig:GW_DM_res}). 
Now, we wonder if \textit{LISA} will be able to non ambiguously detect DM. To do so, we use Bayesian model selection criteria which allow one to quantitatively compare the efficiency of the two different models, the GW $m^\mathrm{GW}_\mathrm{DM}$ and DM $m^\mathrm{DM}_\mathrm{DM}$ models, on the same dataset $d_\mathrm{DM}$. More precisely, we use the Bayes factor $\mathcal{B}$ which gives the ratio of the probability of getting the dataset $d_\mathrm{DM}$ under the condition of one of the model, e.g. $m^\mathrm{DM}_\mathrm{DM}$, to the probability of getting $d_\mathrm{DM}$ under the condition of the second model, $m^\mathrm{GW}_\mathrm{DM}$. This is equivalent to the ratio of the evidences (see Eq.~\ref{eq:Bayes_factor}). For each model, the sampler provides us the evidence of the given model.
For the dataset and models under consideration, the Bayes factor $\mathcal{B}$ is 
\begin{align}
    \log \mathcal{B} &= \log p(d_\mathrm{DM}|m^\mathrm{DM}_\mathrm{DM}) - \log p(d_\mathrm{DM}|m^\mathrm{GW}_\mathrm{DM})= \mathcal{O}(10^3) \gg 1\, ,
\end{align}
i.e given the data that contains a simulated signal, the DM model $m^\mathrm{DM}_\mathrm{DM}$ is largely preferred compared to the GW model.
Therefore, a priori, \textit{LISA} is able to make the difference between one GB and scalar DM. This suggests that if a scalar DM field exists with coupling and oscillation frequency in \textit{LISA} range, \textit{LISA} will be able to uncorrelate it from single galactic binaries.
In the following section, we try to answer the opposite question, namely if it is possible that a galactic binary is seen as DM by \textit{LISA}, i.e if both GW and DM models efficiently fit a GB data.

\subsection{Can a galactic binary be misinterpreted as scalar dark matter ?}

We now focus on the opposite situation compared to the previous section, namely we use a dataset from a GB which we fit with a DM model.
For this situation, the results are shown in Table ~\ref{tab:DM_GW}.
Here, the $v$ parameter denotes the velocity of the wave, for the GW data, it corresponds to the speed of light, while the DM model is constrained by the DM velocity distribution, so it is driven towards $\sim 10^{-3}\: c$. 

Computing the RNR of the best fit DM model on GW data, one finds 
\begin{subequations}
\begin{align}
    \sqrt{\mathrm{RNR}^\mathrm{DM, GW}_A} &\approx 572 \, \\
     \sqrt{\mathrm{RNR}^\mathrm{DM, GW}_E} &\approx 682 \, .
\end{align}
\end{subequations}
The RNR values are still much larger than 1, and of the same order of magnitude as the GW data SNR Eq.~\eqref{eq:SNR_GW_data}, implying that the model is inefficient\footnote{Note that, here, we used the galactic velocity distribution as a prior, but for completeness, we can also relax this velocity constraint, i.e use an uniform prior in $[0,c]$. In this case, the best fit presents a much higher velocity (of order $10^7$ m/s), but the resulting RNR is of the same order of magnitude as before $\sim 500$.}.
The Bayes factor in this case is
\begin{align}
    \log \mathcal{B} &= \log p(D_\mathrm{GW}|M^\mathrm{GW}_\mathrm{GW}) - \log p(D_\mathrm{DM}|M^\mathrm{DM}_\mathrm{GW}) = \mathcal{O}(10^5) \gg 1 \, ,
\end{align}
which indicates that the GW model is by far preferred, given the GW dataset that we use. Therefore, we can conclude that \textit{LISA} is able to distinguish between scalar DM and one single GB.
\\
\noindent
\begin{minipage}{\textwidth}
  \begin{minipage}[b]{0.4\textwidth}
    \captionof{table}{On the left column, GW injected parameters. On the middle and right sides, we show respectively the DM model best fit and the mean and standard deviation of the full distribution of posteriors.}\label{tab:DM_GW}
    \end{minipage}
    \hfill
    \begin{minipage}[b]{0.55\textwidth}
    \centering
    \begin{tblr}{vlines, colspec={cccc}}
        \hline
        & \SetCell[c=1]{}$D_\mathrm{GW}$ & \SetCell[c=1]{} $M^\mathrm{DM}_\mathrm{GW}$ & \SetCell[c=1]{} Mean $\pm$ 1 $\sigma$ \\
        \hline
        \hline
        $\varepsilon$ $(10^{-5})$ & $2.396$& $1.989$ &$2.006\pm 0.014$\\
        \hline
        $f$ (mHz) & $4.215$ & $4.215$ & $4.215 \pm 0.000$\\
        \hline
        $\beta$ (rad) & $0.817$ & $-0.101$ & $-0.098 \pm 0.004$ \\
        \hline
        $\lambda$ (rad) & $5.149$ & $0.806$ & $0.804 \pm 0.005$\\
        \hline
        $v$ $(10^6)$ (m/s) & $299.8$ & $1.014$ & $1.003 \pm 0.010$ \\
        \hline
        $\Phi$ (rad) & $4.890$ & $5.521$ & $5.521 \pm 0.007$\\
        \hline
        $\dot f$ (aHz/s) & $55.91$ & $-$ & $-$\\
        \hline
        $\iota$ (rad) & $1.047$ & $-$ & $-$\\
        \hline
        $\Psi$ (rad) & $0.785$ & $-$ & $-$\\
        \hline
    \end{tblr}
  \end{minipage}
\end{minipage}

\subsection{What about two galactic binaries ?}

In the beginning of this chapter, when discussing about the DM and GW transfer functions, we mentioned that the two main differences between DM and GW are the polarization and the velocity of the wave, i.e the former is effectively non propagating at the scale of TDI, while the latter travels at speed $c$. These are the reasons why \textit{LISA} can a priori make the difference between both signals, as we have seen in the previous sections. We can wonder if, by considering the signals emitted by multiple GB simultaneously, this could change the story, i.e if \textit{LISA} could be misled and interpret this effective signal as scalar DM.  

It is not possible to construct a dipolar signal from one or multiple quadrupolar signals, therefore the polarizations will still differ. However, one can imagine situations where two GB with opposite sky localizations emit GW with same frequency and amplitude, which therefore would create a standing GW. In such (hypothetical) situations, both scalar DM wave and standing GW would not propagate at the time scale of one TDI combination (at leading order).

Let us now derive the transfer function of such system.
Assuming same GW frequency, same amplitude $\mathcal{A}$, phase $\Phi$ and $\imath=\dot f =0$ for both binaries, but opposite source localization $\hat k$, the transfer function of the two binaries combined is (following Eq.~\eqref{eq:TDI2_GW_TF}) 
\begin{subequations}
\begin{align}
    &\mathcal{T}^\mathrm{Double \: GW}_X(\omega) = 8\sin\left(\frac{\omega L}{c}\right)\sin\left(\frac{2\omega L}{c}\right)\Re\left[i\hat h^\mathrm{SSB}_{ij}e^{-i\left(\frac{6\omega L}{c}-\Phi\right)}\sum_{\ell=2,3} w_\ell \frac{\hat n^i_{1\ell}\hat n^j_{1\ell}}{1-(\hat n_{1\ell}\cdot \hat k)^2}\:\times \right.\, \\
    &\left.\left(\cos(\vec k \cdot \vec x_1)\sin\left(\frac{\omega L}{c}\right) - (\hat n_{1\ell}\cdot \hat k)\left(\sin(\vec k \cdot \vec x_1)\cos\left(\frac{\omega L}{c}\right)-\sin\left(\vec k \cdot \vec x_1 - \frac{\omega L \hat n_{1\ell}\cdot \hat k}{c}\right)\right)\right)\right]\, \nonumber,
\end{align}
i.e a standing wave, as expected. When $\omega L/c \ll 2\pi$, this reduces to
\begin{align}
    |\mathcal{T}^\mathrm{Double \: GW}_X(\omega)| &\approx 16\left(\frac{\omega L}{c}\right)^3\left|\hat h^\mathrm{SSB}_{ij} \left(\hat n^i_{13}\hat n^j_{13}-\hat n^i_{12}\hat n^j_{12}\right)\cos(\vec k \cdot \vec x_1)\right| \, .
\end{align}
\end{subequations}
This transfer function is similar to the one of a single GW (up to a factor 2 as expected), and with extinction positions when $\hat k$ is orthogonal to $\hat x_1$. Note that the standing wave wavelength does still correspond to $k_\mathrm{GW} = \omega/c$, while in the case of DM, the de Broglie wavelength of the fields is $10^3$ larger (since $\lambda_\mathrm{DM} = 2\pi /|\vec k_\mathrm{DM}| = 2\pi c^2/\omega v_\mathrm{DM} \sim 10^3  \lambda_\mathrm{GW}$). If we assume that those GWs have the smallest frequency in \textit{LISA} band, i.e $f=10^{-4}$ Hz, $\lambda_\mathrm{GW} \gg |\vec x_1|$, such that the double GW are seen as an homogeneous standing wave by \textit{LISA}, as the scalar DM field. \\
\noindent
\begin{minipage}{\textwidth}
  \begin{minipage}[b]{0.49\textwidth}
    We artificially generate a signal from the superposition of two GBs with opposite sky localizations. In Table ~\ref{tab:double_GW}, we show the parameters of only one of them, as well as the scalar DM best fit. The SNR is
    \begin{subequations}
    \begin{align}
    \sqrt{\mathrm{SNR}^\mathrm{2 \: GW}_A} &\approx 251 \, \\
    \sqrt{\mathrm{SNR}^\mathrm{2 \: GW}_E} &\approx 120 \, ,
    \end{align}
    while the RNR are 
    \begin{align}
        \sqrt{\mathrm{RNR}^\mathrm{DM, 2 \: GW}_A} &\approx 243 \, \\
        \sqrt{\mathrm{RNR}^\mathrm{DM, 2 \: GW}_E} &\approx 119 \, .
    \end{align}
    \end{subequations}
    Therefore, it seems the model does not fit correctly the data. As we have not implemented multiple binaries fit, we cannot compare with a model of two GBs.  
    \end{minipage}%
    \hfill
    \begin{minipage}[b]{0.49\textwidth}
    \centering
    \begin{tblr}{vlines, colspec={cccc}}
        \hline
        & \SetCell[c=1]{} $D_\mathrm{2 \: GW}$ & \SetCell[c=1]{} $M^\mathrm{DM}_\mathrm{2 \: GW}$  \\
        \hline
        \hline
        $\varepsilon$ $(10^{-4})$ & $7.106$& $9.446$ \\
        \hline
        $f$ (mHz) & $0.121$ & $0.121$ \\
        \hline
        $\beta$ (rad) & $1.046$ & $0.482$\\
        \hline
        $\lambda$ (rad) & $0.431$ & $3.699$\\
        \hline
        $\Phi$ (rad) & $5.936$ & $3.290$\\
        \hline
        $v$ $(10^6)$ (m/s) & $299.8$ & $1.149$\\
        \hline
        $\imath$ (rad) & $\pi/2$ & $-$ \\
        \hline
        $\Psi$ (rad) & $0$ & $-$ \\
        \hline
        $\dot f$ (aHz/s) & $0$ & $-$ \\
        \hline
    \end{tblr}
    \captionof{table}{On the left column, we show the GW injected parameters in the simulation for the double GW fit. On the right, we show the DM model best fit.}     
    \label{tab:double_GW}
  \end{minipage}
\end{minipage} 
Therefore, even with such two GB, one should not expect degeneracy between DM and GW signals as seen by \textit{LISA}. This conclusion is also strengthen by the footnote $13$, where we stated that when trying to model GW data by scalar DM, even when relaxing the prior on the velocity, i.e with a best fit velocity close to the speed of light, the DM model still poorly reflects the GW data. This suggests that the main signature which allows one to make the difference between DM and GW is the polarization. This is why combining several GW of rank $2$ polarization does not mimic the effect of a scalar DM with a rank $1$ polarization.

\section{\label{sec:delta_varepsilon}Realistic limit on the sensitivity of \textit{LISA} to ultralight dark matter couplings}

As mentioned in Section ~\ref{sec:model_DM_signal_DM}, the small velocity of the DM wave implies a large width of the $\varepsilon$ parameter posterior distribution. This broadening of the distribution due to this correlation implies that for the specific realization of the data considered in Section ~\ref{sec:model_DM_signal_DM}, the sensitivity of the detector to the coupling $\varepsilon$ decreases compared to a situation where $\varepsilon$ is decorrelated from the other parameters (i.e when $v_\mathrm{DM}$ is fixed). 
In this section, we derive analytically the width of the $\varepsilon$ distribution, as function of the frequency, and we check for its consistency for the specific realization of the data used in the simulation. Then, in Chapter ~\ref{chap:sens_experiments}, we will use this analytical result to infer the sensitivity of \textit{LISA} to ULDM couplings for any frequency in the \textit{LISA} band, taking into account this correlation.

\subsection{Analytical likelihood}

As it was discussed in Section ~\ref{sec:Generation_signal_LISA}, we use the TDI $A,E$ combinations to fit a model onto our dataset. As shown in Eq.~\eqref{eq:TDI_A_noise_PSD}, the two combinations have the same noise PSD, such that only the signal will differ between the two.
In the following, we will work using continuous Fourier transforms (and not DFT), and we will express Fourier transform quantities with a hat.
We start by expressing the (complex) Fourier signal of the $A$ combination as a function of parameters $(\varepsilon^D,f^D,\beta^D,\lambda^D,v^D_\mathrm{DM},\Phi^D)$ at frequency $\omega^D$\footnote{The signal is purely monochromatic (up to a small correction, see next paragraph), and therefore we will be interested at the likelihood at a unique Fourier bin, which for simplicity, we assume is exactly the same as the data frequency $f^D=\omega^D/2\pi$.}
\begin{subequations}
\begin{align}
    &\hat d^\mathrm{DM}_A(\omega^D) = \frac{\sqrt{16 \pi G \rho_\mathrm{DM}}v^\mathrm{D}_\mathrm{DM} \varepsilon^D}{\omega^D c^2}\left(\frac{|\mathcal{T}^\mathrm{DM}_Z|e^{-i\left(\frac{4\omega^D L}{c}+\vec k^D \cdot \vec x_3 - \Phi^D \right)}-|\mathcal{T}^\mathrm{DM}_X|e^{-i\left(\frac{4\omega^D L}{c}+\vec k^D \cdot \vec x_1 - \Phi^D \right)}}{\sqrt{2}}\right) \, \\
    &= \frac{16\sqrt{8 \pi G \rho_\mathrm{DM}}v^\mathrm{D}_\mathrm{DM} \varepsilon^D}{\omega^D c^2}\sin \left(\frac{\omega^D L}{c}\right)\sin \left(\frac{2\omega^D L}{c}\right)\sin^2\left(\frac{\omega^D L}{2c}\right)e^{-i\left(\frac{4\omega^D L}{c}-\Phi^D\right)}\,\\
    &\left(\left(\hat n_{12}\cdot \hat e^D_v\right)e^{-i\vec k^D \cdot \vec x_3} - \left(\hat n_{23}\cdot \hat e^D_v\right)e^{-i\vec k^D \cdot \vec x_1}\right) \, \nonumber \\
    &\approx \frac{16\sqrt{8 \pi G \rho_\mathrm{DM}}v^\mathrm{D}_\mathrm{DM} \varepsilon^D}{\omega^D c^2}\left(\hat n_{12}-\hat n_{23}\right)\cdot \hat e^D_v \sin \left(\frac{\omega^D L}{c}\right)\sin \left(\frac{2\omega^D L}{c}\right)\sin^2\left(\frac{\omega^D L}{2c}\right)\label{eq:TDI_A_data_DM}\,\\
    &e^{-i\left(\frac{4\omega^D L}{c}-\frac{\omega^D v^\mathrm{D}_\mathrm{DM}|\vec x_\mathrm{AU}|\cos(\beta^D)}{c^2}-\Phi^D\right)} \, \nonumber \,
\end{align}
where we have used the transfer function of $X$ Eq.~\eqref{eq:amp_TF_X2}, which we also use for the $Z$ combination by the simple $\hat n_{23} \rightarrow \hat n_{12}$ and $\vec x_1 \rightarrow \vec x_3$ changes. At the last line, we approximated $\vec k^D \cdot \vec x_1 \approx \vec k^D \cdot \vec x_\mathrm{AU} + \mathcal{O}(kL)$ (as $kL \ll 1$ for all DM frequencies of interest, as discussed previously), where $|\vec x_\mathrm{AU}| \approx 1.5 \times 10^{11}$ m is the mean Earth-Sun distance, and similarly for $\vec k^D \cdot \vec x_3$. Therefore, both dot products are equal at leading order. In addition, we assume that spacecrafts are close to the ecliptic plane such that the dot product $\hat k^D \cdot \vec x_\mathrm{AU}$ evolves with time and can be approximated by $\sim -|\vec x_\mathrm{AU}|\cos(\beta^D)\cos(\omega_E t -\lambda^D)$, where $\omega_E$ is Earth rotation frequency around the Sun. As $\omega_E \ll \omega$ and in particular $\omega_E$ is outside of the \textit{LISA} band, the oscillation can be neglected because it will only produce a small broadening of the Fourier peak at frequency $f=\omega/2\pi$, as mentioned previously. In a similar manner, the signal of the $E$ combination reads
\begin{align}\label{eq:TDI_E_data_DM}
    \hat d^\mathrm{DM}_E(\omega^D) &= \frac{16\sqrt{8 \pi G \rho_\mathrm{DM}}v^\mathrm{D}_\mathrm{DM} \varepsilon^D}{\omega^D c^2}\frac{3\hat n_{13}\cdot \hat e^D_v}{\sqrt{3}} \sin \left(\frac{\omega^D L}{c}\right)\sin \left(\frac{2\omega^D L}{c}\right)\sin^2\left(\frac{\omega^D L}{2c}\right)\,\\
    &e^{-i\left(\frac{4\omega^D L}{c}-\frac{\omega^D v^\mathrm{D}_\mathrm{DM}|\vec x_\mathrm{AU}|\cos(\beta^D)}{c^2}-\Phi^D\right)} \, \nonumber ,
\end{align}
i.e it differs from Eq.~\eqref{eq:TDI_A_data_DM} just by the geometric factor, at leading order $((\hat n_{12}-\hat n_{23})\cdot \hat e^D_v)/\sqrt{2} \rightarrow 3(\hat n_{13} \cdot \hat e^D_v)/\sqrt{6}$.
\end{subequations}
Similarly, in Fourier space, the model for both combinations reads 
\begin{subequations}
\begin{align}
    \hat m^\mathrm{DM}_A(\omega^M) &=\frac{16\sqrt{8 \pi G \rho_\mathrm{DM}}v^M_\mathrm{DM} \varepsilon^M}{\omega^M c^2}\left(\hat n_{12}-\hat n_{23}\right)\cdot \hat e^M_v\sin \left(\frac{\omega^M L}{c}\right)\sin \left(\frac{2\omega^M L}{c}\right)\sin^2\left(\frac{\omega^M L}{2c}\right)\,\nonumber \\
    &e^{-i\left(\frac{4\omega^M L}{c}-\frac{\omega^M v^M_\mathrm{DM}|\vec x_\mathrm{AU}|\cos(\beta^M)}{c^2}-\Phi^M\right)}\, \label{eq:TDI_A_model_DM} \\
    \hat m^\mathrm{DM}_E(\omega^M) &=\frac{16\sqrt{8 \pi G \rho_\mathrm{DM}}v^M_\mathrm{DM} \varepsilon^M}{\omega^M c^2}\left(\sqrt{3}\hat n_{13}\cdot \hat e^M_v\right)\sin \left(\frac{\omega^M L}{c}\right)\sin \left(\frac{2\omega^M L}{c}\right)\sin^2\left(\frac{\omega^M L}{2c}\right)\,\nonumber \\
    &e^{-i\left(\frac{4\omega^M L}{c}-\frac{\omega^M v^M_\mathrm{DM}|\vec x_\mathrm{AU}|\cos(\beta^M)}{c^2}-\Phi^M\right)}\, \label{eq:TDI_E_model_DM},
\end{align}
\end{subequations}
i.e they are the same as Eqs.~\eqref{eq:TDI_A_data_DM} and \eqref{eq:TDI_E_data_DM}, but where all quantities with $M$ superscript are the unknown parameters. 

Finally, we define the standard deviation of the TDI combinations $A,E$ noise at frequency $f$ as
\begin{align}
    \hat \sigma_{A,E}(f) &= \sqrt{\int^{f+\frac{1}{2T_\mathrm{obs}}}_{f-\frac{1}{2T_\mathrm{obs}}} df N_{A,E}(f)} \approx \sqrt{\frac{N_{A,E}(f)}{T_\mathrm{obs}}} \equiv \hat \sigma(f)\, ,
\end{align}
where the $N_{A,E}$ are given by Eq.~\eqref{eq:TDI_A_noise_PSD}). At the second to the last equality, we assumed that the noise level is constant over the frequency bandwidth of interest, which is $1/T_\mathrm{obs}$ wide, for a monochromatic signal.

Let us now write the (monochromatic) joint Gaussian likelihood in Fourier space as (from Eq.~\eqref{eq:likelihood_time_domain_general_LISA})
\begin{align}\label{eq:likelihood_DM_general}
    &\log \mathcal{L}(f) = \kappa  -\frac{\left|\hat d^\mathrm{DM}_A(f)-\hat m^\mathrm{DM}_A(f)\right|^2+\left|\hat d^\mathrm{DM}_E(f)-\tilde m^\mathrm{DM}_E(f)\right|^2}{2\hat \sigma^2(f)}  \, ,
\end{align}
where $f$ denotes the Fourier frequency and $\kappa$ is a normalization constant independent of the model, and therefore will not play a role in the following analysis. In the following, we will be interested in the joint likelihood of $\varepsilon$ and $v_\mathrm{DM}$ parameters. In practice, 1) we assume that all the other parameters are individually uncorrelated with the rest (see Fig.~\ref{fig:cornerplot_DM_DM}, where one can notice that the parameters that we want to marginalize over are not correlated with any other parameter); and 2) we use a Laplace approximation, which approximates the posterior distribution by a Gaussian around the best fit value, and we approximate the best fit value by the injected value\footnote{This is a valid approximation if the data contains no noise, which is the case. This approximation can also be confirmed by Table ~\ref{tab:DM_DM}.}. Therefore, the marginalization consists essentially in taking $(f^M,\beta^M,\lambda^M,\Phi^M) = (f^D,\beta^D,\lambda^D,\Phi^D)$, such that the velocity-coupling joint likelihood becomes
\begin{subequations}
\begin{align}
    &\log \mathcal{L}(v^M_\mathrm{DM},\varepsilon^M) =  -\frac{\left|\hat d^\mathrm{DM}_A(f)-\hat m^\mathrm{DM}_A(f)\right|^2+\left|\hat d^\mathrm{DM}_E(f)-\hat m^\mathrm{DM}_E(f)\right|^2}{2\hat \sigma^2(f)} \, \\
    &= -\frac{T_\mathrm{obs}}{2N_{A,E}(f)}\mu_\mathrm{DM}\left(\frac{16\sqrt{8 \pi G \rho_\mathrm{DM}}}{2 \pi f c^2}\right)^2 \sin^2\left(\frac{2 \pi f L}{c}\right)\sin^2 \left(\frac{4 \pi f L}{c}\right)\sin^4\left(\frac{2 \pi f L}{2c}\right)\, \label{eq:likelihood_DM} \\
    &\left(\left(v^D_\mathrm{DM} \varepsilon^D\right)^2+\left(v^M_\mathrm{DM} \varepsilon^M\right)^2-2v^D_\mathrm{DM} \varepsilon^Dv^M_\mathrm{DM} \varepsilon^M\cos\left(\frac{2\pi f |\vec x_\mathrm{AU}|\cos(\beta)}{c^2}\left(v^D_\mathrm{DM}-v^M_\mathrm{DM}\right)\right)\right) \, \nonumber,
\end{align}
with
\begin{align}
    \mu_\mathrm{DM} &= \left(\left(\hat n_{12}-\hat n_{23}\right)\cdot \hat e_v\right)^2+\left(\sqrt{3}\hat n_{13} \cdot \hat e_v\right)^2 \, ,
\end{align}
\end{subequations}
where we dropped the first term of Eq.~\eqref{eq:likelihood_DM_general} since it is a normalization factor, where we consider a monochromatic signal at frequency $f$ and where we dropped the $D$ and $M$ superscripts for the marginalized parameters (i.e when $p^M = p^D$). 
\begin{figure}
    \centering
    \includegraphics[width=0.6\textwidth]{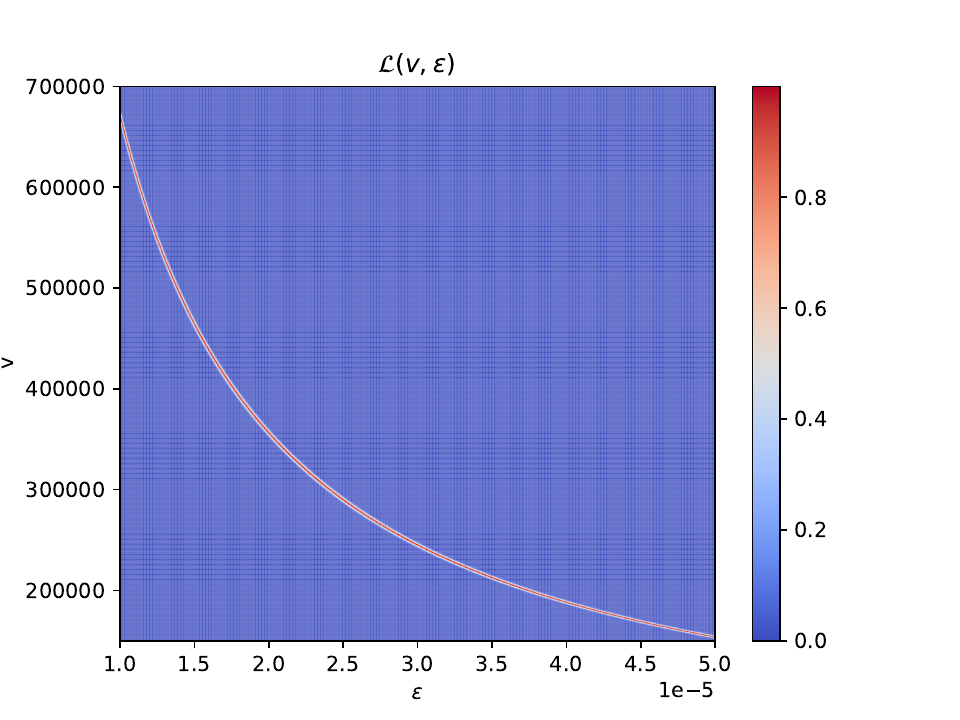}
    \caption{Joint likelihood $\mathcal{L}(v^M_\mathrm{DM},\varepsilon^M)$ for the data parameters shown in Table ~\ref{tab:DM_DM}.}
    \label{fig:joint_likelihood_v_eps}
\end{figure}
By using the true value of parameters $\varepsilon^D, f, \beta, v^D_\mathrm{DM}$ from Table ~\ref{tab:DM_DM}, one can obtain the joint likelihood as function of $\varepsilon^M, v^\mathrm{M}_\mathrm{DM}$. This is shown in Fig.~\ref{fig:joint_likelihood_v_eps}, where we set numerically $\mu_\mathrm{DM}\sim 4.70$\footnote{This factor is time dependent and evolves with the orbit. Numerically, we can recover the time evolution of the dot product between the DM wind direction and any \textit{LISA} arm, and we find that $\left(\hat n_{12}\cdot \hat e_v\right)-\left(\hat n_{23}\cdot \hat e_v\right) \sim 1.29 +0.43\cos(\omega_E t+ \varphi)$ and $\hat n_{13}\cdot \hat e_v \sim 0.79 -0.03\cos(\omega_E t+ \varphi')$, where $\omega_E, \varphi,\varphi'$ are respectively the Earth rotation frequency around the Sun and two irrelevant phases. As we are interested in frequencies $\omega \gg \omega_E$, we assume that the second term is constant and therefore $\left(\hat n_{12}\cdot \hat e_v\right)-\left(\hat n_{23}\cdot \hat e_v\right) \sim 1.29 +0.43 \sim 1.72$ and $\sqrt{3}(\hat n_{13}\cdot \hat e_v) \sim 1.32$, such that the sum of the squares is $\sim 4.70$.}. We recover the banana-shaped likelihood as in Fig.~\ref{fig:cornerplot_DM_DM}. 

\subsection{Coupling posterior width}

From this joint likelihood, we can compute the joint posterior distribution which reads
\begin{align}\label{eq:joint_posterior}
    &\mathcal{P}(v^M_\mathrm{DM},\varepsilon^M) = \frac{\mathcal{L}(v^M_\mathrm{DM},\varepsilon^M)\pi(v^M_\mathrm{DM})\pi(\varepsilon^M)}{\int dv^M_\mathrm{DM} d\varepsilon^M \mathcal{L}(v^M_\mathrm{DM},\varepsilon^M)\pi(v^M_\mathrm{DM})\pi(\varepsilon^M)} \, ,
\end{align}
where $\pi(v^M_\mathrm{DM}),\pi(\varepsilon^M)$ are respectively the prior distribution on the velocity, given by Eq.~\eqref{DM_vel_distrib_scalar} and the prior on the coupling, which is uniform. The denominator ensures that the posterior distribution integrates to 1 over the whole parameter space. 
Then, to obtain the posterior distribution of the $\varepsilon$ parameter, we marginalize over $v^M_\mathrm{DM}$ i.e 
\begin{align}\label{eq:posterior_varepsilon}
    \mathcal{P}(\varepsilon^M) &= \frac{\int dv^M_\mathrm{DM} \mathcal{L}(v^M_\mathrm{DM},\varepsilon^M)\pi(v^M_\mathrm{DM})\pi(\varepsilon^M)}{\int dv^M_\mathrm{DM} d\varepsilon^M \mathcal{L}(v^M_\mathrm{DM},\varepsilon^M)\pi(v^M_\mathrm{DM})\pi(\varepsilon^M)} \, .
\end{align}
From this, one can compute the mean $\mu_\varepsilon$ and standard deviation $\sigma_\varepsilon$ of $\mathcal{P}(\varepsilon^M)$ by numerical integration. For the parameters that are correctly fitted (i.e the frequency $f$, the ecliptic latitude $\beta$), we use the true values for both data and model parameters (we neglect the small difference between the true values and the best fit values, see Table ~\ref{tab:DM_DM}), and we do the same for the velocity $v^D_\mathrm{DM}$ and the coupling $\varepsilon^D$. In practice, using \textit{Mathematica}, we numerically integrate $\mathcal{P}(\varepsilon^M)$ over $\varepsilon^M$ with a global adaptative method, where the region of integration is cut in various sub regions in which the numerical integration is performed. The boundaries of the integral correspond to the prior bounds used in the simulation ($[10^{-6},10^{-4}]$) and we find
\begin{subequations}\label{eq:mean_std_varepsilon}
\begin{align}
    \mu_\varepsilon &= \int d \varepsilon^M \varepsilon^M \mathcal{P}(\varepsilon^M) \sim 2.69 \times 10^{-5} \,  , \\
    \sigma_\varepsilon &= \sqrt{\int d \varepsilon^M (\varepsilon^M- \mu_\varepsilon)^2 \mathcal{P}(\varepsilon^M)} \sim 1.17 \times 10^{-5} \, .
\end{align}
\end{subequations}
As it can be noticed from Table ~\ref{tab:DM_DM}, the relative error on the mean and standard deviation is $\sim 10\%$ and $\sim 36\%$, compared to what is found in the simulation.
The main reason for this discrepancy is that low DM velocities $0 \leq v^M_\mathrm{DM} \lesssim 1.5 \times 10^5$ m/s seem to be disfavoured by data (see Fig.~\ref{fig:cornerplot_DM_DM}), while this velocity range is permitted by our prior. As a consequence, "large" values of coupling $\varepsilon^M$ are not allowed in the analysis using \textit{Nessai} since one still requires $v^M_\mathrm{DM} \varepsilon^M \sim v^D_\mathrm{DM} \varepsilon^D$ for both amplitudes to match. Therefore, the mean and standard deviation get shifted to smaller values than expected, as we have calculated it above. However, we argue that this velocity cut-off is an artefact of the sampler and is inconsistent with the data that is provided. First, the minimum value of velocities on the posterior distribution given by the simulation depends on the prior that we use for the coupling $\varepsilon$\footnote{We always use a uniform prior distribution, but the minimum and maximum allowed values are somewhat arbitrary.}. Second, there is, to our knowledge, no physical reason why the low velocities would not be permitted by the data. Indeed, as mentioned previously, the velocity is not well constrained by the data because the phase induced by the propagation is too small to be detected, and therefore this induced a cut-off of large velocities. However, any arbitrary small velocity $v_s$ should be allowed (as long as $\pi(v_s) \neq 0$ and that we can associate a "large" coupling $\varepsilon_l$ such that $v_s \varepsilon_l \sim v^D_\mathrm{DM} \varepsilon^D$ again). 

In Chapter ~\ref{chap:sens_experiments}, where we will derive \textit{LISA} sensitivities to scalar ULDM fields, we will use the previous simple semi-analytical method to derive $\sigma_\varepsilon$ for all frequencies in \textit{LISA} range.

\section{Conclusion}

The conclusions of this chapter are two-fold. The first one is that we managed to answer the question asked at the beginning of the chapter, namely, are GB and scalar ULDM signals degenerate for \textit{LISA}, i.e is it possible to model accurately a GW signal with scalar DM and vice versa ? It seems that it is not the case, i.e through a heliocentric orbit of one year, \textit{LISA} is able to decorrelate the two signals. This is encouraging with regard to the possible detection of DM by \textit{LISA}. The second conclusion is that, as we have seen in Section ~\ref{sec:model_DM_signal_DM}, the fact that the ULDM field is non relativistic makes it non propagating as seen by \textit{LISA}, especially at low frequency\footnote{As we shall see in Chapter ~\ref{chap:sens_experiments}, even at higher frequencies, when we consider a signal strength such that SNR = 1, the velocity is still badly constrained by data.}. Therefore, the velocity of the DM wave is essentially invisible for \textit{LISA}. The amplitude of the signal being proportional to the DM-SM coupling, but also to the DM velocity, this means that \textit{LISA} is in reality sensitive to the product of the two parameters, but not to one or the other individually. In other words, if one assumes that the velocity is not fixed, but is a free parameter of the theory, the sensitivity of \textit{LISA} to the coupling will decrease due to its correlation with the velocity. In Chapter ~\ref{chap:sens_experiments}, we will derive two sensitivity curves for \textit{LISA} : a standard sensitivity curve where the coupling is assumed to be uncorrelated with the rest of the parameters (i.e we assume the velocity is fixed at $v_\mathrm{DM} \sim 10^{-3}\: c$), and a second one, where we take into account those correlations, i.e where we keep the velocity as a free parameter.
The former has already been derived for one given dilaton-SM coupling in \cite{Yu23}, but the latter, which is more realistic, has never been considered in the literature.

\chapter{\label{chap:axion_photon_LISA}Search for vacuum birefringence and dichroism in \textit{LISA}}

\begin{figure}[h!]
    \centering
    \includegraphics[width=\textwidth]{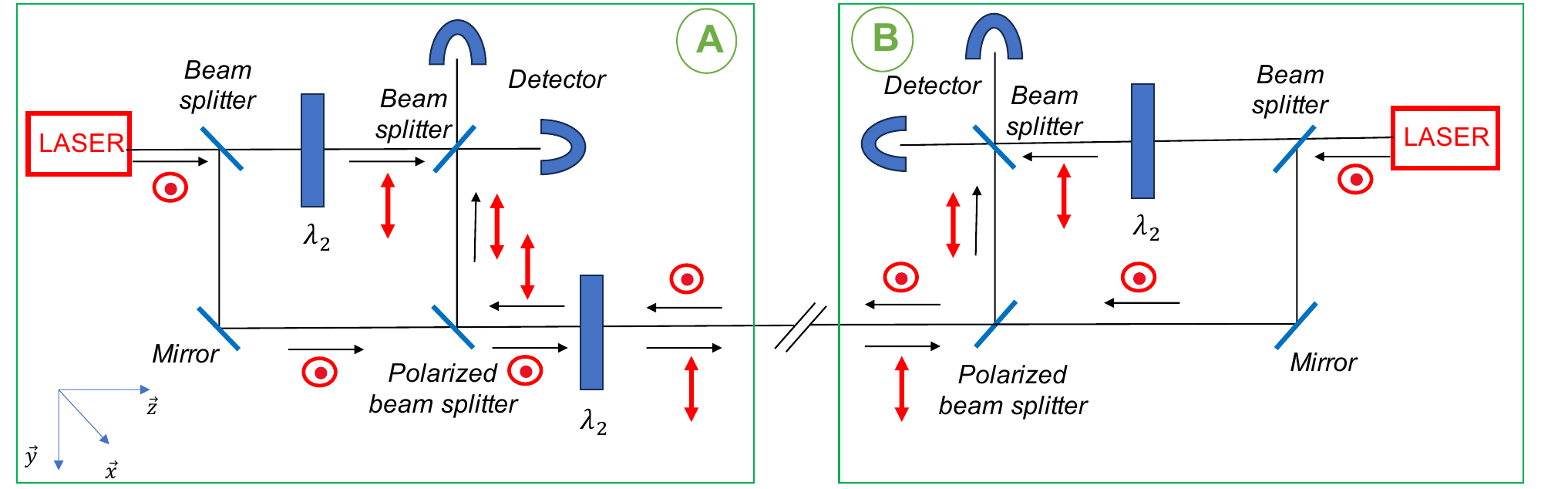}
    \caption{Simplified optical benches in \textit{LISA} spacecrafts. The two optical benches $A$ and $B$ are inside two different spacecrafts (shown as the light green cube) and exchange light between each other. The optical elements are shown in blue, the light polarization in red and the direction of propagation with black arrows. Inside each of these optical benches, an initial "horizontal" linear polarization is produced by the laser and the association of the optical elements inside each bench fulfill all the necessary requirements for the interferometers to work (see text.).}
    \label{fig:opti_bench_LISA}
\end{figure}

As we discussed it in Chapter ~\ref{axion_photon_coupling}, the axion-photon coupling can be probed by its effect on polarization of light. In this section, we study the sensitivity of \textit{LISA} to such effects. 

We first describe the light polarization in \textit{LISA} optical benches. For our purposes, we are interested in the polarization of light produced and detected inside the spacecrafts and travelling in vacuum between the spacecrafts. The current (simplified) optical bench is shown in Fig.~\ref{fig:opti_bench_LISA}, from \cite{Bayle23}, where we only take into account the interference between laser beams of the local spacecraft with a distant one. We assume for simplicity that the light wave is produced with "horizontal" linear polarization along the $x$-axis, is travelling along the $z$ direction and arrives on the detector with "vertical" polarization after passing through the half-wave plate. In terms of light polarization, several requirements are necessary for the detection of GW in \textit{LISA}.

Each spacecraft contains two optical benches, as shown in Fig.~\ref{fig:opti_bench_LISA}. Inside one given optical bench (say $A$ in spacecraft $1$), two beams are interfering : the one produced by the local laser of the given optical bench and the one which has been produced by one of the optical benches of another spacecraft ($B$ in spacecraft $2$) and which has been travelling in free space. We require the two beams to have the same polarization, when they enter the interferometer inside optical bench $A$ (or at least that they do not have opposite linear polarizations), otherwise the interference between the two beams is zero.

\begin{figure}[h!]
\begin{minipage}{\textwidth}
  \begin{minipage}[b]{0.5\textwidth}
    As leaving one spacecraft, one small portion of the local beam can be reflected on it, come back inside and create an additional noise in the interferometer. Assuming that the local beam is at frequency $f_\ell$ while the distant beam (from the second spacecraft) has slightly different frequency $f_\ell + \delta f$, due to standard Doppler from S/C differential velocity, and frequency planning, see \cite{Heinzel24}, the reflected component of the local beam of frequency $f_\ell$ could interfere with itself but only creating a DC component. However, it could also interfere with the distant beam, and create a noise at $\delta f$. In addition, the motion of the reflecting component at the edge of the spacecraft would induce a time dependent phase on the reflected beam, which would mimic a Doppler effect induced by the GW, and therefore, this would greatly alter the measurement.
    To overcome this issue, we require the outgoing polarization to be different than the incoming one. Then, the polarized beam splitter selects only the polarization of interest, i.e the one of the incoming beam.
    It can be shown easily that the current optical benches in Fig.~\ref{fig:opti_bench_LISA} fulfill those requirements. 
    The \textit{LISA} constellation is depicted in Fig.~\ref{fig:LISA_const}. Two optical benches are present in each spacecraft which permits six individual interference measurements at each time. Following our arbitrary choice on light polarization in Fig.~\ref{fig:opti_bench_LISA}, one can see that two different linear polarizations travel in space : vertical polarization travels clockwise while horizontal polarization travels counterclockwise.
    \end{minipage}
    \hfill
    \begin{minipage}[b]{0.45\textwidth}
    \centering
    \includegraphics[width=0.8\textwidth]{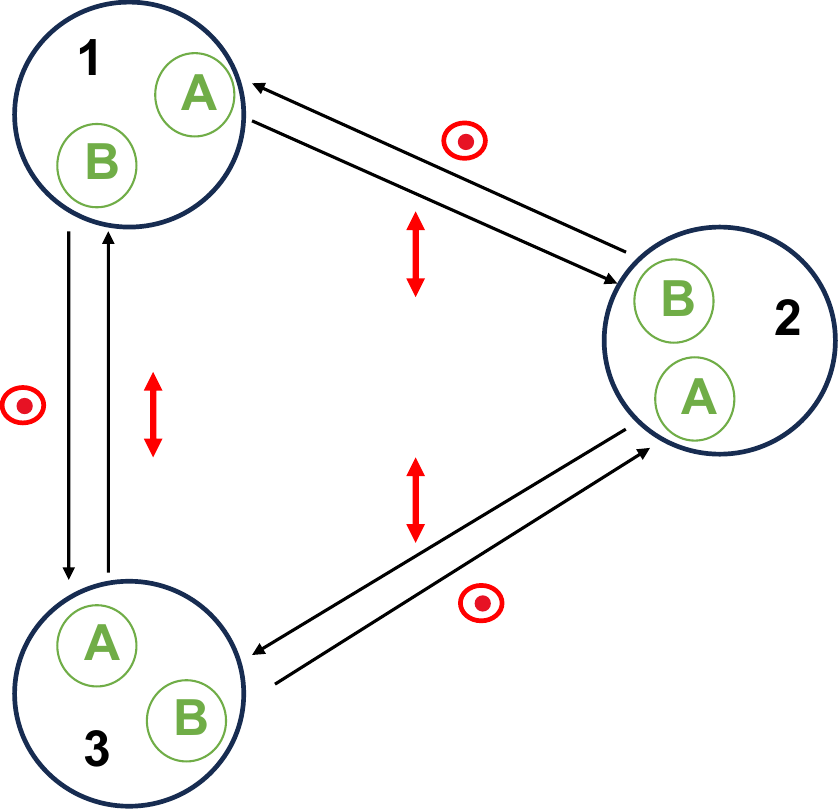}
    \caption{\textit{LISA} constellation involves three spacecrafts, denoted $1,2,3$, each containing two optical benches as described in Fig.~\ref{fig:opti_bench_LISA} denoted by "A" and "B". Six different light signals are sent between the different spacecrafts to form the interferometer; three vertically polarized beams are propagating clockwise and the other three horizontally polarized beams are propagating counterclockwise.}
    \label{fig:LISA_const}
  \end{minipage}
\end{minipage}
\end{figure}

\section{Birefringence}

We first study the possibility of \textit{LISA} to detect axion-photon coupling through vacuum birefringence as described in Eq.~\eqref{delta_c_axion_photon}.
In the last section, we presented the standard design of \textit{LISA} optical benches with linear polarized light travelling in free space. In this section, we consider a non standard design of the optical benches to check if the search of axion-photon coupling in \textit{LISA} is possible. 

\subsection{Jones formalism}

For this, in addition to the two previous requirements for \textit{LISA} to work, we require the light polarization to be circular in free space, i.e when travelling from one spacecraft to another, in order to see any phase shift due to the long propagation between the spacecrafts. Indeed, the axion field only impact the phase velocity of circularly polarized light and not linearly polarized one. To do so, we will still consider an initial linear polarization but we will introduce quarter-wave plates, or other optical elements, inside the original optical benches shown in Fig.~\ref{fig:opti_bench_LISA}.

We can model the system using Jones formalism for polarization states of light and optical elements which is very convenient when treating interference phenomena \cite{Collett05}. This formalism discomposes the light polarization on a basis formed of the two linear polarizations. The polarization state is modelled using a $2\times1$ matrix while the optical elements are represented by $2\times2$ matrices. 

As shown in Fig.~\ref{fig:opti_bench_LISA}, we assume the light beam to propagate along the $z$-axis with an initial polarization along the $x$-axis, that we call "horizontal" polarization. Then, its Jones representation is \cite{Collett05}
\begin{align}
    |H\rangle &= \begin{pmatrix}
        1 \,\\
        0 
    \end{pmatrix} \, .
\end{align}
The other linear polarization along $y$-axis, that we will call "vertical" is represented as 
\begin{align}
    |V\rangle &= \begin{pmatrix}
        0 \,\\
        1
    \end{pmatrix} \, .
\end{align}
A half-wave plate adds a phase $\phi=\pi$ to the light polarization parallel to its slow axis\footnote{The principle of a wave plate is that it presents two perpendicular directions with respective refractive index $n_f, n_s$ with $n_f<n_s$, that we call respectively fast and slow axis \cite{Hecht02}. This leads to a phase shift between the light polarization along the fast axis and the one along the slow axis.}. Its Jones matrix representation is \cite{Collett05}
\begin{align}
    \hat \lambda_2(\theta) &= e^{-i\frac{\pi}{2}}\begin{pmatrix}
        \cos^2 \theta - \sin^2\theta & 2 \sin\theta \cos \theta \,\\
        2 \sin\theta \cos \theta & \sin^2\theta-\cos^2\theta 
    \end{pmatrix} \equiv \begin{pmatrix}
        \cos(2\theta) & \sin(2\theta) \,\\
        \sin(2\theta) & -\cos(2\theta) 
    \end{pmatrix} \, ,
\end{align}
up to an irrelevant total phase, where $\theta$ is the angle between the horizontal axis (the $x$-axis) and the fast axis of the plate. Similarly, a quarter-wave plate introduces a phase shift of $\pi/2$ between the two linear polarized states and can be represented as \cite{Collett05}
\begin{align}
    \hat \lambda_4(\theta) &= \begin{pmatrix}
        \cos^2(\theta)-i\sin^2(\theta)& (1+i)\sin(\theta)\cos(\theta) \,\\
        (1+i)\sin(\theta)\cos(\theta) & \sin^2(\theta)-i\cos^2(\theta) 
    \end{pmatrix}\, ,
\end{align}
up to a total phase. Note that if the half or quarter wave plate is rotated such that its fast or slow axis is along the initial linear horizontal or vertical polarization of light (i.e $\theta=0$ or $\theta=\pi/2$), there is no effect, i.e the input and output polarization are the same (up to an irrelevant phase). Note also that the determinant of these matrices is $1$, i.e they preserve energy, or in other words, they have no absorption.
In the Jones formalism, right and left circular polarizations are represented as \cite{Collett05}
\begin{subequations}
\begin{align}
    |R\rangle &= \frac{1}{\sqrt{2}}\begin{pmatrix}
        1 \,\\
        i 
    \end{pmatrix} \, \\
    |L\rangle &= \frac{1}{\sqrt{2}}\begin{pmatrix}
        1 \,\\
        -i 
    \end{pmatrix} \, .
\end{align}
\end{subequations}
We now introduce horizontal, vertical, left and right polarizers, which respectively transmit only horizontal, vertical, left and right polarizations of the input, defined as \cite{Fymat71}
\begin{subequations}
\begin{align}
    \hat C_H &= \begin{pmatrix}
        1 & 0\,\\
        0 & 0 
    \end{pmatrix} \, \\
    \hat C_V &= \begin{pmatrix}
        0 & 0\,\\
        0 & 1 
    \end{pmatrix} \, \\
    \hat C_L &= \frac{1}{2}\begin{pmatrix}
        1 & i\,\\
        -i & 1 
    \end{pmatrix} \, \\
    \hat C_R &= \frac{1}{2}\begin{pmatrix}
        1 & -i\,\\
        i & 1 
    \end{pmatrix} \, .
\end{align}
\end{subequations}
In practice, the right polarizer is made of the combination $\hat \lambda_4(\pi/4)\hat C_H \hat \lambda_4(-\pi/4)$ (and for the left polarizer, one replaces $\hat C_H$ by $\hat C_V$ or exchange the two quarter-wave plates).
Finally, a mirror reverses the direction of one of the two linear polarizations while letting the other one unchanged. Its representation is \cite{Fymat71}
\begin{align}
    \hat M = \begin{pmatrix}
        -1 & 0 \,\\
        0 & 1 
    \end{pmatrix}\, ,
\end{align}
therefore it is the "horizontal" polarization that gets flipped. 

It can be shown easily that the quarter and half-wave plates commute, implying that a set of optical elements containing $n$ half-wave plates, $m$ quarter-wave plates can be described as the matrix multiplication $\hat \lambda_2^n \hat \lambda_4^m$. However, the mirror and polarizers do not commute with the other optical elements. 

\subsection{Requirements for both gravitational waves and axion detection}

Between the polarized beam splitter and the end of the optical bench, we will model the set of optical elements of $n$ half-wave plates, $m$ quarter-wave plates and $\ell$ polarizers (of any kind) that we add by $\hat S(\theta,\theta')$, where $\theta(\theta')$ is the angle between the fast axis of the half-wave (quarter-wave) plates and the horizontal axis. We will denote this set of elements with superscript $A$ or $B$ for the optical bench $A$ or $B$, which are respectively associated with angles $\theta_A(\theta_B),\theta'_A(\theta'_B)$, i.e we note $\hat S^{(A)}(\theta_A,\theta'_A)$ and $\hat S^{(B)}(\theta_B,\theta'_B)$. 

The first requirement (noted $R_1$) is that the incoming polarization from space can interfere with the local beam, and is not fully rejected by the polarized beam splitter. 
For an initial $|H\rangle$ polarization state for the beam on both spacecrafts, the polarization of the beam after leaving the optical bench $A$ and $B$ is respectively $\hat S^{(A)}(\theta_A,\theta'_A)|H\rangle$ and $\hat S^{(B)}(\theta_B,\theta'_B)|H\rangle$. Then, after travelling in free space, each light beam enters the other spacecraft and goes through the set of optical elements but with opposite propagation direction. In such a case, light sees the fast axis with opposite angle compared to horizontal plane such that the total polarization of the beam becomes respectively 
\begin{subequations}
\begin{align}\label{eq:LISA_birefringence_cond1_1}
    \hat S^{(B)}(-\theta_B,-\theta'_B)\hat{S}^{(A)}(\theta_A,\theta'_A) &|H\rangle \,\\
    \hat S^{(A)}(-\theta_A,-\theta'_A)\hat{S}^{(B)}(\theta_B,\theta'_B) &|H\rangle \, .
\end{align}
\end{subequations}
$R_1$ implies that both polarizations Eq.~\eqref{eq:LISA_birefringence_cond1_1} are not a pure "horizontal" linear polarization (see Fig.~\ref{fig:opti_bench_LISA}), i.e  
\begin{align}
  R_1 \Rightarrow \left\{ 
\begin{array}{rcl} 
 \hat S^{(B)}(-\theta_B,-\theta'_B)\hat{S}^{(A)}(\theta_A,\theta'_A) |H\rangle &\neq |H\rangle \,\\
    \hat S^{(A)}(-\theta_A,-\theta'_A)\hat{S}^{(B)}(\theta_B,\theta'_B) |H\rangle &\neq |H\rangle \, ,
\end{array} \right.\label{eq:LISA_birefringence_cond1}
\end{align}
up to a constant phase.

The second requirement ($R_2$) for the optical interferometer to work is that the reflected component of light is the same as the initial one, such that it is rejected by the beam splitter, i.e we require that after travelling inside the set $\hat S(\theta,\theta')$, getting reflected on the mirror and travelling back inside $\hat S(\theta,\theta')$ (but with opposite propagation direction, as before), the polarization is "horizontal". This means
\begin{align}\label{eq:LISA_birefringence_cond2}
  R_2 \Rightarrow \left\{ 
\begin{array}{rcl}    
\hat S^{(A)}(-\theta_A,-\theta'_A)\hat{M} S^{(A)}(\theta_A,\theta'_A)  |H\rangle &= |H\rangle \, \\
\hat S^{(B)}(-\theta_B,-\theta'_B)\hat{M} S^{(B)}(\theta_B,\theta'_B)  |H\rangle &= |H\rangle \, ,
\end{array} \right.
\end{align}
still up to a constant phase.

Finally, the additional requirement ($R_3$) to be sensitive to the axion field is that at least one of the two light polarizations in outer space has to be circularly polarized. Following the constellation representation of Fig.~\ref{fig:LISA_const}, if we assume the light propagating clockwise is right circularly polarized, then, the light propagating counterclockwise can be anything else, i.e linearly polarized or left polarized state\footnote{If it is linearly polarized or left polarized, the beatnote between the two beams will oscillate at the axion frequency, but the amplitude will be twice as big in the second case.}. Mathematically, 
\begin{align}\label{eq:LISA_birefringence_cond3}
 R_3 \Rightarrow \left\{ 
\begin{array}{rcl}      
    \hat{S}^{(A)}(\theta_A,\theta'_A) |H\rangle &= |R\rangle \,\\
    \hat{S}^{(B)}(\theta_B,\theta'_B) |H\rangle &\neq |R\rangle \, ,
\end{array} \right.
\end{align}
up to a constant phase.

\subsection{Solution}

\begin{figure}[h!]
    \centering
    \includegraphics[width=\textwidth]{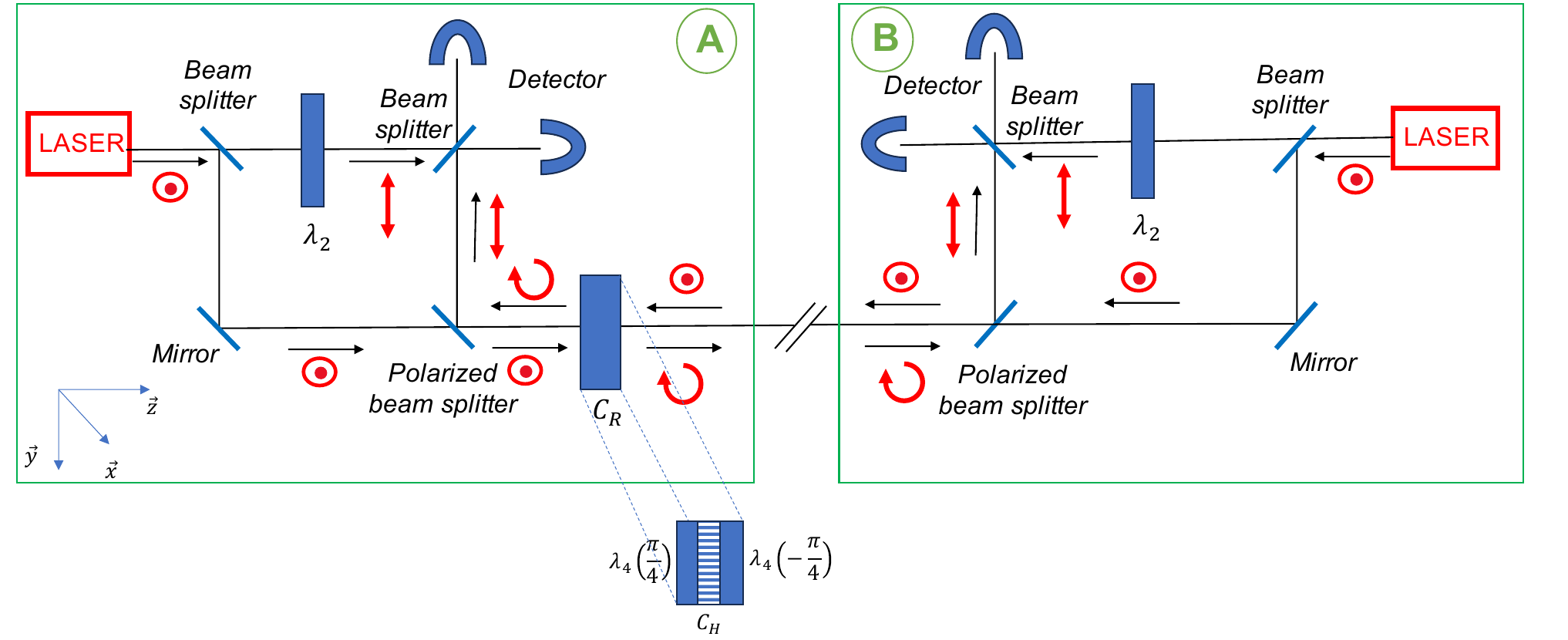}
    \caption{Modified optical bench inside \textit{LISA} spacecraft, compared to the original one presented in Fig.~\ref{fig:opti_bench_LISA}, in order to be sensitive to vacuum birefringence induced by the axion-photon coupling. The half-wave plate at the end of the optical bench $A$ is now replaced with a right polarizer which is a combination of a quarter-wave plate, a horizontal polarizer and a second quarter-wave plate, which are oriented conveniently. This configuration produces a right polarized light at the output of optical bench $A$, and fulfill all the requirements for the light interferometer to work (see text).}
    \label{fig:opti_bench_LISA_modified}
\end{figure}

One can show that a solution to all these requirements is 
\begin{subequations}
\begin{align}
    \hat S^{(A)} &= \hat C_R \, \\
    \hat S^{(B)} &= \mathds{1}_2 \, ,
\end{align}
\end{subequations}
where the latter is the identity matrix. The solution is therefore to replace the last $\hat \lambda_2$ in the optical bench $A$ (see Fig.~\ref{fig:opti_bench_LISA}), by a $\hat \lambda_4(\pi/4)\hat C_H \hat \lambda_4(-\pi/4)$ combination, i.e a first quarter-wave plate oriented with $\theta=\pi/4$, then a horizontal linear polarizer and then a second quarter wave plate oriented with $\theta=-\pi/4$, as it is shown in Fig.~\ref{fig:opti_bench_LISA_modified}.
We show explicitly how it resolves the various requirements. Starting with $R_3$, we have
\begin{subequations}
\begin{align}
    \hat S^{(A)} |H\rangle &= |R\rangle \, \\
    \hat S^{(B)} |H\rangle &= |H\rangle \neq |R\rangle \, ,
\end{align}
\end{subequations}
and therefore, in free space, one polarization is right circularly polarized and the other linearly polarized such that the axion field impacts only one of the two beams (see Fig.~\ref{fig:opti_bench_LISA_modified}). 

Then, looking at $R_1$, we have
\begin{subequations}
\begin{align}
    \hat S^{(B)}\hat{S}^{(A)} |H\rangle &= |R\rangle \neq |H\rangle \,\\
    \hat S^{(A)}\hat{S}^{(B)} |H\rangle &= |R\rangle \neq |H\rangle  \, ,
\end{align}
\end{subequations}
i.e the light coming from the optical bench $A$ can interfere in the interferometer in optical bench $B$ and vice-versa. Indeed, the light leaving optical bench $A$ is right circularly polarized, arrives in optical bench $B$ and since a circular polarization is a linear superposition of the two linear polarizations, the vertical component is reflected towards the interferometer, as shown in Fig.~\ref{fig:opti_bench_LISA_modified}. The other beam coming from optical bench $B$ is in the "horizontal" polarization state in vacuum, arrives on the right polarizer of optical bench $A$, gets right circularly polarized and similarly as above, its vertical component gets reflected by the polarized beam splitter inside the interferometer (see Fig.~\ref{fig:opti_bench_LISA_modified}).

Finally, let us make sure that no parasitic component of light gets inside the interferometers after getting reflected at the edge of the spacecraft (requirement $R_2$). First, as we have not modified anything in the optical bench $B$, we have
 \begin{align}
     \hat S^{(B)}\hat M \hat{S}^{(B)}|H\rangle &= |H\rangle \, ,
 \end{align}
 i.e the reflected component is in the "horizontal" polarization state, and therefore transmitted by the polarized beam splitter.
 Then, it is easy to show that
 \begin{subequations}
 \begin{align}
     \hat C_R \hat M \hat C_R &= 0 \, ,
 \end{align}
such that 
\begin{align}
     \hat S^{(A)}\hat M \hat{S}^{(A)}|H\rangle &= 0 \, ,
 \end{align}
 \end{subequations}
i.e the entire power reflected at the edge of optical bench $A$ is absorbed by the right polarizer and therefore does not jeopardize the functioning of the interferometer. This can be intuitively understood by the fact that after reflection, the right circularly polarization becomes left circularly polarized and goes back inside the right polarizer, whose role is to extract and transmit only the right polarization of the input. Therefore, nothing from the left polarized parasitic light goes through.

In this section, we have shown that a simple modification of \textit{LISA} optical benches can make it sensitive to vacuum birefringence. Note that this modification leads to a decrease of power of light that arrives at each interferometer by a factor $4$. In the next section, we will derive explicitly the signal induced in \textit{LISA} TDI combinations, neglecting the reduction in power. 

Note that other non standard designs of the optical benches exist. Indeed, while we only looked at the phase velocity difference between left and right circular polarizations of light, this birefringence effect can also be seen as the oscillation of the direction of a linear polarization around its nominal direction. Then, starting with a linear polarization (say $|H\rangle$) and adding a polarized beam-splitted such that the small $|V\rangle$ component is separated from the rest, one could detect such oscillation by photon-counting. This is essentially the detection idea in e.g. \cite{Nagano19, Yu23, Ejlli22}. However, in the case of \textit{LISA}, the laser power arriving at each interferometer from the distant spacecraft is small (of order of hundreds of picoWatt \cite{Petiteau08}), and since the oscillation of direction of polarization is $\mathcal{O}(g_{a\gamma}) \ll 1$, the number of photons arriving on the detection port will be extremely suppressed.

\subsection{Expected Doppler effect in \textit{LISA}}

In this section, we derive the Doppler effect in \textit{LISA} induced by the axion-photon coupling. As a reminder, this coupling modifies the phase velocity of left and right circularly polarized light as derived in Eq.~\eqref{delta_c_axion_photon}. As derived in Section ~\ref{fibers_phase_shift} of Chapter ~\ref{optical_exp_axion}, this leads to a phase shift of a circularly polarized light travelling in vacuum for a length $L$ compared to an unaffected linearly polarized beam\footnote{In reality, this coupling affects the two linear polarization states, i.e a linear state slightly oscillates into the other one and vice-versa. Using e.g. an initial pure "horizontal" polarization state, one could see such effect using a vertical polarizer and counting the photons that go through due to the rotation of the polarization plane. However, different dispersion relation between the two linear states arises at second order in the perturbation.} of the form\footnote{The factor $2$ difference with Eq.~\eqref{eq:phase_fiber_axion_photon} comes from the fact that in Section ~\ref{fibers_phase_shift}, we assumed the amplitude of the phase shift of e.g. the right circularly polarized compared to the left one, while here we compare one circular polarization with a linear one.}
\begin{align}\label{eq:phase_shift_axion_photon_LISA}
        \Delta \phi(t) &= \frac{\sqrt{16\pi G \rho_\mathrm{DM}} E_P g_{a\gamma}}{\omega_a c} \sin\left(\frac{\omega_a L}{2c}\right)\sin\left(\omega_a\left(t-\frac{L}{2c}\right)+\Phi\right) \equiv \Delta \phi_\mathrm{single \: link}
\end{align}
where $\omega_a$ is the axion frequency, $g_{a\gamma}$ is the axion-photon coupling and $E_P$ is the reduced Planck energy. Therefore, following Fig.~\ref{fig:LISA_const_modified}, we have 
\begin{subequations}
    \begin{align}
         \Delta \phi_{12} &= \Delta \phi_{23} = \Delta \phi_{31} = 0 \, \\
         \Delta \phi_{13} &= \Delta \phi_{32} = \Delta \phi_{21} = \Delta \phi_\mathrm{single \: link} \, ,
     \end{align}
\end{subequations}
where the $ij$ subscript denote the arm pointing from spacecraft $j$ to spacecraft $i$. \newline
\noindent
\begin{figure}[h!]
\begin{minipage}{\textwidth}
    \begin{minipage}[b]{0.4\textwidth}
    \centering
    \includegraphics[width=0.8\textwidth]{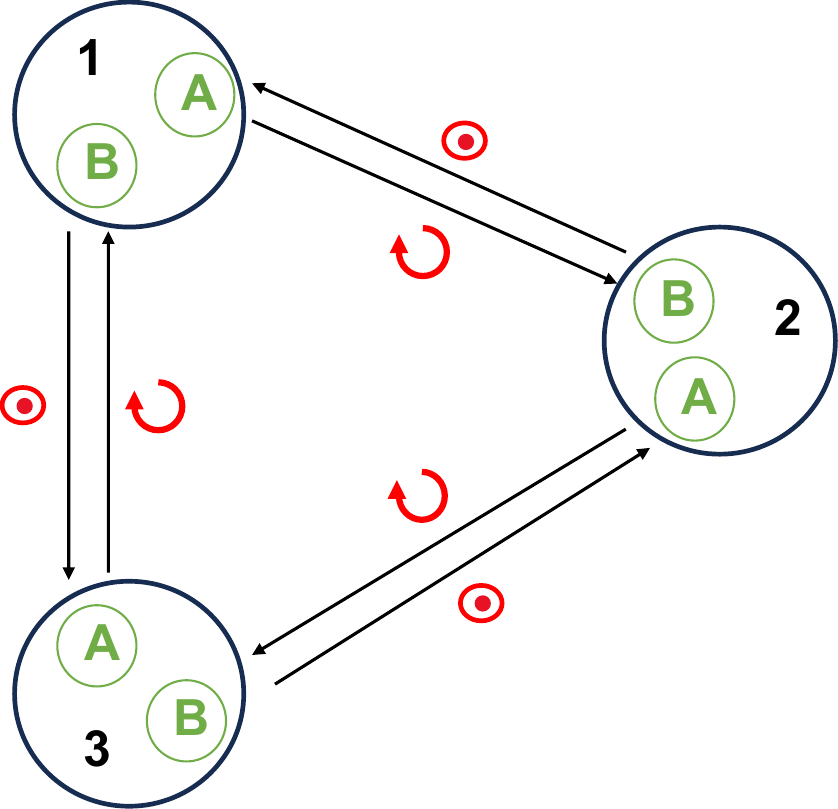}
    \caption{\textit{LISA} proposed modified constellation for the search of vacuum birefringence, as a consequence of Fig.~\ref{fig:opti_bench_LISA_modified}. In this framework, only the arms where light is circularly polarized are impacted by the axion field, and get phase shifted (see text).}
    \label{fig:LISA_const_modified}
    \end{minipage}
    \hfill
    \begin{minipage}[b]{0.55\textwidth}
    Then, using Eq.~\eqref{eq:Doppler_phase_shift_link}, we find the single-link Doppler shifts as
    \begin{subequations}\label{eq:Doppler_axion_photon}
    \begin{align}
    &y_{12} = y_{23} = y_{31}= 0 \, \\
    &y_{13}(t) = y_{32}(t) = y_{21}(t) = \frac{\sqrt{16\pi G \rho_\mathrm{DM}} E_P g_{a\gamma}}{2\pi \nu_0 c} \times \,\nonumber\\
    & \sin\left(\frac{\omega_a L}{2c}\right)\cos\left(\omega_a\left(t-\frac{L}{2c}\right)+\Phi\right) \, ,
    \end{align}
    \end{subequations}
    where $\nu_0$ is the laser frequency. Note that, throughout this derivation, we neglect the propagation phase of the field, and it should be a valid approximation, because as it was discussed in Chapter ~\ref{chap:LISA_DM}, this would add a common phase to all spacecrafts \textit{plus} a small contribution $\propto k_a L \ll 1$, for all axion frequencies of interest.
    \end{minipage}
\end{minipage}
\end{figure}

This configuration suggests that the optimized TDI combination to consider is the Sagnac $\alpha$ one, which, in the constant and equal arm length approximation, reads (using Eq.~\eqref{eq:Sagnac_combination})
\begin{align}
    \alpha_2 &= -\left(1-e^{-\frac{3i\omega_a L}{c}
    }\right)\left(y_{13}-y_{12}+e^{-\frac{i\omega_a L}{c}}\left(y_{32}-y_{23}\right)+e^{-\frac{2i\omega_a L}{c}}\left(y_{21}-y_{31}\right)\right) \, .
\end{align}
The amplitude of the signal at Fourier frequency $\omega$ is then 
\begin{align}\label{eq:signal_amp_axion_photon_Sagnac}
    |s_\mathrm{DM}(\omega)| &= \frac{2\sqrt{16\pi G \rho_\mathrm{DM}} E_P g_{a\gamma}}{2\pi \nu_0 c}\sin^2\left(\frac{3\omega L}{2c}\right)\, .
\end{align}
In Chapter ~\ref{chap:sens_experiments}, we will use this signal to infer the sensitivity of \textit{LISA} to $g_{a\gamma}$.

\section{Dichroism}

As discussed in Section ~\ref{axion_photon_coupling}, in presence of an external magnetic field, the polarisation of light parallel to the magnetic field gets absorbed due to the photon-axion conversion in this direction. This implies that for incoming linear polarized light, the polarization gets elliptical in presence of magnetic field.

The \textit{LISA} constellation orbiting around the Sun, it will be immersed in the solar magnetic field, therefore we can wonder if one could see such dichroic effects using \textit{LISA}.

As it was discussed in the previous section, each light beam is polarized linearly, and depending on the direction of propagation, the polarization is either along the $x$ or $y$ axis (see Fig.~\ref{fig:LISA_const}). Therefore, in the presence of axions and considering a constant magnetic field in the $x-y$ plane, both polarizations get affected, i.e both components in the direction of the magnetic field get a relative phase shift compared to the other component (perpendicular to the magnetic field). Therefore, comparing the two polarization states together, the phase shift is twice as large compared to Eq.~\eqref{eq:dichroism_phase_general}, i.e
\begin{align}\label{eq:phase_ellipt_LISA}
    \phi(L) &=\frac{\hbar c}{\mu_0}\frac{g^2_{a\gamma}B^2_0}{2 q^2}\left(qL-\sin(qL)\right) \, .
\end{align}
For our estimates in Chapter ~\ref{chap:sens_experiments}, we will neglect the orbit of the constellation, and we will simply make the optimistic assumption that the magnetic field orientation is the same along the whole orbit. We will also make the assumption that the axion wind is parallel to the photon propagation, i.e $q=|\vec k_a- \vec k_\gamma| = |k_a - k_\gamma|$.

\section{Conclusion}

In this chapter, we have seen how to search for the axion-photon coupling in \textit{LISA} using its effect on birefringence and dichroism. We have shown that a slight modification of the current optical benches inside \textit{LISA} spacecrafts would make \textit{LISA} an axion-photon coupling detector. In Chapter ~\ref{chap:sens_experiments}, we will show the expected sensitivity of \textit{LISA} to this coupling is very competitive with existing bounds.

In terms of dichroism, one can make estimate of the phase shift that light acquires when moving in the solar magnetic field following Eq.~\eqref{eq:phase_ellipt_LISA}, but as we shall see in Chapter ~\ref{chap:sens_experiments}, such phase shift is a constant of time, and therefore it will be invisible.

\clearpage
\pagestyle{plain}
\printbibliography[heading=none]
\clearpage
\pagestyle{fancy}

\part{Sensitivity estimates}
\chapter{\label{chap:exp_summary}Experimental considerations}

In this chapter, we will review the various experimental parameters that impact the theoretical signals presented in the previous chapters, and that for all experiments under consideration. 
In particular, for most of these experiments, we will estimate their main characteristics, their noise levels and their integration time.
Then, this will allow us to compute their expected sensitivity to a given coupling, as shown in Eq.~\eqref{coupling_constraint_general_DM}, which we will do in the next chapter (Chapter ~\ref{chap:sens_experiments}).

\section{\textit{DAMNED} and optical fibers}\label{sec:DAMNED_exp}

We first discuss the sensitivity of an already existing experiment, located at SYRTE, based on an optical cavity, \textit{DArk Matter from Non Equal Delays} (\textit{DAMNED}) \cite{Savalle21}. In the dilatonic framework developed in Section ~\ref{dilaton_pheno}, we have seen that the fine structure constant $\alpha$ and the electron mass $m_e$ oscillate at different space time locations, according to the phase value of the dilatonic field. Those oscillations imply a variation of the Bohr radius $a_0 = \hbar/ c m_e \alpha$ and a variation of length of objects accordingly. \textit{DAMNED} aims at detecting those oscillations by using an optical cavity and measuring the tiny variations of the length of the cavity through an additional phase shift of the laser. 

As an experiment aiming at measuring length variations of an optical cavity, \textit{DAMNED} could be sensitive to the vacuum birefringence effect from the axion-photon coupling on an optical cavity presented in Chapter ~\ref{optical_exp_axion}, whose corresponding signal is shown in Eq.~\eqref{eq:phase_final_DAMNED}. Its sensitivity to such coupling will be obtained in Section ~\ref{sec:DAMNED_sens_axion_photon}. 

For the experimental parameters of \textit{DAMNED}, we follow \cite{Savalle21}. An infrared laser of wavelength $\lambda = 1.542 \:  \mu$m is sent inside an optical cavity of length $\ell = 15.42$ cm\footnote{such that the condition $\ell = n \lambda, n \in \mathbb{N}$ is fulfilled.} and of finesse $\mathcal{F}=8 \times 10^5$, such that $r\sim 1- 2 \times 10^{-6}$.
We assume the experimental noise to be white between roughly 10 kHz and 1 MHz with phase noise PSD $S_\phi(f) \sim 10^{-8}$ rad$^2$/Hz \cite{Savalle19}. This implies that the experiment will be sensitive at axion masses between approximately $4\times 10^{-11}$ and $4\times 10^{-9}$ eV. The experiment lasted twelve days \cite{Savalle19}. For our estimates, we will assume that the experiment runs for 1 year, i.e $T_\mathrm{obs} \sim 3.15 \times 10^7$ s.

For optical fibers, we will use measurements on a french 86 km urban link \cite{Jiang08} whose phase noise PSD is estimated at $S_\phi \sim 10^{-4}$ rad$^2$/Hz from $1$ to $10^3$ Hz. We will also assume multiple measurements of the phase for a total integration time of one year. Similarly as \textit{DAMNED}, optical fibers could detect the axion-photon coupling through vacuum birefringence effect, as shown in Chapter ~\ref{optical_exp_axion}, with its corresponding signal in Eq.~\eqref{eq:phase_fiber_axion_photon}. Its expected constraint on such coupling will also be obtained in Section ~\ref{sec:DAMNED_sens_axion_photon}. 

\section{Rydberg atoms in a microwave cavity}

We now turn to the experiment involving Rydberg atoms inside a microwave cavity for the detection of DP-photon coupling $\chi$ presented in Chapter ~\ref{chap:Rydberg_exp_DP}. We recall that in this experiment, we aim at measuring the quadratic Stark effect, i.e the change in the transition frequency of atoms $\Delta \nu$, which is proportional to the electric field squared $E^2$. We first discuss the main sources of noise in this experiment, and then the various numerical parameters that will be used for the estimation of the sensitivity of the experiment.

\subsection{Statistical measurement noise} \label{sec:stat-noise}

The first source of noise limiting the sensitivity of the experiment is the statistical noise related to the measurement of the frequency shift experienced by Rydberg atoms under the perturbation from an external electric field. 
We denote the PSD measurement noise of $E^2$ as $S_{E^2}$, which translates into a minimal detectable power of the total field inside the cavity of 
\begin{align}\label{eq:Emin}
     E^2_\mathrm{min} = \sqrt{\frac{2S_{E^2}}{\sqrt{T_\mathrm{obs} \tau(\omega_U)}}} \, ,
\end{align}
where $T_\mathrm{obs}$ is the individual integration time and $\tau(\omega_U)$ is the coherence time of the field in the case where $T_\mathrm{obs} \gg \tau(\omega_U)$, as discussed in the end of Chapter ~\ref{chap:Rydberg_exp_DP}.

This means, that for SNR = 1, the sensitivity on $\chi$ for a mass of the DP $m_U$ corresponding to an angular frequency $\omega_U$ can be computed by equaling Eqs.~(\ref{eq:Etot2}) and \eqref{eq:Emin}, which leads to 
\begin{align}
\left[\chi(\omega_U)\right]_\mathrm{stat} =& \frac{E^2_\mathrm{min}} {S(\omega_U,\omega_A;\rho_\mathrm{DM}, X_A;L,r)} = \frac{\sqrt{2S_{E^2}}}{(T_\mathrm{obs}\tau(\omega_U))^{\frac{1}{4}}S(\omega_U,\omega_A;\rho_\mathrm{DM}, X_A;L,r)}\, . 
\label{eq:chi_stat}
\end{align}

\subsection{\label{Systematic_Rydberg}Amplitude fluctuation of the applied field}

The main systematic identified for this experiment comes from fluctuations of the amplitude of the applied electromagnetic field.  Indeed, the principle of the experiment consists in measuring oscillations of the electric field intensity at the center of the cavity. Amplitude fluctuations of the injected electromagnetic field will mimic such a signal and can jeopardize the results of the experiment. In this section we assume that the main source of fluctuations of the field inside the cavity are fluctuations of the power of the signal that is fed into the cavity, i.e. relative intensity noise (RIN) of the signal generator.

We model the amplitude of the injected electric field by including a stochastic component, i.e. replacing the previously considered constant $\vec X_A$ by
\begin{align}
    \vec X_A \rightarrow \vec X_A\left[1+\int d\omega_0 \frac{\Delta X_A(\omega_0)}{X_A}\cos(\omega_0 t + \phi_0)\right] \, .
    \label{stochastic_noise}
\end{align}
In this expression, $\Delta X_A\left(\omega_0\right)$ is a stochastic contribution modeling the spectral amplitude of the noise characterized by the RIN PSD denoted $S_\mathrm{RIN}(\omega_U)$. Since RIN is related to fluctuation of power inside the cavity, we must relate its PSD to the intensity fluctuation $\Delta X_a$. At first order in the fluctuation $\Delta X_A$, the injected power $P$ inside the cavity is
\begin{subequations}
\begin{align}
    P(\omega_0) = \left(X_A + \Delta X_A(\omega_0) \right)^2 = X^2_A + 2X_A \Delta X_A(\omega_0)  + \mathcal{O}((\Delta X_A)^2) \equiv P_0 + \Delta P(\omega_0)  \, ,
\end{align}
and the RIN PSD is defined as 
\begin{align}
    \frac{\Delta P(\omega_0) }{P_0}&\propto \sqrt{\frac{2S_\mathrm{RIN}(\omega_0)}{T_\mathrm{obs}}} \rightarrow \frac{\Delta X_A(\omega_0) }{X_A} = \sqrt{\frac{S_\mathrm{RIN}(\omega_0)}{2T_\mathrm{obs}}} \label{eq:PSD_noise}
\end{align}
Typically, the RIN of frequency generators in the microwave domain (GHz frequencies of interest here) is characterized by a flicker noise, see e.g. \cite{Rubiola}, such that we can parametrize its PSD as 
\begin{align}
    S_\mathrm{RIN}(\omega) = \frac{P_\mathrm{RIN}}{\omega} \, ,
    \label{eq:flicker_noise}
\end{align}
\end{subequations}
where $P_\mathrm{RIN}$ is dimensionless.

Let us now show how the $\Delta X_A$ fluctuations can produce an harmonic signal in $\left|\vec E\right|^2$ of angular frequency $\Delta \omega=\omega_A - \omega_U$, i.e. mimic the searched signal of Eq.~(\ref{signal_stark}). 

We will work to leading order in $\Delta X_A/X_A$ and in particular neglect terms that are $\mathcal O\left(\left(\frac{\Delta X_A}{X_A}\right)^2\right)$ and $\mathcal O\left(\frac{\Delta X_A}{X_A}\frac{X_\mathrm{DM}}{X_A}\right)$. Considering the modification of applied field amplitude Eq.~\eqref{stochastic_noise}, the fluctuation $\Delta X_A$ will only be considered at frequencies $\omega_0$ producing a noise in the electric field squared at angular frequency $\Delta \omega$. By simply computing the electric field squared component $\propto X_A \Delta X_A$, we find that these angular frequencies are $\omega_0 = \{\Delta \omega; 2\omega_A -\Delta \omega \}$. Considering the RIN as a flicker noise characterized by a PSD of the form Eq.~\eqref{eq:flicker_noise}, the fluctuation amplitude at $\omega_0=\Delta \omega$ will be multiple order of magnitudes larger than its amplitude at $\omega_0=2\omega_A-\Delta \omega \sim 2 \omega_A$ ($\omega_A \sim$ GHz, while $\Delta \omega \leq$ kHz). For this reason, in the following, we will only consider the fluctuation at frequency $|\Delta X_A(\omega_0=\Delta \omega)|$ such that the amplitude of the applied field Eq.~\eqref{stochastic_noise} becomes 
\begin{align}
    \left(\vec X_A+\Delta \vec X_A(\Delta \omega)\cos(\Delta \omega t+\phi_0)\right)\cos(\omega_A t + \phi_A) \, .
    \label{stochastic_noise_single}
\end{align}

In Appendix ~\ref{ap:RIN_amplitude}, we show how to derive the RIN contribution to the total electric field oscillating at frequency $\Delta \omega$. This reads
\begin{align}
\left[E^2(\omega_U,\omega_A)\right]_\mathrm{RIN}&= \sqrt{\frac{P_\mathrm{RIN} N(\omega_U,\omega_A)}{8 T_\mathrm{obs} \Delta \omega}} \cos(\Delta \omega t + \varphi)\, ,
\label{general_E_power_noise}
\end{align}
where $\sqrt{N(\omega_U,\omega_A)}$ corresponds to the noise amplification factor by the cavity, quadratic in $X_A$ and whose expression is also given in Appendix ~\ref{ap:RIN_amplitude}. 

The RIN contribution to $E^2$ from Eq.~(\ref{general_E_power_noise}) will limit the sensitivity of the experiment to values of $\chi$ that makes the signal from Eq.~(\ref{eq:Etot2}) larger than the systematic (i.e. larger than Eq.~(\ref{general_E_power_noise})). In other words, the RIN will limit the sensitivity of the experiment to values of $\chi$ that are larger than
\begin{equation}\label{eq:chi_RIN}
    \left[\chi(\omega_U)\right]_\mathrm{RIN} = \frac{\sqrt{\frac{P_\mathrm{RIN} N(\omega_U,\omega_A)}{8 T_\mathrm{obs} \Delta \omega}}}{S(\omega_U,\omega_A;\rho_\mathrm{DM}, X_A;L,r)} \, ,
\end{equation}
where the function $S$ is defined in Eq.~(\ref{eq:Etot2}). Note that this limit is linear in $X_A$, the amplitude of the applied electric field.

Let us discuss shot noise. For a general detector, light is received as a random flux of photons whose probability density follows the Poisson distribution \cite{Vinet}. In our case, it can also be viewed as an amplitude fluctuation of the electric field squared, as seen by Rydberg atoms, thus is analogous to the RIN. Here, we compute an estimate of the shot noise PSD and we show that it is negligible compared to the RIN PSD. Let us start with the typical number of photons received by the atoms at interval $1/f_s$ as the ratio of laser energy and photon energy
\begin{subequations}
\begin{align}
    N_\gamma &= \frac{E_\mathrm{laser}}{E_\gamma} = \frac{1}{\hbar \omega_A}\int dS_\mathrm{laser} \frac{X^2_A \epsilon_0 c}{2 f_s} \, ,
\end{align}
where we integrate over the laser waist, given by the mode radius $r_L$.
The variance on the number of photons received is also $N_\gamma$ (from Poisson distribution) and therefore the shot noise PSD normalized by the number of photons is 
\begin{align}
    S_\mathrm{SN} &= \frac{\sigma^2_{N_\gamma}}{N^2_\gamma f_s} =\frac{1}{N_\gamma f_s} = \frac{2\hbar \omega_A}{X^2_A \epsilon_0 c \pi r^2_L} 
\end{align}
\end{subequations}
Using experimental parameters that we will discuss in the following (with $r_L=0.01 $ m), one can show easily that $S_\mathrm{SN} \ll S_\mathrm{RIN}$, therefore we will neglect this contribution.

\subsection{\label{sec:opti_param}Optimum choice for applied field amplitude $X_A$}

While the signal Eq.~\eqref{eq:Etot2} is linear in applied field amplitude $X_A$, increasing $X_A$ would also increase the systematic effect $\sqrt{N(\omega_U,\omega_A)}$, as it is quadratic in $X_A$, while leaving the statistical effect unchanged. This suggests an optimum value of $X_A$ at each applied frequency, and this is what we want to find in this section.

The sensitivity of the experiment at a given angular frequency $\omega_U$ relies on the signal amplitude Eq.~\eqref{eq:Etot2} but also on the limiting noise. Combining Eq.~\eqref{eq:chi_stat} and Eq.~\eqref{eq:chi_RIN}, this is simply
\begin{align}
\left[\chi(\omega_U)\right]_\mathrm{limit} &= \frac{\sqrt{\frac{2S_{E^2}}{\sqrt{\tau(\omega_U)}}+\frac{P_\mathrm{RIN}N(\omega_U,\omega_A)}{8\Delta \omega \sqrt{T_\mathrm{obs}}}}}{T^{\frac{1}{4}}_\mathrm{obs}S(\omega_U,\omega_A;\rho_\mathrm{DM}, X_A;L,r)} \,,
    \label{eq:chi_limit}
\end{align}
where we used the usual quadratic sum of uncertainties since the two contributions are uncorrelated.

The maximum angular frequency difference $\Delta \omega$ corresponds to, from Section \ref{bandwidth_exp}, half the sampling frequency ($=\pi f_s$). To better understand the sensitivity of the experiment, we will simplify the expressions of $S(\omega_U,\omega_A;\rho_\mathrm{DM}, X_A;L,r)$ and $N(\omega_U,\omega_A)$ in Eqs.~\eqref{eq:Etot2}, ~\eqref{general_E_power_noise} and \eqref{noise_amp}, considering $r = 1 - \epsilon$, $\epsilon \ll 1$ and $\omega_U \sim \omega_A$\footnote{Even in the case where $\Delta \omega = \pi f_s$, we can assume $\omega_U \sim \omega_A$ for our estimates, as we have $f_s \lessapprox \frac{\omega_U}{Q} \approx \omega_U(1-r) \ll \omega_U$, for $f_s$ and $Q$ that we will choose.}. In that case, 
\begin{subequations}
\begin{align}
    N(\omega_U) &\approx \frac{X^4_A\epsilon^2}{\cos^4(\frac{\omega_U L}{2c})}\, , \label{syst_noise_approx}\\
    S(\omega_U) &\approx \frac{\beta c X_A \sqrt{2\mu_0 \rho_\mathrm{DM}}\sqrt{\epsilon}}{\sqrt{2}\cos^2(\frac{\omega_U L}{2c})}\left(1+\cos\left(\frac{\omega_U L}{2c}\right)\right)\,,\label{signal_approx}
\end{align}
\end{subequations}
at lowest order in $\epsilon$.
As expected, we see from \eqref{syst_noise_approx} that the two uncertainties in the numerator of \eqref{eq:chi_limit}) depend differently on $X_A$ (the statistical uncertainty is independent of $X_A$, the RIN contribution is quadratic in $X_A$), while the signal strength in the denominator is linear in $X_A$. This suggests an 'optimum' value of $X_A$ such that Eq.~\eqref{eq:chi_limit} is minimum,
\begin{subequations}
\begin{align}
    \frac{d\chi(\omega_U)}{dX_A} &= 0\,\\
   \Rightarrow X_A(\omega_U) &\approx \sqrt[4]{\frac{16S_{E^2} \Delta\omega}{ \epsilon^2 P_\mathrm{RIN}}\sqrt{\frac{T_\mathrm{obs}}{\tau(\omega_U)}}}\left|\cos\left(\frac{\omega_U L}{2c}\right)\right| \approx \sqrt[4]{\frac{16\pi f_s S_{E^2}\sqrt{\omega_U T_\mathrm{obs}}}{10^3 \epsilon^2 P_\mathrm{RIN}}}\left|\cos\left(\frac{\omega_U L}{2c}\right)\right|\, ,\label{optimum_amplitude}
\end{align}
\end{subequations}
where we used Eqs.\eqref{syst_noise_approx},\eqref{signal_approx}. We considered the maximum angular frequency shift between the DM and applied frequencies $\Delta \omega =\pi f_s$, in order to have the best sensitivity on $\chi(\omega_U)$. Note that this equation to approximate the optimum value of $X_A$ is not valid for angular frequencies $\omega_U L/c = 2\pi + 4n\pi, n \in \mathbb{N}$ and $\omega_U L/c = \pi + 2n\pi, n \in \mathbb{N}$. In the first case, the signal decreases significantly (see Eq.~\eqref{signal_approx}) and the experiment becomes insensitive to DM, while in the second case, Eq.~\eqref{optimum_amplitude} would indicate to apply $X_A=0$, which would automatically set the signal to 0, at first order in $\chi$. In this case, one must relax the constraint $\omega_U = \omega_A$ and compute the exact value of $X_A$ with the exact expressions of noise and signals, when $\omega_A$ corresponds to a mode of the cavity.

From Eq.~\eqref{optimum_amplitude}, we can express the sensitivity of the experiment $\chi(\omega_U)$ as
\begin{align}
\chi(\omega_U) &\approx \frac{\sqrt{2}\left|\cos\left(\frac{\omega_U L}{2c}\right)\right|}{1+\cos\left(\frac{\omega_U L}{2c}\right)}\left(\frac{P_\mathrm{RIN}S_{E^2}}{10^3\pi f_s }\right)^{\frac{1}{4}}\frac{\left(\omega_U T_\mathrm{obs}^{-3}\right)^{\frac{1}{8}}}{\beta c\sqrt{2\mu_0 \rho_\mathrm{DM}}}\, ,
\label{optimum_chi}
\end{align}
for $\frac{\omega_U L}{c} \neq \pi + 2\pi n$, $n \in \mathbb{N}$. 
This is a simplified expression that provides an approximate evaluation of the optimal sensitivity of the experiment.

When we will evaluate the sensitivity of the experiment in Section ~\ref{sens_Rydberg_DP}, we will assume that the applied field amplitude $X_A$ is modified each time its angular frequency $\omega_A$ is shifted, so that the condition Eq.~\eqref{optimum_amplitude} is always fulfilled. However, we will use the full expressions of $S(\omega_U,\omega_A;\rho_\mathrm{DM}, X_A;L,r)$ and $N(\omega_U,\omega_A)$ in Eqs.~\eqref{eq:Etot2} and \eqref{general_E_power_noise}.

\subsection{Experimental parameters}

Now, we discuss in details the numerical values of the various experimental parameters. They are all summarized in Table \ref{tab:Table_microwave_rydberg}.

First, we consider the random polarized case for the DP polarization direction, i.e $\beta=1/\sqrt{3}$ (see Eq.~\eqref{beta_DP}). Some already existing constraints come from experiments with this assumption \cite{SHUKET,SquAD,WISPDMX,Tokyo1,Tokyo2,FUNK,Qualiphide,FAST}.

Let us remind that the observational scheme considered here consists in electric signals of angular frequency $\omega_A$. For each injected electric field we perform a measurement of duration $T_\mathrm{obs}$ and then shift the angular frequency of the applied field by $\pi f_s$. Each measurement of duration $T_\mathrm{obs}$ provides constraints in the frequency range $\{\omega_A-\pi f_s; \omega_A+\pi f_s\}$ in steps of $2\pi/T_\mathrm{obs}$.
We will consider an individual measurement duration of $\Tobs= 60$ s. This duration is arbitrary, but we list two important considerations for this choice. We require $\Tobs$ to be large for best sensitivity (see Eq.~\eqref{eq:chi_limit}), but short enough to allow scanning a large range of DM frequencies in a reasonable amount of time ($\leq \mathcal{O}$(month)).

Now, let us discuss the sampling frequency. In \cite{Bridge2016}, the measurement of the transition probability for a given frequency lasts about 300 $\mu$s. At least three such measurements are necessary to fully determine the resonance (amplitude, width and centre frequency), which implies a maximum sampling rate of about 1~kHz, which is the maximum value we will assume. In principle the process could be faster with higher laser power and/or non-destructive techniques. In regular dispersive measurements, the photon scattering rate per atom, i.e the rate at which an atom absorbs and re-emits incident photons, is high implying that after a single detection, the atom is too hot, and no longer trapped for a second detection. In that situation, the measurement is said to be destructive. An alternative method is a non-destructive measurement, based on a low photon scattering rate, meaning that a single atom can be used for multiple measurement. This non-destructive process has already been experimentally tested, and is based on a differential dispersive measurement \cite{Vallet}. As a consequence, it is not necessary to produce new Rydberg atoms for each frequency measurement, implying that high ($>$ 1 kHz) sampling frequency is feasible. We will thus consider two scenarios for our order of magnitude estimates of the experimental sensitivity. One with a ``modest'' sampling rate of 100~Hz and a second, more optimistic one, with higher sampling at $f_s=1$~kHz. In both cases we will assume a single shot spectroscopic resolution of $\sim$1~kHz for differential polarizabilities of $\Delta\alpha/2h \approx 10^5$~Hz/(V/m)$^2$, corresponding to Rydberg states with principle quantum numbers $n\sim 60-70$ \cite{Millen2011}\footnote{While we are in the regime $T_\mathrm{obs} \gg \tau(f_U)$, note that we did not take into account the stochastic nature of the field which would impact the sensitivity of the experiment. In particular, we are sampling a signal at $10^2-10^3$ Hz, which roughly corresponds to the inverse of the coherence time of the field at GHz frequency (see Eq.~\eqref{coherence_DM}). Expressing the field as a superposition of plane waves, see e.g. \cite{foster:2018aa}, and assuming an intrinsic frequency $f_U=10^9$ Hz, a quick simulation reveals that this frequency broadening decreases the SNR by a factor $\mathcal{O}(1)$ for $f_s=10^3$ Hz, and $\mathcal{O}(10)$ for $f_s=10^2$ Hz. As expected, this suggests that one would need to increase the sampling frequency of the apparatus, in order to be in the regime where $f_s > 1/\tau(f_U)$, to avoid substantial loss in sensitivity.}.

In terms of statistical noise, in \cite{Bridge2016} the reported resolution of the spectroscopy of $n=56$ Rydberg states in Sr is of the order of a few kHz, at a maximum possible sampling rate of 1~kHz (see previous paragraph). Two different sampling rates will lead to different measurement noise PSD and directly affect the sensitivity of the experiment (cf Eq.~\eqref{optimum_chi}).
To put some numbers, assuming a single shot spectroscopic resolution of $\sim$1~kHz \cite{Bridge2016} for differential polarizabilities of $\Delta\alpha/2h \approx 10^5$~Hz/(V/m)$^2$, leads to a noise PSD of the measured electric field power of $S_{E^2}\approx 10^{-6}$~(V/m)$^4$/Hz for $f_s=100$~Hz and $S_{E^2}\approx 10^{-7}$~(V/m)$^4$/Hz for $f_s=1$~kHz. 

Regarding the systematic effect, the amplitude of the flicker noise can be considered to be $P_{\RIN} = 10^{-13}$ (based on ``off the shelf'' components studied in \cite{Rubiola} more than a decade ago) in the modest case, and we assume an improved RIN control for the optimistic case, with an amplitude of $P_{\RIN} = 10^{-15}$. The level of the systematic effect PSD is also impacted by the sampling rate, following Eq.~\eqref{general_E_power_noise}. Moreover, as derived in Section \ref{sec:opti_param}, the approximate sensitivity of the experiment Eq.~\eqref{optimum_chi} scales as $(P_\mathrm{RIN}S_{E^2}/f_s)^{1/4}$, implying that the modest and optimistic scenarii will differ in the sensitivity by a factor $\sim$10. 
The optimum value of the amplitude of the applied field $X_A$ is then derived from Eqs.~ \eqref{eq:chi_limit} (for the minimum value\footnote{The smallest value of $X_A$ corresponds to DM/applied frequencies close to odd modes, from Eq.~\eqref{optimum_amplitude}. In this regime, both approximate amplitudes of noise Eq.~\eqref{syst_noise_approx} and signal Eq.~\eqref{signal_approx} reach infinity, implying that we must consider the real expression of $\chi(\omega)$ derived in Eq.~\eqref{eq:chi_limit} and check for which value of $X_A$, the sensitivity on $\chi$ is the highest.}) and \eqref{optimum_amplitude} (for the maximum value, where we will consider DP frequencies until $\sim 2 \times 10^{10}$ Hz) and all other experimental parameters. Since it depends on the DM Compton frequency, we provide the range of optimal $X_A$, for the modest case
\begin{align}\label{eq:values_Xa_rydberg}
    18 \,\,\mathrm{V/m} &\lesssim X_A \lesssim 2.4 \times 10^5 \ \mathrm{V/m} \, .
\end{align}
It is independent of the sampling frequency since, from Eq.~\eqref{optimum_amplitude}, $X_A(\omega_A) \propto (f_s S_{E^2})^{1/4}$ and $S_{E^2}\propto f^{-1}_s$, but not of the systematic effect level $P_\mathrm{RIN}$.
\begin{table}[ht!]
\centering
\begin{tabular}{c c}
\hline \hline
Parameters & Numerical values \\
\hline \hline
Quality factor $Q$ \cite{SYRTE_fountains} & $10^4$ \\
Mirrors reflectivity $r$ & $1-2 \times 10^{-4}$ \\
Cavity length $L$ & $7.5 \mathrm{\ cm}$\\
Injected field strength $X_A(\omega)$ &  $[18,7.6\times 10^5]$ V/m \\
Sampling frequency $f_s$ & $10^2 \, ; \, 10^3$ Hz\\
Individual measurement time $T_\mathrm{obs}$ & 60 s \\
Range of $f_A=\omega_A/2\pi$ & $[0.5,20.5]$ GHz\\
Range of $\Delta \omega$ & $[2\pi/\Tobs,\pi f_s]$ rad/s \\
Statistical noise PSD $S_{E^2}$ & $ 10^{-4}/f_s \,\, (\mathrm{V/m})^4/\mathrm{Hz}$ \\
Systematic effect PSD $S_\mathrm{RIN}(\omega)$ & $10^{-13}/\omega \,; \, 10^{-15}/\omega$ \\
\hline
\end{tabular}
\caption{Assumed experimental parameters for the experiment involving Rydberg atoms in a microwave cavity for the detection of the DP-photon coupling. Two different numerical values for the same parameter means that one is used for the modest scenario while the other is used for the optimistic one (see text).}
\label{tab:Table_microwave_rydberg}
\end{table}
In Section ~\ref{sens_Rydberg_DP}, we will obtain the expected sensitivity of this experiment to the DP-photon kinetic mixing coupling $\chi$, in both modest and optimistic case, and using all the experimental parameters described above.

\section{\textit{SHUKET}}\label{sec:SHUKET_update}

We now turn to dish antenna experiments for the search of DP-photon $\chi$ coupling, as described in Chapter ~\ref{chap:SHUKET_exp}. We will focus on a practical example to illustrate how the Kirchhoff integral theorem and the overlap of modes affect the expected power received by an antenna from a spherical dish emitter. 
We consider the setup of the \textit{SHUKET} experiment \cite{SHUKET}, which has been used to search for a DP in the frequency range $5-6.8$ GHz using a spherical dish and a horn antenna for the detection. In the analysis of this experiment, following \cite{Horns}, it has been assumed that all the power emitted by the dish is received by the antenna. In this section, we will  present the experimental parameters, in particular we will assume the mean DP frequency $f_U \sim 6$ GHz, and show that the calculations done in Chapter ~\ref{chap:SHUKET_exp} are valid in that case. Then, in the next chapter, Section ~\ref{sec:SHUKET_sens}, we will estimate the power received by the antenna and estimate its implication on a more realistic sensitivity of the experiment on $\chi$. 

The spherical dish used in \cite{SHUKET} has a curvature radius of $R=32$ m and an area of $A_\mathrm{dish} = 1.2$ m$^2$, implying a radius of $r \approx 0.618$ m (see Fig.~\ref{fig:SHUKET_Kirchhoff}). Then, $R \gg r$ (or equivalently the dish-fictional plane distance $a\sim 6 \times 10^{-3}$ m, as shown in Fig.~\ref{fig:Dish_plane_Vinet}, in the case of \textit{SHUKET} is way smaller than $r$, i.e. $a\ll r$) which ensures that the low curvature approximation is valid. 
The expression of the electric field emitted by the dish at its surface is given by Eq.~(\ref{eq:E_out}) where $ \vec Y_{\parallel,D}$ has the form 
\begin{align}
 \vec Y_{\parallel,D}(\theta,\phi) = \vec Y - \left(\vec Y\cdot \hat e_r\right)\hat e_r \approx \begin{pmatrix}
      Y_x \,\\
      Y_y \,\\
      0
  \end{pmatrix} + \mathcal{O}(\theta)\, \label{field_polarization}, 
\end{align}
with $\theta, \phi$ corresponding to a spherical coordinate system centered in dish's curvature center and where we have used the notation $\vec Y=(Y_x,Y_y,Y_z)$ in the cartesian coordinate system depicted in Fig.~\ref{fig:SHUKET_Kirchhoff} and where we used the low curvature approximation at the last step\footnote{The relative error induced by the neglect of $\mathcal{O}(\theta)$ is $\sim$ 1\%.}. This means that at leading order in $\theta$, the polarization of the emitted electric field does not depend on the dish coordinate, and is only polarized in the $x-y$ plane.

For this configuration, the power emitted by the dish is independent of the mass of the DP and is given by Eq.~\eqref{power_emit_dish}\footnote{Note there is a factor 2 discrepancy with what \cite{SHUKET} indicated, due to a calculation error.} 
\begin{align}\label{power_SHUKET}
    P^\mathrm{SHUKET}_\mathrm{dish} &= 1.73 \times 10^{-20} \left(\frac{\chi}{10^{-12}}\right)^2 \mathrm{ \ W} \, .
\end{align}
In Appendix ~\ref{ap:SHUKET_Vinet_approx}, we show how with these experimental parameters, the procedure using the thin optical element approximation, described in Chapter ~\ref{chap:SHUKET_exp} to compute the electric field from the dish to the fictional plane is valid for \textit{SHUKET}. Indeed, we find that this approximation leads to a relative error on the electric field of order $10^{-4}$. Additionally, the calculation in Appendix ~\ref{ap:SHUKET_Vinet_approx} shows that the small curvature approximation and Eq.
~\eqref{field_polarization} are valid in our regime.

In the \textit{SHUKET} experiment, the detection is made by the \textit{Schwarzbeck BBHA-9120-D} antenna, which is sensitive to electric field frequencies from 0.8 to 18 GHz. The datasheet \cite{datasheet_ant} contains several pieces of information that will be necessary for our calculations on the expected amount of power received by the antenna wires, namely the gain and the antenna factor of the antenna as function of those various field frequencies. These quantities have been measured experimentally by the manufacturer.
The antenna consists in a rectangular surface of area $S_\mathrm{phys} = 0.25 \times 0.142$ m$^2$,  with $A=0.25$ m the long length and $B=0.142$ m the small length. This implies that for the fundamental mode of the antenna at frequency $\omega_c$, the effective size of the antenna corresponds to its physical size (in other words $A_\mathrm{eff}(\omega_c)=A$ and $B_\mathrm{eff}(\omega_c)=B$) and therefore the ratio between the long and small sides of the antenna is $R_\mathrm{AB}(\omega_c)=A/B\approx 1.76$.  We will make the assumption that the ratio $R_{AB}$ is independent of the frequency,  i.e $R_\mathrm{AB}(\omega_c) \equiv R_\mathrm{AB} = A_\mathrm{eff}(\omega)/B_\mathrm{eff}(\omega)$, for any frequency $\omega>\omega_c$, such that the effective surface of the antenna at frequency $\omega$ is simply $S_\mathrm{eff}(\omega)= A_\mathrm{eff}(\omega)B_\mathrm{eff}(\omega)=A_\mathrm{eff}^2(\omega)/R_\mathrm{AB}$. Then, using Eq.~\eqref{eq:effective_surface}, the effective long and small lengths of the antenna at the frequency $\omega_U$ are simply given by
\begin{align}
\frac{A_\mathrm{eff}^2(\omega_U)}{R_\mathrm{AB}} &= \frac{e_r\pi G(\omega_U)c^2}{\omega^2} \: \: \: ; \: \: \: B_\mathrm{eff}(\omega_U)=\frac{A_\mathrm{eff}(\omega_U)}{R_\mathrm{AB}} \, .
\end{align}
These two dimensions define the effective surface area over which we need to integrate Eq.~(\ref{eq:overlap_int}) to estimate the output of the antenna.

Therefore, we will use the electric field at the surface of the fictional plane Eq.~\eqref{E_field_complete} to compute the field power received by the antenna. This will be done in details in Chapter ~\ref{chap:sens_experiments} on sensitivities.

\section{\label{sec:exp_param_MICROSCOPE}\textit{MICROSCOPE}}

We now discuss the experimental parameters of \textit{MICROSCOPE} for the detection of ULDM candidates that produce a violation of the UFF, as discussed in Chapter ~\ref{chap:Classical_tests_UFF}.
As mentioned in Section ~\ref{sec:exp_test_UFF}, \textit{MICROSCOPE} is a space mission that was launched to test the UFF in space \cite{Microscope17} with a final result on the $\eta$ parameter Eq.~\eqref{eq:eta_MICROSCOPE}.
The full \textit{MICROSCOPE} dataset used to constrain the UFF is made of 17 sessions \cite{Rodrigues22}. The noise PSD of the differential acceleration of the session 404 is given by \cite{Pihan}
\begin{align}
    S_a(f) =  2.2\times 10^{-24} f^{-1} + 2.3 \times 10^{-17} f^{4} \: \mathrm{\left(m/s^2\right)^2/Hz}\, ,
    \label{PSD_MICROSCOPE}
\end{align}
for frequencies between $10^{-5}$ Hz and 0.3 Hz. The $f^{-1}$ slope noise comes from thermal effects of the gold wire connecting the test masses to the cage, while the high frequency noise in $f^4$ is the second derivative of the position measurement white noise \cite{Microscope17}.

The UFF signal is modulated by spinning the satellite at a frequency chosen to minimize the noise, $f_\mathrm{EP} \sim 3$ mHz. Note that $f_\mathrm{EP}$ is in reality a linear combination of the spin frequency $f_\mathrm{spin}$ and orbital frequency $f_\mathrm{orb}$, with $f_\mathrm{spin} \gg f_\mathrm{orb} \sim 1.5 \times 10^{-4}$ Hz \cite{Microscope17} for session 404. Three different spin frequencies exist depending on the session \cite{Berge23}, such that one should take both contributions into account for the determination of $f_\mathrm{EP}$ for other sessions. The bucket frequency of this acceleration noise corresponds approximately to $f_\mathrm{bucket} \sim 30$ mHz \cite{Pihan,Microscope17}. 

To take into account the contribution of the full experiment, i.e the 17 different sessions, we make the hypothesis that all of them have the same level of acceleration noise Eq.~\eqref{PSD_MICROSCOPE} while we consider the orbital period to be constant $T_\mathrm{single \: orbit} \approx 5946$ s \cite{Rodrigues22}. This hypothesis is sufficient for the rough sensitivity analysis we do here, but will have to be revisited when doing a complete \textit{MICROSCOPE} data analysis in search for DM candidates.

Using the total number of orbits of the experiment, i.e $N_\mathrm{orbits}= 1362$ \cite{Rodrigues22}, we find that the total integration time is $T_\mathrm{int} = T_\mathrm{single \: orbit} \times N_\mathrm{orbits} \equiv 8.1 \times 10^6$ s ($\approx 94$ days).

As an experiment measuring the differential acceleration between two test masses, the corresponding theoretical signal is given by Eq.~\eqref{delta_a_UFF_osc} projected onto the sensitive axis of the experiment (see Eq.~\eqref{Eotvos_param}).
In Appendix ~\ref{ap:dot_product_AI}, we compute the projection of the galactic velocity onto the sensitive axis of \textit{MICROSCOPE}, and we find that $\hat e_v \cdot \hat e_\mathrm{meas} \approx 0.85 \cos(\omega_\mathrm{spin} t + \psi)$, where $\omega_\mathrm{spin} = 2\pi f_\mathrm{spin}$ and $\psi$ is an irrelevant phase.

\section{Atom interferometers}

Similarly as for \textit{MICROSCOPE}, we have shown in Chapter ~\ref{chap:AI} that the sensitivity of some AI experiments to ULDM fields are proportional to the dot product between the laser pulse direction and the galactic velocity (see e.g. Eqs.~\eqref{eq:MZ_delta_phase_shift} and \eqref{eq:phase_new_prop}). In Appendix ~\ref{ap:dot_product_AI}, we derive the various projections of the galactic velocity onto the laser pulses directions of those experiments, which depend on their location on Earth, but also on their integration time. The AI experiments under consideration are located at Stanford, USA, Fermilab, USA and Oxford, UK. In Appendix ~\ref{ap:dot_product_AI}, we find $\hat e_v \cdot \hat e_\mathrm{kick}\Big|_\mathrm{Stanford} \approx 0.43$, $\hat e_v \cdot \hat e_\mathrm{kick}\Big|_\mathrm{Oxford} \approx 0.99$ and $\hat e_v \cdot \hat e_\mathrm{kick}\Big|_\mathrm{Fermilab} \approx 1.00$.

\subsection{Stanford Tower}\label{sec:stanford}

The most stringent constraint on the E\"otv\"os parameter $\eta$ obtained from atom interferometry experiment is achieved using a Bragg atom interferometry experiment in the Stanford Tower. In this experiment, the relative acceleration of freely falling clouds of two isotopes of Rubidium ($^{85}$Rb and $^{87}$Rb) is measured \cite{Stanford20}. The differential interferometric phase measurement between these two inteferometers leads to a constraint on $\eta$ given by \cite{Stanford20}
\begin{subequations}
\begin{align} \label{equ:eta_Stanford}
\eta = (1.6 \pm 5.2)\times 10^{-12}\, .
\end{align}
Both atoms are launched upwards inside the Stanford Tower for an interferometric sequence of total duration $2T=1.91$ s, corresponding to the total time for the atoms to fall back \cite{Stanford20}.

Double diffraction interferometry is performed using two lasers  both resonant with the $|5^{2}S_{1/2}\rangle \rightarrow | 5^{2}P_{3/2}\rangle$ $^{87}$Rb transition, i.e. their wavevectors are $k_1 \approx k_2 \approx 2\pi/\lambda$, with $\lambda = 780$ nm. Three different momentum splittings have been used: $\{4\hbar k, 8\hbar k, 12 \hbar k\}$. All are in agreement with no EP violation \cite{Stanford20}. In the following, we will consider the highest momentum transfer since this will enhance the signal searched for, which corresponds to using an effective wavevector $k_\mathrm{eff}=24\pi/\lambda$ in Eq.~\eqref{eq:MZ_delta_phase_shift}.

The resolution per shot on the differential acceleration is $\sigma_{\Delta a} = 1.4 \times10^{-11}g$ \cite{Stanford20}, with $g = 9.81$ m/$s^2$ the Earth gravitational acceleration on ground. We assume white noise (see below) with a corresponding acceleration noise PSD
\begin{align}
    S_{\Delta a} = \frac{2\sigma^2_{\Delta a}}{f_s}
    \approx \left(7.7 \times 10^{-11}\ g\right)^2 \ \mathrm{Hz}^{-1} \, ,
\end{align}
with $f_s$ the sampling frequency of the experiment, defined as $f_s = 1/T_\mathrm{cycle}$ where $T_\mathrm{cycle} = 15$ s is the duration of one interferometric sequence, including atom preparation, launch and free fall \cite{Stanford20}. 
The raw measurement in this experiment is actually a differential phase shift $\Delta \Phi$ which is related to the differential acceleration between the atoms $\Delta a$ through $\Delta \Phi = k_\mathrm{eff}T^2 \Delta a$ \cite{Storey}. Therefore, we can infer the PSD of the phase shift using
\begin{align}\label{phase_acceleration_noise}
    \sqrt{S_\Phi} = k_\mathrm{eff}T^2\sqrt{S_{\Delta a}} \, .
\end{align}

The final uncertainty of this differential acceleration measurement is mostly limited by electromagnetic effects, coming from the Bragg lasers and non homogeneous magnetic field \cite{Stanford20}. Knowing the final uncertainty of the differential measurement \eqref{equ:eta_Stanford} we derive the total ``effective'' experiment time $T_\mathrm{int}$ under our white noise assumption i.e. assuming that individual experimental cycles are uncorrelated. Then, the number of cycles $N$ can be derived as
\begin{align}
    N = \frac{1}{g^2}\left(\frac{\sigma_{\Delta a}}{\sigma_\eta}\right)^2 \approx 77 \, ,
\end{align}
and
\begin{align}
    T_\mathrm{int} = N\times T_\mathrm{cycle} \approx 1148 \mathrm{\: s} \, .
\end{align}
\end{subequations}

As a Bragg differential AI, the Stanford experiment is sensitive to ULDM fields that produce oscillations of rest mass and transition frequency of atoms, whose corresponding signal is presented in Eq.~\eqref{eq:MZ_delta_phase_shift}.

\subsection{\textit{AION-10}}\label{sec:AION-10}

The Atom Interferometer Observatory and Network (\textit{AION}) is an experimental program to search for ULDM and gravitational waves in the $10^{-1}-10$ Hz range using atom interferometry \cite{Badurina20}. \textit{AION-10} is a 10 meter-long single-photon atom gradiometer instrument which will use $^{87}\mathrm{Sr}$ atoms and that will be built in Oxford \cite{Badurina22}. Contrary to the other experiments previously introduced, \textit{AION-10} will operate in a distant future, assuming a much better control on noise than current experiments.
As a gradiometer, \textit{AION-10} will be sensitive to the signal presented in Eq.~\eqref{eq:phase_gradio}.
Following \cite{Badurina22}, we will use the following experimental parameters ultimately envisaged for \textit{AION-10}:
\begin{subequations}\label{AION_exp_param}
\begin{align}
    T_\mathrm{int} &= 10^8 \mathrm{\: s} \, , \\
    T &= 0.74 \mathrm{\: s} \, , \\
    S^\mathrm{AION}_\Phi(f) &= 10^{-8} \mathrm{\: rad^2/Hz} \, , \\
    n &= 1000 \, , \\
    \omega_\mathrm{Sr} &= 2.697 \times 10^{15} \mathrm{\: rad/s} \, , \\
    \Delta r &= 4.86 \mathrm{\: m} \, , \\
    L &= 10 \mathrm{\: m} \, ,
\end{align}
\end{subequations}
respectively the interrogation time, the free evolution time, the gradiometer phase noise PSD, the number of LMT kicks, the optical transition frequency used, the gradiometer separation, and the total size of the baseline. Note that, while a $10$ m facility is used, which would imply a free fall time of $2T = \sqrt{8L/g} \sim 2.86$ s, i.e $T\sim1.43$ s, \cite{Badurina22} computed the optimum free evolution time as function of the other experimental parameters, in particular considering a non-zero separation between the two interferometers $\Delta r \neq 0$, reason for the difference.

\subsection{\textit{SPID}}\label{sec:SPID}

We now focus on the \textit{SPID} variation, which can also probe oscillations of rest mass and transition frequency of atoms through the signal presented in Eq.~\eqref{eq:phase_new_prop} (Section ~\ref{sec:SPID_prop_pres}). 

We consider first the experimental parameters from one running mode of \textit{MAGIS-100}. They are listed in \cite{Abe21}, see their Fig. 3. Their noise levels are much lower than current experiments, similar to \textit{AION-10}. \textit{MAGIS-100} will use two isotopes of Strontium ($^{87}$Sr and $^{88}$Sr) in a 100-meter baseline. The transition under consideration in \cite{Abe21} is $|5^1 S_0\rangle \rightarrow |5^3 P_1\rangle$ with frequency $\omega_\mathrm{Sr} = 2.73 \times 10^{15}$ rad/s. In terms of acceleration noise level and order of LMT, we consider the upgraded parameters, i.e 
\begin{subequations}
    \begin{align}
        \sqrt{S^\mathrm{MAGIS}_a}(f) &= 6 \times 10^{-17} \mathrm{g/}\sqrt{\mathrm{Hz}} \, , \\
        n &= 1000 \, .
    \end{align}
In addition, for a $L =100$ m high tower, the free-fall time of atoms is given by $2T=\sqrt{8L/g}$, which implies $T\sim 4.5$ s. Following Eq.~\eqref{phase_acceleration_noise}, the phase noise PSD is 
\begin{align}
    S^\mathrm{MAGIS}_\Phi(f) &= \left(\frac{n \omega_\mathrm{Sr} T^2}{c}\right)^2 S_a(f) \sim 10^{-8} \: \mathrm{rad^2/Hz}  \, ,
\end{align}
\end{subequations}
i.e a similar phase noise PSD as \textit{AION-10}. Finally, the full integration time of the experiment corresponds to 1 year of observation, $T_\mathrm{int} \approx 3.16 \times 10^7$ s.

Another reason why we consider the \textit{SPID} setup is to directly compare the expected sensitivity on axion and dilaton couplings of two experiments with the exact same experimental parameters but operating a gradiometer on one hand and the \textit{SPID} variation on the other hand. For this matter, we will compare the sensitivity of the current version of \textit{AION-10}, i.e a gradiometer, with a \textit{SPID} setup using the same noise levels. In this case, in addition to the phase noise, we also include the EOM noise (which we show to be negligible in the following), which is used to shift the frequency of the laser to account for the isotope shift ($\sim 1$~GHz) as discussed in Sec.~\ref{sec:SPID_prop_pres} and depicted in Fig.~\ref{fig:isotope_AI}. An alternative scheme would be to phase-lock two lasers with a $\sim 1$~GHz frequency offset. In either case, this setup generates an additional frequency fluctuation $\sigma_f(t)$ in the second beam, which comes from the EOM internal noise and from the phase noise of the GHz reference frequency. Whilst the raw EOM phase noise PSD can be approximated as $S^\mathrm{GHz}_\Phi(f) \approx 10^{-13} \left(f/\mathrm{Hz}\right)^{-2}\, \mathrm{rad^2/Hz}$ \cite{Barke15,Hartnett2012,Xie2017a}, it enters only differentially in the \textit{SPID} setup at times separated typically by $L/c$. The resulting PSD is altered by a factor $2\left(1-\cos\left(2 \pi f L/c\right)\right) \approx (2\pi f L/c)^2$, since $2\pi f L/c \ll 1$ for all frequencies $f$ of interest. Since this noise is uncorrelated with the gradiometer phase noise, the total phase noise PSD of the \textit{SPID} setup is simply
\begin{align}
S^\mathrm{SPID}_\Phi(f)&=S^\mathrm{GHz}_\Phi(f)\left(\frac{2\pi f L}{c}\right)^2 + S^\mathrm{AION}_\Phi(f) \approx S^\mathrm{AION}_\Phi(f) \, ,
\end{align}
For the remaining experimental parameters, such as the total time of experiment, the flight time of atoms and the number of LMT kicks, we will assume the same values as \textit{AION-10}, see Eq.~\eqref{AION_exp_param}.
We will also consider the use of four different pairs of isotopes of alkaline-Earth neutral atoms commonly used as optical clocks (see Table ~\ref{tab:alk_isotope_freq}) and we will assess the impact of the choice of the atoms on the ULDM sensitivity.

\section{\textit{LISA}}

In Chapter ~\ref{chap:LISA_DM}, we already discussed most of the experimental parameters affecting \textit{LISA} sensitivity to DM fields. \textit{LISA} one-link phase noise PSD is given in Eq.~\eqref{eq:noise_PSD_LISA}. 
The sensitivity of a given TDI $O$ combination to a coupling $\varepsilon$ at frequency $f$ is given by \cite{Babak21}
\begin{align}\label{eq:sens_LISA_O_TDI}
    S_O(f) &= \frac{N_O(f)}{\mathcal{T}^2_O(f)} \, .
\end{align}
The amplitude of the transfer functions of TDI $X_2$ combination is (at leading order in $v_\mathrm{DM}/c$ and using Eq.~\eqref{eq:amp_TF_X2})
\begin{subequations}
\begin{align}
    \left|\mathcal{T}^\mathrm{DM}_X(f)\right| &\approx 16\sin \left(\frac{2\pi f L}{c}\right)\sin \left(\frac{4\pi f L}{c}\right) \left(\hat n_{23}\cdot \hat e_v\right)\sin^2\left(\frac{\pi f L}{c}\right) \, .
\end{align}
From this, one can compute both $A,E$ transfer functions (using Eq.~\eqref{eq:AET_TDI})
\begin{align}
\left|\mathcal{T}^\mathrm{DM}_A(f)\right| &\approx \frac{16}{\sqrt{2}}\left|\sin \left(\frac{2\pi f L}{c}\right)\sin \left(\frac{4\pi f L}{c}\right) \left(\left(\hat n_{12} - \hat n_{23}\right)\cdot \hat e_v\right)\right|\sin^2\left(\frac{\pi f L}{c}\right) \, \\
\left|\mathcal{T}^\mathrm{DM}_E(f)\right| &\approx \frac{48}{\sqrt{6}}\left|\sin \left(\frac{2\pi f L}{c}\right)\sin \left(\frac{4\pi f L}{c}\right) \left(\hat n_{13} \cdot \hat e_v\right)\right|\sin^2\left(\frac{\pi f L}{c}\right) \, \\
\left|\mathcal{T}^\mathrm{DM}_{AE}(f)\right| &\approx \frac{16}{\sqrt{2}}\left|\sin\left(\frac{2\pi f L}{c}\right)\sin \left(\frac{4\pi f L}{c}\right)\right| \sqrt{\left(\left(\hat n_{12} - \hat n_{23}\right)\cdot \hat e_v\right)^2 + \left(\sqrt{3} \hat n_{13} \cdot \hat e_v\right)^2}\sin^2\left(\frac{\pi f L}{c}\right) \label{eq:joint_AE_TF}\, ,
\end{align}
where at the last line, we define the $A,E$ joint transfer function, which we used for our analysis of the realistic limit on sensitivity of \textit{LISA}, in Section ~\ref{sec:delta_varepsilon}.
\end{subequations}
The noise transfer functions of the $A,E$ combinations are defined in Eq.~\eqref{eq:TDI_A_noise_PSD}, and the Sagnac $\alpha$ noise transfer function is \cite{Hartwig23}\footnote{In \cite{Hartwig23}, only the first generation noise transfer functions are presented, so in order to compute the second generation noise transfer function for the Sagnac $\alpha$ combination, one needs to add a factor $(2\sin(3\omega L/2c))^2$, using Eq.~\eqref{eq:Sagnac_combination}. In addition, the modification of the optical benches of \textit{LISA} that we propose to be able to be sensitive to vacuum birefringence implies a decrease in the amplitude of light (or equivalently in the number of photons $N_\gamma$) that reaches each interferometer by a factor $2$ because light gets partially absorbed by the right polarizer, see Fig.~\ref{fig:opti_bench_LISA_modified}. Since the OMS noise includes a contribution from shot noise whose PSD $\propto 1/N_\gamma$, we have 
conservatively added a factor $2$ to the whole OMS contribution.}
\begin{align}\label{eq:TDI_Sagnac_noise}
    N_\alpha(f) &= 4\sin^2\left(\frac{3 \pi f L}{c}\right)\left(12S_\mathrm{oms}(f) +4\left(3-2\cos\left(\frac{2\pi f L}{c}\right)-\cos\left(\frac{6\pi f L}{c}\right)\right)S_\mathrm{acc}(f)\right) \, .
\end{align}

In the following, we will make our estimates assuming a fixed arm length $L = 2.5 \times 10^9$ m, and two different integration times. For the estimates related to Chapter ~\ref{chap:LISA_DM} on searches for time oscillating rest mass, we will be using the parameters used in the simulation : we will use one year of observation time $T_\mathrm{obs}\sim 3.15 \times 10^7$ s, and we will compute the sensitivity of the $A,E$ TDI combinations together, i.e we will be using the joint transfer function Eq.~\eqref{eq:joint_AE_TF} (the noise transfer functions are the same for both combinations, see Eq.~\eqref{eq:TDI_A_noise_PSD}). For the estimates related to Chapter ~\ref{chap:axion_photon_LISA} on searches for vacuum birefringence, we will be assuming four years of integration time, $T_\mathrm{obs} \sim 1.26 \times 10^8$ s, which is the expected mission duration \cite{Colpi24} and we will compute the sensitivity of the Sagnac combination, as it is the optimized combination to use for this matter, as we discussed it in Chapter ~\ref{chap:axion_photon_LISA}.

\chapter{\label{chap:sens_experiments}Sensitivity results on ultralight dark matter couplings}

In this chapter, we will present the main results of this thesis, i.e the sensitivity estimates of the experiments presented in Chapters ~\ref{optical_exp_axion}, \ref{chap:Rydberg_exp_DP}, \ref{chap:SHUKET_exp}, \ref{chap:Classical_tests_UFF}, \ref{chap:AI}, \ref{chap:LISA_DM} and \ref{chap:axion_photon_LISA} with the experimental parameters and noises presented in Chapter ~\ref{chap:exp_summary}.

\section{Dilaton-SM couplings $d_i's$}

\begin{figure}[h!]
    \centering
    \includegraphics[width=\textwidth]{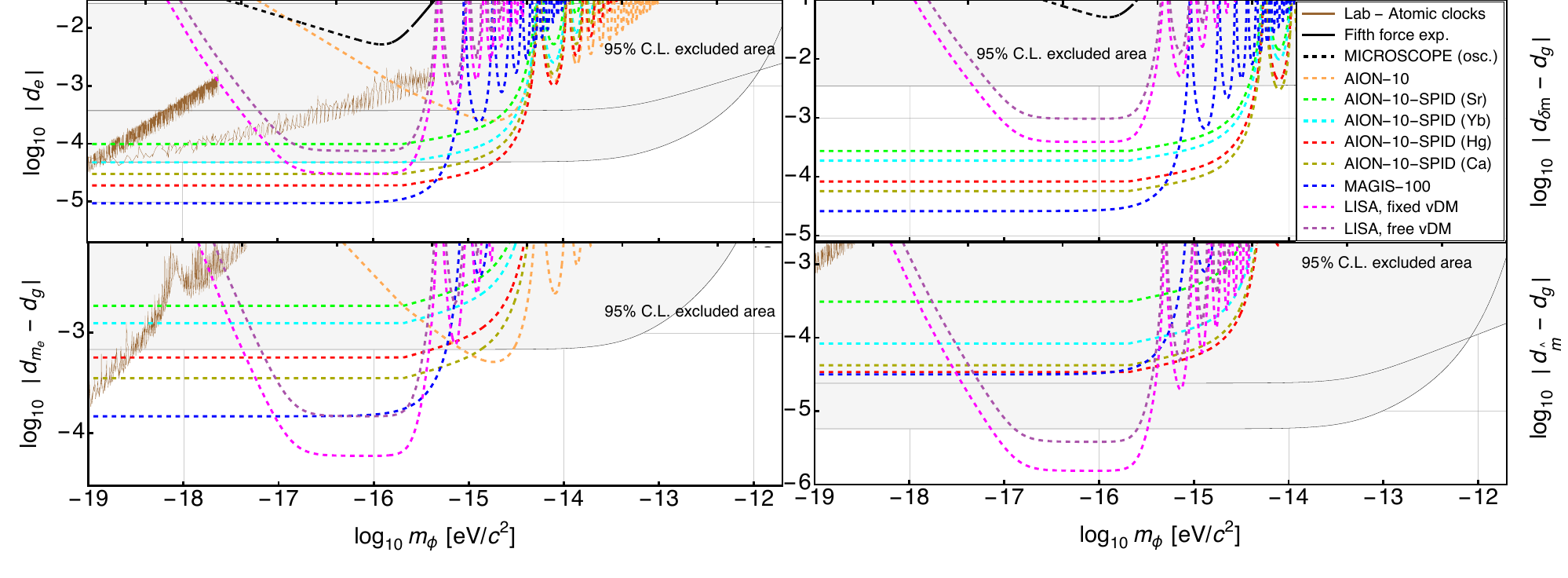}
    \caption{Current constraints on all the dilatonic couplings of interest in this thesis : $d_e$ (top left), $d_{m_e}-d_g$ (bottom left), $d_{\delta\hat m}-d_g$ (top right), $d_{\hat m}-d_g$ (bottom right) from \cite{Microscope22,Wagner12,Hees16,Kennedy20}, with 95\% confidence level (shown in light grey background). All these existing constraints are shown in solid lines. The expected sensitivity of all the experiments considered in this thesis are shown in dashed lines (with 68\% detection threshold). The sensitivity of \textit{MICROSCOPE} is shown in black, and denoted "MICROSCOPE (osc.)". The expected sensitivity of \textit{AION-10} is shown in orange dashed line while the expected sensitivity of \textit{AION-10} using the \textit{SPID} variation, noted "AION-10-SPID" are respectively shown in green, light blue, red and gold dashed lines, depending on the isotope pair used. The expected sensitivity of \textit{MAGIS-100} is shown in dark blue. Finally, the expected sensitivity of \textit{LISA} is shown in pink and purple respectively when the DM velocity is assumed as a fixed and free parameter (see text).}
    \label{fig:full_dilaton_constraints}
\end{figure}

In Fig.~\ref{fig:full_dilaton_constraints}, we show the existing most stringent laboratory constraints on the dilatonic couplings : Torsion balances \cite{Wagner12} and \textit{MICROSCOPE}\footnote{The best \textit{MICROSCOPE} sensitivity comes from the static term of the field, i.e when considering a fifth force generated by Earth on the test masses, while we only focused on the oscillatory term in this thesis.} \cite{Microscope22}, which are denoted as "Fifth force exp.", and hyperfine and optical clocks \cite{Hees16, Kennedy20} denoted as "Lab-Atomic clocks".

\subsection{\textit{MICROSCOPE}}

\textit{MICROSCOPE}'s sensitivity to the oscillating dilaton field is given by the projection of the differential acceleration Eq.~\eqref{eq:delta_a_dil_lab_frame} to the sensitive axis of measurement, i.e
\begin{align}\label{eq:delta_a_MICRO}
    |\Delta \vec a(t)| &= \frac{\sqrt{16 \pi G\rho_\mathrm{DM}}v_\mathrm{DM}}{c}\left|\left([Q^\mathrm{Pt}_M]_d-[Q^\mathrm{Ti}_M]_d\right)\hat e_v \cdot \hat e_\mathrm{meas.}(t)\Big|_\mathrm{\mu SCOPE}\right| \, .
\end{align}
The dot product is computed in Appendix ~\ref{ap:dot_product_AI} and the mass charges $[Q^\mathrm{Pt}_M]_d, [Q^\mathrm{Ti}_M]_d$ are defined in Eq.~\eqref{partial_dil_mass_charge} and the various partial dilatonic charges are shown in Table ~\ref{dilatonic_charge_table}. The sensitivity is then computed using Eq.~\eqref{coupling_constraint_general_DM} and the experimental parameters of Section ~\ref{sec:exp_param_MICROSCOPE}.
The associated sensitivity of \textit{MICROSCOPE} to the various dilatonic couplings is shown in black dotted line in Fig.~\ref{fig:full_dilaton_constraints} (denoted "MICROSCOPE (osc.)").
As mentioned above, this sensitivity is not competitive with \textit{MICROSCOPE}'s ability to detect a fifth force induced by the dilaton field, as it is shown in black full line in Fig.~\ref{fig:full_dilaton_constraints}.

\subsection{\label{sec:AI_dilaton_SM}Atom interferometers}

For the various atom interferometric schemes, we use Eqs.~\eqref{eq:MZ_delta_phase_shift}, \eqref{eq:phase_gradio} and \eqref{eq:phase_new_prop} and we express respectively the amplitude of the phase shift between atomic species A and B in differential two-photon transition AI, gradiometers and the \textit{SPID} setup, with respectively \textit{AION-10} and \textit{MAGIS-100} experimental parameters. The experimental parameters of these various experiments are discussed in Sections ~\ref{sec:stanford}, \ref{sec:AION-10}, \ref{sec:SPID}. Then, the different phase shifts read 
\begin{subequations}
\begin{align}
&\Delta \Phi^\mathrm{Stanford}_\mathrm{^{87}Rb,^{85}Rb} \approx \frac{4\sqrt{16 \pi G \rho_\mathrm{DM}}v_\mathrm{DM}k_\mathrm{eff}}{\omega^2_\phi c}\Big|\left([Q^\mathrm{^{87}Rb}_M]_d-[Q^\mathrm{^{85}Rb}_M]_d\right) \hat e_v \cdot \hat e_\mathrm{kick}\Big|_\mathrm{Stanford}\Big|\sin^2\left(\frac{\omega_\phi T}{2}\right) \, , \\
&\Delta \Phi^\mathrm{AION-10}_\mathrm{^{87}Sr} \approx \frac{4 n \omega^0_\mathrm{^{87}Sr} \Delta r \sqrt{16 \pi G \rho_\mathrm{DM}}[Q^\mathrm{^{87}Sr}_\omega]_d}{\omega_\phi c}\sin^2\left(\frac{\omega_\phi T}{2}\right) \, , \\
&\Delta \Phi^\mathrm{MAGIS-100}_\mathrm{^{88}Sr,^{87}Sr} \approx \frac{4 n \omega_0 \sqrt{16 \pi G\rho_\mathrm{DM}}v_\mathrm{DM}}{\omega^2_\phi c^2}\left|\left([Q^\mathrm{^{88}Sr}_M]_d-[Q^\mathrm{^{87}Sr}_M]_d\right)\hat e_v \cdot \hat e_\mathrm{kick}\Big|_\mathrm{Fermilab}\right|\sin^2\left(\frac{\omega_\phi T}{2}\right) \, , \\
&\Delta \Phi^\mathrm{AION-10-SPID}_\mathrm{AB} \approx \frac{4 n \omega_0 \sqrt{16 \pi G\rho_\mathrm{DM}}v_\mathrm{DM}}{\omega^2_\phi c^2}\left|\left([Q^A_M]_d-[Q^B_M]_d\right)\hat e_v \cdot \hat e_\mathrm{kick}\Big|_\mathrm{Oxford}\right|\sin^2\left(\frac{\omega_\phi T}{2}\right) \, ,
\label{Amp_phase_shift_dilaton}
\end{align}
\end{subequations}
where the $[Q]_d$ charges contain partial dilatonic mass and/or frequency charges defined in Eqs.~\eqref{partial_dil_mass_charge} and \eqref{partial_dil_freq_charge}. Note that for the \textit{SPID} variation, we let the atomic species $A,B$ free because we will derive expected sensitivities for various isotope pairs.

In Fig.~\ref{fig:full_dilaton_constraints}, the sensitivity of the Stanford tower experiment to the various dilatonic couplings is not visible, because it is less sensitive than the best existing constraints (for example, it reaches $d_e \sim 2 \times 10^3$ at best, around $m_\phi c^2 \sim 10^{-16}$ eV, almost 8 orders of magnitude above the constraint from \textit{MICROSCOPE}).
As it was discussed in \cite{Badurina22}, \textit{AION-10} would improve the current constraint on the $d_{m_e}-d_g$ coupling, over a mass range approximately between $7 \times 10^{-16}$ eV and $4\times 10^{-15}$ eV.

With the same experimental parameters as \textit{AION-10}, operating the \textit{SPID} setup (denoted "\textit{AION-10-SPID}" in Fig.~\ref{fig:full_dilaton_constraints}) would improve the current constraints on various dilatonic couplings. By respectively using Ca and Hg isotopes pairs, the constraint on $d_e$ would be improved by a factor $\sim 2$ and 3 respectively, over a mass range covering four orders of magnitude (approximately from $10^{-19}$ eV to $10^{-15}$ eV), compared to \textit{MICROSCOPE}, the current best constraint in this mass range. Regarding the $d_{m_e}-d_g$ coupling, the use of Ca and Hg isotopes respectively would improve the best constraints, by a factor 2.5 and 1.5 respectively over the same mass range. Finally, all pairs of isotopes presented would improve the current best constraint on $d_{\delta m}-d_g$ by one order of magnitude, depending on the isotope pair used, over a mass range covering more than 4 orders of magnitude  (from $10^{-19}$ eV to $3 \times 10^{-15}$ eV). 

One can notice also that operating both gradiometers and \textit{SPID} at \textit{AION-10} would give complementary sensitivities for the search of the $d_{m_e}-d_g$ coupling of the dilaton. Indeed, while the gradiometer would improve the current constrain in the $\sim 7 \times 10^{-16}-3\times 10^{-15}$ eV/$c^2$ mass range, the sensitivity of the \textit{SPID} setup at lower mass is better, as described in the last paragraph. In addition, \textit{SPID} would be more sensitive than \textit{AION-10} to the $d_e, d_{\hat m}-d_g$ and $d_{\delta m}-d_g$ couplings in the full range of masses of interest (i.e lower than $10^{-14}$ eV/$c^2$). 

Regarding the sensitivity of \textit{MAGIS-100} operating the \textit{SPID} setup, one can notice that it would overall give the best constraint at low masses.
One can notice that \textit{MAGIS-100} would reach better sensitivity on most couplings, compared to the \textit{AION-10-SPID} variation. The reason is that for low masses, such that $\omega_\phi T \ll 1$, the signal Eq.~\eqref{Amp_phase_shift_dilaton} is quadratic in the free fall time $T$, and \textit{MAGIS-100} using a much longer baseline than \textit{AION-10}, the free fall time is roughly $6$ times longer for \textit{MAGIS-100} (as mentioned previously, the free fall time in the case of \textit{AION-10} is smaller than what the facility size allows), resulting in increased signal. However, note that we made the calculations assuming three years of integration time for \textit{AION-10} (from \cite{Badurina22}) and the \textit{SPID} variation, but only one year for \textit{MAGIS-100} (from \cite{Abe21}). This is why the difference in sensitivity is roughly one order of magnitude between the two dual-Sr configurations, and not more. Note that compared to the sensitivity curves on dilaton couplings presented in \cite{Abe21}, but using the gradiometer setup of \textit{MAGIS-100}, the curve presented in this paper would constrain a larger DM mass range. Indeed, while the former has a peak sensitivity around a mass $10^{-15}$ eV and quickly loses sensitivity for lower masses, the \textit{SPID}-like setup of \textit{MAGIS-100} would have a constant sensitivity of the same order of magnitude for 3 orders of magnitude of mass ($10^{-19}-10^{-16}$ eV).

\subsection{\textit{LISA}}\label{sec:LISA_sens_dilaton}

In Chapter ~\ref{chap:LISA_DM}, we have seen that despite the fact that \textit{LISA} can probe dilaton-SM couplings through oscillation of the rest mass of the freely falling test masses, the DM velocity cannot be constrained accurately and this leads to large uncertainty on such couplings. Now, we might wonder what would be the effect of such uncertainty in the regions of the parameter space where \textit{LISA} can probe the smallest couplings, i.e when SNR=1.

Following Eq.~\eqref{eq:geo_factor_DM_GW}, the sensitivity of the combination $A,E$ on the coupling $\varepsilon$ without correlation with other parameters is (at SNR=1)
\begin{align}\label{eq:LISA_varepsilon_general}
    \varepsilon_\mathrm{min}(f) &= \frac{1.5 \times 2\pi f c^2}{\sqrt{16\pi G \rho_\mathrm{DM}}v_\mathrm{DM}}\sqrt{\frac{S_{A,E}(f)}{T_\mathrm{obs}}} \: \: \: \mathrm{or} \: \: \: \varepsilon_\mathrm{min}(f) = \frac{2\pi f c^2}{\sqrt{16\pi G \rho_\mathrm{DM}}v_\mathrm{DM}}\sqrt{\frac{S_{A,E}(f)}{\sqrt{T_\mathrm{obs}\tau(f)}}}\, ,
\end{align}
depending on if $T_\mathrm{obs}$ is smaller or larger than $\tau(f)$ (see Eq.~\eqref{coupling_constraint_general_DM}). Now, we estimate the sensitivity of the experiment, taking into account the correlation between the coupling and the DM velocity (see Section ~\ref{sec:delta_varepsilon}). When SNR=1, the likelihood Eq.~\eqref{eq:likelihood_DM} reduces to\footnote{At frequencies where $T_\mathrm{obs} < \tau(f)$, the noise in Eq.~\eqref{eq:likelihood_DM} scales as $\sqrt{N_A/\sqrt{T_\mathrm{obs}\tau(f)}}$, and we will obtain the same result.}
\begin{align}\label{eq:likelihood_DM_SNR1}
    \log \mathcal{L}(f,\varepsilon^M) &= -\frac{1}{2}\left(1+\left(\frac{v^M_\mathrm{DM} \varepsilon^M}{v^D_\mathrm{DM} \varepsilon_\mathrm{min}(f)}\right)^2-2\frac{v^M_\mathrm{DM}\varepsilon^M}{v^D_\mathrm{DM} \varepsilon_\mathrm{min}(f)} \cos\left(\frac{2\pi f |\vec x_\mathrm{AU}|\cos(\beta)}{c^2}\left(v^D_\mathrm{DM}-v^M_\mathrm{DM}\right)\right)\right) \, ,
\end{align}
where the parameters with superscript $D$ mean true values injected in the simulation, and the ones with superscript $M$ mean model values. Similarly as what was done in Section ~\ref{sec:delta_varepsilon}, we will now compute numerically the standard deviation on the posterior distribution of $\varepsilon^M$, in order to have a new sensitivity estimate of \textit{LISA}. 
\begin{figure}
\centering
\begin{minipage}{.5\textwidth}
  \centering
  \includegraphics[width=\textwidth]{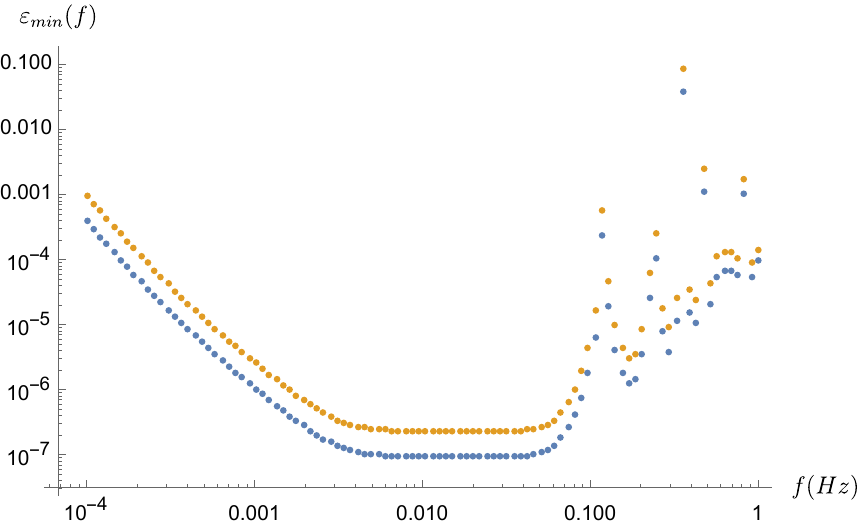}
\end{minipage}%
\begin{minipage}{.5\textwidth}
  \centering
  \includegraphics[width=\textwidth]{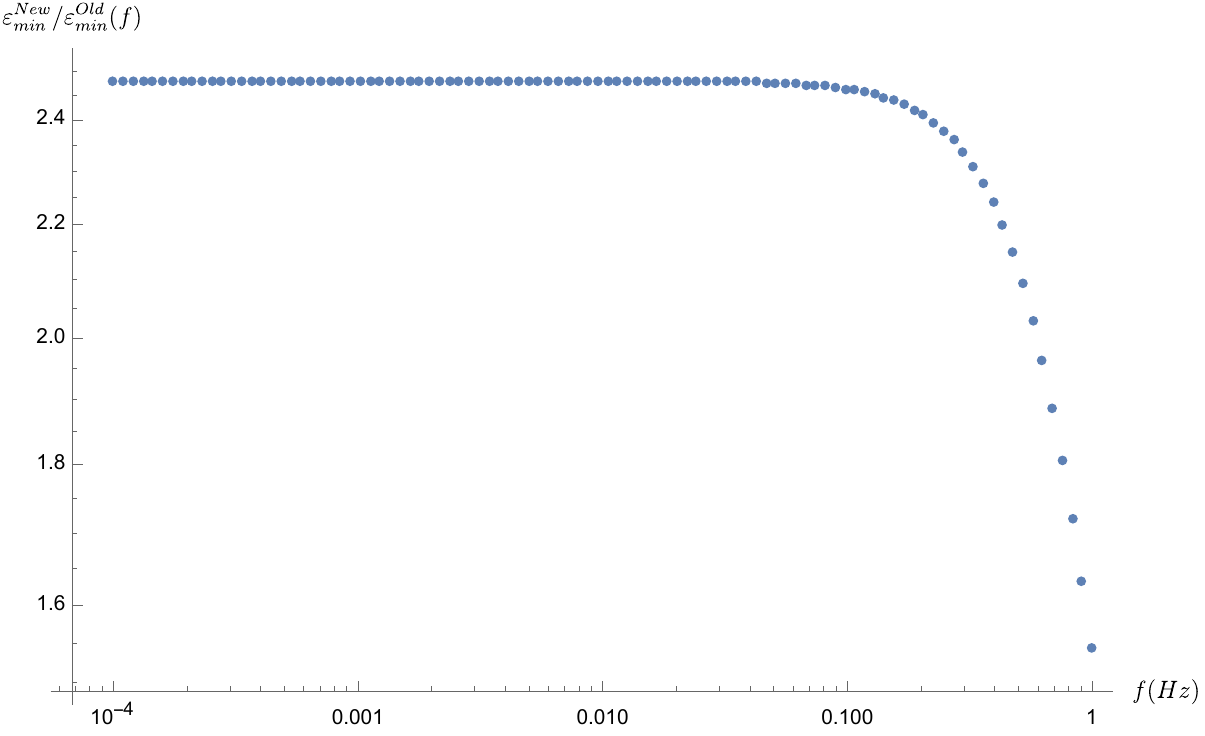}
\end{minipage}
\caption{On the left, we show the sensitivity of \textit{LISA} to the general coupling $\varepsilon_\mathrm{min}$ when the velocity is assumed fixed (blue curve) and when it is a free parameter (orange curve). One can notice a deterioration of the sensitivity when the velocity is a free parameter of the Bayesian analysis. On the right, we show the ratio of the orange to the blue curves. At low frequency, the ratio is $\sim 2.5$, and when the frequency becomes $f>10^{-1}$ Hz, this ratio starts decreasing because the argument of the cosine in Eq.~\eqref{eq:likelihood_DM_SNR1} becomes important (see text.).}
\label{fig:varepsilon_LISA_new}
\end{figure}

In the following, we consider the same specific values of $\beta^D, v^D_\mathrm{DM}$ as in Chapter \ref{chap:LISA_DM} (as a reminder, we assume the mean values of velocity and sky localization from the velocity distribution Eq.~\eqref{DM_vel_distrib}, i.e $(v^D_\mathrm{DM},\beta^D) = (10^{-3}c,1.046 \: \mathrm{rad})$).
In order to infer $\sigma_{\varepsilon_\mathrm{min}}(f)$, the uncertainty on the posterior distribution of $\varepsilon_\mathrm{min}$ at a given frequency $f$, we numerically integrate the marginalized likelihood Eq.~\eqref{eq:likelihood_DM_SNR1} over $v^M_\mathrm{DM}$, following Eq.~\eqref{eq:posterior_varepsilon} and using an uniform prior on $\varepsilon^M$ with (arbitrary) bounds $[0,100 \: \varepsilon_\mathrm{min}(f)]$. This gives us the posterior distribution $\mathcal{P}(\varepsilon^M(f))$ which we again integrate over $\varepsilon^M$ (as in Eq.\eqref{eq:mean_std_varepsilon}) to obtain $\sigma_{\varepsilon_\mathrm{min}}(f)$.

In the left panel of Fig.~\ref{fig:varepsilon_LISA_new}, we show sensitivities of \textit{LISA} to $\varepsilon$ when $v_\mathrm{DM}$ is fixed, i.e from Eq.~\eqref{eq:LISA_varepsilon_general} and when $v_\mathrm{DM}$ is a free parameter, i.e from the width of the posterior distribution Eq.~\eqref{eq:posterior_varepsilon} (but with the likelihood given in Eq.~\eqref{eq:likelihood_DM_SNR1}). One can notice that at low masses, letting $v_\mathrm{DM}$ free decreases the sensitivity by a factor $\sim 2.5$ compared to the fixed $v_\mathrm{DM}$ case (right panel of Fig.~\ref{fig:varepsilon_LISA_new}). This difference is entirely determined by the cosine term in Eq.~\eqref{eq:likelihood_DM_SNR1}. Indeed, as mentioned in Chapter ~\ref{chap:LISA_DM}, it is the accuracy on the determination of the velocity $v^M_\mathrm{DM}$ which impacts the width of the coupling $\varepsilon^M$ posterior distribution. In particular, in order to maximize the likelihood, one needs to get the $\log \mathcal{L}$ Eq.~\eqref{eq:likelihood_DM_SNR1} as close to $0$ as possible, i.e one needs the cosine term to be close to $1$. This means that the argument of the cosine
\begin{subequations}
\begin{align}
    \frac{2\pi f |\vec x_\mathrm{AU}|\cos(\beta)}{c^2}\left(v^D_\mathrm{DM}-v^M_\mathrm{DM}\right) \sim \frac{10^3 \Delta v}{c} \left(\frac{f}{1 \: \mathrm{Hz}}\right)  \, ,
\end{align}
must be close to $0$, where $\Delta v = v^D_\mathrm{DM}-v^M_\mathrm{DM}$. We can consider that the accuracy on the velocity parameter starts to increase when $\Delta v \leq \sigma_v$, where $\sigma_v$ is the width of the velocity prior distribution. This implies that the accuracy on the coupling starts to increase for frequencies
\begin{align}
\frac{10^3 \sigma_v f}{c} \slashed{\ll} 2\pi &\Rightarrow f \slashed{\ll} 10 \: \mathrm{Hz} \, ,
\end{align}
i.e when $f\gtrapprox 0.1$ Hz. This is what is shown on the right panel of Fig.~\ref{fig:varepsilon_LISA_new}. 
\end{subequations}

Since $\varepsilon = [Q^\mathrm{TM}_M]_d$, and $[Q^\mathrm{TM}_M]_d = \sum_i [Q^\mathrm{TM}_{M,i}]_d \: d_i/\sqrt{2}$ (from Eq.~\eqref{partial_dil_mass_charge}), the sensitivity of \textit{LISA} to the coupling $d_i$ reads
\begin{align}
    d_i &= \frac{\sqrt{2}\sigma_{\varepsilon_\mathrm{min}}}{[Q^\mathrm{TM}_{M,i}]_d} \, .
\end{align}

Using Eq.~\eqref{eq:sens_LISA_O_TDI} together with Eqs.~\eqref{eq:TDI_A_noise_PSD} and \eqref{eq:joint_AE_TF}, we can now derive the sensitivity curves. In Fig.~\ref{fig:full_dilaton_constraints}, we show respectively in pink and purple the sensitivity of \textit{LISA} to the different dilatonic $d_i$ couplings when the velocity is assumed as a fixed or free parameter. Similarly as for the $\varepsilon$ coupling shown in Fig.~\ref{fig:varepsilon_LISA_new}, there is an approximate factor $\sim 2.5$ difference. One can notice that \textit{LISA} will be competitive with the other futuristic experiments for all dilatonic couplings. In particular on the $d_{m_e}-d_g, d_{\hat m} -d_g$ couplings, it will reach unconstrained regions of the parameter space, improving respectively fifth force constraints by a factor $\sim 10, \sim 5$ over one order of magnitude of mass ($\sim 10^{-17} \: \mathrm{eV}-10^{-16} \: \mathrm{eV}$), in the case of fixed velocity. The reason it is much more sensitive to this coupling than AI experiments comes from the fact that AI experiments mainly use isotopes $I1,I2$, while \textit{LISA} makes use of the finite light travel time with a single atomic species. Isotopes have very close quark contribution to their rest mass, and therefore the differential mass charge $\Delta [Q_{M,\hat m}]_d = [Q^\mathrm{I1}_{M,\hat m}]_d - [Q^\mathrm{I2}_{M,\hat m}]_d $ is very small, and the corresponding sensitivity decreases significantly.

\section{Axion-gluon coupling $f^{-1}_a$}

\begin{figure}[h!]
    \centering
    \includegraphics[width=\textwidth]{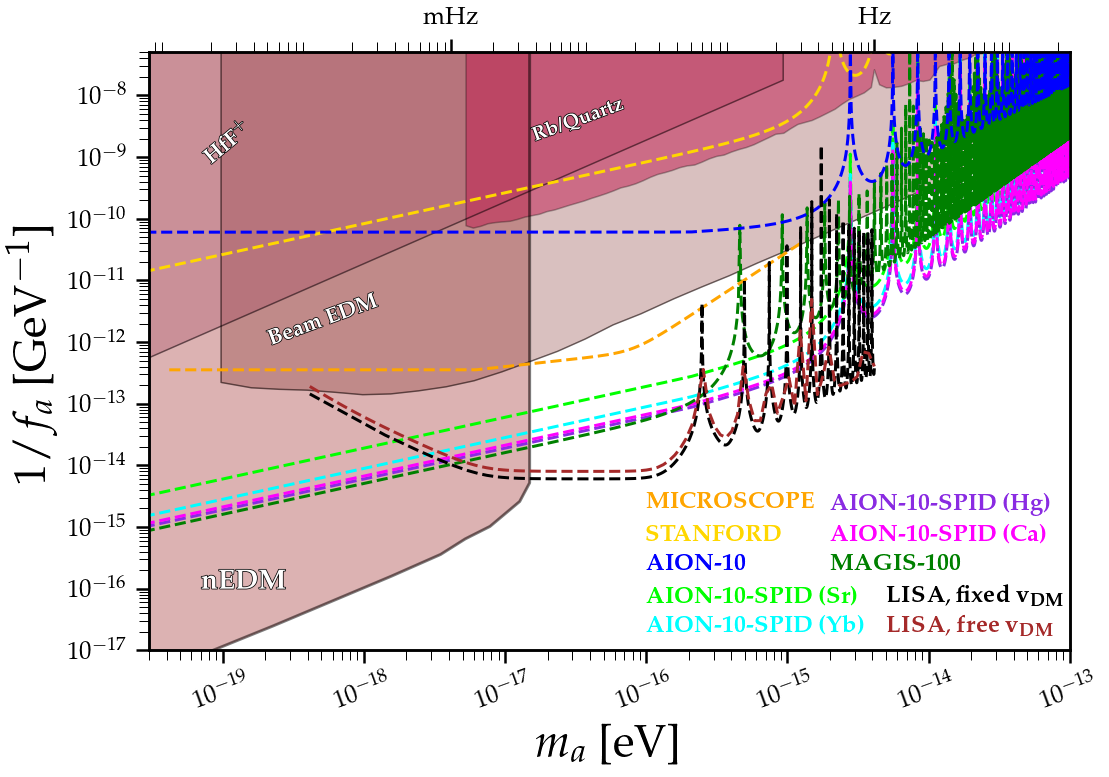}
    \caption{Current lab constraints on $1/f_a$ axion coupling \cite{Beam_EDM,HfF,Rb_quartz,nEDM} (from \cite{AxionLimits}) (the constraint from \cite{Rb_quartz} has been rescaled for consistent value of local DM energy density) (shown in solid lines). The new sensitivity estimates resulting from this thesis are shown in dashed lines, i.e from Stanford, \textit{AION-10} and \textit{MAGIS-100} atom interferometry experiments, EP classical test from \textit{MICROSCOPE} and \textit{LISA}. The expected sensitivity of the \textit{AION-10-SPID}-like experiment is shown in four different colors, each using four different pairs of isotopes, denoted "AION-10-SPID".}
    \label{fig:constraints_fa}
\end{figure}
The current best laboratory constraints on the axion-gluon coupling are from \cite{Beam_EDM,HfF,Rb_quartz,nEDM} and are shown in solid red in Fig.~\ref{fig:constraints_fa}. 

\subsection{\label{sec:MICRO_sens_axion_gluon}\textit{MICROSCOPE}}

Similarly as in the previous section, we can express the amplitude of the differential acceleration between two test-masses A and B as (using Eqs.~\eqref{eq:delta_a_axion_lab_frame})
\begin{align}
|\Delta \vec a(t)| &= \frac{16 \pi G \rho_\mathrm{DM} v_\mathrm{DM} E^2_P}{f^2_a\omega_a c^2}\left|\left([Q^\mathrm{Pt}_M]_a-[Q^\mathrm{Ti}_M]_a\right)\hat e_v \cdot \hat e_\mathrm{meas.}(t)\Big|_\mathrm{\mu SCOPE}\right| \, .
\label{Amp_delta_acc_axion}
\end{align}
The axionic mass charges $[Q^\mathrm{Pt}_M]_a$ and $[Q^\mathrm{Ti}_M]_a$ can be found in Table ~\ref{axionic_charge_table}. Similarly as in the previous section, the sensitivity to the axion-gluon coupling is then computed using Eq.~\eqref{coupling_constraint_general_DM} and the experimental parameters of Section ~\ref{sec:exp_param_MICROSCOPE}.
In Fig.~\ref{fig:constraints_fa}, the expected sensitivity of \textit{MICROSCOPE} is shown by the orange full line. This curve presents two breaking point frequencies. The first one $f\sim 1$ mHz corresponds to half of $f_\mathrm{spin} \sim 3$ mHz. As discussed in Appendix ~\ref{ap:dot_product_AI}, we consider two different frequency regimes, depending on whether the signal frequency is higher or lower than $f_\mathrm{spin}$. In the axion case, the signal Eq.~\eqref{eq:delta_a_axion_lab_frame} oscillates at twice the axion field frequency, therefore the breaking point is $f_\mathrm{spin}/2$. The second breaking point arising at $f \sim 15$ mHz corresponds to half of the bucket frequency of the acceleration noise PSD, for the same reason as above. Note that there is no breaking point frequency associated with the coherence time of the field because it would arise at a frequency larger than the bandwidth of the noise PSD, at around 500 mHz.
As it can be seen from this curve, a complete re-analysis of \textit{MICROSCOPE}'s data would enable to constrain a new region of the parameter space, over approximately two orders of magnitude in mass, compared to existing laboratory experiments. 

\subsection{Atom interferometers}

For the atom interferometry experiments, we do the same procedure as in Section ~\ref{sec:AI_dilaton_SM} and we find 
\begin{subequations}\label{Amp_phase_shift_axion}
\begin{align}
&\Delta \Phi^\mathrm{Stanford}_\mathrm{^{87}Rb,^{85}Rb} \approx \frac{16 \pi G  \rho_\mathrm{DM} E^2_P v_\mathrm{DM}k_\mathrm{eff}}{\omega^3_a f^2_a c^2}\Big|\left([Q^\mathrm{^{87}Rb}_M]_a-[Q^\mathrm{^{85}Rb}_M]_a\right)\hat e_v \cdot \hat e_\mathrm{kick}\Big|_\mathrm{Stanford}\Big|\sin^2(\omega_aT)  \label{Amp_phase_shift_MZ_axion} \, , \\
&\Delta \Phi^\mathrm{AION-10}_\mathrm{^{87}Sr} \approx \frac{32 \pi G \rho_\mathrm{DM} E^2_P n \omega^0_A \Delta r [Q^\mathrm{^{87}Sr}_\omega]_a}{\omega^2_a f^2_a c^3}\sin^2(\omega_aT) \, , \\
&\Delta \Phi^\mathrm{MAGIS-100}_\mathrm{^{88}Sr,^{87}Sr} \approx \frac{16 \pi G \rho_\mathrm{DM} E^2_P n \omega_0 v_\mathrm{DM}}{\omega^3_a f^2_a c^3}\left|\left([Q^\mathrm{^{88}Sr}_M]_a-[Q^\mathrm{^{87}Sr}_M]_a\right)\hat e_v \cdot \hat e_\mathrm{kick}\Big|_\mathrm{Fermilab}\right|\sin^2(\omega_aT) \, , \\
&\Delta \Phi^\mathrm{AION-10-SPID}_\mathrm{AB} \approx \frac{16 \pi G \rho_\mathrm{DM}E^2_P n \omega_0 v_\mathrm{DM}}{\omega^3_a f^2_a c^3}\left|\left([Q^A_M]_a-[Q^B_M]_a\right)\hat e_v \cdot \hat e_\mathrm{kick}\Big|_\mathrm{Oxford}\right|\sin^2(\omega_aT) \, .
\end{align}
\end{subequations}
In Fig.~\ref{fig:constraints_fa}, we present the sensitivity of the Stanford Tower \cite{Stanford20}, \textit{AION-10} experiment \cite{Badurina22}, \textit{MAGIS-100} experiment \cite{Abe21} and the \textit{SPID} AI setup with \textit{AION-10} experimental parameters, denoted ``AION-10-SPID''.

One can notice that \textit{MICROSCOPE} is approximately two to three orders of magnitude more sensitive than Stanford. This is consistent considering that the signal is quadratic in $1/f_a$, that \textit{MICROSCOPE} constrains the E\"otv\"os parameter $\eta$ three orders of magnitude better than the Stanford experiment and that the difference in axionic mass charges of species used in the experiments is 2 orders of magnitude larger for MICROSCOPE. 

Contrary to the gradiometer setup of \textit{AION-10} which is almost insensitive to the axion-gluon coupling (at leading order, the sensitivity is proportional to the axionic frequency charge of the optical transition of Sr, which is 0), the \textit{SPID} variation or \textit{AION-10} would have the largest sensitivity to this coupling compared to existing laboratory experiments. Such an experiment would improve the current lab constraint \cite{Beam_EDM} by 2 orders of magnitude over a  mass range of 4 orders of magnitude  ($10^{-17}-10^{-13}$ eV). In this mass range, the \textit{MAGIS-100} experiment, which uses the same setup has also a very interesting sensitivity on the coupling, which is comparable to the one of \textit{AION-10} in the \textit{SPID} setup. 

\subsection{\textit{LISA}}

Similarly as for the dilatonic couplings, \textit{LISA} will be able to probe the axion-gluon coupling through Doppler effects Eq.~\eqref{eq:geo_factor_DM_GW}. In the same way as what we did for the dilatonic couplings in Section ~\ref{sec:LISA_sens_dilaton}, we will derive two sensitivity curves. The first one is computed assuming no correlations between parameters, where the constraint on the dimensionless coupling $E_P/f_a$ is given by
\begin{align}
    \left(\frac{E_P}{f_a}\right)^2_\mathrm{min}(f) &= \frac{1.5 \times (2\pi f)^2 c^3}{8\pi G \rho_\mathrm{DM} v_\mathrm{DM}}\sqrt{\frac{S_{A,E}(2f)}{T_\mathrm{obs}}} \: \: \: \mathrm{or} \: \: \: \left(\frac{E_P}{f_a}\right)^2_\mathrm{min}(f) = \frac{(2\pi f)^2 c^3}{8\pi G \rho_\mathrm{DM}v_\mathrm{DM}}\sqrt{\frac{S_{A,E}(2f)}{\sqrt{T_\mathrm{obs}\tau(f)}}}\, ,
\end{align}
taking into account the fact that the signal has a different normalization amplitude (from Eq.~\eqref{eq:Doppler_pseudoscalar_DM}) and that it oscillates at twice the axion frequency.
The second sensitivity curve will take into account the correlations between parameters. The signal being quadratic in $1/f_a$, the lack of precision on the galactic velocity induces an uncertainty 
\begin{subequations}
\begin{align}
    \sigma_\varepsilon \equiv \sigma_{E^2_P/f^2_a}(f) \, .
\end{align}
Note that here, we choose a dimensionless random variable parameter $(E_P/f_a)^2$, for consistency with our analysis involving the dimensionless parameter $\varepsilon$.
As we are interested in the uncertainty on $E_P/f_a$, we make the assumption that, at the frequency $f$, the random variable $E_P/f_a(f)$ follows a normal distribution with zero mean\footnote{Here, we are interested in computing an upper limit on the coupling in case of no detection, therefore, we can assume that the mean of the distribution of such coupling is $0$.} and deviation $\sigma_{E_P/f_a}(f)$ such that we have 
\begin{align}
    \sigma_{E^2_P/f^2_a}(f) &= \sqrt{2}\sigma^2_{E_P/f_a}(f) = \sqrt{2}E^2_P\sigma^2_{1/f_a}(f)\, \\
    \rightarrow \sigma_{1/f_a}(f) &= \frac{1}{2^{1/4}}\frac{\sqrt{\sigma_{E^2_P/f^2_a}(f)}}{E_P} \label{eq:delta_fa_LISA} \, ,
\end{align}
\end{subequations}
where we used the fact that $E_P$ is constant. To get a curve of $\sigma_{1/f_a}(f)$ as function of the frequency, we perform the same iteration as with the dilatonic couplings, i.e we compute the posterior distribution of the $\varepsilon_\mathrm{min} = (E_P/f_a)^2_\mathrm{min}$ variable, where $\varepsilon_\mathrm{min}$ is computed following \eqref{eq:LISA_varepsilon_general}.

In Fig.~\ref{fig:constraints_fa}, we show the sensitivity of \textit{LISA} to $1/f_a$ coupling when the DM velocity is fixed at $v_\mathrm{DM}$ (\textit{LISA}, fixed $v_\mathrm{DM}$) and when the velocity is taken as a free parameter, i.e for $\sigma_{1/f_a}$ as given by Eq.~\eqref{eq:delta_fa_LISA} (\textit{LISA}, free $v_\mathrm{DM}$). One can notice a slighter difference in sensitivity between the two curves, compared to the dilaton case, and this is because the axion signal is quadratic in the coupling while the dilaton signal is linear. At axion masses between $10^{-17}$ and $10^{-14}$ eV, \textit{LISA} would be the most sensitive experiment to the axion-gluon coupling, in particular around $10^{-16}$ eV, where the sensitivity is improved by a factor $\sim$ 3 compared to the best AI experiments (\textit{MAGIS} and \textit{AION-10-SPID}).

\section{Axion-photon coupling $g_{a\gamma}$}

\begin{figure}[h!]
    \centering
    \includegraphics[width=0.85\textwidth]{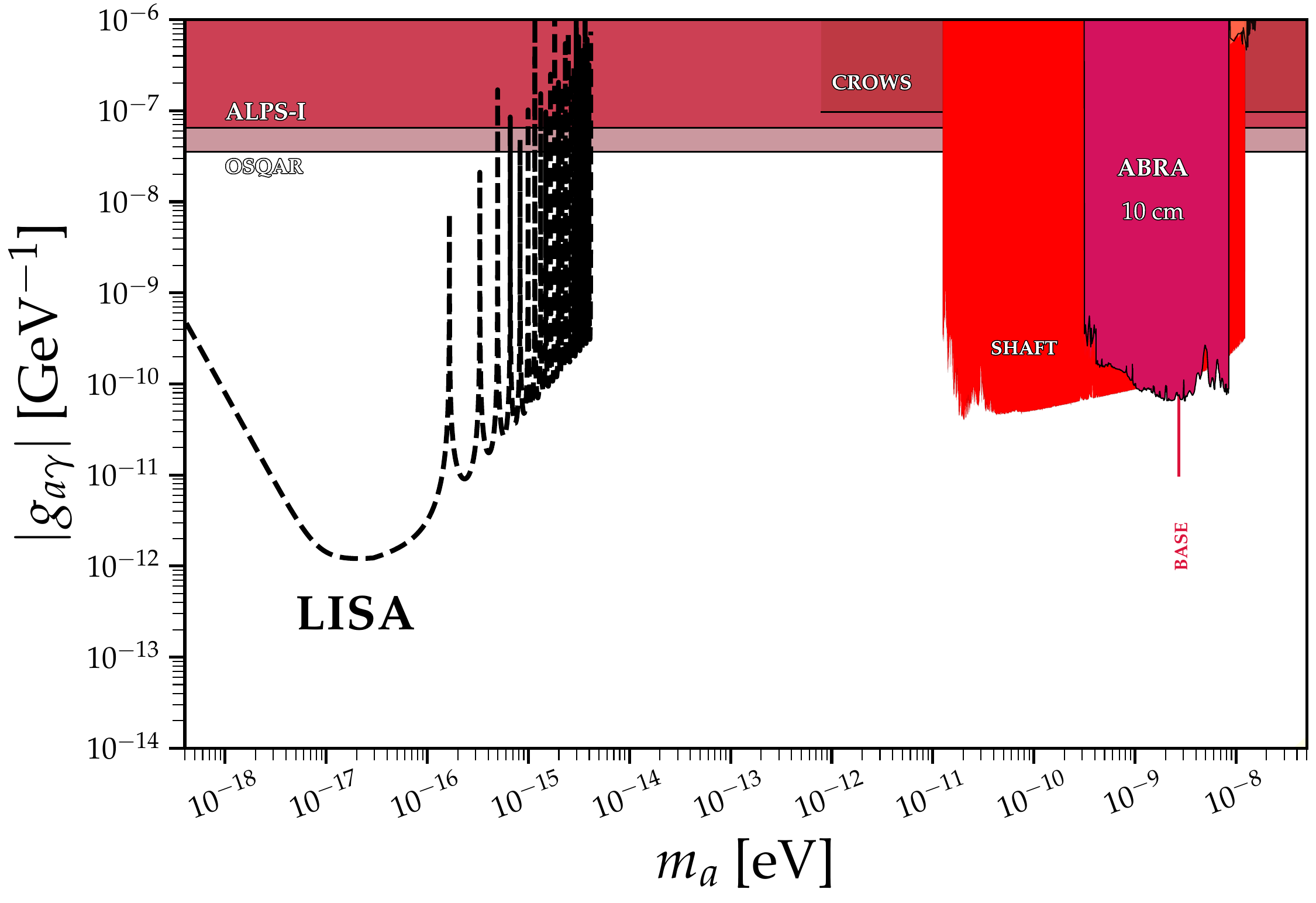}
    \caption{Current laboratory constraints on the axion-photon coupling $g_{a\gamma}$ (from \cite{AxionLimits}). The expected sensitivity of \textit{LISA} through vacuum birefringence is shown in black.}
    \label{fig:LISA_axion_photon}
\end{figure}

In Fig.~\ref{fig:LISA_axion_photon}, we show in red the current best laboratory constraints on the axion-photon coupling, from \cite{ADMX:2010,ADMX:2018_1,ADMX:2019,ADMX:2021,ADMX_Sidecar:2018,Capp:2020_1,Capp:2020_2,Capp:2022_1,CAPP:2020_3,Capp:2022_2,Capp:2022_3,HAYSTAC:2018,HAYSTAC:2020,Organ,Alesini:2019,Alesini:2020,Alesini:2022,Ouellet:2018beu,Salemi:2021gck,ADMX:2021nhd,Crisosto:2019fcj,Adair:2022rtw,Kim:2023vpo,Yang:2023yry,CAPP:2024dtx,Yoon:2022gzp,Devlin:2021fpq,Quiskamp:2023ehr,McAllister:2017lkb,HAYSTAC:2023cam,QUAX:2023gop,Gramolin:2020ict,DePanfilis,Wuensch:1989sa,Hagmann:1996qd,Hagmann}.

\subsection{\label{sec:DAMNED_sens_axion_photon}\textit{DAMNED} and optical fibers}

\begin{figure}[h!]
\begin{minipage}{\textwidth}
  \begin{minipage}{0.55\textwidth}
    Therefore, both experiments in their current form are not competitive. For \textit{DAMNED}, this is because the signal resonances happen at modes of the cavity, where $\omega_a \ell/c = n\pi, \: n\in \mathbb{N}$, which correspond to GHz frequencies, while the apparatus is sensitive to much lower frequencies, from $10$ kHz to $1$ MHz, as we described in Chapter ~\ref{chap:exp_summary}. One could overcome this by modulating the signal with another periodic signal of frequency comparable to the axion one, e.g. with a rotating waveplate. Note that even in the case where the resonances of the cavity would be visible, the associated sensitivity would reach $g_{a\gamma} \sim 10^{-5}$ GeV$^{-1}$, which is still far from the best existing constraints in this mass range (see Fig.~\ref{fig:LISA_axion_photon}).
    \end{minipage}
    \hfill
    \begin{minipage}{0.4\textwidth}
    \centering
    \includegraphics[width=\textwidth]{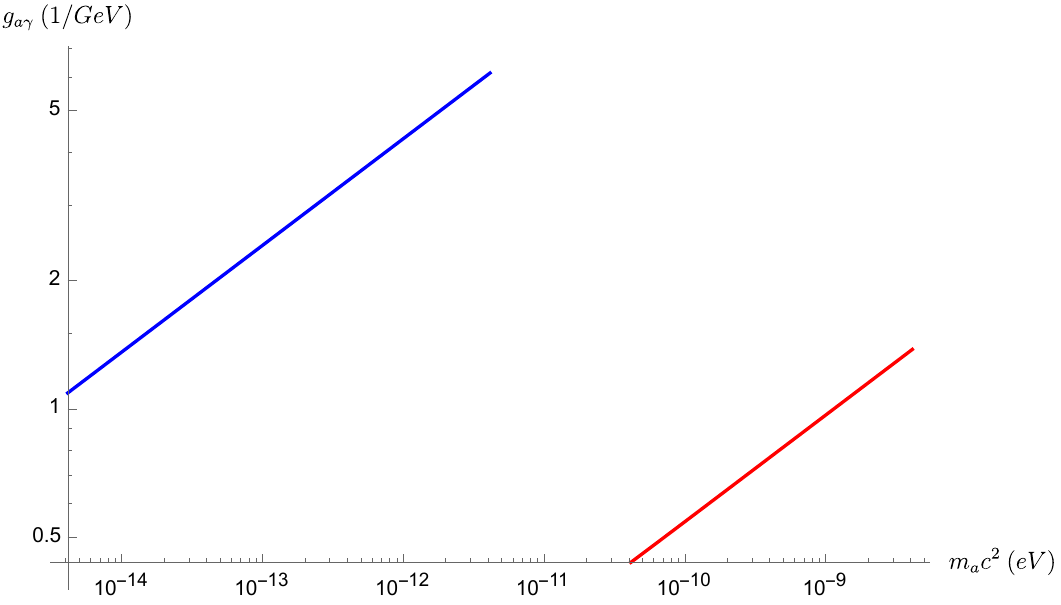}
    \caption{Expected sensitivity of \textit{DAMNED} (red curve) and a 86 km long fiber (blue curve) to $g_{a\gamma}$. Compared to already existing experiments (see Fig.~\ref{fig:LISA_axion_photon}), they are not competitive.}
    \label{fig:constraint_axion_photon}
  \end{minipage}
\end{minipage}
\end{figure}

The respective phase shift induced by the axion-photon coupling in \textit{DAMNED} or the optical fiber can be found in Eqs.~\eqref{eq:phase_final_DAMNED} and \eqref{eq:phase_fiber_axion_photon} respectively. All the experimental parameters of interest are described in Section ~\ref{sec:DAMNED_exp}. As for the other sensitivity curves, using Eq.~\eqref{coupling_constraint_general_DM}, we compute the corresponding sensitivity on the axion-photon coupling $g_{a\gamma}$ of \textit{DAMNED} and the optical fiber, which are respectively presented in red and blue in Fig.~\ref{fig:constraint_axion_photon}. As it can be noticed, the smallest coupling that they can reach is $\mathcal{O}(10^{-1}, 1)$ GeV$^{-1}$ respectively, which is very far away from the best existing constraints in this range (which reach $\sim 10^{-8}$ GeV$^{-1}$, see e.g. Fig.~\ref{fig:LISA_axion_photon}).
\newpage
\subsection{\textit{LISA}}

As mentioned in Chapter ~\ref{chap:axion_photon_LISA}, \textit{LISA} could be sensitive to vacuum birefringence induced by the axion-photon coupling, by a slight modification of its optical benches. Here, we estimate the sensitivity of \textit{LISA} to such coupling using the signal of the Sagnac $\alpha$ TDI combination  $|s_\mathrm{DM}(f)|$ from Eq.~\eqref{eq:signal_amp_axion_photon_Sagnac}. We first compute the signal one-sided PSD at frequency $f$
\begin{align}
    S_s(f) &= \left(\frac{\sqrt{8 \pi G \rho_\mathrm{DM}}E_P g_{a\gamma}}{\pi \nu_0 c}\right)^2\sin^4\left(\frac{3\pi f L}{c}\right) \, ,
\end{align}
bearing in mind that the signal is oscillating at frequency $f$, such that the amplitude of the signal power is $|s_\mathrm{DM}(f)|^2/2$. 
Using the Sagnac $\alpha$ noise transfer function defined in Eq.~\eqref{eq:TDI_Sagnac_noise} and Eq.~\eqref{coupling_constraint_general_DM}, we can derive the expected sensitivity of the Sagnac combination to $g_{a\gamma}$. This is what is shown in black in Fig.~\ref{fig:LISA_axion_photon} with four years of data. One can notice that \textit{LISA} would improve the current laboratory constrain on $g_{a\gamma}$ for axion masses between $\sim 4\times 10^{-19}$ to $4 \times 10^{-15}$ eV, by several orders of magnitude. In particular, between $10^{-17}$ and $10^{-16}$ eV, \textit{LISA} would reach $g_{a\gamma} \sim 10^{-12}$ GeV$^{-1}$, improving by more than 4 orders of magnitude the current best laboratory constraint at these masses.
\newline
We now discuss quickly dichroism effects. As a reminder, in the case of \textit{LISA}, the phase shift (between the elliptical polarization and the initial linear one) induced by the axion-photon coupling $g_\mathrm{a\gamma}$ depends on the travelled distance $L$ inside a constant magnetic field $B_0$ as 
\begin{align}
    \phi(L) &=\frac{\hbar c}{\mu_0}\frac{g^2_{a\gamma}B^2_0}{2 q^2}\left(qL-\sin(qL)\right) \, .
\end{align}
In \textit{LISA}, the laser beams are operating with a wavelength $\lambda = 1064$ nm \cite{Petiteau08}, which means the photon energy is $E_\gamma \approx 1$ eV. As it was discussed in Chapter ~\ref{chap:axion_photon_LISA}, the axion and photon have equal energy for the photon-axion conversion to work in a static magnetic field, i.e $E_\gamma = \hbar k_\gamma c =\sqrt{(k_a c)^2+(m_ac^2/\hbar)^2}= \hbar \omega_a$. This means that if the axion field is DM, the mass of the field is roughly equal to its Compton frequency (using Eq.~\eqref{eq:compton_freq_mass_corresp_DM}), implying that \textit{LISA} is essentially only sensitive to DM mass of $1$ eV. In addition, the $q$ parameter reduces to
\begin{align}
    q &= |k_a - k_\gamma | = \frac{\omega_a}{c}\left|\frac{v_\mathrm{DM}}{c}-1\right| \approx \frac{\omega_a}{c} \equiv \frac{m_a c}{\hbar} \, ,
\end{align}
i.e it corresponds to the wavevector of light, and we can approximate the phase shift by
\begin{align}
     \phi(L)_\mathrm{DM} &\approx \frac{\hbar c}{\mu_0}\frac{g^2_{a\gamma}B^2_0 L}{2 q} \approx 2.5 \times 10^{-35} \left(\frac{g_{a\gamma}}{10^{-10} \: \mathrm{GeV^{-1}}}\right)^2 \: \mathrm{rad}\, ,
\end{align}
since $qL \gg 1$ and where we used a solar magnetic field of $B_0 \sim 3.5$ nT at 1 astronomical unit \cite{Grieder01}. 
For demonstration, let us relax the constraint that axion is DM, i.e we only permit the existence of the axion-photon coupling. In that case, axion can be relativistic, which implies that $\hbar \omega_a \gg m_ac^2$ (we are now sensitive to any axion mass much below 1 eV), and 
\begin{align}
    q &= \left|\sqrt{\left(\frac{\omega_a}{c}\right)^2-\left(\frac{m_a c}{\hbar}\right)^2}-k_\gamma \right| \approx \frac{(m_ac^2)^2 k_\gamma}{2E^2_\gamma} \, ,
\end{align}
where we expanded the square root in small $m_a c^2 /\hbar \omega_a$. For comparison, assuming an axion mass $m_a c^2 = 1$ neV, such that $qL \ll 1$, the phase shift becomes
\begin{align}
    \phi(L)_\mathrm{\overline{DM}} &\approx \frac{\hbar c}{12\mu_0}g^2_{a\gamma}B^2_0 qL^3 \approx 3.4 \times 10^{-22} \left(\frac{g_{a\gamma}}{10^{-10} \: \mathrm{GeV^{-1}}}\right)^2 \: \mathrm{rad} \, .
\end{align}
This phase shift does not oscillate with time, and therefore this effect will a priori not be visible by \textit{LISA}. 

\section{Dark photon-photon coupling $\chi$}

\begin{figure}[h!]
    \centering
    \includegraphics[width=\textwidth]{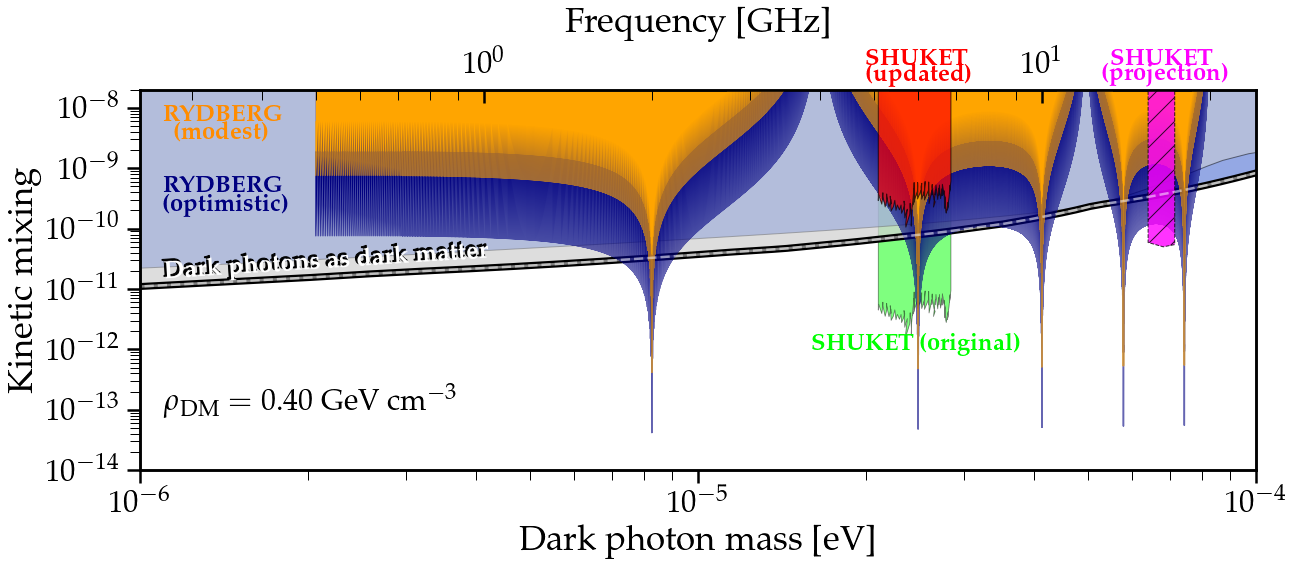}
    \caption{Dark photons-photons kinetic mixing coupling $\chi$ parameter space, with cosmological constraints from CMB (from \cite{AxionLimits,Caputo}). The expected sensitivity of the experiment involving Rydberg atoms in a microwave cavity, with "optimistic" parameters of the system described in Table ~\ref{tab:Table_microwave_rydberg} is shown in blue, denoted "RYDBERG (optimistic)", while the sensitivity with the "modest" parameters of Table ~\ref{tab:Table_microwave_rydberg} is shown in orange, denoted "RYDBERG (modest)" (see text). In green is the original constraint on the kinetic mixing parameter from the \textit{SHUKET} experiment obtained in \cite{SHUKET}. In red is shown the updated constraint using the same experimental parameters and data but considering a realistic modeling of the experiment in the analysis, following Eq.~\eqref{ratio_AF_SHUKET}.  The sensitivity curve with optimized experimental parameters is shown in dashed orange.}
    \label{fig:DP_chi_full_constraint}
\end{figure}

\subsection{\label{sens_Rydberg_DP}Rydberg atoms in microwave cavity}

The sensitivity of the experiment involving Rydberg atoms inside a microwave cavity for the detection of the kinetic mixing coupling $\chi$, obtained considering all experimental parameters described in Table ~\ref{tab:Table_microwave_rydberg} and respectively with $\{f_s=1$ kHz, $P_\mathrm{RIN}=10^{-15}\}$ and $\{f_s=100$ Hz, $P_\mathrm{RIN}=10^{-13}\}$ is presented by the blue and orange curves of Fig.~\ref{fig:DP_chi_full_constraint}. One can clearly see the sensitivity peaks arising from the cavity's odd resonances. At these frequencies, the applied field amplitude $X_A$ takes the smallest value (see Eq.~\eqref{eq:values_Xa_rydberg}), as shown in Eq.~\eqref{optimum_amplitude}, in order to minimize Eq.~\eqref{general_E_power_noise}. Eq.~\eqref{optimum_amplitude} works well for frequencies far from odd resonances. However, on those odd resonances, this approximate equation cannot be used as discussed previously. Instead one should use the exact expressions of signal and noise to optimize Eqs.~\eqref{eq:chi_limit}. As an example, when the applied field frequency corresponds exactly to the first odd resonance of the cavity, i.e $\omega_A L =\pi c$ and $\omega_U = \omega_A + \pi f_s$ the optimum amplitude of $X_A$ is $\sim 18 $~V/m, whose corresponding experimental sensitivity is $\chi \sim 10^{-13}$ in the modest scenario, as shown in Fig.~\ref{fig:DP_chi_full_constraint}.
Additionally, one can notice the presence of specific frequencies where this sensitivity decreases significantly, the experiment is almost insensitive to these DM frequencies. As discussed in the previous chapter, from the approximate expression of the signal contribution Eq.~\eqref{signal_approx}, we have $S(\omega_U,X_A) \simeq 0$ for $\frac{\omega_U L}{c} = 4\pi + 2 \pi n $, $n \in \mathbb{N}$ accounting for the loss of sensitivity.  
In both scenarii presented here (modest and optimistic), one can see from Fig.~\ref{fig:DP_chi_full_constraint} that the experiment setup proposed here would improve the current constraint on the coupling $\chi$ compared to cosmological and astrophysical observations (light grey curve, from CMB \cite{Arias}).

If one decides to run this experiment aiming at unconstrained regions of the exclusion plot, it would take approximately five days of data-taking to cover the mass range from 7 $\mu$eV to 10 $\mu$eV, while around 35 days would be needed to cover the mass range from 35 $\mu$eV to 60 $\mu$eV, assuming no dead time between the $T_\mathrm{obs}=60$~s observation runs. More realistically, reserving say 50\% of the total experimental time for manipulation of the atoms and applied field, the total duration increases by a factor two, which is still very reasonable.

With the appropriate set of parameters, in particular the applied field amplitude $X_A$ following Eq.~\eqref{optimum_amplitude}, both sources of noise, systematic and statistical, are equal in amplitude. This means that, in the search for high sensitivity of the experiment, the optimum choice of $X_A$ is not to increase it as much as possible to maximise the signal. Even though the signal is linear in $X_A$, the systematic uncertainty is quadratic in $X_A$, as stated at the end of Section \ref{Systematic_Rydberg}, implying a loss of sensitivity if the experimenter decides to apply too much power inside the cavity.

If the level of intensity fluctuations (RIN) of the applied field could be reduced e.g. by stabilizing the power using low noise intensity measurements \cite{Orpheus}, the applied field and/or the quality factor of the cavity could be increased leading to an increase of the signal whilst keeping the contribution from the RIN below that of the measurement noise in Eq.~\eqref{eq:chi_limit}. This way, the optimistic curve presented in Fig.~\ref{fig:DP_chi_full_constraint} would be achievable. 

\subsection{\label{sec:SHUKET_sens}\textit{SHUKET}}

In the previous chapter, we have shown that one can use the results of Chapter ~\ref{chap:SHUKET_exp} for the \textit{SHUKET} experiment, in order to infer an updated sensitivity curve. This is what we do in this section. We remind that the idea here is to model more accurately the experiment, taking into account diffraction and mode-matching effects.

\subsubsection{Propagation of the field from the fictional plane to the antenna}\label{sec:shuket_prop}

The electric field induced by the dish at the location of the antenna is provided by Eq.~(\ref{E_field_complete}), which is not solvable analytically. In order to simplify it and get an analytical expression, we will make different approximations:
\begin{itemize}
    \item  The distance between the fictional plane and the antenna needs to be much larger than the typical size of the dish, i.e. $\left|\Delta z\right| \gg \rho' \leq r$ (For \textit{SHUKET}, $\left|\Delta z\right| = R-a \sim R = 32$ m, while $r \sim 0.618$ m). 
    \item  The distance between the fictional plane and the antenna needs to be much larger than the typical size of the antenna, i.e. $\left|\Delta z\right| \gg \rho$ (For \textit{SHUKET}, the largest dimension of the antenna is $A = 0.25$ m $\ll R$).
    \item  The last approximation is known as the far field approximation (FFA). For the DM Compton frequency under consideration, i.e $f_U=6$ GHz, $kL \sim k |\Delta z| \gg 1$ and we can safely neglect the factor $-1$ in Eq.~(\ref{E_field_complete}).
\end{itemize} 

The first two approximations simplify the distance $L$ between any point on the fictional plane $(\rho',\phi',R-a)$ and any point on the antenna (whose center is located at the origin of our coordinate system) $(\rho,\phi,0)$ to
\begin{align}\label{eq:L_low_curv}
L &\approx |\Delta z| + \frac{\rho^2+\rho'^2-2\rho\rho'\cos(\phi-\phi')}{2|\Delta z|} \, .
\end{align}

Using this expression of $L$ as well as the FFA, we can express Eq.~\eqref{E_field_complete} as 
\begin{subequations}
\begin{align}
   \vec U^\mathrm{SHUKET}_\mathrm{dish}(\rho, \Delta z) &\approx \frac{\omega^2_U\chi e^{i\Phi(\rho,\Delta z)}}{\Delta zc}\begin{pmatrix}
Y_x\\
Y_y\\
0
\end{pmatrix} \int_0^r d\rho' \rho' e^{-i\varphi(\rho',\Delta z)}J_0\left(\frac{k\rho\rho'}{|\Delta z|}\right) \, ,\label{E_field_FFA} 
\end{align}
where $J_0$ is the Bessel function of the first kind of order 0 and where the integral is performed over the radius of the fictional plane which closes the dish, where the dependence on the angle $\phi$ disappeared by spherical symmetry and with
\begin{align}
    \varphi(\rho',\Delta z)&= \frac{k\rho'^2}{2}\left(\frac{1}{R}+\frac{1}{\Delta z}\right)\,\\
    \Phi(\rho,\Delta z) &= k\left(\frac{r^2}{2R} - \Delta z - \frac{\rho^2}{2\Delta z}\right) \, .
\end{align}
\end{subequations}
One can find an analytical solution for the last integral in the case where $z_\mathrm{ant}=0$. We expand the exponential inside the integral of Eq.~\eqref{E_field_FFA}, since the argument is much smaller than 1. Indeed, for the parameters of \textit{SHUKET}, we have $\Delta z = -R+a$ (i.e $z_\mathrm{ant}=0$) and 
\begin{align}
    \frac{k\rho'^2}{2}\left(\frac{1}{R}+\frac{1}{\Delta z}\right) &\approx \frac{k \rho'^2 a}{2R^2}< \frac{k r^2 a}{2R^2} = 1.4 \times 10^{-4}\,
\end{align}
Then, the integrand of Eq.~\eqref{E_field_FFA} becomes
\begin{align}
    \rho'(1-i\epsilon \rho'^2)J_0\left(\frac{k\rho\rho'}{|\Delta z|}\right)
\end{align}
at first order in the small parameter $\epsilon \rho'^2$, where $\epsilon=k a/2R^2$. Then, the integral is analytically calculable and the electric field reads 
\begin{subequations}
\begin{align}
    &\vec U^\mathrm{SHUKET}_\mathrm{dish}(\rho,\Delta z) \approx \frac{r\omega_U\chi}{\rho} \begin{pmatrix}
    Y_x \\
    Y_y \\
    0
    \end{pmatrix}
    e^{i\Phi(\rho)}\left(J_1\left(x\right)-i\frac{ar \Delta z}{2\rho R^2}\left(2 J_2\left(x\right)-x J_3\left(x\right)\right)\right) \, \label{E_field_expansion_bessel} ,
\end{align}
where $x=rk\rho/|\Delta z|$. One can verify easily that, with the set of parameters considered, the second term in Eq.~\eqref{E_field_expansion_bessel} containing the Bessel functions of order $2$ and $3$ is smaller by a factor $\sim 10^{6}$ compared to the other term $\propto$ the Bessel function of order $1$. Therefore, we can simplify the expression of the field as
\begin{align}\label{E_field_final}
    \vec U^\mathrm{SHUKET}_\mathrm{dish}(\rho,\Delta z) \approx \frac{r \omega_U \chi}{\rho}e^{i\Phi(\rho)}\begin{pmatrix}
Y_x\\
Y_y\\
0
\end{pmatrix}J_1\left(\frac{rk\rho}{|\Delta z|}\right) \, ,
\end{align}
\end{subequations}
with $J_1$ the Bessel function of the first kind of order 1. 
 
At the center of the curvature radius of the dish is located a polarized horn-antenna of physical surface $S_\mathrm{phys}=0.25 \times 0.142 \ \mathrm{m}^2$.
Numerical integration of the power from the electric field from Eq.~\eqref{E_field_final} over the physical antenna surface leads to
\begin{align}
    P_\mathrm{int}&=\int dS_\mathrm{phys}\frac{\epsilon_0 |U^\mathrm{SHUKET}_\mathrm{dish}(\sqrt{x^2+y^2})|^2 c}{2} \approx 2.85 \times 10^{-22} \Big(\frac{\chi}{10^{-12}}\Big)^2 \mathrm{ \ W} \, \label{S_rec_geom} ,
\end{align}
where we assumed an emission from a plane surface in the random polarization scenario, as in Eq.~\eqref{power_SHUKET}. The ratio of the power emitted by the dish that crosses physically the antenna and the total power emitted by the dish in the \textit{SHUKET} experiment (Eq.~\eqref{power_SHUKET}) is 
\begin{equation}
Q = \frac{P_\mathrm{int}}{P^\mathrm{SHUKET}_\mathrm{dish}} \approx\frac{2.85 \times 10^{-22}}{1.73 \times 10^{-20}} \approx 1.6\: \mathrm{\%} \, .
\end{equation}
Note that, in the geometrical optics approximation, i.e where diffraction effects are neglected, one would obtain $Q=100\: \mathrm{\%}$.
It is interesting to note that the ratio of the physical antenna's surface to the dish's surface is $\sim 3\%$, meaning that there is actually no focus of the field generated by the dish on the antenna.

This result means that, considering simply the propagation of the field from the dish to the antenna using Kirchhoff integral, the majority of the electromagnetic power is lost through diffraction, and the antenna is only able to detect a small amount of energy emitted by the dish.  
One can note that the usual criteria for diffraction effects to be negligible $d_\mathrm{dish} = 2r \gg \lambda$ is not fulfilled in this system, as the proportionality factor between the two parameters is $\sim$ 25, which explains this lack of focus.

As a cross-check of our calculations, integrating Eq.~\eqref{S_rec_geom} over the infinite antenna plane gives
\begin{align}
P_\mathrm{int, full} &\approx 1.73 \times 10^{-20} \Big(\frac{\chi}{10^{-12}}\Big)^2 \mathrm{ \ W} ,
\end{align}
which corresponds to the full power emitted $P^\mathrm{SHUKET}_\mathrm{dish}$ from Eq.~\eqref{power_SHUKET}. This provides some confidence in our set of approximations, made throughout the derivation.

\subsubsection{Detection of the field using a horn antenna}\label{sec:shuket_detection}

In the previous section, we showed that most of the power emitted by the dish is already lost through propagation of the field from the dish to the antenna. As described in Sec.~\ref{sec:detection_field}, there is still a second step to consider before predicting the exact amount of energy generated by the antenna: overlap integral between incident and antenna modes. 

\paragraph{Computation using the modes overlap}
As mentioned in Sec.~\ref{sec:detection_field}, to predict the energy generated by the horn antenna, i.e the overlap of modes, we need to consider the effective surface of the antenna $S_\mathrm{eff}$ at the frequency we are interested in ($f_U$ = 6 GHz). Furthermore, we also need to find an analytical expression for the mode of the field emitted by the dish $\vec M_\mathrm{dish}(x,y)$, such that it is possible to perform the mode overlap integral Eq.~\eqref{eq:overlap_int}.

To be able to compute the integral overlap of modes, we need an analytic expression of the field at coordinate $(\rho,z)$, which is provided by Eq.~\eqref{E_field_final}.
Then, from Eqs.~\eqref{eq:E_mode_dish} and \eqref{E_field_final}, we can separate the mode of the dish $\vec{M}_\mathrm{dish}$ from the constant amplitude $V_\mathrm{dish}$ expressed in Eq.~\eqref{eq:Vdish}, such that the mode of the dish at coordinates $(x,y,0)$ is
\begin{align}\label{dish_mode_SHUKET}
\vec{M}_\mathrm{dish}(x,y) &= \sqrt{\frac{3}{2\pi}} \frac{1}{\rho\left|\vec Y\right|}J_1\left(\frac{rk\rho}{R-a}\right)\begin{pmatrix}
    Y_x\\
Y_y\\
0
\end{pmatrix} \, ,
\end{align}
with $\rho=\sqrt{x^2+y^2}$. Then, using Eqs.~\eqref{eq:mode_ant}, \eqref{ratio_powers} and \eqref{dish_mode_SHUKET} and assuming the polarization of the DP to be randomly distributed, the ratio of receiving to emitted powers is simply
\begin{align}
\gamma_\mathrm{Overlap} &= \left(\int dS_\mathrm{eff} \vec M_\mathrm{ant}\cdot  \vec M_\mathrm{dish}\right)^2 \approx 5.8 \times 10^{-4} e_r\, ,
\label{ratio_overlap}
\end{align}
with $m_{TE_{10}} \approx 25.6$ m$^{-1}$ has been estimated from Eqs.~(\ref{eq:m_value}) and (\ref{eq:effective_surface})  using the antenna gain $G(f_U)=11.86$ dBi at $f_U=6$ GHz provided in the antenna datasheet and which has been measured experimentally by the manufacturer.
For this numerical value of ratio, we assumed an axial detection, with $Y_x = Y_y$ in Eq.~\eqref{dish_mode_SHUKET}.
Assuming $e_r=1$, i.e no loss inside the antenna, this result means that only 0.06\% of the emitted power is actually transmitted to the antenna wires, and therefore detectable.

\paragraph{Computation using the antenna factor}

As explained in Sec.~\ref{sec:antenna_factor_gen}, if the  electric field emitted by the dish is seen as a plane wave by the antenna or equivalently if the mode of this electric field is approximately constant over the effective surface of the antenna, we can use another method to derive the output of the experiment. 

From Eq.~\eqref{dish_mode_SHUKET}, it can be shown that the dish polarization mode is approximately constant over the effective long length of the antenna ($\sim 7$ cm) (with a deviation of $\sim 0.3\%$), with a value of $M^y_\mathrm{dish}(x,y=0) \approx 0.48 \: \mathrm{m}^{-1}$.

For a Horn antenna with internal  $R_0=50 \ \Omega$ resistance (which is typically the case for the \textit{Schwarzbeck BBHA-9120-D} antenna), the antenna factor at frequency $f_U$ is given by \cite{Mclean02}
\begin{align}
\mathrm{AF}(f_U) &= \frac{9.73 \ f_U}{c \sqrt{G(f_U)}} = 49.7 \mathrm{\ m}^{-1} \, ,
\label{AF_gain}
\end{align}
which is consistent with the value given in the antenna datasheet \cite{datasheet_ant}\footnote{Since AF(dB)=$20\log_\mathrm{10}$(AF) \cite{Balanis05}.}
Then, using Eq.~\eqref{ratio_AF}, one can compute the ratio of received to emitted powers as 
\begin{align}
\gamma_\mathrm{AF} &\approx 3.6 \times 10^{-4} \, .
\label{ratio_AF_SHUKET}
\end{align}
Considering that $e_r=1$, Eqs.~\eqref{ratio_overlap} and \eqref{ratio_AF_SHUKET} disagree by a factor of approximately 1.5. This means that our assumption of no loss inside the antenna is most likely wrong, and we can artificially consider $e_r \sim 0.62$ such that both methods coincide.
Therefore, the second result, obtained using the antenna factor, Eq.~\eqref{ratio_AF} is probably more realistic, as the different parameters have been experimentally measured. 

Combining the results from Section ~\ref{sec:SHUKET_update}, we can now reevaluate the constraints on $\chi$ obtained in the \textit{SHUKET} experiment \cite{SHUKET}. In Fig.~\ref{fig:DP_chi_full_constraint}, we show how both effects (diffraction and mode overlap) affect the sensitivity of the \textit{SHUKET} experiment. In green, denoted "SHUKET (original)" is the original sensitivity curve presented in \cite{SHUKET}. Since the power received by the antenna wires is quadratic in $\chi$ coupling, the ratio Eq.~\eqref{ratio_AF_SHUKET} leads to a loss in $\chi$ of an approximate factor 53. Making the assumption that this loss factor is roughly the same over all DM frequencies to which \textit{SHUKET} is sensitive, this leads to an updated sensitivity curve, shown in red in Fig.~\ref{fig:DP_chi_full_constraint}, denoted "SHUKET (updated)".

\subsection{Optimized run of \textit{SHUKET}}

Following the full derivation described above, it is possible to find some optimized experimental configuration such that the power received by the antenna is maximized. The experimental parameters that can easily be modified are the emitter-detector distance $L$ and the optimized DM frequency $f_U$ to search for with this setup\footnote{This requires an antenna appropriate for this frequency.}.

The frequency of the background DM field has several impacts on the power received by the antenna. First, following Eq.~\eqref{E_field_complete}, the electric field amplitude received at coordinates $(\rho,\Delta z)$ (obtained using the Kirchhoff integral) depends highly on the frequency of the field. It can be shown numerically from Eq.~\eqref{E_field_FFA} that the focusing of the electric field improves with  the DM frequency, which is  a consequence of diffraction effects that become non negligible for low frequency, i.e. when $f_U d_\mathrm{dish}/c = 2f_Ur/c \: \slashed{\gg} \: 1$. In particular, as already discussed in Chapter ~\ref{chap:SHUKET_exp}, for the size of dish and the frequency bandwidth of the horn antenna used in \textit{SHUKET}, diffraction effects are always non negligible (i.e.  $f_U d_\mathrm{dish}/c \: \slashed{\gg} \: 1$). As a consequence, the optimal location for the horn antenna is not the center of curvature of the dish and this location becomes frequency dependent. In addition, the antenna factor (or equivalently the antenna gain) is also highly dependent on the frequency of the measured electric field. As it was shown in the previous section, the overlap integral of polarization modes contribute almost equally to the total power loss than the one from the propagation of the field. Therefore, the optimal distance $L$ between the dish emitter and the horn antenna detector where the maximum field power is transmitted is non trivial and depends on the frequency.

The goal of this section is to explore the parameter space to find the optimal frequency $f$ and distance $L$ such that the efficiency coefficient $\gamma_\mathrm{AF}$ from Eq.~\eqref{ratio_AF} is maximized. To do so, we use Eq.~\eqref{E_field_FFA} with unknown parameter $\Delta z$\footnote{Note that in order to use this equation, we must restrict ourselves to dish-horn distance $\left|\Delta z \right|\gg r$.} and Eq.~\eqref{eq:Vdish} to find the mode of the field emitted by the dish. 

\begin{figure}[h!]
\begin{minipage}{\textwidth}
  \begin{minipage}{0.49\textwidth}
    We first consider only the value of the mode at $\rho=0$, and then we show that for the optimized parameters, the mode is indeed constant over the effective size of the antenna, such that the method can be used. Additionally, we interpolated the antenna datasheet \cite{datasheet_ant} to infer the value of the antenna factor as function of the frequency $\mathrm{AF}(f)$. Then, Eq.~(\ref{ratio_AF}) is used to estimate numerically the efficiency coefficient  
    \begin{align}\label{eq:gamma_AF_opt}
    \gamma(f, \Delta z)_\mathrm{AF} &= \frac{Z_0 M^2_\mathrm{dish}(\rho=0,\Delta z,f)}{2R_0\mathrm{AF}(f)^2} \, .
    \end{align}
    The behavior of this efficiency coefficient as a function of the DP frequency and of the distance between the dish and the receiver is shown in  Fig.~\ref{fig:3D_plot_gamma}. One can notice the increase of $\gamma$ for small frequencies and short distances, which is mainly driven by the behavior of the antenna factor. Indeed, even though the loss through diffraction effects is larger at lower frequencies, the $\gamma$ parameter also takes into account the mode matching of the antenna, which is larger at those frequencies. 
    \end{minipage}
    \hfill
    \begin{minipage}{0.49\textwidth}
    \centering
    \includegraphics[scale=0.58]{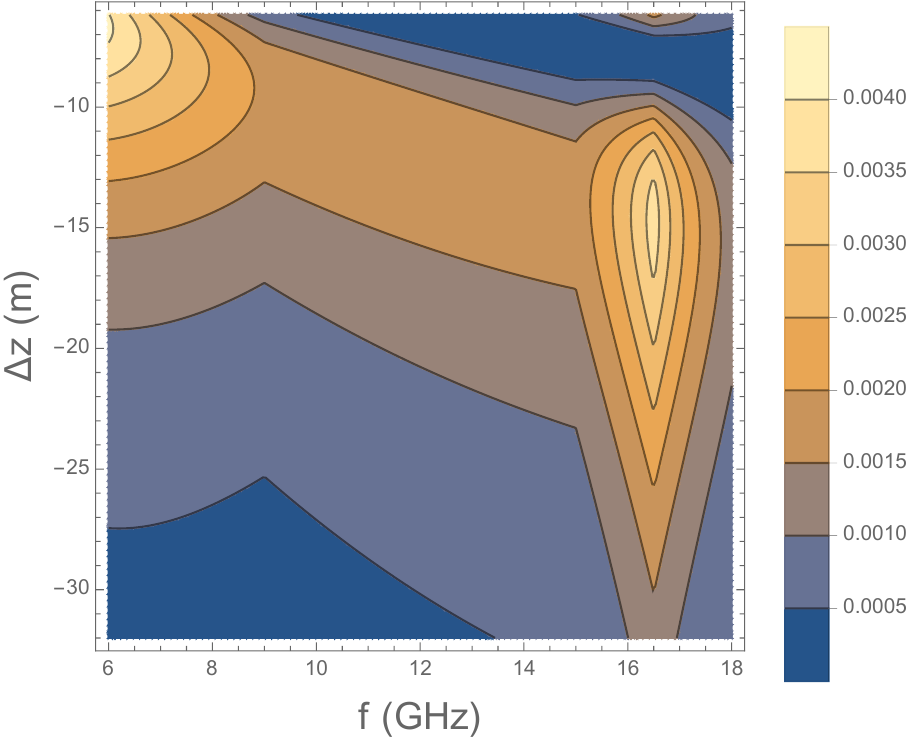}
    \caption{Efficiency coefficient $\gamma(f,\Delta z)_\mathrm{AF}$ as function of the frequency $f$ and $\Delta z$, in the frequency range $f \in [6, 18]$ GHz, of the horn antenna \cite{datasheet_ant} and for distances $|\Delta z| \gg r$. The efficiency coefficient increases for low frequencies and short distances and presents a local maximum around $(f \sim 16.5 \: \mathrm{GHz}, \Delta z \sim -15 \: \mathrm{m})$.}
    \label{fig:3D_plot_gamma}
  \end{minipage}
\end{minipage}
\end{figure}
The expected output signal from the \textit{SHUKET} experiment using the optimized experimental parameters can be computed using the same procedure and approximations as the ones presented in Sec.~\ref{sec:shuket_prop} at the exception of one approximation that is no longer valid. Indeed, Eq.~(\ref{E_field_final}) is obtained by assuming that $\Delta z \sim R = 32$ m, which is no longer the case for the optimized distances, as it is shown in Fig.~\ref{fig:3D_plot_gamma}.  
We now derive an analytic expression of the electric field emitted by the dish in the case of such optimized distances. We will concentrate on one particular optimized frequency $f_O$.
We consider the optimized parameters for DM frequency $f_O=16.5$ GHz and dish-antenna distance $\Delta z_O \approx -14.8$ m, the exponent in the integrand of Eq.~\eqref{E_field_FFA} can be comparable to 1, therefore the same Taylor expansion as previously is not possible. Instead, we expand the Bessel function as 
\begin{align}\label{Bessel0_expansion}
    J_0\left(x\right) = \sum_{m=0}^\infty \frac{(-1)^m x^{2m}}{m!4^m\Gamma(m+1)}
\end{align}
In our case, $x=|k_O\rho\rho'/\Delta z_O|$, $k_O = 2\pi f_O/c$. The maximum value of $\rho$ depends on the effective size of the antenna at frequency $\omega_O = 2\pi f_O$ following Eq.~\eqref{eq:effective_surface}. From the antenna gain $G(\omega_O) = 16.87$ dBi, the maximum value of $x$ is $x_\mathrm{max} \approx 0.30$. In the range $[0,x_\mathrm{max}]$, one can easily show that the relative error on $J_0$ by only taking the first two terms of the sum Eq.~\eqref{Bessel0_expansion} is very small $\sim 10^{-4}$, which indicates that these two terms are sufficient to describe the Bessel function in our system. Therefore, the integrand of Eq.~\eqref{E_field_FFA} becomes
\begin{subequations}
\begin{align}
    \rho'e^{-i\varphi(\rho',\Delta z_O)}\left(1-\left(\frac{k_O\rho\rho'}{2\Delta z_O}\right)^2\right) \, ,
\end{align}
since
\begin{align}
    \sum_{m=0}^1 \frac{(-1)^m x^{2m}}{m!4^m\Gamma(m+1)} = 1 -\frac{x^2}{4} \, .
\end{align}
\end{subequations}
Then, the analytic integration is doable and gives
\begin{align}
&\vec U^\mathrm{Opti}_\mathrm{dish}(\omega_O,\Delta z_O) = \frac{\chi \omega_O R e^{ik_0\Delta z_0}}{2(\Delta z_O)(R+\Delta z_O)^2}\begin{pmatrix}
Y_x \\
Y_y \\
0 
\end{pmatrix}\left(\left(e^{i\Phi'}-1\right)(k_O\rho^2 R-2i\Delta z_O(R+\Delta z_O))-\right.\,\nonumber\\
&\left.\frac{ir^2k^2_O\rho^2}{2}\left(1+\frac{R}{\Delta z_O}\right)\right) \, ,
\end{align}
with $\Phi'=k_Or^2(R+\Delta z_O)/2R\Delta z_O$. Note that from this expression, by setting $\Delta z_O = -R+a$, we recover the first order expression of the field Eq.~\eqref{E_field_final}, as expected.
From this expression, one can show that the mode associated to this electric field is roughly constant over the effective size of the antenna ($\sim 1\%$ variation), therefore both methods presented in the Sec.~\ref{sec:detection_field} can work to compute the relative power received by the antenna. We find
\begin{subequations}
    \begin{align}
        \gamma_\mathrm{Ov.} &\approx 5.1 \times 10^{-3} \,\\
        \gamma_\mathrm{AF} &\approx 3.2 \times 10^{-3} \, \label{ratio_opti}.
    \end{align}
\end{subequations}
Notice that in the same way as for \textit{SHUKET} parameters, the two results differ by an approximate factor 1.5 difference, as expected. Comparing both values with the ones presented in Sec.~\ref{sec:shuket_detection}, one finds an approximate factor 9 improvement in power received. 
Note that this order of magnitude agrees with the one reached at the local maximum in Fig.~\ref{fig:3D_plot_gamma}.
The efficiency coefficient obtained for this optimized experimental setup is one order of magnitude larger than the one obtained using the initial set of parameters, Eq.~\eqref{ratio_AF_SHUKET}. Compared to the original 53 loss factor in $\chi$, this optimization leads to a loss of $\sim 17$ on the coupling $\chi$ compared to the initial approximation of $P_\mathrm{rec} = P_\mathrm{dish}$.  

We use the same procedure to obtain a curve of efficiency coefficient as function of the frequency, where, for each frequency $f$, we always assume the optimized dish-horn distance $\Delta z_O$ such that $\gamma(f,\Delta z_O)$ is maximized. Then, we apply this frequency-dependent efficiency coefficient to the original constrain on the mixing parameter $\chi$ set by \textit{SHUKET} \cite{SHUKET}, in order to obtain an estimate of the sensitivity of a new \textit{SHUKET}-like experiment using these optimized frequencies and distances. However, as can be seen in \cite{SHUKET}, the \textit{SHUKET} constraint on $\chi$ depends on the frequency, even though the assumed signal is frequency independent. As explained in \cite{SHUKET}, this is due to the frequency dependent gain of the power amplifier connecting the horn antenna to the spectrum analyzer. Since we can assume that a new run of \textit{SHUKET} would operate another amplifier with higher gain, we will consider the highest constraint of the original experiment as a basis for our new estimate.

Such projection is shown by the magenta hashed curve in Fig.~\ref{fig:DP_chi_full_constraint}. One can notice that over the $15.5-17.3$ GHz DM frequency range, this new run with optimized parameters would improve the current constraint on the kinetic mixing parameter $\chi$, compared to CMB, shown in light grey.

\subsection{New experimental run of \textit{SHUKET}}

Motivated by the optimization procedure presented in the previous section, a new run of the experiment was performed at IRFU, CEA Saclay, with the same dish and horn antennas used for the original measurement campaign \cite{SHUKET}. In order to ease the data taking, it was decided that the dish-horn distance would not be modified for each frequency, but would stay fixed. In addition, the signal searched for is contained in the frequency range $f \in [8;18]$ GHz, which corresponds to the high frequency band of sensitivity of the horn antenna. Then, one needs to find the optimized single distance $\Delta z_O$ such that for all frequencies $f \in [8;18]$ GHz, 
\begin{align}
    \sum_f \gamma(f,\Delta z_O) \, ,
\end{align}
is maximized. Numerically, we find $\Delta z_O \sim 12.5$ m. Keeping the distance fixed decreases a little the maximum potential sensitivity of the experiment, but keeping in mind that the final constraint on $\chi$ is $\propto \sqrt{\gamma(f,\Delta z)}$, this loss of sensitivity is marginal. However, this eases greatly the experimental process, as the antenna should not be re calibrated at each measurement.   

The full data acquisition covers $8-18$ GHz frequency band, and is done following \cite{Caputo}. In short, the setup is said to be "axial", i.e it is only sensitive to one polarization direction, which corresponds to the antenna mode direction $\vec M_\mathrm{ant}$, which in practice is pointing towards the Zenith. In order to maximize the signal in case of fixed polarization scenario\footnote{In the random polarization scenario, the field has a random polarization at each coherence time, such that for experiment lasting much longer, the average is simply $\beta_\mathrm{random}=\langle\cos^2(\theta)\rangle$ over a sphere, where $\theta$ is the angle between the polarization direction and the axis of measurement, as it was computed in e.g. Eq.~\eqref{power_SHUKET} from Eq.~\eqref{power_emit_dish}, with $\beta_\mathrm{random}=2/3$. In the fixed polarization scenario, the polarization direction is unknown but stays fixed at all times. In this case, it is the daily modulation of the signal due to Earth rotation that can help increasing the signal. Indeed, for a very short experimental time, the experiment probes only one specific direction and a conservative estimate yields $\beta_\mathrm{fixed} \ll 1$ for most experimental setups \cite{Caputo}. However, if the experimental time increases significantly ($\mathcal{O}$(1 day)), the experiment effectively averages over many different polarizations and therefore $\beta_\mathrm{fixed} \rightarrow \beta_\mathrm{random}$ \cite{Caputo}.}, it is found that the optimized data-taking for such experiment is to make $3$ individual measurements and wait for a third of a sidereal day between each measurement \cite{Caputo}. In practice, each acquisition lasts for $T=200$s, and corresponds to a frequency window of $10$ MHz. Then, the three acquisitions are summed to be able to average over the polarization.
Since the total frequency interval is $10$ GHz (from $8$ to $18$ GHz), the total acquisition time is $600 \times 1000 = 6 \times 10^5$s.

\begin{figure}[h!]
    \centering
    \includegraphics[width=0.8\textwidth]{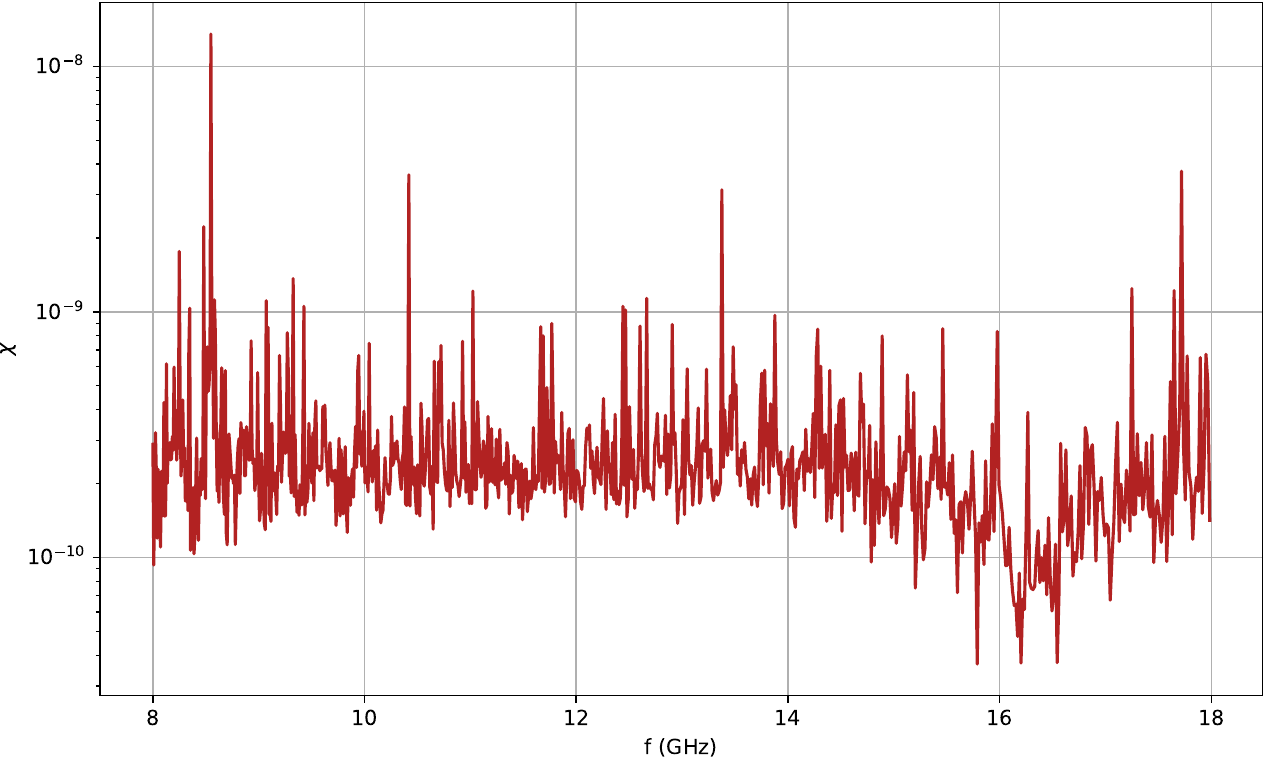}
    \caption{Preliminary constraint (95\% confidence level) of a second run of \textit{SHUKET} to the kinetic mixing coupling $\chi$.}
    \label{fig:new_constrain_SHUKET}
\end{figure}

Next, the background noise in the data is estimated using a Savitzky-Golay filter with a 1 MHz bandwidth (which is much larger than the expected width of the signal $\sim 10$ kHz, from Eq.~\eqref{eq:freq_DM_std} and using the fact that we are looking at DP frequencies around tens of GHz.). In short, this filter smooths out the data by using moving averages \cite{Savitzky64}.
The ratio of the original data (potential signal + background noise) with the noise, given by the filter, follows a Gaussian distribution centered around $1$. Then, each individual Fourier peak which is above 5$\sigma$ in this distribution is identified and analyzed individually to see if it looks like a DM signal. A preliminary analysis was done, where the DM signal is modelled with a Gaussian distribution with a fixed width which is $\sim 10^6$ smaller than the Fourier frequency of the peak (still following Eq.~\eqref{eq:freq_DM_std}), but with variable amplitude and offset.
No DM signal is found in the data, and therefore, the uncertainty on the amplitude is converted into an upper limit on the power received from the hypothetical DP. Then, using Eqs.~\eqref{power_SHUKET} and \eqref{eq:gamma_AF_opt}, this is converted to an upper limit on the kinetic mixing coupling $\chi$.  
The preliminary constraint on $\chi$ from this new run is shown in Fig.~\ref{fig:new_constrain_SHUKET}. The sensitivity estimate around $16$ GHz corresponds quite accurately to the theoretical prediction shown in magenta in Fig.~\ref{fig:DP_chi_full_constraint}. Over a large frequency band ($\sim 10-18$ GHz), this preliminary curve reaches unconstrained regions of the parameter space, compared to CMB, see Fig.~\ref{fig:DP_chi_full_constraint}. One can notice some modulation of the sensitivity in Fig.~\ref{fig:new_constrain_SHUKET}, which comes from the value of the gain amplifier which depends on the frequency.

\section{Discussion}

In this chapter, we have derived the expected sensitivities of various experiments, both already existing and futuristic, to various ULDM couplings to Standard Model fields. In particular, following this theoretical work, a new experimental run of \textit{SHUKET} was performed, with preliminary competitive results. 
We have shown that despite the existing strong constraints on those couplings, many experiments that will be conducted in the next few years will be able to reach unconstrained regions of the parameter space, and therefore have a chance to detect a positive signal, which will, without any doubt, be a major discovery. 

\clearpage
\pagestyle{plain}
\printbibliography[heading=none]

\pagestyle{plain}

\part{Conclusion}
General relativity and the Standard Model of particle physics are regarded today as the most successful theories in modern physics. Despite their incredible success, their inherent formalism differences make their unification into a single theory, the \textit{theory of everything}, currently impossible. Many believe that these theories are therefore incomplete, and theorists around the world are determined to find a quantum theory of gravity. This would certainly provide answers to unresolved mysteries arising in both theories, such as the existence (or not) of dark matter and its nature. Indeed, while dark matter was first revealed almost a hundred years ago, its true microscopic nature is still unknown, making it one of the biggest puzzles in modern physics. However, one must remember that it could still be an artifact of an incomplete theory of gravitation.

Among many other candidates, ultralight fields present an undeniably elegant solution to the dark matter problem. As discussed in Chapter~\ref{DM_pheno}, they induce various effects on matter, such as variation of constants of Nature, modification of classical Maxwell's equations, appearance of equivalence-principle violating accelerations, or production of electromagnetic fields. This kind of phenomenology can be probed accurately with quantum precision technologies, which, over the past $\sim$ thirty years, have reached unprecedented progress in the control of matter and light. These quantum sensors include, for example, atomic clocks, electromagnetic cavities, atom interferometers, and optical interferometers. We have seen in different parts of this thesis how each of these probes can be used to detect ultralight dark matter fields.

SYRTE, as one of the world's leading laboratories in time metrology, possesses some of the most accurate atomic clocks and atom interferometers (see e.g. \cite{Guena12,Fang16,Abgrall15,Farah14}), and is therefore greatly involved in ultralight dark matter searches \cite{Hees16,Savalle21}. SYRTE also has significant expertise in data analysis of dark matter and gravitational wave laboratory experiments. It is in this environment that this thesis took place, making possible the investigation of many different experiments for the search of ultralight dark matter.

In terms of phenomenology, the main focus of this thesis is on 1) the violation of the equivalence principle induced by couplings between a (pseudo-)scalar dark matter field, the dilaton and the axion, and various Standard Model sectors \cite{Gue:2024onx}, 2) the modification of Maxwell's equations via the introduction of a pseudo-scalar field, the axion, coupled to electromagnetism, and 3) the production of an electromagnetic field in vacuum from the coupling of a vector dark matter candidate, the dark photon, to electromagnetism \cite{Gue23,Gue:2024gws}.

The first phenomenological aspect is related to the spacetime variation of constants of Nature, such as the fine structure constant or the electron mass, which, as a consequence, induces an oscillation of intrinsic properties of bodies, such as their rest mass, and the transition frequency of atoms. The amplitude of such oscillations being atomic species-dependent, two bodies of distinct composition would accelerate at different rates, leading to a violation of the weak equivalence principle. We extensively studied how \textit{MICROSCOPE}, as a classical test of the equivalence principle, and atom interferometers, as its quantum equivalent, can probe dark matter fields \cite{Gue:2024onx}. We found that \textit{MICROSCOPE} could reach unprecedented sensitivity to axion-gluon coupling compared to already existing laboratory experiments, which warrants a corresponding detailed analysis of \textit{MICROSCOPE} data to search for a potential signal. Additionally, we showed that diverse atom interferometric setups, despite being similar in terms of noise levels, have contrasting sensitivities to dark matter fields. In particular, we showed that dual-isotope interferometers, where both interferometers are spatially overlapped, have much more potential than gradiometers, where the two interferometers are stacked at different altitudes. 
Even though it is natural to consider experiments involving different atomic species to detect oscillations that are species-dependent, experiments using a single species can still be competitive due to the finite speed of light. This is the case with large space-based gravitational wave detectors, such as \textit{LISA}. Indeed, the movement of test masses caused by dark matter would generate Doppler shifts on light signals exchanged between the spacecrafts. We investigated the possibility of the degeneracy between this signal and a monochromatic signal from a galactic binary, and if it is realistic to expect making the distinction between the two, using a Bayesian approach and a full year of realistic orbits of the spacecraft. We found that the detector will be able to distinguish signals from scalar dark matter and monochromatic gravitational waves. Moreover, due to the non-relativistic nature of the dark matter field, the detector cannot probe accurately its propagation velocity. As a result, by allowing this velocity to be a free parameter in the Bayesian model, the sensitivity of the experiment to dark matter couplings decreases due to correlations.

The second phenomenological aspect of interest in this thesis is vacuum birefringence induced by the coupling between axions and photons. Considering an electromagnetic plane wave, the axion field couples to the magnetic polarization of light and, by its inherent oscillation, causes the light polarization to oscillate as well. If the light polarization is further decomposed into the two circular polarization states, one finds that the two states have different phase velocities, i.e., vacuum becomes birefringent. Considering an electromagnetic cavity in which circularly polarized light is sent in, we showed that this change in phase velocity of light is, mathematically speaking, equivalent to a change in the cavity's length, leading to a phase shift of light at the output of the cavity. We derived the sensitivity of an optical cavity and a tens-of-kilometer-long fiber link to such an effect, and we showed it is not competitive with current bounds. We then turned to \textit{LISA} again. The current version of \textit{LISA}'s optical benches operates with linearly polarized light only. We showed that a slight modification of these benches would 1) produce circularly polarized light and 2) be permitted in the setup for the search of gravitational waves. We showed that such a modification would make \textit{LISA} the most sensitive laboratory experiment to the axion-photon coupling at low masses, by five orders of magnitude.

Finally, we concentrated on laboratory probes of a small electromagnetic field induced by the kinetic mixing coupling between dark photons, a vector dark matter candidate, and photons. This electric field oscillates at the Compton frequency of the DM field, and we have been interested in experiments probing the GHz frequency region. 
We first proposed an innovative way of detecting this electric field by using Rydberg atoms inside a microwave cavity. In short, the idea of the experiment is 1) to use the cavity as a resonator for this small electric field; 2) to inject a strong electromagnetic field inside the cavity at a close frequency from the DM Compton one, in order to produce a slowly oscillating beatnote between the two fields; 3) to use Rydberg atoms to measure this slow component by the quadratic Stark effect it produces on them. The innovative concept of this experiment is twofold: first, the signal in our scheme is only linear in the small kinetic mixing coupling, while most experiments aim at detecting the field power, which is quadratic in the coupling; second, instead of using antennas for the measurement of the signal, we proposed to use atoms. We showed that such an experiment would reach competitive constraints on the kinetic mixing coupling compared to already existing experiments \cite{Gue23}. 
We also studied another kind of experiment to detect this electric field, which involves a dish antenna, which reflects the small electric field and focuses it onto a horn antenna located at the curvature center of the dish \cite{Horns, SHUKET}. On one hand, we investigated the effects of diffraction in this type of experiment from an analytical standpoint, making use of the Kirchhoff integral theorem in the low-curvature dish limit. On the other hand, we estimated the impact of mode-matching inside the horn antenna. We showed that the expected signal intensity can be significantly reduced compared to usual estimates. Our method was applied to the re-interpretation of the \textit{SHUKET} experiment data, the results of which were shown to be degraded by a factor of approximately fifty due to both diffraction and mode-matching \cite{Gue:2024gws}. This analytical method allowed optimizing some experimental parameters to gain sensitivity in future runs. In particular, following this theoretical modeling, a second run of \textit{SHUKET} was recently performed, with a competitive constraint on the kinetic mixing coupling.

This thesis combines several studies that can be useful to the dark matter search community in the near future: proposals of innovative experimental schemes (Chapter ~\ref{chap:Rydberg_exp_DP}), improved modeling of already existing and future experiments (Chapters ~\ref{chap:SHUKET_exp}, \ref{chap:AI}, and \ref{chap:LISA_DM}), and demonstration of the potential of existing and future experiments to probe unpredicted couplings (Chapters ~\ref{chap:Classical_tests_UFF} and \ref{chap:axion_photon_LISA}). It focuses on the theoretical modeling of such experiments and their corresponding sensitivity estimates, but not on the analysis of data produced by such experiments. As a simplifying assumption, the ultralight dark matter fields are considered monochromatic, while in reality, they have a finite coherence time. In situations where the time of observation of the experiment falls below this coherence time, they should be modeled as a stochastic superposition of plane waves oscillating at slightly different frequencies. We believe that this should not significantly change the various sensitivity estimates made in this thesis, but as soon as data analysis is concerned, one should take this stochastic effect into account.

The outlooks opened by this thesis are large and extend from the re-analysis of existing data (such as \textit{MICROSCOPE} or \textit{SHUKET}), the proposal of new experiments to search for ultralight dark matter, and the development of international and futuristic projects (\textit{AION-10}, \textit{MAGIS-100} and \textit{LISA}) that will search for ultralight dark matter. 

\newpage
\chapter*{\centering Acknowledgments}

I genuinely enjoyed these years of PhD, and I owe it to many people I would like to thank.
\newline
I will start by thanking deeply my two PhD advisors, Peter Wolf and Aurélien Hees. I am extremely grateful to have been their PhD student and to have enjoyed their supervision. I really appreciated their pedagogy, open-mindedness and humility, despite their experience. Like many other scientists who know them personally, I am still amazed by their intuition, their rigor and the (very) broad spectrum of knowledge they have. One day, I hope to become a scientist as complete as they are. 
\newline
There are many other people who have had a very positive impact on me throughout these three years. First, I would like to offer a special thanks to Marc Lilley and Robin Corgier for helping me in the preparation of my defense, but more generally for their guidance all along my PhD. I am also thankful to Adrien Bourgoin, Samy Aoulad-Lafkih and Martin Staab for all their pieces of advice, scientifically, but not only. Then, I would like to thank all the members of the theory and metrology team of SYRTE for their support: Pacôme Delva, Christophe Le Poncin-Lafitte, Marie-Christine Angonin, Miltos Chatzinikos and Christopher Aykroyd. 
In addition, I really appreciated the help from all the people I collaborated with, and I thank them for their physical insights: Jérôme Lodewyck and Rodolphe Le Targat from SYRTE, and Etienne Savalle, Laurent Chevalier and Pierre Brun from CEA Saclay.
More generally, I thank all the people from SYRTE with whom I exchanged directly or indirectly during these three years. 
\newline
I would like to express my gratitude to all the members of my PhD jury, Franck Pereira Dos Santos, Diego Blas, Hartmut Grote, Clare Burrage, Antoine Petiteau and Marianna Safronova, for having accepted reading the manuscript and attending the defense. I am very glad to have a jury of such quality and I can only hope to have the opportunity to discuss more with them on those dark matter searches in the future.
\newline
Finally, I am grateful to all my friends and family, especially my parents for their unfailing support in my quest of becoming a researcher in fundamental physics. It has been a long journey, full of obstacles, but they always believed in me, and for this, I thank them with all my heart.

\appendix
\part{Appendix}

\chapter{\label{ap:cavities}Electric fields inside a cavity}
\section{\label{ap:phase_shift_cavity}Phase shift induced by vacuum birefringence}

In this section, we explicitly derive Eq.~\eqref{eq:phase_final_DAMNED}, the phase of light measured at the output of a cavity, in the framework of vacuum birefringence induced by the axion-photon coupling.
We follow closely the derivation of the phase explicited in \cite{savallePhD}. We start by initializing the input electric field
\begin{equation}
\vec \epsilon(t) = \vec \epsilon_0 e^{-i\omega_0 t} \, ,
\end{equation}
with amplitude $\vec \epsilon_0$ and frequency $\omega_0$. We assume this field enters the cavity by one of the sides of the cavity. Note that in the following, all measurable electric fields are implicitly given by the real part only. When entering the cavity, the field gets transmitted by the cavity with a coefficient $\sqrt{1-r^2}$, where $r$ is the mirror reflectivity of the cavity and becomes
\begin{equation}
\vec E_0(t) = \sqrt{1-r^2} \vec \epsilon_0 e^{-i\omega_0 t} \, .
\end{equation}
After two reflections on the mirrors' surfaces, the electric field at the location of the first mirror is
\begin{equation}
\vec E_1(t) = r^2 \vec E_0(t-\tau_\leftrightarrow) = r^2 \vec E_0(t)e^{2i\frac{\omega_0}{c}\ell}e^{2i\frac{\omega_0}{c}\delta l(t-\frac{\ell}{c})}\, ,
\end{equation}
where we used Eq.~\eqref{eq:tau_ell} to express the time variation of length of the cavity from the round trip $\tau_\leftrightarrow$. After $n$ round trips, the electric field inside the cavity can be written as
\begin{equation}
\vec E_n(t) = r^{2n} \vec E_0(t) e^{2in\frac{\omega_0}{c}\ell} e^{2i\frac{\omega_0}{c}\sum_{j=1}^{n} \delta \ell(t-(2j-1)\frac{\ell}{c})}
\end{equation}
at first order in the perturbation $\delta \ell(t)$. The total electric field transmitted outside the cavity (by the cavity's opposite wall, compared to where the input electric field enters) is then
\begin{align}
\vec E^\mathrm{transmit}_\mathrm{tot}(t) &= \sum^\infty_{n=0} E^\mathrm{transmit}_n (t)= \frac{\sqrt{1-r^2} \vec E_0(t)}{r^2}\sum_{n=1}^{\infty}r^{2n}e^{2in\frac{\omega_0}{c}\ell}e^{2i\frac{\omega_0}{c}\sum_{j=1}^{n}\delta \ell^{(j)}} \, .
\end{align}
Since we are interested in the field transmitted to the second mirror (not the one from which the field initially enters the cavity), we explicitly add a factor $\sqrt{1-r^2}$ to account for the transmission to the second mirror and we remove a factor $r^2$, since in that case, the field is not reflected on the second mirror and then on the first one on his way back. This is also why we sum over the fields from $n=1$. We also defined $\delta \ell^{(j)} = \delta \ell(t-(2j-1)\frac{\ell}{c})$.
We now use the fact that the parameter $\delta \ell^{(j)}$ is small ($\mathcal{O}(g_{a\gamma}$)), allowing us to expand the exponential at first order in $\delta \ell^{(j)}$, i.e 
\begin{subequations}
\begin{align}
&\sum_{n=1}^{\infty}r^{2n}e^{2in\frac{\omega_0}{c}\ell}e^{2i\frac{\omega_0}{c}\sum_{j=1}^{n}\delta \ell^{(j)}}=\sum_{n=1}^{\infty}r^{2n}e^{2ink_0\ell}+2ik_0\sum_{n=1}^{\infty}r^{2n}e^{2ink_0\ell}\sum_{j=1}^{n}\delta \ell^{(j)} \,\\
&= \sum_{n=1}^{\infty}r^{2n}e^{2ink_0\ell}+2ik_0\left(\delta \ell^{(1)}\sum_{n=1}^{\infty}r^{2n}e^{2ink_0\ell}+\delta\ell^{(2)}\sum_{n=2}^{\infty}r^{2n}e^{2ink_0\ell}+...\right) \,\\
&= \left(r^2e^{2ik_0\ell}+2ik_0r^2e^{2ik_0\ell}\delta\ell^{(1)}+2ik_0r^4e^{4ik_0\ell}\delta\ell^{(2)}+...\right)\sum_{n=0}^{\infty}r^{2n}e^{2ink_0\ell}\,\\
&=r^2e^{2ik_0\ell}\left(1+\frac{2ik_0}{r^2}\sum_{n=1}^{\infty}\delta \ell^{(n)}r^{2n}\right)\sum_{n=0}^{\infty}r^{2n}e^{2ink_0\ell}\,\\
&=\frac{r^2e^{2ik_0\ell}}{1-e^{2ik_0\ell}r^2}e^{\frac{2ik_0}{r^2}\sum_{n=1}^{\infty}\delta \ell^{(n)}r^{2n}}
\end{align}
\end{subequations}
where $k_0=\omega_0/c$ and where at the last line, we inverted the Taylor expansion, such that the total transmitted electric field becomes
\begin{align}
\vec E^\mathrm{transmit}_\mathrm{tot}(t) &= \frac{\sqrt{1-r^2} \vec E_0(t) e^{2ik_0\ell}}{1-e^{2ik_0\ell}r^2}e^{\frac{2ik_0}{r^2}\sum_{n=1}^{\infty}\delta \ell^{(n)}r^{2n}}
\end{align}
We recover the usual form of electromagnetic plane wave where the amplitude $\vec A(t)$ and phase $\phi(t)$ are respectively given by
\begin{subequations}
\begin{align}
\vec A(t) &= \frac{(1-r^2) e^{2ik_0\ell}}{1-e^{2ik_0\ell}r^2}\vec \epsilon_0 \,\\
\phi(t)&=\omega_0 t - \frac{2k_0}{r^2}\sum_{n=1}^{\infty}\delta \ell^{(n)}r^{2n} \, .
\end{align}
\end{subequations}
When the light source is locked on a mode of the cavity, i.e $k_0 \ell = \pi m$, $m\in \mathbb{N}$, the transmitted amplitude is equal to the input amplitude.  
Using Eq.~\eqref{delta_l}, the phase shift between the transmitted field and the input oscillates at the axion frequency $\omega_a$ whose amplitude is 
\begin{subequations}
\begin{align}
|\Delta \phi(t)| &= \left|\frac{2 k_0}{r^2} \sum_{n=1}^{\infty}r^{2n}\delta \ell^{(n)}\right|\,\\
&=\frac{2\sqrt{16\pi G \rho_\mathrm{DM}} E_P g_{a\gamma}}{\omega_a c}\sin^2\left(\frac{\omega_a \ell}{2c}\right)\frac{\left|\cos(\omega_a t-\frac{2\omega_a \ell}{c}+\Phi)-r^2\cos(\omega_a t+\Phi)\right|}{1-2r^2\cos\left(\frac{2\omega_a \ell}{c}\right)+r^4}\,\\
&\approx \frac{2\sqrt{16\pi G \rho_\mathrm{DM}} E_P g_{a\gamma}}{\omega_a c\sqrt{1-2r^2\cos\left(\frac{2\omega_a \ell}{c}\right)+r^4}}\sin^2\left(\frac{\omega_a \ell}{2c}\right) \, .
\end{align}
\end{subequations}

\section{\label{ap:amp_field_cavity}Amplitude of the square of the total electric field inside a cavity}

In this appendix, we derive the amplitude of the electric field power inside a cavity Eq.~\eqref{eq:Etot2} in the context of the experiment presented in Chapter ~\ref{chap:Rydberg_exp_DP}.
From Eq.~\eqref{general_E_power}, we can write the signal amplitude as
\begin{subequations}
\begin{align}
  &\sqrt{\left(\vec A(\omega_A)\cdot \vec C(\omega_U)+\vec B(\omega_A)\cdot \vec D(\omega_U)\right)^2+{\left(\vec B(\omega_A)\cdot \vec C(\omega_U)-\vec A(\omega_A)\cdot \vec D(\omega_U)\right)^2}}\,\\
  &\equiv X_A X_\mathrm{DM}\beta\sqrt{\left( A'(\omega_A)^2+B'(\omega_A)^2\right)\times\left(C'(\omega_U)^2+ D'(\omega_U)^2\right)}
\end{align}
where 
\begin{align}
    \beta &= \hat e_\mathrm{DM} \cdot \frac{\vec X_A}{X_A} \, ,
\end{align}
\end{subequations}
is the projection of the polarization of the DP field on the polarization of the injected electric field, such that
$\vec X_A \cdot \vec X_\mathrm{DM} = \vec X_A \cdot \vec X_\mathrm{DM,\parallel} = X_A X_\mathrm{DM} \beta$ and where the prime quantities are defined as  
\begin{subequations}
\begin{align}
A'(\omega_A) &\equiv \frac{\sqrt{1-r^2}\left(1+r\right)\cos\left(\frac{\omega_AL}{2c}\right)}{1+2r\cos(\frac{\omega_AL}{c})+r^2}\, , \\
B'(\omega_A) &\equiv \frac{\sqrt{1-r^2}\left(1-r\right)\sin\left(\frac{\omega_AL}{2c}\right)}{1+2r\cos(\frac{\omega_AL}{c})+r^2}\, , \\
C'(\omega_U) &\equiv 1 +\frac{2(1+r)\cos(\frac{\omega_U L}{2c})}{1+2r\cos(\frac{\omega_U L}{c})+r^2}\, , \\
D'(\omega_U) &\equiv \frac{2(1-r)\sin(\frac{\omega_U L}{2c})}{1+2r\cos(\frac{\omega_U L}{c})+r^2}\, ,
\end{align}
\end{subequations}
i.e the polarizations are factorized from the amplitude functions $\{\vec A, \vec B, \vec C, \vec D\}$. The signal amplitude can be easily simplified to 
\begin{align}
    \frac{\sqrt{1-r^2}X_AX_\mathrm{DM}\beta}{\sqrt{1+2r\cos(\frac{\omega_A L}{c})+r^2}}\sqrt{1+4\frac{1+(1+r)\cos(\frac{\omega_U L}{2c})}{1+2r\cos(\frac{\omega_U L}{c})+r^2}} \, .
\end{align}

\section{\label{ap:RIN_amplitude}Amplitude fluctuation inside a cavity}

In this section, we show how to express the contribution in field power from the amplitude fluctuation of the applied field $\Delta X_A$ Eq.~\eqref{general_E_power_noise}.

We use the same procedure as the one described in Section ~\ref{Total_E_field_DP} to compute the RIN contribution to the total electric field at the center of the cavity. Starting from its expression in Eq.~\eqref{stochastic_noise_single}, the first contribution of noise reads
\begin{align}
\left[\vec E^\mathrm{0}_A\right]_\mathrm{RIN}(x=0,t) &= \Re \left[\sqrt{1-r^2}\left(\frac{\Delta \vec{X}_A(\Delta \omega)}{2}\left(e^{-i(\omega_+t-k_+\frac{L}{2}+\phi_+)}+e^{-i(\omega_-t-k_-\frac{L}{2}+\phi_-)}\right)\right)\right]\, ,
\end{align}
where we used the trigonometric identity $2\cos(A)\cos(B) = \cos(A+B)+\cos(A-B)$. Here, the only angular frequency noise we are interested in is $\omega_0 = \Delta \omega$, hence $k_\pm = k_A \pm \Delta k$, $\Delta k = \Delta \omega/c$ and $\Delta \vec X_A$ is evaluated at $\Delta \omega$. After getting reflected on the other boundary, the second contribution is
\begin{align}
\left[\vec E^\mathrm{1}_A\right]_\mathrm{RIN}(x=0,t) &= \Re\left[-\sqrt{1-r^2}re^{ik_AL}\left(\frac{\Delta \vec{X}_A(\Delta \omega)}{2}\left(e^{-i(\omega_+t-k_+\frac{L}{2}+\phi_+)}+e^{-i(\omega_-t-k_-\frac{L}{2}+\phi_-)}\right)\right)\right]\, , 
\end{align}
and after an infinite number of round trips N, the full RIN contribution on the electric field inside the cavity is
\begin{subequations}
\begin{align}
&\left[\vec E^\mathrm{tot}_A\right]_\mathrm{RIN}(x=0,t) = \sum_{n=0}^{N=+\infty} \vec E^{n}_\mathrm{RIN}(x=0,t) \,\\
&=\sqrt{1-r^2}\Delta \vec X_A(\Delta \omega)\left(\Re \left[e^{-i(\omega_+t+\phi_+)}\frac{e^{i\frac{k_+L}{2}}}{1+re^{ik_+L}} + e^{-i(\omega_-t+\phi_-)}\frac{e^{i\frac{k_-L}{2}}}{1+re^{ik_-L}}\right]\right)\, \\
&= \frac{\Delta X_A(\Delta \omega)}{2 X_A}\sum_{i=\pm}\left(\vec A(\omega_i)\cos(\omega_it+\phi_i)+\vec B(\omega_i)\sin(\omega_it+\phi_i)\right) \, \\
&=\frac{\Delta X_a(\Delta \omega)}{2X_A}\left( \vec A(2\omega_A-\omega_U)\cos([2\omega_A-\omega_U]t+\phi_+) +\vec A(\omega_U)\cos(\omega_U t+\phi_-) +\nonumber \right.\\
& \left.\vec B(2\omega_A-\omega_U)\sin([2\omega_A-\omega_U]t+\phi_+) +\vec B(\omega_U)\sin(\omega_U t+\phi_-)\right)
\end{align}
\end{subequations}
where the functions $\vec A$ and $\vec B$ have already been defined in Eqs.~\eqref{eq:A_B} and where we used $\omega_\pm = \omega_A \pm \Delta \omega$ at the last line, and where $\phi_\pm=\phi_A \pm\phi_0$. This expression has to be directly added to the noise-less contribution from Eq.~\eqref{full_field_applied_simplified}.

Then, the RIN contribution to the square of the total electric field at the center of the cavity can be obtained by multiplying the last equation with Eq.~\eqref{full_field_applied_simplified}, as we work at first order in $\Delta X_A/X_A$. Keeping only the terms oscillating at angular frequency $\Delta \omega$, the RIN contribution to $E^2$ is given by
\begin{subequations}
\begin{align}
&\left[E^2(\omega_U,\omega_A)\right]_\mathrm{RIN}=\frac{\Delta X_A(\Delta \omega)}{2X_A}\left(\left(A(\omega_A)\left[A(2\omega_a-\omega_U)+A(\omega_U)\right]+B(\omega_A)\left[B(2\omega_A-\omega_U)+B(\omega_U)\right]\right)\right.\nonumber \\
&\left.\cos(\Delta \omega t+\phi_0) +\left(A(\omega_A)\left[B(2\omega_A-\omega_U)-B(\omega_U)\right]+B(\omega_A)\left[A(\omega_U)-A(2\omega_A-\omega_U)\right]\right)\sin(\Delta \omega t+\phi_0)\right)  \label{full_E_power_noise}\\
& = \frac{\Delta X_A (\Delta \omega)}{2X_A}\sqrt{N(\omega_U,\omega_A)} \cos(\Delta \omega t + \varphi)= \sqrt{\frac{P_\mathrm{RIN} N(\omega_U,\omega_A)}{8 T_\mathrm{obs} \Delta \omega}} \cos(\Delta \omega t + \varphi)\, ,
\end{align}
\end{subequations}
where we have used Eqs.~\eqref{eq:PSD_noise} and \eqref{eq:flicker_noise}. In this expression, the functions $A$ and $B$ are the norm of $\vec A$ and $\vec B$ and $\varphi$ is an irrelevant phase dependent on $A$ and $B$. 

The function $\sqrt{N(\omega_U,\omega_A)}$ correspond to the noise amplification factor by the cavity, quadratic in $X_A$ and whose expression is given by 
\begin{subequations}
\begin{align}
\mathrm{N}(\omega_U,\omega_A) &\equiv \mathrm{N}_1(\omega_U,\omega_A)+\mathrm{N}_2(\omega_U,\omega_A)+\mathrm{N}_3(\omega_U,\omega_A)\, ,
\end{align}
where
\begin{align}
\mathrm{N}_1(\omega_U,\omega_A) &\equiv A(\omega_A)^2\left(\left(A(2\omega_A-\omega_U)+A(\omega_U)\right)^2+\left(B(2\omega_A-\omega_U)-B(\omega_U)\right)^2\right)\, , \\
\mathrm{N}_2(\omega_U,\omega_A) &\equiv B(\omega_A)^2\left(\left(A(2\omega_A-\omega_U)-A(\omega)\right)^2+\left(B(2\omega_A-\omega_U)+B(\omega_U)\right)^2\right)\, , \\
\mathrm{N}_3(\omega_U,\omega_A) &\equiv 4A(\omega_A)B(\omega_A)\left(A(2\omega_A-\omega_U)B(\omega_U)+B(2\omega_A-\omega_U)A(\omega_U)\right)\, .
\label{noise_amp}
\end{align}
\end{subequations}
\chapter{\label{ap:pos_vel_atoms_AI}Motion of atomic wavepackets inside an atom interferometer}
In this appendix, we compute rigorously the wavepackets position and velocity along the $\pi/2-\pi-\pi/2$ interferometer, which will be used for the computation of the phase shift observable in Chapter ~\ref{chap:AI}.

We start by the calculation of the position and velocity of the atom at the end of the trajectory portion 1 in Fig.~\ref{Mach-Zehnder_perturbed}. From Eq.~\eqref{EoM_atom_AI}, the atom does not undergo any kick velocities from the laser pulses, hence after a time T, its equations of motion read 
\begin{subequations}
    \begin{align}
        \vec x^{(1)}_A(t=T) &= \vec v_\mathrm{DM}\left(T-\frac{[Q^A_M]_d X_\mathrm{DM}}{\omega_\phi}[\sin(\omega_\phi T+\Phi)-\sin(\Phi)- \omega_\phi T\cos(\Phi)]\right)\,\\
       \vec v^{(1)}_A(t=T) &= \vec v_\mathrm{DM}\left(1-[Q^A_M]_d X_\mathrm{DM}\left[\cos(\omega_\phi T+\Phi)-\cos(\Phi)\right]\right) \, .
    \end{align}
\end{subequations}
On the other hand, the wavepacket on portion 2 in Fig.~\ref{Mach-Zehnder_perturbed} has undergone a kick velocity with amplitude $v_\mathrm{kick,0}+\delta v_\mathrm{kick}(t=0)$, following Eq.~\eqref{kick_modif}, with 
\begin{align}
\delta \vec v_\mathrm{kick}(t=0) = \vec v_\mathrm{kick,0}\left([Q^L_\omega]_d-[Q^A_M]_d\right)X_\mathrm{DM}\cos(\Phi) \, ,
\end{align}
with $\vec v_\mathrm{kick,0}$ along an arbitrary direction $\hat x$, hence
\begin{subequations}
\begin{align}
    \vec x^{(2)}_A(t=T) &= (\vec v_\mathrm{DM}+\vec v_\mathrm{kick,0})T - X_\mathrm{DM}\left(2\frac{[Q^A_M]_d}{\omega_\phi}\sin\left(\frac{\omega_\phi T}{2}\right)\cos\left(\frac{\omega_\phi T}{2}+\Phi\right)(\vec v_\mathrm{DM} +\vec v_\mathrm{kick,0})+\right.\, \nonumber \\
    &\left.T ([Q^A_M]_d\vec v_\mathrm{DM}-[Q^L_\omega]_d \vec v_\mathrm{kick,0})\cos(\Phi)\right)\, , \\
    \vec v^{(2)}_A(t=T) &= \vec v_\mathrm{DM}+\vec v_\mathrm{kick,0} - X_\mathrm{DM}\left([Q^A_M]_d \cos(\omega_\phi T+\Phi)(\vec v_\mathrm{DM} +\vec v_\mathrm{kick,0})+\right. \, \nonumber \\
    &\left.\left([Q^A_M]_d\vec v_\mathrm{DM} +[Q^L_\omega]_d \vec v_\mathrm{kick,0}\right)\cos(\Phi)\right) \, ,
    \end{align}
\end{subequations}
At time T, both wavepackets undergo a kick velocity with opposite direction, such that their momenta states are exchanged. Following Eq.~\eqref{kick_modif}, this means that at the end of portion 1, the atom undergoes a kick velocity of amplitude $v_\mathrm{kick,0}+\delta v_\mathrm{kick}(t=T)$, with
\begin{align}
    \delta \vec v_\mathrm{kick}(t=T)=\vec v_\mathrm{kick,0}\left([Q^L_\omega]_d-[Q^A_M]_d\right)X_\mathrm{DM}\cos(\omega_\phi T+\Phi) \, ,
\end{align}
with same direction $\hat x$, hence, at the end of the portion 3, its coordinates read
\begin{subequations}
\begin{align}
    &\vec x^{(3)}_A(t=2T) = (2\vec v_\mathrm{DM}+\vec v_\mathrm{kick,0})T+X_\mathrm{DM}\left(\frac{[Q^A_M]_d}{\omega_\phi}\left(\vec v_\mathrm{DM}(\sin(\Phi)+2\omega_\phi T\cos(\Phi))+\right.\right.\,\\
    &\left.\left.\vec v_\mathrm{kick,0}\sin(\omega_\phi T+\Phi)-(\vec v_\mathrm{DM}+\vec v_\mathrm{kick,0})\sin(2\omega_\phi T+\Phi)\right)+[Q^L_\omega]_d \vec v_\mathrm{kick,0}T \cos(\omega_\phi T + \Phi)\right) \, \nonumber , \\
    &\vec v^{(3)}_A(t=2T) = \vec v_\mathrm{DM}+\vec v_\mathrm{kick,0} - X_\mathrm{DM}\left([Q^A_M]_d \left((\vec v_\mathrm{DM}+\vec v_\mathrm{kick,0})\cos(2\omega_\phi T+\Phi)-\vec v_\mathrm{DM}\cos(\Phi)\right)+\right.\,\nonumber \\
    &\left.[Q^L_\omega]_d \vec v_\mathrm{kick,0}\cos(\omega_\phi T + \Phi)\right)\, .
    \label{v3_2T}
\end{align}
\end{subequations}
The atom at the end of the portion 2 undergoes a kick velocity in the other direction, i.e of amplitude $v_\mathrm{kick,0}+\delta v_\mathrm{kick}(t=T)$, but with opposite direction (-$\hat x$), compared to the previous laser kicks, hence  
\begin{subequations}
\begin{align}
    &\vec x^{(4)}_A(t=2T) = (2\vec v_\mathrm{DM}+\vec v_\mathrm{kick,0})T-X_\mathrm{DM}\left(\frac{[Q^A_M]_d}{\omega_\phi}\left(\vec v_\mathrm{kick,0}\sin(\omega_\phi T+\Phi)+\vec v_\mathrm{DM}\sin(2\omega_\phi T+\Phi)-\right.\right.\, \nonumber \\
    &\left.\left.(\vec v_\mathrm{DM}+\vec v_\mathrm{kick,0})\sin(\Phi)-2\vec v_\mathrm{DM}\omega_\phi T\cos(\Phi)\right)-[Q^L_\omega]_d \vec v_\mathrm{kick,0} T \left(\cos(\omega_\phi T +\Phi)-2\cos(\Phi)\right)\right)\, , \\
    &\vec v^{(4)}_A(t=2T) = \vec v_\mathrm{DM}\left(1 + 2 X_\mathrm{DM} [Q^A_M]_d\sin(\omega_\phi T)\right)\sin(\omega_\phi T + \Phi) + 2X_\mathrm{DM} [Q^L_\omega]_d \vec v_\mathrm{kick,0}\sin\left(\frac{\omega_\phi T}{2}\right)\,\nonumber\\
    &\sin\left(\frac{\omega_\phi T}{2}+\Phi\right)\, ,
\end{align}
\end{subequations}
at the end of portion 4.

Since at the end of portion 3, the wavepacket is in the excited state, we must take into account an additional kick at time $t=2T$ of amplitude $v_\mathrm{kick,0}+\delta v_\mathrm{kick}(t=2T)$ and direction -$\hat x$ to this wavepacket, in order to put it back to the ground state. Since, we assume the detection to be immediately after the kick, the position of the wavepacket will not be impacted by this additional kick. The final velocity of this wavepacket read
\begin{align} 
    \vec v^{(4')}_A(2T) &= \vec v_\mathrm{DM}\left((1 - X_\mathrm{DM}[Q^A_M]_d\left(\cos(2\omega_\phi T+\Phi)-\cos(\Phi)\right)\right)- \, \nonumber\\
    &X_\mathrm{DM}[Q^L_\omega]_d \vec v_\mathrm{kick,0}\left(\cos(2\omega_\phi T + \Phi)-\cos(\omega_\phi T +\Phi)\right) \, ,
    \label{v_D2}
\end{align}
resulting in a different velocity compared to the other wavepacket ($\vec v^{(4)}_A(2T)$). 

\chapter{\label{ap:dot_product_AI}Projection of galactic velocity onto sensitive axes of various experiments}
In this appendix, we will compute the factors that are orientation dependent in the various acceleration or phase shifts observables, presented in Chapters ~\ref{chap:Classical_tests_UFF} and \ref{chap:AI}. 

The observable signatures depend on the orientation of the experimental setup with respect to the initial velocity of the atoms or tests masses, which corresponds at leading order to the galactic velocity $v_\mathrm{DM}$. Indeed, for classical tests of the universality of free fall between two macroscopic bodies, one measures $\Delta \vec a \propto \vec v_\mathrm{DM}$ (see Eq.\eqref{delta_a_UFF_osc}) projected onto the sensitive axis of the instrument. Therefore, the signal depends on the projection of the velocity of the bodies in the galactocentric frame with the sensitive axis of the experiment. Similarly, all the AI experiments depend on the scalar product between the initial galactocentric velocity of the atomic clouds and the direction of the velocity kick undergone in the interferometric schemes, i.e. on $\hat e_v \cdot \hat e_\mathrm{kick}$, see Eqs.~\eqref{eq:MZ_delta_phase_shift} and \eqref{eq:phase_new_prop}.

As mentioned in Chapter ~\ref{chap:LISA_DM}, the direction of the DM velocity in the galactic halo points towards $\alpha_\mathrm{DM},\delta_\mathrm{DM}=310.36\degree \mathrm{E},45.28\degree \mathrm{N}$ \cite{vanLeeuwen07} in the equatorial frame.

\section{Atom interferometers}

All the AI-based experiments operate at constant location, \textit{loc}, on Earth with longitude $\lambda_\mathrm{loc}$ and latitude $\phi_\mathrm{loc}$. We assume the velocity kick to be directed vertically, i.e. in the Earth's reference frame $\hat e_\mathrm{kick}=\left(\cos (\lambda_\mathrm{loc})\cos (\phi_\mathrm{loc}),\sin (\lambda_\mathrm{loc})\cos (\phi_\mathrm{loc}),\sin (\phi_\mathrm{loc})\right)$. As first approximation, we consider the declination as equivalent to the terrestrial latitude, i.e $\delta_\mathrm{DM}\approx \phi_\mathrm{DM}$, such that the dot product $\hat e_v \cdot \hat e_\mathrm{kick}$  is simply given by $\cos(\phi_\mathrm{loc})\cos(\phi_\mathrm{DM})\cos(\lambda_\mathrm{loc}-\lambda_\mathrm{DM}(t))+\sin(\phi_\mathrm{loc})\sin(\phi_\mathrm{DM})$. $\lambda_\mathrm{DM}(t)$ is the longitude of $\alpha$ Cygni at the time $t$ of the experiment. Indeed, while $\alpha_\mathrm{DM}$ is fixed, the former follows the Earth rotation with frequency $\omega_E \sim 7 \times 10^{-5}$ Hz and is therefore time dependent
\begin{subequations}
\begin{align}
    \lambda_\mathrm{DM}(t)&=\omega_E t+\varphi \, ,
\end{align}
where the phase $\varphi$ corresponds to the longitude of $\alpha$ Cygni at the origin of time reference considered. 
For short experiments with an experimental time  much smaller than a day (i.e. relevant for Stanford's experiment, see Sec~\ref{sec:stanford}), the dot product is roughly constant and depends therefore on the exact time of the day when the experiment was conducted. In order to infer a sensitivity estimate, we will only consider the mean value of the dot product which is given by $\sin(\phi_\mathrm{loc})\sin(\phi_\mathrm{DM})$, i.e
\begin{align}\label{dot_product_DM_axis_1}
    \hat e_v \cdot \hat e_\mathrm{kick}\Big|_\mathrm{Stanford} &\approx 0.43 \, .
\end{align}
For longer time experiments, like the \textit{SPID} variation, the dot product evolves with time, such that the signal evolves as 
\begin{align}
    s(t) &\propto \hat e_v \cdot \hat e_\mathrm{kick} \cos(\omega_\mathrm{DM}t +\Phi)= [\cos(\phi_\mathrm{loc})\cos(\phi_\mathrm{DM})\cos(\omega_E t +\varphi)+\,\nonumber\\
    &\sin(\phi_\mathrm{loc})\sin(\phi_\mathrm{DM})] \cos(\omega_\mathrm{DM}t +\Phi) , 
\end{align}
i.e the Earth rotation modulates the signal at frequency $f_E=\omega_E/2\pi$. We will make two approximations for our estimates. For all the DM frequencies of interest $\omega_\mathrm{DM}$, the Earth rotation is a slowly varying function, because $\omega_E < \omega_\mathrm{DM}$. Therefore, we will assume that the signal will manifest itself by a single peak at frequency $f_\mathrm{DM}= \omega_\mathrm{DM}/2\pi$ in Fourier space. In addition, we will be interested only by the maximum value of the dot product, which is only a function of the different latitudes.
\begin{align}
    \hat e_v \cdot \hat e_\mathrm{kick} = \mathrm{Max}\left(|\cos(\phi_\mathrm{loc}\pm \phi_\mathrm{DM})|\right) \, .
\end{align}
For the various locations under consideration in this thesis, we have
\begin{align}\label{dot_product_DM_axis_2}
\hat e_v \cdot \hat e_\mathrm{kick}\Big|_\mathrm{Oxford} &\approx 0.99 \, , \\
\hat e_v \cdot \hat e_\mathrm{kick}\Big|_\mathrm{Fermilab} &\approx 1.00 \, ,
\end{align}
where we assume the \textit{SPID} variation operates at the same location as \textit{AION-10}, i.e Oxford \cite{Badurina20} for consistent comparison. 

\section{\textit{MICROSCOPE}}

For \textit{MICROSCOPE}, the axis of measurement is alongside the test masses cylinders' longitudinal symmetry axis \cite{Microscope17}. The orbital motion of the satellite around Earth is sun-synchronous, which means that the orientation of the orbital plane evolves with time with an annual period. As it was mentioned in Section ~\ref{sec:exp_param_MICROSCOPE}, the measurements are distributed on 17 different sessions, each of them lasting $T_\mathrm{int}/17 \sim 5$ days on average. For such duration, we can assume the orbital plane to be fixed during each session. In addition, the satellite spins around an axis that is orthogonal to the axis of measurement with angular frequency $\omega_\mathrm{spin}$. In total, this means the dot product can be written as 
\begin{align}
    \hat e_v \cdot \hat e_\mathrm{meas.}(t)\Big|_\mathrm{\mu SCOPE} &= A(t) \cos(\omega_\mathrm{spin} t + \psi) \, ,
\end{align}
where $A(t)$ depends on the orientation of the orbital plane, i.e is fixed for one session but changes from one session to another, and with $\psi$ an irrelevant phase.
Using \textit{MICROSCOPE}'s publicly available data, we estimated numerically the coefficient $A(t)$ for every session and we find that its value oscillates between 0.71 and 1. For our estimates, we will consider its mean value, i.e $|\hat e_v \cdot \hat e_\mathrm{meas.}| \approx 0.85$. In conclusion, the signal presented in Eq.~\eqref{delta_a_UFF_osc} is modulated by the factor
\begin{align}
    \hat e_v \cdot \hat e_\mathrm{meas.}(t)\Big|_\mathrm{\mu SCOPE} &\approx 0.85 \cos(\omega_\mathrm{spin} t + \psi) \label{dot_product_DM_MICRO} \, .
\end{align}
\end{subequations}
Thus, the observable signal in the \textit{MICROSCOPE} experiment is oscillating at the combination of the DM frequency and the spin frequency, i.e. has harmonics at the two frequencies $\omega_\mathrm{DM} \pm \omega_\mathrm{spin}$. Therefore, for DM frequencies in Eq.~\eqref{delta_a_UFF_osc} such that $\omega_\mathrm{DM} \ll \omega_\mathrm{spin}$, we will consider that the signal in \textit{MICROSCOPE} oscillates at $\omega_\mathrm{spin}$, whereas if $\omega_\mathrm{DM} \gg \omega_\mathrm{spin}$, we will make our estimate with the signal oscillating at $\omega_\mathrm{DM}$. As explained in Section ~\ref{sec:exp_param_MICROSCOPE}, $\omega_\mathrm{spin}$ depends on the measurement session since three different spinning frequencies have been used during the full mission. They differ by a factor 5 roughly \cite{Berge23}. For our rough sensitivity estimates, we use the data from session 404 as a basis, therefore we will assume a constant $\omega_\mathrm{spin} \sim 18.4$~mrad/s.

\chapter{\label{ap:fastDM_fastGB}Fast likelihood for scalar ultralight dark matter}
In this appendix, we derive (in time domain) the slowly oscillating component of TDI $X$ combination of a scalar ULDM signal, that will be used for the DM model in \textit{FastDM} in Section ~\ref{sec:ULDM_GW_discriminate}. 

We will make the detailed calculations for a (pure scalar) DM signal, and then we generalize to GW. 
Let us start again with the one-link response function induced by scalar DM Eq.~\eqref{eq:Doppler_scalar_DM} at time of reception of the photon of the receiver $r$ $t_r$, when it is emitted at time $t_e=t_r-L/c$ by emitter $e$ 
\begin{subequations}\label{eq:Doppler_DM}
\begin{align}
y^\mathrm{DM}_{re}(t_r) &= \left(\hat n_{re} \cdot \hat e_v\right)\frac{\sqrt{16 \pi G \rho_\mathrm{DM}}v_\mathrm{DM}[Q_M]_d}{\omega_\phi c^2}\Re\left[e^{i\phi_r}- e^{i\phi_e}\right] \,\\
&=\left(\hat n_{re} \cdot \hat e_v\right)\frac{\sqrt{16 \pi G \rho_\mathrm{DM}}v_\mathrm{DM}[Q_M]_d}{\omega_\phi c^2}\Re\left[e^{i\phi_r}\left(1- e^{-2i\alpha^\mathrm{DM}_{re}}\right)\right]\, \\
&=2\left(\hat n_{re} \cdot \hat e_v\right)\frac{\sqrt{16 \pi G \rho_\mathrm{DM}}v_\mathrm{DM}[Q_M]_d}{\omega_\phi c^2}\sin(\alpha^\mathrm{DM}_{re})\Re\left[ie^{i\phi_r}e^{-i\alpha^\mathrm{DM}_{re}}\right]\, \\
&\equiv Y_{re}(t_r) e^{i\phi_r} \, ,
\end{align}
\end{subequations}
where
\begin{subequations}
\begin{align}
\phi_r  &= \omega_\phi t_r- \vec k_\phi \cdot \vec x_r(t_r) + \Phi\, \\
\phi_e  &= \omega_\phi t_e  -\vec k_\phi \cdot \vec x_e(t_e)  + \Phi  \, ,
\end{align}
and therefore
\begin{align}
    2\alpha^\mathrm{DM}_{re}(t_r) = \phi_r-\phi_e &= \omega_\phi\left((t_r-t_e)-\frac{\hat k \cdot \hat n_{re} |\vec v_\mathrm{DM}|}{c^2}\left|\vec x_r(t_r)-\vec x_e(t_e)\right|\right) \, \\
    &=\frac{\omega_\phi \left|\vec x_r(t_r)-\vec x_e(t_e)\right|}{c}\left(1 - \frac{\hat k \cdot \hat n_{re} |\vec v_\mathrm{DM}|}{c}\right) + \mathcal O(\mathrm{Shapiro}) \, ,
\end{align}
\end{subequations}
where we neglect the impact of the Shapiro delay on the signal.
The expression of $\alpha^\mathrm{DM}_{re}$ is therefore given by
\begin{equation}\label{eq:alpha_DM}
    \alpha^\mathrm{DM}_{re}(t_r) =\frac{\omega_\phi \left|\vec x_r(t_r)-\vec x_e(t_e)\right|}{2c}\left(1 - \frac{\hat k \cdot \hat n_{re} |\vec v_\mathrm{DM}|}{c}\right) \, .
\end{equation}

Let us now decompose the one-link response function into a slowly evolving part and a fast oscillating part. Let us  consider the observation time baseline $T_\mathrm{obs}$ such that the corresponding Fourier frequencies  are given by integer times $1/T_\mathrm{obs}$. Let us consider the Fourier frequency $q_0$ which is the closest to $f_\phi$, i.e.
\begin{equation}\label{eq:q}
	q_0= \mathrm{round}\left[f_\phi T_\mathrm{obs}\right]/T_\mathrm{obs} \, 
\end{equation}
where $\mathrm{round }\left[x\right]$ provides the closest integer to $x$. One can rewrite Eq.~(\ref{eq:Doppler_DM}) as
\begin{subequations}
\begin{equation}
    y^\mathrm{DM}_{re}(t_r) = \Re\left[\mathcal Y^\mathrm{DM}_{re}(t_r) e^{2\pi i q_0 t_r}\right] \, ,
\end{equation}
where 
\begin{equation}\label{eq:mathcal_Y_DM}
    \mathcal Y^\mathrm{DM}_{re} = 2i\left(\hat n_{re} \cdot \hat e_v\right)\frac{\sqrt{16 \pi G \rho_\mathrm{DM}}v_\mathrm{DM}[Q_M]_d}{\omega_\phi c^2}\sin(\alpha^\mathrm{DM}_{re})e^{-i\beta^\mathrm{DM}_{re}}\, ,
\end{equation}
with
\begin{align}
    \beta^\mathrm{DM}_{re}&=\alpha^\mathrm{DM}_{re}+2\pi q_0 t_r - \phi_r  \,\nonumber\\
    &= \alpha^\mathrm{DM}_{re} - \Phi + 2 \pi (q_0-f_\phi) t_r  + \frac{2\pi f_\phi}{c}\frac{|\vec v_\mathrm{DM}|}{c}\hat k\cdot \vec x_r(t_r)  \, . \label{eq:beta_DM}
\end{align}
\end{subequations}
One can do the exact same job for the GW signals starting from Eq.~\eqref{eq:Doppler_GW} to find
\begin{subequations}
\begin{align}
    \mathcal Y^\mathrm{GW}_{re} &= \frac{\mathcal{A}}{2\left(1-\hat n_{re} \cdot \hat k_\mathrm{GW}\right)}\sin(\alpha^\mathrm{GW}_{re})\Re\left[i\hat h^\mathrm{SSB}_{ij}(\imath,\Psi)\hat n^i_{re} \hat n^j_{re}e^{-i\beta^\mathrm{GW}_{re}}\right]\,\label{eq:mathcal_Y} \\ 
    \alpha^\mathrm{GW}_{re} &= \pi\left(f_\mathrm{GW} + \dot f_\mathrm{GW}\xi_r\right)|\vec x_r(t_r)-\vec x_e(t_e)|\left(1-\hat k \cdot \hat n_{re}(t_r)\right) \, \label{eq:alpha} \\
    \beta^\mathrm{GW}_{re} &= \alpha^\mathrm{GW}_{re} + \Phi_\mathrm{GW} + 2 \pi (q_0-f_\mathrm{GW}) t_r + \frac{2\pi f_\mathrm{GW}}{c}\hat k\cdot \vec x_r(t_r) -\pi \dot f_\mathrm{GW}\xi^2_r \, ,  \label{eq:beta}
\end{align}
\end{subequations}
where $\xi_r=t_r-\hat k \cdot \vec x_r/c$.

We now quickly show how to express TDI combinations in this framework. We will focus on the case of constant and equal arm-length $L$. In TDI, we encounter quantities such as
\begin{align}
    y_{ij}\left(t-\frac{L}{c}\right) = & Y_{ij}\left(t-\frac{L}{c}\right) e^{i\phi_i\left(t-\frac{L}{c}\right)}\, .
\end{align}
The quantity $Y_{ij}$ are slowly evolving with time such that $Y_{ij}(t-L/c)\approx Y_{ij}(t)$. For GW, this assumption holds if the GW amplitude slowly evolves with time, if the frequency drift of the GW is small and if the spacecraft velocity is small compared to the speed of light. For DM, this assumption holds for the same reasons, in particular, the amplitude of the wave is constant (in the approximation where the time of integration is smaller than the coherence time), and $Y_{ij}$ depends on the spacecraft velocity, which is small compared to $c$. Taking DM as an example, we can write
\begin{subequations}
\begin{align}
    \phi_i\left(t-\frac{L}{c}\right)&=\omega_\phi t-\frac{\omega_\phi L}{c} -\vec k_\phi \cdot \vec x_i\left(t-\frac{L}{c}\right) +\Phi\, \\
    & =\omega_\phi t-\frac{\omega_\phi L}{c} -\vec k_\phi \cdot \vec x_i(t) + \Phi + \mathcal{O}(v_j/c \times L/x_i) \,\\
    &\approx \phi_i(t) - \frac{\omega_\phi L}{c} \, ,
\end{align} 
\end{subequations}
where $v_j$ is the velocity of the emitting spacecraft. This leads to
\begin{align}
    y^\mathrm{DM}_{ij}\left(t-\frac{L}{c}\right) &\approx \Re\left[Y^\mathrm{DM}_{ij}(t) e^{i\phi_i(t)}e^{-\frac{i\omega_\phi L}{c}}\right] =\Re\left[\mathcal Y^\mathrm{DM}_{ij}(t)e^{-\frac{i\omega_\phi L}{c}} e^{2\pi i q_0 t}\right] \, .
\end{align}
Applying recursively this relationship leads to
\begin{equation}
    D_{ijk}y^\mathrm{DM}_{kl}(t) = \Re\left[\mathcal Y^\mathrm{DM}_{kl}(t)e^{-\frac{2i\omega_\phi L}{c}} e^{2\pi i q_0 t}\right] \, ,
\end{equation}
and so on. Using this result, the first generation TDI combination can be written as 
\begin{subequations}
\begin{equation}
    X^\mathrm{DM}_1(t)=\Re\left[\mathcal X^\mathrm{DM}_1 (t) e^{2\pi i q_0 t}\right] \, ,
\end{equation}
and $\mathcal X^\mathrm{DM}_1$ is slowly evolving with time and is given by (using Eq.~\eqref{eq:TDI_X_gen}) 
\begin{align}
\mathcal X^\mathrm{DM}_1 &= \left(1-e^{\frac{-2i\omega_\phi L}{c}} \right) \left[\mathcal Y^\mathrm{DM}_{13} -\mathcal Y^\mathrm{DM}_{12} + e^{\frac{-i\omega_\phi L}{c}}\left(\mathcal Y^\mathrm{DM}_{31}-\mathcal Y^\mathrm{DM}_{21}\right)\right] \nonumber \\
 &= 2i \sin \left(\frac{\omega_\phi L}{c}\right)e^{\frac{-i\omega_\phi L}{c}} \left[\mathcal Y^\mathrm{DM}_{13} -\mathcal Y^\mathrm{DM}_{12} + e^{\frac{-i\omega_\phi L}{c}}\left(\mathcal Y^\mathrm{DM}_{31}-\mathcal Y^\mathrm{DM}_{21}\right)\right]\, .\label{eq:X1_sl}
\end{align}
\end{subequations}
Similarly the second generation TDI can be written as 
\begin{subequations}
\begin{equation}
    X^\mathrm{DM}_2(t)=\Re\left[\mathcal X^\mathrm{DM}_2 (t) e^{2\pi i q_0 t}\right]\, ,
\end{equation}
with the slowly evolving part
\begin{align}
\mathcal X^\mathrm{DM}_2 = &\left(1-e^{\frac{-2i\omega_\phi L}{c}}-e^{\frac{-4i\omega_\phi L}{c}}+e^{\frac{-6i\omega_\phi L}{c}}\right) \left[\mathcal Y^\mathrm{DM}_{13} -\mathcal Y^\mathrm{DM}_{12} + e^{\frac{-i\omega_\phi L}{c}}\left(\mathcal Y^\mathrm{DM}_{31}-\mathcal Y^\mathrm{DM}_{21}\right)\right]\, .
\end{align}
Since $1-x-x^2+x^3=(1+x)(1-x)^2$, the previous expression becomes simply
\begin{align}
     \mathcal X^\mathrm{DM}_2 = &\left(1+e^{\frac{-2i\omega_\phi L}{c}}\right)\left(1-e^{\frac{-2i\omega_\phi L}{c}}\right)^2\left[\mathcal Y^\mathrm{DM}_{13} -\mathcal Y^\mathrm{DM}_{12} + e^{\frac{-i\omega_\phi L}{c}}\left(\mathcal Y^\mathrm{DM}_{31}-\mathcal Y^\mathrm{DM}_{21}\right)\right] \nonumber\\
    = &\left(1+e^{\frac{-2i\omega_\phi L}{c}}\right)\left(1-e^{\frac{-2i\omega_\phi L}{c}}\right) \mathcal X^\mathrm{DM}_1 =2 i \sin \left(\frac{2\omega_\phi L}{c}\right) e^{\frac{-2i\omega_\phi L}{c}}\mathcal X^\mathrm{DM}_1\, . \label{eq:X2_sl}
\end{align}
\end{subequations}
We have now expressed the second generation TDI combination in time domain induced by scalar DM as function of a slowly evolving part (the $\mathcal{X}$ factors) and a fast oscillating part, oscillating at the closest Fourier bin to the wave Compton frequency.
As mentioned previously, this formulation is very convenient because in Fourier domain, the signal is simply the product of the Fourier transform of $\mathcal{X}$, which is numerically fast to obtain, and the Fourier transform of the $\exp(2\pi i q_0 t)$, which has an analytical solution (see \cite{Cornish07}).
These expressions are directly implemented in \textit{FastGB} and \textit{FastDM} softwares for the fit of both GB and scalar ULDM data.
\chapter{\label{ap:SHUKET_Vinet_approx}Propagation of the electric field from the dish to the fictional plane in \textit{SHUKET}}
The calculation presented in Section ~\ref{sec:propag_field} is based on the thin optical element approximation, which is valid only for low curvature dish $R\gg r$, see the discussion in Section ~\ref{sec:dish_plane}. 
In this appendix, we want to check if Eq.~\eqref{Vinet} is valid in the \textit{SHUKET} setup, in the case of emission from standing DP electric field. To do so, we propose the following "reverse engineering-like" method : 
\begin{enumerate}
\item Consider the field on the dish using boundary conditions Eq.~\eqref{eq:E_out}, i.e $\vec U_D(\vec x) = i\chi \omega_U \vec Y_{\parallel, D}(\vec x)$,
\item Compute the field at each point $\vec x'$ on the plane closing the dish $\vec U_P(\vec x')$ from Eq.~\eqref{Vinet},
\item Compute numerically the field on the dish $\vec U^\mathrm{test}_D(\vec x)$ from $\vec U_P(\vec x')$ using Kirchhoff integral theorem Eq.~\eqref{E_Dirichlet},
\item Compute the relative error between $\vec U_D(\vec x) $ and $\vec U^\mathrm{test}_D(\vec x)$.
\end{enumerate}

We consider the plane located at $z'=R-a$, then the field at a point $(\rho', \phi',z')$ on the plane going towards the dish, located at $z \geq z'$, is given by
\begin{align}
\vec U_\mathrm{P\rightarrow D}(\rho',\phi',z') &= i\chi \omega_U e^{-ikf(\rho')} \vec Y_{\parallel, D}(\rho',\phi',f(\rho')+z') \, .
\label{E_plane_Vinet}
\end{align}
Notice the change of sign on the wavevector $k$, compared to Eq.~\eqref{Vinet} as we are now considering emission towards the dish (positive $z$ axis). In addition, we aim at computing the field reflected by the dish, while Eq.~\eqref{Vinet} gives the incident field. By boundary conditions, we assume that both incident and reflected are equal, up to a sign, hence the positive sign in front of Eq.\eqref{E_plane_Vinet}.

Using Eqs.~\eqref{E_Dirichlet}, \eqref{Green_plane} and \eqref{E_plane_Vinet}, the field in a point on the receiving surface, i.e, the dish, of coordinates $(\rho,\phi,z)$ is given by 
\begin{subequations}
\begin{align}
\vec U^\mathrm{test}_D(\rho,\phi,f(\rho)+z') &= -\frac{i\chi \omega_U f(\rho)}{2\pi}\int_0^r  d\rho' \rho'e^{-ikf(\rho')}\int_0^{2\pi}d\phi' \frac{ikL-1}{L^3}e^{ikL}\vec Y_{\parallel, D}(\rho',\phi', f(\rho')+z') \,\\
&\approx -\frac{i\chi \omega_U f(\rho)}{2\pi}\begin{pmatrix}
Y_x\,\\
Y_y\,\\
0
\end{pmatrix}\int_0^r  d\rho' \rho'e^{-ikf(\rho')}\int_0^{2\pi} d\phi' \frac{ikL-1}{L^3}e^{ikL} \, ,
\end{align}
\end{subequations}
where the difference of $z$ positions $\Delta z$ between the fictional plane and the dish from Eq.~\eqref{Green_plane} is now positive and corresponds exactly to $f(\rho)$, $\Delta z=f(\rho) = (r^2-\rho^2)/2R > 0$.
Then, the relative error between $\vec U^\mathrm{test}_D$ and $\vec U_D \approx i \chi \omega_U \left(Y_x,Y_y,0\right)^T$ (from Eq.~\eqref{field_polarization}) is 
\begin{subequations}
\begin{align}\label{error_Vinet}
    &\epsilon(\rho,\phi) \approx \left|\frac{\vec U^\mathrm{test}_D - \vec U_D}{\vec U_D}\right|\approx \left|-\frac{f(\rho)}{2\pi}\left(\int_0^r  d\rho' \rho'e^{-ikf(\rho')}\int_0^{2\pi} d\phi' \frac{ikL-1}{L^3}e^{ikL}\right) - 1 \right| \, .
\end{align}
\end{subequations}
\begin{figure}[h!]
\begin{minipage}{\textwidth}
  \begin{minipage}{0.65\textwidth}
    \centering
    \includegraphics[width=\textwidth]{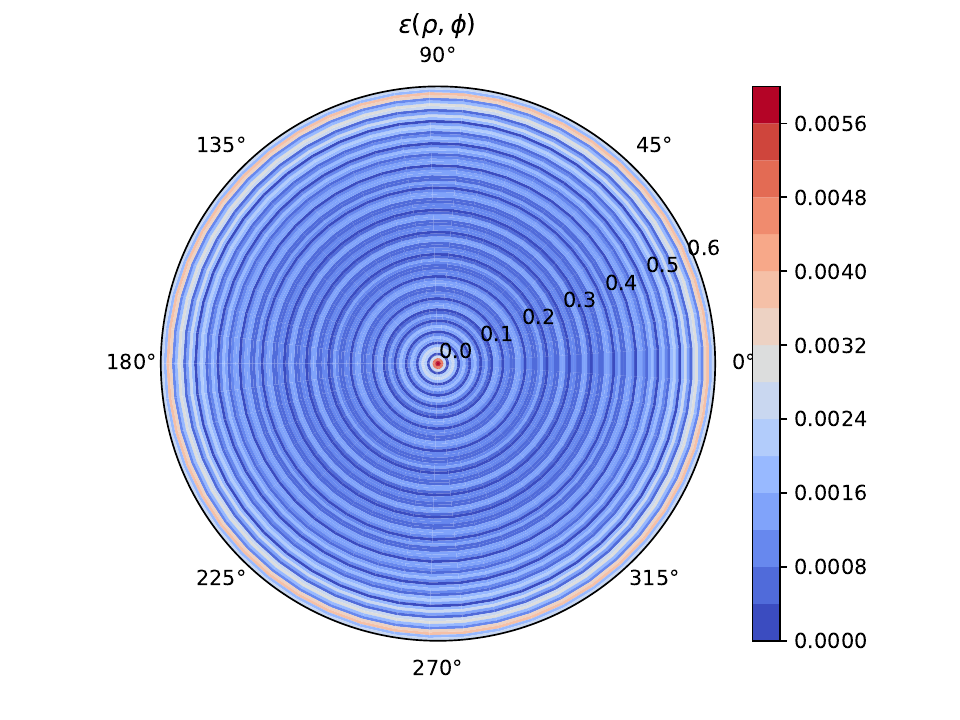}
    \end{minipage}
    \hfill
    \begin{minipage}{0.32\textwidth}
    \caption{Numerical computation of $\epsilon(\rho,\phi)$ from Eq.~\eqref{error_Vinet} for various dish positions $(\rho,\phi)$. Over the full surface, it is of the order of the \textperthousand. The dish antenna is located inside the non-radiative near field region of the fictional plane (i.e the distance $L<0.62 \sqrt{(2r)^3/\lambda}$). One can notice the interference pattern with a typical period $\sim 0.05$ m, which corresponds to the wavelength of emission.}
    \label{fig:err_relat_Vinet}
  \end{minipage}
\end{minipage}
\end{figure}
Fig.~\ref{fig:err_relat_Vinet} shows the numerical computation of this relative error as function of the position $(\rho, \phi)$ on the dish. One can notice that the error is less than 1\%, implying that Eq.~\eqref{Vinet} can be safely used for the propagation of the electric field from the dish in the \textit{SHUKET} system, with negligible error. Note that the error $\epsilon(\rho,\phi)$ is computed only for points on the dish with radial coordinates $\rho < r$, as the circle of points belonging to the dish with coordinates $(r,\phi,z_\mathrm{plane}), \phi \in [0,2\pi[$ belongs to the fictional plane as well, and Kirchhoff integral is only valid for the computation of the field at reception points outside the emission surface. Therefore, one can deduce that the calculations presented in \cite{Vinet} are valid for \textit{SHUKET}.

\end{document}